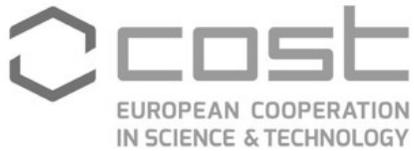
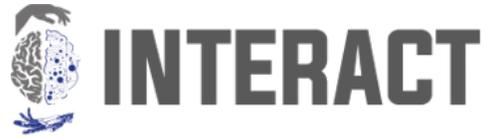

# COST Action INTERACT WG2 Whitepaper

Editors: Alister Burr, Ana Garcia Armada, Carsten Smeenk, Yang Miao
(all names in alphabetical order)

December 11, 2024

COST CA20120 INTERACT
Working Group 2 (Signal Processing and Localisation)

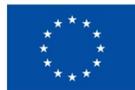

# Contents

# List of contributors

Abdelali Arous, 1.1
Abuu Bakari Kihero
Ainur Ziganshin
Agnieszka Czapiewska, 1.8
Alda Xhafa, 2.1, 3.2
Alessio Fascista, 2.4
Alejandro Castilla 3.3
Alehandro Lopez Escudero 3.3
Alicja Olejniczak
Alister Burr, 1.3
Ana Garcia Armada, 1.1, 1.2, 1.3, 1.7, 3.2
Ana Moragrega, 2.3
Armin Dammann, 2.6
Artan Salihu, 2.1
Carlos Alvarez, 2.1
Carmillo Gentile 3.3
Carsten Smeenk, 3.1, 3.2
Christian Gentner, 2.6
Claire Goursaud, 1.4
Daniel Egea, 2.1, 3.2
Davide Scazzoli, 2.2
Dragana Bajic, 1.2
Dinh-Thuy Phan-Huy
Ebubekir Memisoglu, 3.2
Emanuel Staudinger, 2.6
Emmanuele Peschiera, 1.3
Emil J. Khatib, 2.1
Eneko Iradier
Erik Leitinger, 2.1
Eugenio Moro
Fabio Broghammer, 2.6
Federica Fieramosca, 3.4
Flor de Guadalupe Ortiz Gomez
Fran Fabra, 2.1, 3.2
Francesco Linsalata
Francois Rottenberg, 1.3
Fredrik Tufvesson, 2.1
Gonzalo Seco-Granados, 2.1, 3.2
Guillermo Diaz 3.5
Guillaume Villemaud
Grega Morano, 2.6
Hamza Haif, 1.1
Hanadi Salman, 1.7
Haoqiu Xiong 3.5
Hadi Alidoustaghdam 3.1
Heraldo Cesar Alves Costa, 3.4
Huseyin Arslan, 1.1, 1.3, 3.2, 1.7
Iker Sobron 3.5
Inaki Eizmendi 3.5
Inigo Bilbao
Iratxe Landa 3.5
Jan Sýkora, 12
Jean Marie Gorce
Jiteng Ma, 2.1
José A. del Peral-Rosado, 2.6
José A. López-Salcedo, 2.1, 3.2
Juraj Machaj, 2.1
Ladislav Polak, 2.3
Laurent Clavier 1.4
Lianet Mendez-Monsanto, 1.1, 1.2, 1.3, 1.7, 3.2
Liza Afeef, 1.3
Manuel Castillo Cara, 2.1, 2.3
Manuel Velez 3.5
Marcin Hoffmann 1.5
Marek Simka 2.3
Markus Hofer 3.3
Markus Ulmschneider, 2.1
Markus Wirsing
Maurizio Magarini 2.2
Mehmet Kemal Ozdemir, 2.3
Minseok Kim 3.3
Miroslav Hutar, 3.5
Mostafa Rahmani, 1.6
Nopphon Keerativoranan 3.5
Narcis Cardona 3.3
Onur Günlü, 1.7
Pawel Kryszkiewicz 1.5
Pawel Sroka 1.5
Peter Brida, 2.1, 3.5
Piotr Rajchowski, 3.5
Raquel Barco, 2.1
Reiner Thomae 3.1, 3.3
Rizqi Hersyandika 3.2
Robert Pöhlmann, 2.6
Ronald Raulefs
Ruisi He 3.3
Salah Eddine Zegrar
Saul Fenollosa
Saw James Myint 3.3
Siwei Zhang, 2.6
Sobia Jangsher, 1.4
Sofie Pollin 3.2, 3.5
Stefan Schwarz 2.1
Stefano Savazzi 3.4
Steffen Schieler 3.3
Shanglin Yang
Syed Najaf Haider Shah, 3.2
Tomaz Javornik, 2.1
Thomas Feys, 1.3
Thomas Zemen 3.3
Vasile Bota, 1.2
Vitaly Skachek, 1.2
Werner G. Teich, 1.2
Xuesong Cai, 2.1
Xuhong Li, 2.1
Yang Miao 3.1, 3.2, 3.3, 3.5
Zhiyu Wang, 2.1
Zhixiang Zhao, 3.2
Zhuangzhuang Cui 3.5

# Foreword

The upcoming next generation of wireless communication is anticipated to revolutionize the conventional functionalities of the network by adding sensing and localization capabilities, low-power communication, wireless brain computer interactions, massive robotics and autonomous systems connection. Furthermore, the key performance indicators (KPI) expected for the 6G of mobile communications promise challenging operating conditions, such as user data rates of 1 Tbps, end-to-end latency of less than 1 ms, and vehicle speeds of 1000 km/h.

This evolution needs new techniques, not only to improve communications, but also to provide localization and sensing with an efficient use of the radio resources. The goal of Interact's Working Group 2 is to design novel physical layer technologies that can meet these KPI, by combining the data information from statistical learning with the theoretical knowledge of the transmitted signal structure. Waveforms and coding, advanced multiple-input multiple-output (MIMO) and all the required signal processing, in sub-6-GHz, millimeter-wave bands and upper-mid-band, are considered while aiming at designing these new communications, positioning and localization techniques. This White Paper summarizes our main approaches and contributions.

The document is divided in three parts. Part I is focused on signal processing for communications, with the aim of satisfying the challenging requirements for the evolution of wireless communications systems. New ideas regarding waveforms, channel coding, massive MIMO, and massive user access are proposed, considering hardware impairments, operation within new radio access network (RAN) architectures and the implications on security.

Part II deals with signal processing applied to localization. Several algorithms and techniques are explained that are suitable to provide localization information in different scenarios. Localization strategies using radio systems are proposed, including cellular and other signals of opportunity, as well as with emerging technologies such as reconfigurable intelligent surfaces (RIS).

Part III offers our view and proposals about integrated communication and sensing (ISAC). We start by defining our understanding about this novel terminology and explain strategies for efficiently managing the radio resources jointly for communications and sensing. Channel models and measurements, parameter estimation are explained and Interact's proposals are exemplified for the particular cases of sensing in wireless fidelity (WiFi), cellular systems, and using ambient backscattering.

# 1  Signal Processing for Communications

In this section we discuss the fundamental physical layer technologies for wireless communications looking towards the next generation, beginning with the most fundamental: the waveforms used for data transmission. Here we consider alternative domains beyond the frequency domain used in orthogonal frequency division multiplexing (OFDM) in fourth generation (4G) and 5G, including orthogonal time frequency space (OTFS) and chirp-based waveforms, among others. We discuss the transform domains upon which they are based and their relative advantages and disadvantages. As research moves forward to 6G it is timely to consider alternative more flexible waveforms, given that no single universally optimal waveform is likely to be found.

We next consider channel coding, reviewing options that can approach the Shannon bound. This includes variants of low-density parity-check code (LDPC), polar and turbo codes, but also others such as spinal and rateless codes which have advantages in certain applications. We further describe decoding methods including some based on "noise guessing" ("guessing random additive noise decoding" (GRAND)) and machine learning (ML). The network view of channel coding is also increasingly important, including physical layer network coding (PLNC) and distributed coding.

The spatial domain has become increasingly important in 5G, and this trend is only likely to accelerate in 6G. Here we address trends in MIMO and massive multiple-input multiple-output (mMIMO), including non-coherent massive MIMO, hybrid beamforming, energy-saving techniques, and distortion-aware processing, and also cell-free massive MIMO (CFmMIMO) or distributed MIMO (D-MIMO), which effectively implements the Network MIMO concept and is likely to play a significant role in 6G.

6G is likely to serve applications requiring massive access with a high density of wireless devices, where global coordination may not be feasible. We address some of the challenges this gives rise to, and proposed methods such as non-orthogonal multiple access (NOMA), rate splitting multiple access (RSMA), and alternative interference modelling and handling methods.

We then consider some of the impairments that can arise in wireless communication systems, such as- peak-to-average power ratio (PAPR) and non-linearities, self-interference, phase noise, and the problems posed by ultra-wideband (UWB) technology. A number of approaches for modelling and mitigating these affects are discussed.

In recent years a new approach called open radio access network (O-RAN) has been introduced, creating a more flexible and open architecture for the RAN. A feature of this is disaggregation of the physical layer between separate radio units and distributed units, between which digitized signals are carried on fronthaul connections. Here we introduce the O-RAN concept in general, but we focus in particular on the effects of disaggregation on physical layer performance and also on fronthaul compression.

A very important topic for any communication system is its security and its vulnerability to attack. We review potential threats and conventional cryptographic methods, and then consider physical layer security methods for a wide range of systems and applications.

Finally we briefly consider a rather different wireless communication scenario: underwater communications. This is certainly wireless communications, but we consider the use of hydroacoustic waves rather than electromagnetic waves, and the challenges that arise from this.

## 1.1  Waveforms

It is clear that a paradigm shift in the waveform design, i.e. in the physical layer (PHY) is needed to accommodate the new challenging applications while considering, in addition to other factors, the complexity, energy and spectral efficiencies, MIMO and full-duplex compatibility.

This section provides an overview of the current candidate waveforms for 6G, first by introducing some fundamental definitions about waveforms regarding their design for future systems, then by describing the state of the art and categorizing them based on various aspects, and finally by providing some thoughts on possible 6G waveform designs for future communications.

### 1.1.1  Waveform Definition

A waveform determines how a wireless signal is propagated over the physical resources (time and frequency). This involves designing *what*, *where*, and *how* to transmit. Fundamentally, a waveform defines the structure of the grid where the data/symbols are arranged, the pulse shape used to represent the data at those points, and the multiple access that allows resource sharing among different users [SZA24]. The aspects involved in designing a waveform are as follows:

**Representation from the lattice perspective**: One of the fundamentals of designing a waveform relies on *what* to transmit. This is, the population of the data lattice. Usually, this data lattice is directly placed over the physical resource grid, i.e. the time-frequency (TF) lattice points, as was the case for long term evolution (LTE) and the 5G, which are based on the OFDM. However, both domains can be decoupled, which leads to an additional degree of freedom.

First, on the one hand, the data or logical lattice simply refers to the mathematical construct used to arrange the raw symbols, such as a quadrature amplitude modulation (QAM), in a multidimensional space. Beyond time and frequency, there are many other possible logical domains that can be exploited to improve system performance, such as delay, Doppler, code, user, angle, etc. On the other hand, the physical resource grid is characterized by PHY parameters, mainly time and frequency. A precoding/spreading step are needed to convert the logical lattice to the physical grid. The TF grid is characterized by the following:

- The minimum spacing in frequency, $\Delta f$, which determines the symbols duration $T = \frac{1}{\Delta f}$.

- The minimum sampling rate, $T_s$, which determines the total bandwidth according to $B = \frac{1}{T_s}$.

- The maximum time duration of the frame $T_{frame} = NT$ containing $N$ symbols being transmitted with the minimum frequency spacing.

**Related transformations**: In order to convert the symbols in the logical lattice to a signal in the time domain to be able to transmit, a transformation is needed. Depending on the logical domain, different mother transforms are used. The simplest example would be the use of the discrete Fourier transform (DFT) for OFDM. For example, other waveforms that are also defined directly in the TF, such as the orthogonal chirp division multiplexing (OCDM) or the affine frequency division multiplexing (AFDM) waveforms use the discrete Fresnel transform (DFnT) and discrete affine Fourier transform (DAFT), respectively, which achieve some advantages over the effective channel. On the other side, OTFS multiplexes the information in the delay-Doppler (DD) domain and transfers information symbols from the DD to the TF via the inverse symplectic finite Fourier transform (ISFFT), i.e. DFT along the delay axis and inverse discrete Fourier transform (IDFT) along the Doppler axis. Apart from the TF or DD, some designs are based on the delay-sequency and the delay-scale domains, such as the orthogonal time sequency multiplexing (OTSM) and orthogonal delay scale space (ODSS) waveforms, respectively. They are based on the Walsh-Hadamard transform (WHT) and discrete Mellin transform (DMnT), respectively. These transforms have different impacts on the effective channel. Details about the relationship among the main domains and their impact over the effective channel can be found in [SZA24, ZYX+].

**Input-output relation, channel estimation and equalization complexity**: The input-output relationship refers to the relationship between the received and transmitted signals in any domain, depending on the effective channel and other impairments. Based on the waveform design, the channel estimation and equalization processes will be different. For example, one of the main advantages of OFDM is the low complexity frequency domain one-tap equalizer in static frequency-selective channels, which is lost under time-varying effects. Although, for example, OTFS is robust in high-mobility and presents channel sparsity under specific conditions, the channel estimation and equalization steps are still computationally expensive and require a high pilot overhead. Other waveforms, such as constant envelope-orthogonal frequency division multiplexing (CE-OFDM), and its version for high-mobility frequency modulated-orthogonal frequency division multiplexing (FM-OFDM), require channel estimation in the time domain. Aside from the benefits of waveforms in scenarios with high mobility, constant envelope, or other advantages, it is paramount to consider the overhead and computational costs associated with channel estimation and equalization processes. It's critical to determine the domain in which these tasks can be performed most efficiently, as this will have a significant impact on system performance and may ultimately dictate the choice of waveform.

**Energy focus, pulse shaping, leakage/out-of-band (OOB) effects**: The waveform design also involves considering energy focus, windowing/filtering, pulse shaping and leakage/OOB effects. For example, orthogonal delay-Doppler division multiplexing (ODDM) also multiplexes the information in the DD domain as OTFS, but ODDM uses a orthogonal transmission pulse in the DD domain, and therefore, it outperforms OTFS in terms of OOB emission and bit error rate (BER) by achieving perfect coupling between the modulated signal and the DD channels [ZYX+].

### 1.1.2 State of the Art

This section provides an overview of the state of the art waveform designs for 5G and 6G. We categorize them based on the common transform properties, while highlighting recent work on each category that deals with waveform design, channel estimation, and equalization techniques that are proposed to improve the performance of these waveforms.

1. **OFDM and its variants**:

   (a) **OFDM**: It is undoubtedly one of the most popular waveforms in the history of wireless communications (not only cellular, but also WiFi). The popularity of OFDM is due to its near optimal performance in frequency selective channels. Using a cyclic prefix (CP), this waveform is able to operate successfully under the effects of multipath channels. The low complexity single-tap frequency domain equalizer, improved spectral efficiency, and MIMO support for improved data rates and reliability are among its main advantages. However, OFDM performance degrades significantly in time-varying, high-mobility scenarios such as high-speed trains, vehicle-to-everything (V2X), and low Earth orbit (LEO) satellites. Because OFDM packs the frequency-domain subcarriers so close together, any synchronization error will adversely affect the system. The mobility and resulting Doppler spread leads to loss of orthogonality. In addition, it suffers from large sidelobes in the frequency domain and high PAPR.

   (b) **5G candidates**: For 5G, several candidates were proposed based on the original OFDM with fixed numerology (uniform lattice) that was used in LTE. Some of them were based on an alternative pulse shape, such as filter bank multicarrier (FBMC). However, the ultimate choice was using the same pulse shape but with the multi-numerology approach (non-uniform lattice). All of these propositions multiplexed the information symbols in the TF domain, unlike most of the 6G candidates, which explore the use of other logical domains.

   (c) **Constant envelope**: To overcome the disadvantages of high PAPR, one of the main waveforms, adopted as optional for the uplink in LTE and 5G is the DFT-spread-OFDM (DFT-s-OFDM) or single-carrier frequency division multiple access (SC-FDMA). Similarly, CE-OFDM has been proposed [TAP+08], which is shown to achieve good performance in dense multipath with the use of cyclic prefix transmission in conjunction with a frequency-domain equalizer. An overhead-free channel estimation technique is proposed in [CHFGGTA23]. However, all of these proposals are not suitable for high-mobility scenarios. For this reason, FM-OFDM was proposed as a constant envelope alternative robust to time-varying channels. The channel estimation is performed in the time domain, and a block-equalizer is proposed in [HA23].

2. **Delay-Doppler waveforms**:

   Recently, waveform design exploiting the DD domain has gained significant popularity, with OTFS being the focal point of discussion. The primary driver behind this trend is its effectiveness in high-mobility scenarios and its suitability for ISAC. Furthermore, OTFS facilitates the application of novel MIMO communication designs, for both point-to-point MIMO and multi-user MIMO scenarios. Next, we outline the most relevant waveforms utilizing the DD domain, providing a historical recap of the journey leading to OTFS [LJY+]:

   (a) **OTFS (pre-coded OFDM)**: Early implementations of OTFS utilized an overlay on OFDM, employing a two-stage process involving the 2D symplectic finite Fourier transform (SFFT). While this approach offers compatibility with OFDM standards, it fails to fully exploit the unique properties of OTFS enabled by the Zak transform, which bridges DD and TF domains. However, this implementation is constrained by the OFDM structure, leading to a pulse-discontinuity issue due to the spread of each DD domain symbol across multiple TF domain pulses. This discontinuity can result in significant OOB emissions, impacting system performance. In response, various OTFS variants have been proposed to address this challenge.

   (b) **ODDM**: The ODDM modulation, introduced in [LY22b], utilizes a realizable orthogonal basis constructed through staggered multi-tone modulation, offering superior performance over OTFS in terms of OOB emissions and BER. This superiority stems from achieving perfect coupling between the modulated signal and the DD channels. Additionally, the authors proposed delay-Doppler multicarrier (DDMC) modulation, employing delay-Doppler orthogonal pulses (DDOP), which demonstrated effective DD orthogonality through periodically extended root raised cosine pulses or a Nyquist pulse train [LY22a]. Furthermore, a low-complexity pulse shaping framework on the DD plane was presented in [BF].

(c) **OTFS 2.0 (Zak transform)**: Recently, researchers have emphasized the importance of the Zak transform in OTFS, originally introduced by J. Zak in solid state physics [Jan88] and later extended to signal processing. This mathematical tool elucidates the physical relationship between TF and DD domains. Compared to OTFS implementations based on OFDM transceivers, Zak transform-based implementations offer reduced complexity and clearer physical insights. OTFS transmissions based on the Zak transform typically do not require overlay with OFDM, hence termed as one-stage implementations. A discrete Zak transform (DZT)-based OTFS realization was proposed in [LAW22], but suffers degraded performance under insufficient DD resolutions (termed "fractional delay and Doppler"). In [Moh21], a Zak transform-based OTFS implementation was presented, showing that rectangular windows can achieve the perfect DD orthogonality. This approach has been extended to OTFS 2.0 modulation in [MHCC22, MHCC23], highlighting DD domain information transmission based on the Zak transform. OTFS 2.0 employs intentional precoding of information symbols to satisfy the quasi-periodicity property of the Zak transform, and apply DD domain shaping pulses at both the transmitter and receiver to convey information, resulting in a mathematically simple input-output relation defined purely in the DD domain. However, practical implementation of DD communications necessitates further study due to real-world signal transmission constraints.

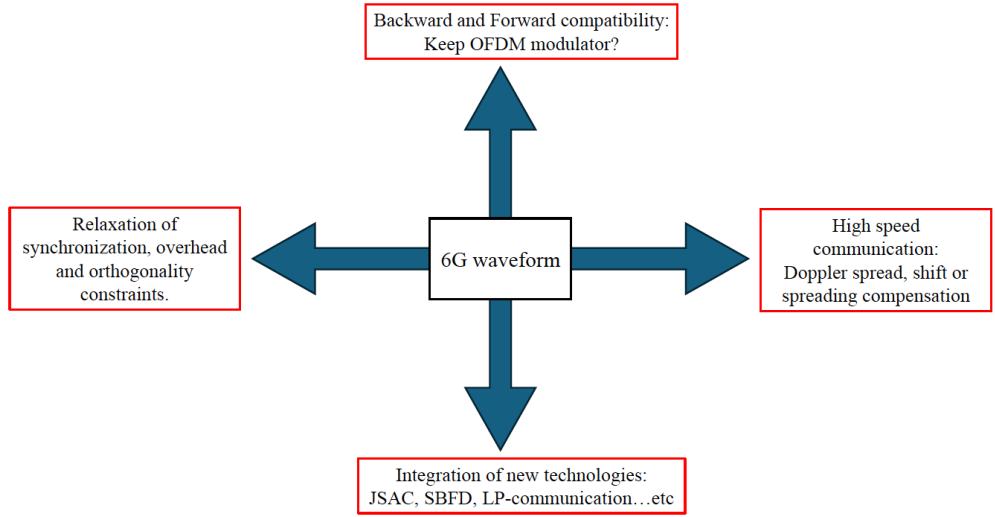

Figure 1: 6G waveforms directions and design criteria.

3. **Other domains apart from TF and DD**:

    (a) **OTSM**: OTSM, a single-carrier modulation scheme, delivers a comparable BER to OTFS. Information symbols are multiplexed in the delay-sequency domain using the WHT, where "sequency" refers to the number of zero-crossings per unit interval. Notably, WHT necessitates only addition and subtraction operations, avoiding the need for multiplicative operations, thus significantly reducing the complexity of OTSM modulation/demodulation compared to OFDM and OTFS [TV21].

    (b) **ODSS**: Unlike narrowband channels, for which time contractions or dilations due to Doppler effect can be approximated by frequency-shifts, the Doppler effect in wideband channels

results in frequency-dependent non-uniform shift of signal frequencies across the band. In [KPM22] authors propose ODSS modulation, which is based on the DMnT and provides enhanced performance in terms of BER in wideband time-varying channels. It multiplexes the symbols in the delay-scale domain.

4. **Chirp-based waveforms**:

   (a) ***OCDM***: it is based on the DFnT, and outperforms OFDM in terms of BER in multipath channels. Nevertheless, OCDM cannot achieve full diversity in general time-varying channels since its diversity order depends on the delay-Doppler profile of the channel [OZ16].

   (b) ***AFDM***: it can achieve full diversity. It is based on the DAFT, which, unlike DFnT, the DAFT is a generalization of the DFT characterized by two parameters that can be adapted to better cope with doubly selective channels [BCKK21].

### 1.1.3 Waveform Vision

The research community continues its search for 6G candidate waveforms. Unlike previous generations such as the third generation (3G) and 4G, which introduced entirely new PHY technologies, 5G retained the OFDM waveform while introducing a flexible numerology to accommodate different services. Despite the undeniable advantages of OFDM in multipath channel propagation, it has specific vulnerabilities such as susceptibility to high mobility, high PAPR, OOB emissions, and others. Meanwhile, the recently proposed 6G candidates outlined in the previous sections outperform OFDM under certain conditions, but lack a universally optimal waveform. Therefore, the adoption of a flexible waveform approach that can adapt to different 6G services without excluding the OFDM seems promising.

In [SZA24], the authors propose a hybrid waveform solution based on the OFDM umbrella. This forward-looking yet backward-compatible framework not only explores different options of OFDM-like waveforms, but also allows the integration of novel alternatives such as the ones discussed in the above sections. Such an approach facilitates a sustainable, gradual, and smooth network upgrade, avoiding the need for immediate large-scale changes that may prove highly undesirable to various stakeholders in the wireless communications industry. However, further analysis and comparisons are needed to determine the specific functioning of this unified framework.

## 1.2 Channel Coding

The goals of 6G in terms of data rates, reliability, latency, energy efficiency, and scalability, regarding the messages lengths, impose challenging requirements to channel coding, [CBMD23]. 6G will change the 5G's applications triangle, i.e., enhanced mobile broadband (eMBB) to next-generation broadband, ultra-reliable and low-latency communications (URLLC) to real-time control, and massive machine type communications (mMTC) to pervasive access. It also is supposed to allow for new types of applications, e.g., artificial intelligence, augmented reality (AR), and holograms, [GKC+23]. Consequently, these goals would require further developments in the domain of channel coding. The error-correcting code (ECC) should be suitable for applications with different KPIs, with wider ranges of the reliability and latency requirements, should have increased flexibility in terms of codeword lengths and coding rates, and should perform energy-efficient encoding/decoding with moderate complexity, while having small sizes of the required chips.

### 1.2.1 State of the Art

**FEC Codes.**

**LDPC codes:** The 5G standards specify the use of protograph-based LDPC codes on traffic channels, since they ensure high reliability, are quite flexible in terms of their code rates, code lengths, and decoding delay, and conveniently support hybrid automatic repeat request (HARQ) [3GP18b]. To meet the envisioned requirements of 6G in terms of reliability, flexibility and implementation, some modified/optimized variants of the LDPC coding were proposed. [LYT+21], analyzes a LDPC decoding algorithm, the check Node self-update (CNSU), which requires less memory and power, and its hardware implementation is shown to be able to provide very high throughput. The convolutional

LDPC code (CC-LDPC) is shown in [ZW20] to exhibit lower error-floor and implementation complexity, and faster decoding convergence, when compared to "classical" block LDPC codes. To decrease the computation involved, the authors of [DVN21] proposed a LDPC protograph-based code to be used in 1-bit analog-to-digital converter (ADC)-based mMIMO transmissions; the paper concludes that this approach overcomes the error-floor of conventional LDPC and might be appropriate for 6G base stations with low resolution. The spatially-coupled LDPC (SC-LDPC) codes are shown to mix the good properties of regular and irregular LDPC block codes, [CDF+14], and achieve the channel capacity when decoded with the belief propagation (BP) algorithm, [KRU13].

**Polar codes:** The performances of polar codes, which are used in the low-delay control channels of 5G, for different codeword lengths are discussed in [SSS13, LT17, MHU17]; the analysis shows they behave well for small lengths code words. Further research, [MHU16], showed that the polar codes don't present an error-floor for decoders based on successive cancellation (SC); moreover, the simulation results presented in [TV15] indicate that polar codes decoded with cyclic redundancy check (CRC) aided SC ensure better BER performance than LDPC codes with the same parameters.

The multi-level polar coded modulation, i.e., combinations of polar codes with high-order modulation schemes, were studied in [SSS13], while bit-interleaved polar coded modulation was studied in [MEKL16]; these two approaches seem to be able to provide increased throughput.

**Concatenated codes:** Starting with the occurrence of turbo codes, several performant types of concatenated coding schemes have been developed, e.g., parallel concatenated code (PCC) [WR61] and serially concatenated code (SCC) [BDM98]. Another powerful type of concatenated codes is the hybrid concatenated code (HCC) [DP97], composed of various component codes; they are a promising option which should be further studied. All of these concatenated codes may be decoded using the turbo approach, based on exchanging iteratively soft information.

**Rateless codes:** Rateless codes are suitable to wireless transmissions due to their strong message-recovery capabilities and to the fact that the transmitter might not need channel state information. Literature presents a large number of their applications in current communications systems, such as cooperative relay networks [LL09], [CC11], IEEE ad-hoc 802.11b networks [NSMH10], and in decreasing the PAPR in OFDM transmitters [JL10].Intensive research has been made to combat the high error-floor of these codes, e.g. [TLSV13]. The use of a rateless space-time code in mMIMO transmissions is proposed in [Alq18] and shown to bring some promising performance improvements. In [BL19] the authors study the use of a fountain-code based mechanism to provide multi-connectivity in millimeter wave (mmWave) mobile networks. The simulation results using a 60GHz mmWave test bed showed an increase of the link reliability and efficiency. [FKRA13] analyzes a combination of a rateless code and an irregular repetition code in mobile edge computing. It is shown that, due to the resiliency of the rateless code and the spatial diversity ensured by the repetition code, this combination decreases the computation and communication latencies. Another approach in using the rateless codes is presented in [SLT+22]. It employs a sophisticated Raptor code in a scheme where the transmitter receives acknowledgment information. The analysis and simulations show that this approach brings significant performance improvements, when compared to the use of a Raptor code and no feedback information.

**Spinal codes:** Another type of codes which was studied in literature are the spinal codes. As shown in [PBS11] and [MWL+21], these codes achieve near Shannon capacity performance, especially for short block codes, allow for a wide range of coding rates and can be easily adapted to time-varying channels, [PIF+12].

**Decoding and other aspects.**

**Unified coding:** To overcome the use of different codes, coding rates, and decoders for the various channels of the future transmission, [GKC+23] proposes a concept of unified coding. It considers the automorphism ensemble decoding (AED) framework and "code agnostic" decoders as potential approaches for unified coding. The AED would allow the dynamic distribution of the decoding resources to ensure reliability, throughput, and various block lengths. The "code-agnostic" decoding would be able to decode (almost) any linear code, which would be very important for practical applications. An unified LDPC decoder is proposed in [WZZ+18].

**Belief Propagation:** The idea of using the BP algorithm to decode other types of channel codes is evaluated in literature, since it provides soft output, thus allowing for iterative decoding. In [CEEtB18] the BP-based decoding of polar codes showed promising results, while the use of BP to the decoding of turbo codes, presented in [KHGW23], suggests the possibility to use the BP approach as a more general method to decode any channel code.

**The Noise Guessing Approach:** The GRAND, [DLM19], uses a different approach for channel decoding. Instead of estimating the correct codewords, which have the same occurrence probabilities, it uses maximum-likelihood-decoding to assess the probabilities of the noise values; then it orders these probabilities and uses them in the descending order. The number of attempts needed is shown to be much smaller than the brute-force search over all possible codewords. An improved version of GRAND, proposed in [Duf21], which uses soft inputs might be the basis for further research of the performance that could be provided by this approach. The Fading-GRAND, used in decoding the random linear code is shown in [AJG22] to significantly outperform the Bose–Chaudhuri–Hocquenghem (BCH) codes using the Berlekamp-Massey decoder, in transmission over Rayleigh-faded channels, while requiring less complex processing and low decoding times. The use of GRAND in conjunction with random linear codes is also shown to provide high rates in the finite-block length regime, [PPV10], [YDKP14], which makes this combination a promising option for practical URLLC.

**Machine Learning:** In the last years the use artificial intelligence (AI)/ML techniques in channel decoding has received more attention. Paper [HRW14] showed that, out of the AI/ML-based channel decoding algorithms, "deep unfolding" seems to bring improvements to the current iterative algorithms with fixed number of iterations. This technique unfolds the algorithm's structure onto several hidden layers of a neural network and introduces some trainable multiplicative weights and bias parameters. Similarly, in [NBB16, WJZ18, VXL18, WCN+21], the Tanner graphs of different LDPC codes are used to structure neural networks, thus improving the performance of the BP-based decoders. Other papers, e.g., [ECEtB19], studied the application of AI/ML to code design, which showed promising results, especially for short codes.

The sequence-based redundancy, [OH17], is an approach which aims at merging the pilot-generation, encoding, and signaling, by using learning encoders which generate an autoencoder structure. A similar receiver structure, performing detection, synchronization, and decoding, is reported in [DCCtB23] to outperform the "classical" systems in transmissions of short messages conventional systems. Paper [JJW+20] also shows that the use AI/ML techniques for channel coding and estimation could improve the performance of the ultra-broadband networks.

**Shaping Loss:** The shaping loss, which is due to the non-optimal probability distribution of transmitted symbols, is another aspect which would increase the transmission's reliability. In [BSS15] and [LZM+16] probabilistic shaping and geometric shaping schemes, respectively, are introduced, while [SAH20] analyzes the joint use of probabilistic and geometric shaping to design constellations with an 'autoencoder'.

**Using CRC:** Another less-redundant approach to increase the transmission reliability of low complexity standards is proposed in [TFP15]. It uses the cyclic redundancy check (CRC) code to correct the received packets, by means of iterative decoding techniques, besides their classical role of detecting the erroneous ones. The paper shows that this approach improved the sensitivity of a Bluetooth low energy (BLE) by 3 dB; moreover, it corrected a significant percentage of the received packets with no additional redundancy, decreasing the latency and saving energy.

**Network view.**

**Distributed coding:** The distributed coding is another technique used in cooperative relay-assisted transmissions. The research in this area is focused on increasing the either coverage, or reliability [MMYZ07], [ZPWL09], or throughput of cellular, emergency and V2X networks, [LYTH23], [WXX+23].

**Network coding:** To cope with the significant increase of the network access density, the physical layer might have to operate over the network, or subnetworks, instead of operating on individual links, in order to apply the signal processing and transmission techniques, thus generating the physical layer network coding approach, [SB18b], which generated significant amount of research in the last years. Paper [EKY23] presents a survey of the network coding schemes, including those using Opportunistic Network Coding Schemes, which are expected to increase the network's data rates and decrease its latency. Paper [DRE+23] analyzes the fixed sliding window random linear network coding (RLNC) approach and shows that it would ensure lower latency in mmWave backhaul network of 6G. The use of a separated RLNC algorithm that uses cooperative control frames is shown in [PC21] to improve the retransmission performance by decreasing the required overheads. The use of high-order modulations in PLNC employing deep neural networks is studied in [PJC21].

### 1.2.2 Research on Channel and Network Coding Conducted in WG2

**FEC Codes.**

**Rateless codes:** A topic studied within WG2 is the employment of the two-level rateless-forward error correction (FEC) coding in several transmission schemes over Rayleigh faded channels. Paper [BS22] analyzes a generic rateless-FEC coding scheme used to perform link adaptation by transmitting the minimum number $N_{aM}$ of rateless symbols needed to ensure a target message non-recovery probability $p_{NR-t}$ for a given set of modulations and FEC-coding rates $(b, R_c)$ at the current signal-to-noise ratio (SNR) value of a Rayleigh block-faded channel. To this end, it first computes the block error rate (BLER) ensured by the FEC code by using an approximation which is shown to provide good accuracy and computes the message non-recovery probability $p_{NR}$ of a generic rateless code with an overhead $\delta$ using a probabilistic approach. Then it derives the expressions of $N_{aM}$ and spectral efficiency $\beta$, and evaluates the performance provided, pointing out the impact of the message and transmission parameters. Finally, it describes and analyzes the performance of a link adaptation algorithm that changes the $(b_i, R_{cj})$ pair and uses the minimum number $N_{aM}$ of rateless symbols needed to provide the highest $\beta$, while ensuring $p_{NR-t}$ at each SNR.

Regarding the $p_{NR}$ vs. SNR performance, the numerical results show that the minimum SNR at which a given $p_{NR-t}$ can be ensured gets smaller as the message length, $m$ symbols, increases and the $k$-bit rateless symbol length decreases. It is also shown that, for a given pair $(b, R_c)$, the $p_{NR}$ vs. SNR curve has a steep "waterfall", e.g., the minimum SNR increases by about 1 dB, while the minimum number of symbols $N_{aM}$ increases slightly, e.g., at SNR = 10 dB, $N_{aM}$ increases from 109 to 111, if $p_{NR-t}$ goes from $10^{-4}$ to $10^{-8}$. Note that the minimum number of correctly FEC-decoded symbols theoretically needed to recover the message is $N_{aM} = m(1 + \delta) = 105$. This property might be very useful in systems that serve applications with various requirements of reliability since the adaptation could be made by slightly modifying the number $N_{aM}$ of rateless symbols to be transmitted. The value of $\beta$ provided by this approach is high due to the high coding rate of the rateless code. The adaptation also has fine granularity since the rateless code's rate $R_r$ has very discrete variations that compensate for the more abrupt variations of the modulation and FEC coding rate.

The employment of the FEC-rateless coding in a generic HARQ algorithm with limited incremental redundancy (IR) is studied in [BV22]. The proposed HARQ-IR algorithm aims at recovering a message composed of $m$ data blocks, each of them containing $k$ bits. This algorithm transmits $N_a$ rateless symbols in the initial attempt and sends $N_r = (2m - N_a)/q$ symbols in each retransmission, $q$ being the number of retransmissions allowed, thus limiting the number of transmitted rateless symbols to $2m$. It derives the expressions of the message non-recovery probability $p_{NR}$, the average number of rounds $A_{avg}$ needed to ensure a target $p_{NR-t}$, the average spectral efficiency $\beta$ and the average coding rates, in an orthogonal frequency division multiple access (OFDMA) generic structure over a Rayleigh block-faded channel. The impacts of the algorithm's parameters, i.e., the number $N_a$ of rateless symbols sent in the initial round, the maximum value of $q$ and the FEC code's coding rate $R_c$, upon performance are evaluated by the numerical results.

This approach ensures very small $p_{NR-t}$ values at low SNRs. The results also show that the increase of $N_a$ leads to smaller average delay and $\beta$, while observing constraints regarding $p_{NR-t}$ and $N_t$, the maximum number of transmitted symbols. Therefore, to meet various delay and reliability requirements, the algorithm should modify adaptively the value of $N_a$ and/or the FEC coding rate and modulation, $R_c$, and $b$. The first option would be rather simple to implement, while implementing both options would require a slightly modified version of the link adaptation (LA) functionality.

The performance of relay-assisted transmissions using concatenated rateless and FEC coding is studied in [BS24]. The paper analyzes the message non-recovery probabilities ($p_{NR}$) and spectral efficiencies provided by two cooperative relaying algorithms, which use both the direct source-destination (S-D) link and the source-relay-destination (S-R-D) path, over Rayleigh-faded channels and employ rateless-FEC coding. Within the repetition redundancy (RR) algorithm, the relay retransmits only the correctly FEC-decoded rateless symbols, while the destination performs soft-combining of the received symbols on both paths before performing the FEC and rateless decodings. Within the IR incremental redundancy algorithm, the relay recovers the source's message, performs a rateless encoding and transmits only a limited number of additional symbols which are FEC-encoded. The IR receiver performs the FEC-decoding and concatenates only the correct symbols from both paths, before performing the rateless decoding.

The paper derives the expressions of the post FEC-decoding block-error, the $p_{NR}$ and spectral efficiencies ensured by the two algorithms. Their performances are compared to the ones provided by the direct S-D transmission and by the two-hop relaying (THR) algorithm, both using the same concatenated FEC-rateless coding, and to the performance of the RR algorithm using only FEC coding with the same global coding rate. The RR algorithm with FEC-rateless coding is shown to provide the smallest $p_{NR}$, out of the studied algorithms, at the expense of smaller $\beta$, compared to the IR algorithm or to the direct transmission.

This preliminary study indicates that the use of the rateless coding, besides the FEC coding, decreases significantly the $p_{NR}$ with a reasonable coding rate. The use of the relay-assisted RR algorithm extends much the coverage area and/or allows for ensuring very low $p_{NR}$ at rather small SNRs, being a promising option for the ultrareliable transmissions. It also shows that the direct transmission and the transmission using the RR algorithm should be used adaptively, according to the SNR values in each frame period, thus bringing a new dimension to the LA functionality, meaning that, besides the dynamic change of the modulation and coding rates, the LA should modify adaptively the type of transmission, i.e., direct or relay-assisted, to ensure a small target $p_{NR}$.

**Splitting Codes:** Paper [Baj23] deals with a less studied theoretical topic, namely the splitting code (SpC). It elaborates on the properties of the splitting codes related to their telescopic scalability. To this end, the paper analyzes the conditions of shortening the SpCs, thus offering increased error correction capability; it also points out that the SpCs could be lengthened. Both scaling types present very fine granularity and do not alter the encoding and error-correcting procedures. The analysis of the SpCs is interesting from a mathematical point of view, offering the possibility to investigate the factorization of large numbers in the context of error correction. The paper also mentions a potential application based on the splitting property of the SpCs [BDZ21], namely a hybrid three-stage incremental HARQ procedure that transmits the code word in the first stage, auxiliary control symbols in the second stage, and retransmits the sub-words detected as incorrect in the third stage. At each stage, error correction can be turned on or off, keeping both the retransmission and residual error rates low.

**Network view.**

**Physical Layer Network Coding:** As continuation of research performed in previous Cost Actions, in the context of Wireless PLNC with hierarchical decode and forward (HDF) relaying strategy systems, paper [HSC23] addresses the problem of asynchronous reception in a 2-source hierarchical MAC (H-MAC). The asynchronous approach is motivated by the difficulty of precise synchronization among different sources in an H-MAC stage. Even when individual link delays are known to the receiver, a simple compensation is not possible due to the non-orthogonal nature of the H-MAC. Compensation at the transmitters would require a channel state feedback channel which is not always available. In some scenarios with various distances between individual nodes, even if the delays were known at the transmitter, a simple time base adjustment would not solve the problem with respect to all involved receivers. The authors assume delays of a fraction of the symbol time duration and propose a delay-independent linear precoding at the transmitter side, together with symbol-spaced and fractionally-spaced processing at the receiver. For each approach, they derive an equivalent channel model and analyze its properties in the case of a root raised cosine modulation pulse. The proposed techniques are numerically evaluated in terms of hierarchical bit error rate (H-BER). The results show that in some cases the asynchronous reception can lead to a performance boost.

In the 5G new radio (NR) standards, the system dynamically determines the modulation and coding scheme (MCS) along with the transport block size (TBS) to ensure maximized data rates, while observing a target BLER, which could be ensured above a specific SNR value, which also determines the required transmit power. The 3GPP TS 23.501 standard [3GP23a] specifies the required BLER for each application. Although the existing literature provides lookup tables [LWE+20, MMR+12, PRHC17], they do not cover the latest standard updates, including modulations up to 1024-QAM, and lack a wide range of MCS combinations. Paper [MMSLMGA23] presents a comprehensive set of BLER vs. SNR curves over AWGN channels, obtained by simulations, in a wide variety of 5G NR MCSs and TBS configurations, including 1024-QAM, using the LDPC codes in accordance with the 3GPP standards. To point out the SNR gain provided by the LDPC codes, it shows the BER curves of the uncoded modulations in similar conditions.

These resources not only contribute to the preliminary design of 5G communication systems, but also play a key role in the real-time MCS selection by the scheduler during communication. Furthermore, this information is essential for integration into system level simulator (SLS) [LWE+20], which use precomputed BLER-SNR tables to derive BLER without performing physical layer processing, thus saving significant computation time.

**Generic approaches.**

**Machine Learning:** In the context of using AI/ML techniques in channel coding, one of the main problems when decoding the FEC codes with neural networks is that the number of codewords (patterns) increases exponentially with the dimension of the code. This makes the training of the network unfeasible for large code dimensions. Paper [TP22] proposes the use of a convolutional neural network (CNN) to decode terminated convolutionnal code (CC). The CNN decoder consists of only convolutional layers and a circular padding is applied. As the CNN is properly matched to the dimension and structural properties (e.g., memory) of the CC, by increasing the number of hidden layers for CCs with greater memory, it can actually learn the internal structure of the CC. This allows to upscale the CNN decoder to match the CCs with the same internal structure, but with larger code word dimensions, i.e., to exhibit the generalization property. A retraining of the weights and bias are not required. It is shown that the upscaled CNN successfully decodes code words it has never been seen during the training process. The simulations presented, for constraints $m \leq 3$, show that the error rate performance of the CNN decoder is close to the optimum performance of a maximum likelihood detection (MLD), the SNR differences being of 0.6 dB at most. Moreover, the SNR gap to MLD remains approximately constant with the increase of the code word length, confirming the generalization property of the proposed decoder.

**Shaping:** The decrease of the shaping loss and average energy per transmitted symbol while maintaining the transmission rate is studied in [BKS24]. The authors study the probabilistic amplitude shaping (PAS), which uses uniformly spaced signal points with different symbol probabilities according to their optimized distribution, the shaping being applied after the channel coding (shaping after coding (SAC)). Since the performance of SAC depends on the joint optimization of the channel code, shaping, and bit-mapping, the authors analyze a configuration composed of a $L = 2000$-bits-long non-binary LDPC code over $GF(2^4)(NB)$ of rate 3/4, a bit-plane coded modulation (BPCM) mapping, and the 4-PAM modulation. The frame error rate (FER) performances of SAC and shaping before coding (SBC) transmissions are compared by simulations to the performances of the same NB-LDPC code, to the QC-LDPC code used in 5G, both in unshaped configurations, and to tangential-sphere (TS) bounds for SAC and SBC methods. The SAC and SBC configurations are shown to ensure an SNR gain of about 0.8 dB, compared to the unshaped configuration using the NB code, and an SNR gain of about 1.2 dB, compared to the unshaped configuration using the QC-LDPC code. Moreover, the two shaped configurations have an SNR loss smaller than 1 dB, compared to their corresponding TS bounds.

### 1.2.3 Research Challenges in Channel and Network Coding

The use of multiple types of channel codes to meet the reliability, latency and codeword length requirements of various use-cases requires excessive processing and power consumption of the transmission systems. Moreover, channel coding in future cellular networks will have to deal with high levels of interference, due to the huge numbers of devices. The design and performance estimation of coding schemes should target the BLER predicted by the Normal Approximation (NA), especially for short codes.

The ultimate research challenge would be the development of a **unified coding** solution, which could be adapted to various types of applications' block lengths requirements and to allow for different decoding algorithms depending on the latency and reliability targets, and available power and computational resources at the receiver.

The construction and decoding of **LDPC** codes that would have better reliability and lower error-floors for short block lengths and require less processing, and the increase of their lifting size to increase the decoder's throughput are other promising areas of research. Also, the study of employment and implementation of spatially-coupled LDPC codes is another interesting area of research.

The use of non binary (NB) codes, e.g. NB-LDPC, could bring some advantages, since their symbols could be mapped directly on the QAM constellations' symbols and the demapping has no information loss, but their employment raises some challenges, i.e., the implementation complexity and compatibility with the existing binary systems.

Similarly, the study of the potential employment of **polar codes** in more types of traffic, besides the control channels, by improving the performance of long polar codes, the development of more powerful and flexible concatenated codes and the application of the BP algorithm to decode the polar and convolutional codes are promising and challenging topics for future research.

The **rateless codes** are also an option which will require a lot of research effort; the construction of rateless codes with lower error-floors, the elaboration of more efficient decoding methods and the study of their applications in mmWave, edge computing and HARQ schemes with selective incremental redundancy are several promising topics.

Another challenge will be the elaboration of high-performance and low-complexity decoders for the **spinal codes**.

The **distributed coding** in relay-assisted transmissions, using either block codes or rateless codes, is another area of research. The research should focus on the optimal distribution of the global redundancy and on its employment in HARQ schemes, to either increase coverage or improve reliability and/or throughput in cellular, emergency and V2X networks.

The use of **physical network coding** will also require more research effort , since it might ensure more efficient ways of managing the increased density of the communications networks.

The use of **GRAND-based decoders** and the use of **CRC bits for error-correction** as well, might be worth studying for applications that impose limited processing and not very severe reliability requirements.

One of the most important challenge for future research is the design and performance evaluation of the **AI/ML-based channel coding algorithms**. It should develop methods of constructing adaptable (quasi)optimal codes for various requirements and design more performant and versatile decoding algorithms, whose performances should be guaranteed for all use-cases of interest. Another research topic would be the use of AI/ML approach in elaborating an autoencoder structure, which would generate jointly the encoding, pilot signals and modulation at the transmitter, and the demodulation, synchronization and decoding at the receiver.

Since coexisting sensing and communication is foreseen in 6G, channel coding has to solve an additional issue, to allow adaptive use in communication-centric sensing and sensing-centric communications, namely, channel coding should balance between data rate and reliability (communications KPIs) and accuracy and resolution (sensing KPIs).

## 1.3 (Massive) MIMO

A common trend in evolving generations of wireless communication systems is the combination of an increasing number of antennas starting from a few in 4G to a massive number in 5G. This trend is continuing its trend in 6G. In this section, we provide an overview of key challenges and opportunities related to mMIMO for future systems. The section is structured in five subsections covering the following topics: CFmMIMO, non coherent mMIMO, hybrid beamforming, energy-saving techniques and finally distortion-aware processing. Each subsection provides a motivation and state-of-the-art review of the related topic, together with the description of INTERACT past and future contributions.

### 1.3.1 Cell-free massive MIMO

The concept of CFmMIMO was introduced in [NAY+15] as a generalisation of centralised mMIMO in which antennas are distributed at access points (APs) across a service area rather than centralised at a single base station per cell. Signals are then processed at a central processing unit (CPU), which is connected to the APs by what are called backhaul connections in [NAY+15], but will be referred to here as fronthaul, in line with O-RAN terminology. The concept is related to distributed antenna systems (DAS), and is also known as distributed massive MIMO (DmMIMO) or more simply D-MIMO.

The CFmMIMO concept however differs from DAS in that the antennas are not simply distributed over a single cell, but over the entire service area. In this sense it abolishes the concept of the cell - hence the term "cell-free" - and thus goes beyond the cellular networking paradigm which has dominated wireless mobile networks for more than 40 years. The big advantage of this is that it also abolishes inter-cell interference and the problem of the disadvantaged cell-edge user. It can be regarded as an implementation of Network MIMO [KFV06] or a more advanced version of coordinated multi point (CoMP) [IDM+11].

However it also introduces new challenges, in particular the scalability problem [BS20]. As envisaged in [NAY+15] APs over the entire service area are connected to a single CPU. This is clearly not scalable as the size of the network increases. The simplest approach to enable scalability would be to divide the service area into distinct, independent regions, each served by a separate CPU. However this effectively reintroduces disadvantaged users located at the edges of these regions, which are then subject to uncoordinated interference from the neighbouring region. To avoid these effects, a user should be served by a number of APs around it, regardless of which region those APs fall into, which requires either that CPUs are connected together, to exchange signals from the APs connected to each one, or that some APs are connected to multiple CPUs. [BS20] describes a proposal based on the former approach, while [Zha23] develops the latter approach under the name decentralised distributed massive MIMO.

As the 6G research agenda has taken shape in the past few years, it has become clear that some version of CFmMIMO features strongly among the research topics currently under development. For example in Europe it features in the programmes of the 6G Flagship, the UK 6G Innovation Centre, and the Hexa-X-II project, and in the USA in the NextG Alliance. However, the term CFmMIMO is not always used: the term D-MIMO has recently become widely accepted in the context of 6G. As such it seems likely to feature in 6G when standardization begins in Third Generation Partnership Project (3GPP) in 2025.

However until now CFmMIMO has been largely an academic concept, and has been analysed based on assumptions which do not generally apply in practical networks. In particular the analysis has assumed flat Rayleigh fading channels using single carrier modulation, while of course in practice the channel is frequency-selective and OFDM is used. It also follows the assumption from mMIMO that orthogonal time-domain pilot sequences are used on the uplink for channel estimation. In many standards, and especially in 5G new radio (5G-NR), the pilots are in the frequency domain, across subcarriers of one or more OFDM symbols in a single physical resource block (PRB). Since there are 12 subcarriers in a PRB, this limits the number of orthogonal sequences to 12, and often there are fewer than that in practice. Further work is required, and will be undertaken in the remainder of the COST INTERACT Action.

### 1.3.2 Non coherent massive MIMO

Traditional communication systems are based on coherent detection schemes that rely on the use of pilots or reference signals for channel estimation and equalization. However, the use of pilots reduces the effective data rate, and the channel estimation and equalization processes can introduce complexity in the transceivers. In typical scenarios, the overhead can be limited if the channel is considered quasi-static and the number of antennas is not very large. On the contrary, for mMIMO, numerous reference signals are required, especially in the multi user (MU) mMIMO scenario with reduced or moderate mobility. Additionally, in general, coherent schemes are not suitable for high mobility scenarios because they require a large number of reference signals to effectively track channel variations. As a result, non-coherent (NC) schemes are gaining interest.

NC schemes combined with mMIMO are able to transmit information without the need for channel estimation and equalization, achieving the same asymptotic performance as the coherent schemes. This eliminates the need for reference signals and reduces the complexity of the transceivers. Several works in the literature show that the NC scheme is flexible and can be integrated into different systems [AH15]-[CHA19]. In [AH15], an NC scheme based on differential M-ary phase shift keying (DMPSK) modulation is proposed, which allows the use of differential detection while taking advantage of the increased number of receiving antennas. The work in [BAZ$^+$18] provides implementation improvements to [AH15]. Finally, in [CHA19], the NC scheme is combined with OFDM. The performance superiority of NC schemes under non-trivial conditions makes it a good candidate for future communications.

To improve the performance of NC schemes, [AH15] proposed to multiplex users in the constellation domain. This means that at the base stations (BS), a joint symbol is received, which is the result of the superposition of all the individual symbols transmitted by each user in the same time-frequency resource, with their different channel effects. It is crucial to properly design the individual constellations of each user so that the BS can distinguish the transmitted information of each user in the joint-constellation. Several works in the literature focus on the design of constellations in a similar context [AH15], [BAZ$^+$18], [LZM19], [XXX$^+$19]. However, they either focused on the single-user case, which were suboptimal solutions for the multi-user case, or they were applied to NC techniques that were not based on the DMPSK and thus show a worse performance for the same number of antennas.

For this reason, [LMCHGAD22] aims to solve the problem of constellation design for the multi-user NC massive single-input multiple-output (SIMO) based on DMPSK in uplink (UL) scenarios. The authors show that a classical analytical approach is intractable and propose two different procedures to solve the problem:

- Gaussian-approximated Optimization (GAO) is proposed. This method obtains the individual user constellation and the bit-mapping policy for each user by assuming that the received joint symbols follow a bivariate Gaussian distribution. Here the problem is divided into two: the first yields the individual constellation of the users that best resembles a QAM joint-constellation, and the second determines the bit-mapping policy for each user.

- Monte-Carlo-based optimization (MCO) method is proposed. It relies on a single problem that gives the individual constellation and bit mapping policy of all users. It consists in evaluating the BER through Monte-Carlo simulations. It outperforms the GAO approach and can be used irrespective of the characteristics of the propagation channel.

The authors propose evolutionary computation (EC) techniques to solve both non-convex approaches, which can be solved offline. The paper presents several novel constellations for different scenarios and configurations of the number of users and constellation sizes. Simulations show that these constellations outperform the current state-of-the-art solutions and that the performance of NC schemes is superior to the coherent 5G for high mobility and low SNR scenarios.

### 1.3.3 Hybrid Beamforming Technique for Efficient MIMO Systems

The increasing demands of current wireless networks for higher energy and spectral efficiency leads to several investigations in MIMO design [LA20]. As UWB and mMIMO systems become more prevalent, managing network resources efficiently has become a critical challenge. Additionally, the beam squint effect, which arises from propagation delays across antenna elements, significantly degrades array gain and system capacity, necessitating innovative solutions to maintain and enhance network performance.

To address these issues, several innovative solutions are proposed. First, a sub-grouped LAS-MIMO architecture combined with a hybrid precoding algorithm is introduced in [AMA22] to reduce the cost and hardware overhead of traditional hybrid MIMO systems. This algorithm controls the architecture to maximize the achievable sum rate of each subgroup by selecting optimal beams from a predefined set and independently calculating phase shifters and digital precoders while using successive successive interference cancellation (SIC) to minimize inter-subgroup interference. Another hybrid precoding technique is introduced in [CAMA22] for further controlling the LAS design.

Additionally, to counter the beam squint effect in ultra-wideband mMIMO systems, a transceiver design incorporating LAS and analog subband filters is proposed in [AKA24]. This design chunks the wideband signal into narrowband groups, managing their squints with an exhaustive search-based switching/precoding mechanism. A simplified, thresholded search-based algorithm further reduces complexity while maintaining performance.

Furthermore, user scheduling and precoding techniques are introduced to exploit the beam squint phenomenon to serve multiple users and improve overall system capacity [KAA22]. These techniques, based on the LAS-mMIMO system, control and utilize beam squinting to serve spatially separated groups of users, significantly enhancing beam gain and capacity.

### 1.3.4 Energy-saving techniques

Sustainable deployments of next-generation mMIMO BS will depend on their overall energy consumption. For this reason, energy-saving mechanisms at the BSs (or "network energy savings for NR", as coined by 3GPP) are attracting attention from both academia and industry. Traditional designs that aim at maximizing communication rate (or related indicators) are being challenged by studies that propose a different configuration of transmission schemes/resources for a fixed set of rate requirements, and repeat this process when the traffic changes. The goal of these load-aware techniques is primarily to decrease the static energy consumption at low and medium loads. Indeed, always-on BSs in current systems are associated with large fixed consumption, for example, due to their large number of active radio frequency (RF) chains, active-mode direct current (DC) supply, and massive baseband processing.

Energy-saving techniques generally operate in four domains, i.e., time (symbol), space (antennas or entire BSs), frequency (subcarriers or carriers), and power, by adapting the number of active symbol slots, antennas/BSs, (sub)carriers, or transmit power levels, respectively. In the space domain, the works [PR23b, PR22] have looked into the design of optimal precoders and number of active antennas in cellular mMIMO systems. Both studies build upon an improved power amplifier (PA) consumption model that considers non-fixed PA efficiency, differently from what is considered in traditional downlink problems. By modeling the PA efficiency as an increasing function of the transmit (output) power since PAs are more efficient close to saturation, the PA consumption turns out to be non-linearly related to the transmit power. In single-carrier systems, minimizing the PAs consumption subject to per-user rate constraints leads to precoders that use a number of antennas proportional to the number of users. When there is only one user, the optimal precoder utilizes only the antenna with the highest channel gain with high transmit power to avoid the use of many antennas with low transmit powers and poor PA efficiencies [PR22]. Different is the case in multicarrier systems with many subcarriers, where the per-user per-subcarrier rate constraints (expressed as per-subcarrier zero inter-user interference constraints) require the use of all the BS antennas when the objective function is the PA consumption. However, by considering a BS consumption model that comprises a fixed term and a consumption scaling with the number of active antennas, it has been proved that when the number of subcarriers goes to infinity, there is a globally optimal number of active antennas, which is tipycally lower than the total number of antennas [PR23b]. Indeed, retrieving the optimal antenna powers generally requires solving a system of fixed-point equations. Asymptotically in the number of subcarriers, this is not necessary as it turns out that the conventional zero-forcing (ZF) is optimal. This precoder leads to a constant transmit power among the active antennas. The number of active antennas can be then retrieved as solution to a quartic equation, and its value illustrates the tradeoff between circuit-limited regime (use as less antennas as possible) and PA-limited regime (use as many antennas as possible). In general, a BS with space-domain energy-saving solutions can consume more than $2.5\times$ less power than a traditional BS without energy-saving solutions [PR23b].

In CFmMIMO systems, where distributed APs jointly serve the users, implementing energy-saving techniques becomes even more critical. The full activation of APs and their antennas is impractical. The work [PMR24] considers a multicarrier CFmMIMO system with spatial correlation at the APs. It is shown that when the number of subcarriers is large, a centralized ZF precoder among a set of active APs (leading to uniform transmit power among the antennas of the active APs) is close-to-optimal in a PA consumption sense. A random matrix theory method is proposed to efficiently retrieve the powers at the APs as dependent only on the second-order channel statistics. This reduces the fronthaul requirements as only the active APs need to communicate with the network central unit. The proposed strategy results in significant savings in network energy consumption (up to a factor of $9\times$) at low loads.

### 1.3.5 Distortion-aware processing

Energy consumption in today's cellular networks is predominantly attributed to the BS [LCRB23]. Within the BS, the PAs are responsible for the majority of the energy consumption, ranging from $\sim 50 - 80\%$ [ABB+22, AGD+11, GHC20]. Moreover, these PAs are designed to operate in the linear regime, leading to very low energy efficiencies, these can be as little as 5% to 30% [AGD+11]. The energy efficiency of these PAs can be increased by operating closer to the saturation point of the PAs. However, this leads to nonlinear distortion of the communication signal. In today's mMIMO systems, which use conventional precoding techniques, this distortion is known to coherently combine in the user direction, which limits the performance [MGEL18a]. In order to overcome this, recent works have looked at incorporating knowledge of the distortion into the precoder design [RCVdP22, RCVdP23, FCVdPR22, FMR23, FVdPR24]. This allows for the suppression of the distortion at the user locations, limiting the distortion that the users experience. In [RCVdP22, RCVdP23], a closed-form solution is obtained for a distortion-aware precoder that is designed for the single-user mMIMO case, taking into account a third-order polynomial amplifier model. This precoder maximizes the SNR under an average power constraint while taking an additional constraint into account which nulls the distortion at the user location. It is shown, that this precoder can null the distortion, at the cost of a small reduction in array gain, which vanishes as the number of antennas increases. When considering multiple users and a polynomial amplifier model that takes into account higher-order polynomial terms, the complexity of the distortion-aware precoding problem increases. In order to tackle this problem, in [FMR23, FVdPR24] a neural network is introduced which learns the mapping between the channel and precoding matrix, taking into account the distortion. This mapping is obtained by learning the weights of a graph neural network in a self-supervised manner, by maximizing the sum rate expression, taking into account distortion. By learning this mapping from channel to precoding matrix, a lot of the complexity is offloaded to the training step, which reduces the online computational complexity of the solution. These methods open perspectives to operate PAs closer to saturation, thereby drastically reducing the power consumption of these amplifiers.

## 1.4 Massive Access

### 1.4.1 Problem description

In a scenario characterized by a high density of interconnected devices, achieving comprehensive global system coordination is impractical. Traffic is usually sporadic and requires the transfer of short packets. To support this, grant-free resource access was introduced in 5G-NR [DNP+22]. Grant-free random access is a medium access control (MAC) protocol to ensure simple and low overhead communication. In grant free, the signatures are preallocated to the users in a way to avoid potential collision. ALOHA, slotted ALOHA are some of the examples of grant free random access schemes. Efficiency of these schemes has been a challenging research direction adopted in most of the early research on this topic. In [QB06], the transmission was done opportunistically and the system throughput was shown to be increased logarithmically with the number of devices.

However, the lack of coordination generates important challenges:

1. Activity detection becomes difficult if transmissions are not *a priori* declared. If the receiver has calculation capabilities, when the number of devices communicating in overlapping resource blocks becomes important, their identification is in itself an issue.

2. An important source of communication degradation is caused by interference. Under such circumstances, interference levels can vary significantly between individual packets, with only their statistical properties estimable prior to transmission. Notably, the variance of the resulting noise due to interference remains unpredictable.

3. Non Orthogonal multiple access: it is not so efficient to handle channel resource with orthogonal methods as it has been traditionally done in cellular networks. On the other hand, sensing before transmitting is not efficient, neither in terms of performance nor in terms of energy. NOMA is then a natural option but raises many challenges.

### 1.4.2 Activity detection

**Open research questions:** In grant-free schemes, each device transmits its information directly, saving time (no additional delay is introduced by the grant procedure) and wireless resources. However, this implies that the base station is able to identify which are the transmitting devices among all the possible ones in the network. This process is known as active user detection (AUD). Although the AUD problem can be written as a multi user detection (MUD) one, there are specificities which prevent to exploit the same tools. We thus need to design new processing techniques dedicated to AUD.

In practice, if we consider a NOMA system with $N \geq SF$ users (with $SF$ the spreading factor), the users activity can be modelled with the set $\mathbf{b} \in \{0,1\}^N$ where $b_i = 1$ corresponds to user $i$ being active (and $b_i = 0$ to user $i$ being inactive). The received signal corresponds to the sum of the contributions from the active users, and is thus given by:

$$\mathbf{y} = \sum_i h_i.b_i.c_i + \mathbf{w} \tag{1}$$

with $c_i$ user's #$i$ code, $h_i$ the channel coefficient, and $w$ the additive white Gaussian noise (AWGN) contribution.

Given the received signal and the set of user codes, the objective of the AUD process is to recover the set of active users $\mathbf{b}$.

The first and simplest solution is to use conventional correlation receiver (CCR) [RSMD19]. This detector correlates the received signal $\mathbf{y}$ to the corresponding code sequences $c_i$. The user is considered active if the correlation exceeds a predefined threshold $T$. Each potential users are thus detected separately. This detector has a low complexity, however, as sequences are not orthogonal, the performances are very poor.

On another side, we can use maximum-likelihood [HWL+20]. A maximum-likelihood detector identifies the most likely active users set $\hat{\mathbf{b}}^{ML}$, given the received signal. It is obtained by searching the active user set that minimizes the distance between its expected contribution and the actual received signal. The maximum-likelihood receiver can be expressed as follows in an AWGN channel and equiprobable activity:

$$\hat{b}^{ML} = \underset{\{b_i\}}{\arg\min} \left\| \mathbf{y} - \sum_{i=1}^{N} h_i.b_i.c_i \right\|^2 \tag{2}$$

The maximum-likelihood solution suffers from a high computation complexity $\mathcal{O}(2^N)$, as it is based on an exhaustive search over all the existing possibilities. Thus, the high complexity of the maximum-likelihood detector makes it unadapted with classical processors when the size of the network increases.

The research community has thus proposed to exploit the fact that the users sporadically access to the channel, to formulate the problem under the sparse signal reconstruction framework. The problem can then be modeled by a standard compressed sensing (CS) problem [ASL19, CSD+17]. Thus, it can be solved by using the greedy-based CS algorithm : orthogonal matching pursuit (OMP) [Sch18, ASSU22]. Meanwhile, approaches such as deep learning (DL) are also investigated [KAS20, DKS+21]. However, these solutions are not satisfying enough. The CS performance decreases when the network size increases and the learning approach requires a highly consuming learning phase.

There is therefore more work to be done on the AUD for NOMA systems, in order to achieve greater reliability without impacting on complexity.

**Some proposals:** Meanwhile, quantum computing has emerged as a promising solution to solve difficult problems with a reduced complexity. Indeed, it exploits a promising quantum attribute, superposition, which permits to handle both 0 and 1 states simultaneously, enabling parallel processing.

One of the most famous quantum algorithms is Grover's algorithm [Gro96]. It was designed to search for specific values (i.e. solutions verifying some constraints given by the problem to solve) in an unsorted database.

In the AUD context, the maximum-likelihood formulation perfectly matches with Grover's algorithm structure. This is thus a promising approach to reduce the complexity while keeping the optimal achieved AUD accuracy. Authors in [BNH14] have implemented Boyer, Brassard, Høyer and Tapp (BBHT) [BBHT98] and Dürr-Høyer algorithm (DHA) [DH96] in the context of MUD with a simple case, where the iterations are upper-bounded by $22.5\sqrt{K/S}$, with $K$ the database size, and $S$ the number of valid solutions. Authors in [HHG21, HHG22], have also compared the classical and quantum performances for several code families for the AUD problem with the original Grover's algorithm implementation, in a specific case. Nevertheless, these works consider Grover's algorithm in a noiseless case. Meanwhile, the maximum-likelihood approach has to be considered to match the AUD case in a noisy scenario. This was done in [NI23], where the authors have adapted Grover's algorithm to find the minimum for power domain NOMA purpose, when the number of solutions is unknown. Additionally, the authors of [XKS+23] have used an improved quantum approximate optimization algorithm (QAOA) for power domain NOMA purposes within the power domain NOMA context.

The next step would be to consider more realistic channel coefficients. To do so, one might consider extensions of Grover such as the quantum minimum searching algorithm (QMSA) family which searches for the minimum value of the database (BBHT and DHA). Another promising approach, is the use a quantum annealing, which permits to solve a problem in a more analog way, but for which the expression of the problem in an appropriate way is more challenging.

### 1.4.3 Interference as a noise

**Open research questions:** The absence of a global system coordination manifests as additional noise caused by interference. Under such circumstances, interference levels can vary significantly between individual packets, with only their statistical properties estimable prior to transmission. Specifying its probability density function (PDF) is an important issue, for instance when deriving the likelihood for designing an optimal receiver.

One of the pioneer contribution is from Middleton [Mid77], who obtained general expressions based on series expansions assuming Poisson distributed interference sources. The expression can be simplified using only significant terms (Gaussian mixtures [HB08], $\epsilon$-contaminated noise [AALM17]). Other works proposed empirical models, justified by observation and improvements in the system performance (Generalized Gaussian [Fio06], Gaussian-Laplace mixtures [BN10] or Cauchy-Gaussian mixture [MJGC17]). Another proposed class of model, which allows to better represent the heavy tail behaviour of the interference, is the $\alpha$-stable distribution. It comes from the spatial distribution of the interfering sources and relies on a theoretical derivation, finding its foundation in stochastic geometry [WPS09, WA12]. In a network, we can express interference as

$$I = \sum_{i \in \Omega} l(d_i).\mathcal{Q}_i, \qquad (3)$$

where $d_i$ is the distance between interferer $i$ and the destination and $l(d)$ the attenuation as a function of the distance; a classical model is $l_{\gamma,\epsilon}(d) = d^{-\gamma}\infty_{r \geq \epsilon}, d \in \mathcal{R}^+$ where $\gamma$ is the channel attenuation coefficient; $\epsilon$ is a guard zone, meaning no interferer can be closer than $\epsilon$ from the receiver; $Q_i$ includes the propagation effects (multipath, shadowing) and the physical layer characteristics; $\Omega$ is the set of interferers. If applied in an *ad hoc* network, an unbounded received power assumption makes the interference fall in the attraction domain of a stable law. This unbounded assumption means taking the limit as $\epsilon \to 0$;

In the case of the interference power, a detailed study has been carried out by Haenggi and Ganti [HG09], who showed that the distribution is heavily dependent on the path loss attenuation coefficient $\gamma$. Again [WPS09], in a network with infinite radius and no guard zone ($\epsilon = 0$), the interference power has the totally skewed $\alpha$-stable distribution, where $\alpha$ depends on $\gamma$.

Interference models remains however dependent on the scenario and generally difficult to use in practical context. Besides the space, frequency and time dependencies are difficult to model.

**Some proposals:** Promising prospects to meet extreme communication requirements in terms of throughput, latency and reliability is the 6G "in-X" sub-networks architecture were first introduced in [BMA20, ABM+20] as highly specialised autonomous cells. The term "X" represents the entity in which the subnetwork cell is deployed, for example, a production module, a robot, a vehicle, a house, or even the human body. Scenarios are characterised by uncoordinated deployments of the subnetworks and a high density of devices.

Creating small subnetworks, even if each subnetwork can coordinate efficiently their own devices, leads to an important intereference between the different subnetworks. This inter-networks interference can be modeled by an exponential mixture models (EMM). A robust estimation process with bootstrapong and a protocol to guarantee the reliability of the communication has been proposed, assuming the set of interfering devices do not change during a packet transmission,

A further reserach direction is to use copula models [Nel99] for the dependence structure in the interference vector. This is especially important when the set of interferer is not constant over a block or when several antennas or frequency bands are considered. The joint PDF of a random vector in $\mathbb{R}^n$, $\mathbf{X} = [X_1, \ldots, X_n]$ is given in the form

$$F(x_1, \ldots, x_n) = C(F_1(x_1), \ldots, F_n(x_n)), \tag{4}$$

where $C : [0,1]^n \to [0,1]$ is called a *copula function*, and $F_i$, $i = 1, \ldots, n$ are the marginal distribution functions. When both the joint and marginal distributions admit density functions, the joint probability density function is then given in the form

$$p_{\mathbf{X}}(x_1, \ldots, x_n) = c(F_1(x_1), \ldots, F_n(x_n)) \prod_{i=1}^n p_{X_i}(x_i). \tag{5}$$

In our interference modeling, the copula exists and is unique and allows to decompose the joint PDF into the product of the marginal densities and the copula $c : [0,1]^n \to \mathbb{R}_+$ which captures dependence between the different components of $\mathbf{X}$. One example in the context of internet of things (IoT) wireless networks is to use the class of $t$-copulas with $\alpha$-stable marginal distributions, consistent with theoretical and empirical analysis [PW10, CPL$^+$20].

### 1.4.4 Resource access and NOMA

**Open research questions:** Orthogonal multiple access (OMA) restricts the number of users that can access the resource. NOMA is a key technology to scale and support a massive number of devices. In this scheme, multiple users can simultaneously transmit their data on the same channel resource either with different power levels (power domain NOMA) or codes (code domain NOMA) to differentiate at the receiving end. In power domain NOMA the user data is superimposed on each other with different power levels based on the channel conditions. On the receiver side SIC is applied to differentiate the data on the receiving end. The research questions are mainly in three categories, ($a$) channel difference between users, ($b$) power allocation and ($c$) SIC. For a downlink PD-NOMA, the superimposed messages of $M$ users is as follows:

$$x = \sum_{m=1}^{M} \sqrt{P_m} q_m \tag{6}$$

where $P_m$ is the power assigned to user $m$ for transmission and $q_m$ is the message of user $m$. On the receiver side, the $m^{th}$ users receives

$$y_m = h_m x + z_m = h_m \sum_{m=1}^{M} \sqrt{P_m} q_m + \mathbf{z}, \tag{7}$$

where $h_m$ is the channel gain from the base station to user $m$, and $\mathbf{z}$ is AWGN.

RSMA is also a non orthogonal transmission scheme and in literature it has shown to outperform NOMA in terms of multiplexing gain [CMSP20]. There are different type of RSMA schemes based on the layering and the decoding mechanisms and 1-layer RSMA is the basic building block of all the variations. In 1-layer RSMA, the common message symbols are decoded followed by SIC operations and then the individual private messages are decoded. Consider a downlink transmission to $M$ users. A total of $M+1$ message streams $\{s_c, s_1, s_2, \ldots, s_M\}$ are created from $M$ messages $\{W_1, W_2, \ldots, W_M\}$. The transmit signal is as follows:

$$x = \alpha_c P_c s_c + \sum_{m=1}^{M} \alpha_m P_m s_m \tag{8}$$

where $\alpha_c$ is the power coefficient for the common message, $\alpha_m$ is the power coefficient for the $m^{th}$ user private message, $s_c$ is the common message stream, $s_m$ is the $m^{th}$ user message stream, $P_c$ is the power for the common message and $P_m$ is the power for the $m^{th}$ user private message. The MLD receiver is used as a benchmark to compare the performance of the receiver designs. In MLD received the common and private message pair are estimated by minimum distance estimation rule.

The implementation of SIC to separate the signals at the receiving end is challenging. It is dependent on the model imperfections such as the channel estimation which deteriorates the performance.

**Some proposals:** In grant free random access, the objective is to maximize the success probability, maximize throughput or minimize the latency in the network.

A future direction is to do mobility aware NOMA user pairing in order to maximize the number of users that are being served by a channel resource. NOMA pairing is presently done based on channel gain difference between users, when users are moving the channel gain differences will vary as well. To this end, the mobility should be a factor in the pairing decision to ensure the pair are not frequently switched.

The practical implementation of SIC is challenging and it is required in both NOMA and RSMA. DL can be employed to separate the signals at the receiver. Using DL the model imperfections can be accounted for in the DL model by learning the non linear features. DL can be employed to design the receiver for RSMA. It can also be implemented to jointly optimize precoding and symbol detection in RSMA [CLDCL23].

## 1.5 Impairments

Wireless communications play a pivotal role in our interconnected world, enabling seamless data exchange across devices. However, several impairments affect the transmitted signal, so the reliability and quality of wireless signals can be downgraded. Some of the most common impairments are typically related to propagation aspects, such as path loss, attenuation, or fading, which have been deeply addressed in [BDE23].

Nevertheless, some impairments are closely related to the receiver architecture and require additional signal processing to overcome them. Indeed, hardware equipment introduces limitations that may non-linearly affect the signals or insert additional noise. This section focuses on describing and analyzing impairments beyond the propagation area.

### 1.5.1 PAPR

A PA is used to amplify the signals before the transmission by an antenna at the transmitter [Ars21]. The PAs have linear and nonlinear power intervals for the amplification. The nonlinear amplification causes in-band and out-of-band interference. When the signal with a high PAPR is amplified without causing distortion, the power backoff technique is generally used in practice. However, this causes power inefficiency and increases the power consumption. Therefore, the solution to avoid PA impairments is to find a way to decrease the PAPR of the signal before passing through the PA at the transmitter. Many methods for PAPR reduction have been developed in the literature [RM13]. Although OFDM has a high PAPR problem, it is a widely used waveform in many wireless standards and uses PAPR reduction methods. On the other hand, sensing systems use waveforms with a constant envelope, such as frequency modulated continuous wave (FMCW).

ISAC is a potential key technology for 6G mobile networks. While ISAC can be performed by communication- and sensing-based waveforms, literature and 6G standardization primarily focus on communication-based waveforms such as OFDM. Consequently, novel PAPR reduction techniques are essential to enable an ISAC system to use communication-based waveforms while concurrently meeting communication and sensing requirements. Here are two scenarios: the first involves using only the communication signal, while the second incorporates separate communication and sensing signals for ISAC. The communication and sensing signals contain random symbols and known sequences [MTOA23]. In the first scenario, the communication signal is optimized to minimize multi-user interference in MIMO systems, adhering to a low PAPR constraint [BC23, HMLN21, CWH$^+$23]. The second scenario involves joint modifications of communication and sensing signals to decrease the PAPR [HHM$^+$22, YWS$^+$23]. The second scenario requires additional resources for sensing functionalities. It is essential to highlight that modifying the signals is done to satisfy both communication and sensing requirements. Therefore, a different method can be developed for different ISAC requirements.

There are a lot of different sensing scenarios with different requirements in ISAC systems that have not been investigated in the literature. Considering these scenarios, new PAPR reduction techniques are required for an excellent trade-off between communication and sensing performance. Unlike available PAPR reduction techniques, the sensing signal characteristics should be utilized innovatively to have low computational complexity and high spectral efficiency in ISAC systems.

### 1.5.2 Non-linearities

Every radio front end is somehow nonlinear. From the PA perspective, when the transmitted signal has high envelope variations (like in OFDM signal), or the transmitted signal is close to the saturation region of the PA (while aiming at high energy efficiency), various samples will undergo different amplification. This will be observed as a distortion of the signal. In practical applications, the nonlinear distortion in transceivers is kept below a certain threshold (e.g., employing the Error Vector Magnitude limit imposed on the transmitter). This allows the designers of algorithms in the RAN to neglect the distortion utilizing linear radio channel model, e.g., fading and white noise.

Next-generation networks will have higher bandwidth, making linear frontend design even harder. At the same time, energy efficiency becomes one of the main design goals. To make it happen, we suggest optimizing the front end during higher-layer algorithms/procedures or making these algorithms aware of the front-end nonlinearity. While this makes the design more difficult, it opens new degrees of freedom, enabling potentially higher spectral and energy efficiency.

The problem of nonlinear PA modeling has been known for years [BHH+07, Gha11], and the models vary from straightforward amplitude distortion models, e.g., Rapp model or soft limiter, to advanced models, e.g., Voltera series. The bigger problem is the influence of nonlinear PA on the performance of wireless systems. While there are some measurement or simulation results, the most valuable ones are analytical models. In the case of the OFDM waveform model of wanted signal and uncorrelated distortion, power spectra densities can be found in [LO14].

A new problem was recently observed in the spatial distribution of nonlinear distortion in the mMIMO system. While in an IID Rayleigh channel, it is commonly assumed that distortion is emitted omnidirectionally [BHKD14], it is shown in [LVdP18] that for some cases, the nonlinear distortion pattern is the same as the spatial distribution of the wanted signal. There is no Signal-to-Distortion gain even if the number of antennas exceeds infinity. In [MGEL18b], a complete continuous-time model for calculating wanted signal and distortion distribution was proposed, although it does not provide a concise model.

The nonlinear amplifier characteristics are currently considered from many angles. First, [TMFPA16] defines an analytical signal to distortion (plus) noise ratio (SDNR) model for OFDM signal distorted by nonlinear PA. This allows the maximization of the link rate by adjusting the transmit power relative to the saturation power of the amplifier. Moreover, there is a potential to utilize the nonlinearity knowledge to improve the reception performance. While there are iterative receivers able to remove nonlinearity distortion [SO21], it is shown that the nonlinear distortion increases the minimum Euclidean distance between transmitted sequences, i.e., a system with nonlinearity can have better reception quality than a fully linear system (without distortion) [GDM13].

First, in [Kry23], some COST INTERACT contributors considered multiple PA nonlinearity (Rapp, soft limiter) and power consumption models (perfect, class A, class B) for maximization of energy efficiency or spectral efficiency of an OFDM transmission. In addition, the nonlinear character of a battery discharge has been considered. It is shown that a different power allocation model is to be used depending on the aim and utilized hardware. Next, the nonlinear distortion in a multicarrier system, depending on the power allocation, has some specific power spectral density in frequency, which can be used for resource allocation. It can be shown that per subcarrier power allocation can achieve higher throughput in a single-user and multi-user scenario. Finally, we have analyzed the spatial distribution of nonlinearity distortion in a mMIMO system under realistic, RayTracer-based channel impulse response. As expected, the nonlinear distortion is different in many locations than in an idealistic iid Rayleigh channel. This confirmed that nonlinear distortion is a significant problem in mMIMO systems. Especially in line-of-sight (LOS) channels and when analog beamforming is used. A distortion-aware power allocation scheme has been proposed for an analog beamforming-based transmitter [HK23].

### 1.5.3 Self-interference

One alternative gaining traction towards 6G is in-band full-duplex (IBFD), which can double the spectral efficiency and reduce the transmission latency [LYM+21]. IBFD is a research trend that has been actively explored for many years for increasing the throughput of wireless communications systems [SSG+14, LSHL15, KLH15].

Due to its benefits, IBFD has been proposed and implemented in various communication systems, such as WiFi [HDL+18], 5G/6G [ZCZ15, AS21], satellite communications [BSZMO15], or Digital Terrestrial Television (DTT) [ZLW+21, LZW+21]. Although the benefits of IBFD systems are noticeable, all potential use cases have a common challenge: canceling the self-interference signal leaking into the receiver antenna from the transmission antenna. From a mathematical perspective, the signal at the receiver in the frequency domain is

$$Y = X_{FWS} \cdot H_{FWS} + X_{LBS} \cdot H_{LBS} + N_0, \tag{9}$$

where $X_{FWS}$ and $X_{LBS}$ are the desired feedforward signals (FWSs) and loopback signals (LBSs), respectively. $H_{FWS}$ and $H_{LBS}$ are the FWS signal and LBS channel responses. $N_0$ is a function representing the thermal noise at the receptor, which is assumed to be AWGN.

In this line, the research community has focused on developing solutions for decreasing/eliminating the impact of self-interfence. Different technique families have also been tested to improve the cancellation performance, which can be classified into propagation domain, analog domain, and digital domain techniques [SSG+14, KPH19, NZQ+18]. In the first group, the loopback signal power is reduced before reaching the receiver. These techniques can be divided into passive and active. The former implies controlling the antenna directions, placements, and separations. As mentioned above, these techniques reach up to 50 dB of cancellation [BMK13, AES13], but the performance highly depends on the application case, and the desired signal can also be mitigated. On the other hand, active propagation domain techniques modify the communication path to obtain more favorable propagation conditions. Beamforming is a well-known example, which combines elements in an antenna array to modify the antenna pattern. Nevertheless, having full knowledge of the environment characterization is critical [IBF+22, IBF+23, BIF+24]. In [ESZS16], the authors present a digital transmit beamforming technique to reduce the loopback signal interference with a gain ranging from 20 to 80 dB obtained from field trials. However, the main drawback of this method is that traditional antennas should be replaced by more advanced radiating systems that would enable beamforming.

Secondly, the analog domain cancellation can be frequency or time-based and present a helpful alternative to reduce the dynamic range requirements of the digital domain cancellation modules [IBM+21]. In [BVAV18], the authors propose a time analog domain cancellation based on a multitap finite impulse response (FIR) filter. Although promising cancellation results were obtained, the performance showed a high-frequency dependency, ranging from 15 to 40 dB between 2.9 and 3.4 GHz. More recently, in [RMDG19], the authors propose combining an RF-tapping technique and an inherent self-interference cancellation technique. The results show cancellation of 80 dB between 2.40 and 2.44 GHz. Nevertheless, the cancellation considerably diminishes in the rest of the frequencies. The frequency dependency conclusion is also recurrent in other works [Mat21, SHS20, DvdBL+14].

Finally, the digital domain techniques refer to channel modeling and digital cancellation [ZLW+21]. In [TKS19], the authors design a least squares (LS) estimation denoised with Wavelet Decomposition (WD). The paper shows that for an SNR of 15 dB, a cancellation of 5 dB can be obtained. Then, in [WWLJ19], the authors propose a variable tap filter based on a Wiener filtering of an adaptive number of taps in the time and frequency domain to deal with the time variability of the loopback channel. Computer simulations showed that 25 dB of cancellation could be obtained when the interference-to-noise ratio reached 50 dB.

Taking advantage of the latest AI development, several authors have tested the cancellation performance of AI-aided solutions. A good example is [KCP21], where the authors design an AI-based analog domain cancellation technique (i.e., a two-tap canceller) based on AI. Also, in [KKK22], a fully connected neural network offers up to 18 dB gain when used to improve a linear estimation to deal with non-linear and dynamic behaviors. Although the results are promising, the knowledge concerning AI-based techniques for pure signal cancellation has not advanced significantly. Moreover, some COST INTERACT contributors have designed a CNN-based cancellation architecture. In particular, a Super Resolution Convolutional Neural Network (SRCNN), usually implemented for super-resolution purposes, is adapted to remove the loopback signal. The results have been compared to a well-performing traditional estimation algorithm (i.e., Wiener filtering) to evaluate the canceling capacity of the proposed method. The evaluation has been carried out under different loopback channel conditions derived from field measurements. Simulations indicate that our proposal outperforms state-of-the-art methods by up to 20 dB.

### 1.5.4 Phase noise

The ever-evolving landscape of mobile communications promises to enhance the user experience and connect our society in an intelligent environment. One of the key technologies for achieving higher capacity is using mmWave in 5G [OK20]. In addition, frequencies higher than 100 GHz are being considered for 6G. However, combining higher frequencies and the OFDM waveform can increase the system's vulnerability to phase noise (PN).

PN originates in the transceiver's local oscillator (LO), which is responsible for the up and down-conversion of signals. Ideally, the LO generates a signal at a specific frequency, which can be viewed as a Dirac delta function in the frequency domain. However, due to circuit imperfections and thermal noise, the signal is generated at the desired frequency and frequencies around it. As a result, the power spectral density (PSD) widens, leading to a loss of orthogonality between subcarriers, namely inter-carrier interference (ICI). In the time domain, this drift causes time-varying deviations in the signal phase. Therefore, the received signal exhibits a common phase error (CPE), manifested as a time-varying rotation of symbols in the constellation [QHN$^+$18]. The 5G NR standardized by the 3GPP uses a PN model based on a multi-pole/zero mask [QHN$^+$18]. Various models for different parameter set oscillators are available in TDoc R1-163984 [3gp] and TR 38.803 Section 6.1.10 [38.17].

To facilitate the pilot-based estimation of PN, the 3GPP introduced the phase tracking-reference signal (PT-RS) in the 5G NR standard [3GP18d]. However, this results in increased pilot overhead, diminishing data efficiency, and potentially complicating resource management at the base station in certain scenarios [QHN$^+$18].

In the literature, numerous works focus on traditional pilot-based PN estimation and compensation in OFDM systems [Arm06, CA08, PRF07, BDD23]. ML-based PN estimation and/or compensation has recently gained popularity [PCS$^+$20, MTL21, MC22, FS20]. In [PCS$^+$20], a pilot-based scheme estimates PN and a deep neural network (DNN) is designed to compensate for it in THz communications. [MTL21] refines traditional pilot-based estimation using a DNN. [MC22] presents a DNN architecture for PN estimation, where the input to the DNN is a sparse time-frequency grid with PN estimation in the pilot positions, and the network completes the grid with the remaining estimations. Finally, [FS20] proposes an ML-based soft-demapper capable of enhancing soft-bit computation in the presence of residual ICI after pilot-based estimation and compensation with PT-RS.

In contrast to the previous ML works, in [SA22], some COST INTERACT contributors propose to use ML to reduce the number of pilots while maintaining the same performance without sacrificing the data rate. This work focuses on the downlink channel using simple ML algorithms with limited computational complexity at user terminals. After equalization, symbols are classified using weighted k-Nearest Neighbors (k-NN) or Gaussian mixture model (GMM). The symbols classified with the highest amplitude are then used to estimate and compensate the CPE, and the symbols can be de-rotated. The results show that ML can eliminate the need for PT-RS for CPE tracking.

Other COST INTERACT contributors focused on the phase correction of the Gaussian minimum shift keying (GMSK) modulation. In particular, data generated in simulations covered the extended pedestrian A (EPA) channel and AWGN noise in the range Eb/N0 = 1-20 dB. The general concept of GMSK detection with an auxiliary phase correction block is presented in Fig. 2.

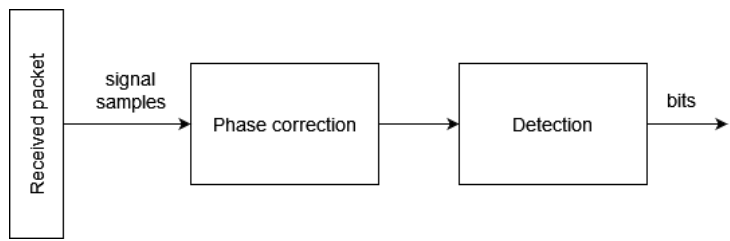

Figure 2: The general concept of GMSK detector.

The phase correction can be realised by the neural network trained on ideal correction reference data (obtained directly from the modulator). After phase correction, the received data can be further processed via the GMSK detector. The structure of the DL model for phase estimation is illustrated in Fig. 3.

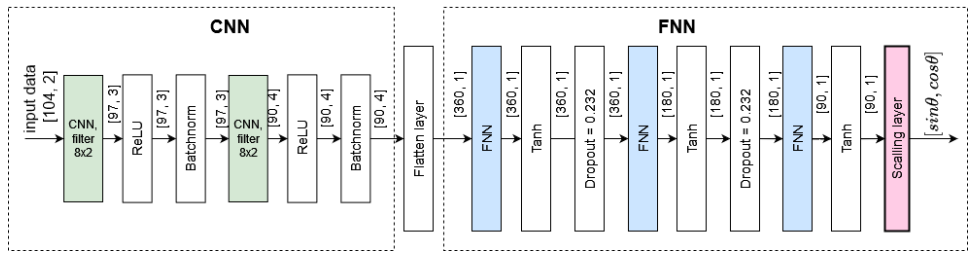

Figure 3: The DL-based phase corrector structure.

The presented model consists of convolutional and feedforward neural networks connected with a flattened layer. The output of the network represents the estimated phase $\theta$ rotation in terms of $sin\theta$ and $cos\theta$ representation. The accuracy of the performed phase correction for exemplary packet and EPA channel outperforms the state-of-the-art methods.

### 1.5.5 Wideband and ultra-wideband challenges and Beam squint issues

Ultrawideband (UWB) technology is expected to play a significant role in enabling innovative applications and addressing various challenges of next-generation wireless networks. UWB signaling is desired mainly due to its ability to provide high data-rate communication in short ranges, facilitate precise localization due to its high time-domain resolution, inherent covertness, which provides robustness against eavesdropping and jamming attacks, robustness against multipath fading, etc. The following discussion highlights some critical challenges that UWB systems may face.

**Spatial-wideband effect:** UWB signaling implementation requires enormous bandwidth, only available in higher frequency bands such as mmWave and THz bands. Owing to the high path loss problem at these higher frequencies, directional/beam-based communication through transceivers with mMIMO antenna arrays configuration is inevitable. However, it has been shown that mMIMO systems implementing UWB signaling suffer from spatial-wideband effects. This effect can be perfectly explained by the Huygens principle, which dictates that in antenna array systems if the incident signal is not perpendicular to the array, the signal received at different array elements will be a slightly delayed version of the original signal. The delay incurred across the elements depends on the inter-element spacing and the signal's angle of arrival (AoA) or angle of departure (AoD). In conventional small-scale MIMO systems with a relatively small number of antennas, the maximum delay across the antenna aperture is usually much smaller than the symbol duration, so its effect can be ignored. However, high-dimensional antenna arrays, such as mMIMO and UWB signaling, can cause the delay to be in the order of or even larger than the symbol duration, leading to a delay squinting effect in the spatial delay domain. This means a significant delay spread is observed across the array even in pure LOS propagation conditions. The delay squinting effect can cause significant inter-symbol interference in the system if not properly considered during guard duration design between consecutive symbols [WGJ+18]. The delay squinting effect also makes the array steering vector frequency-dependent in the angular-frequency domain [AKA24, KAA22]. In multicarrier systems such as OFDM, signals at different subcarriers point to different physical directions. Signals related to such derailed subcarriers might not arrive at the intended receiver or may align with sidelobes or nulls of the receiver's radiation pattern, thereby degrading system capacity [AKA24]. This phenomenon is known as beam squinting. As the array size grows even larger and the beams narrow, the beam squinting transforms into a complete beam-splitting problem whose impact on the system performance is far worse.

**Molecular Absorption and Temporal Broadening Effects in THz Frequencies:** Apart from the conventional path loss problem, channels at high frequencies also exhibit a severe form of frequency-selective absorption known as molecular absorption (MoA) effect. Essentially, some of the frequencies in THz bands align with natural resonance frequencies of the atmospheric contents like Oxygen and water molecules. When excited at their resonance frequencies, these molecules absorb significant energy from the signal, significantly elevating path loss at these particular frequencies. Such frequency-selective MoA effects naturally create small windows of varying bandwidths across THz bands on which the signals can be reliably transmitted. Note that the MoA effect is a function of both frequency and distance. At higher THz frequencies and larger distances between communication nodes, the number of absorption peaks increases, which reduces the widths of the transmission windows even more. This may impact the operation of the UWB signaling. Further analysis of the MoA effect showed that some absorbed energy is re-radiated at the same frequency, considering it a noise source. However, some experimental studies have revealed that the re-radiated energy is highly correlated with the original signal and can be considered its scattered component. This is known as the molecular scattering (MoS) effect. The MoS effect has been found to increase the multipath richness and randomness of the channel [HDHC20].

UWB signal in THz frequencies may also experience temporally dispersive channels even in line-of-sight propagation conditions. The cause of this dispersion is two-fold. First, the signal with bandwidth larger than the transmission windows created by the frequency-selective MoA effect experiences a frequency-selective channel. This frequency selectivity due to MoA manifests itself in time-domain as temporal broadening effect (TBE) where the duration of the received pulse is larger than the transmitted pulse duration [HBA14]. Second, the frequency-dependent refractive index of the atmospheric contents, such as water vapor and Oxygen molecules, induces different group velocities to different portions of the UWB signal, which causes the received signal to spread in time [SEO21]. Group velocity- and MoA-induced temporal dispersion can lead to inter-symbol interference and system capacity degradation if poorly handled.

**Worse-than-Rayleigh fading channels in UWB Channels:** UWB signals have been observed to experience a severe form of fading, referred to as worse-than-Rayleigh fading (WRF), which exhibits a higher probability of deep fade compared to the conventional Rayleigh fading channels [KFSA23]. This is simply because the UWB signal has a high temporal resolution, which implies that only a small number of multipath componentss (MPCs) fall within an interval of a resolution bin. Accordingly, in this case, the central limit theorem no longer holds, invalidating the Rayleigh fading process assumption. Additionally, with only a few MPCs in each delay bin, the chances of having a total destructive superposition of these MPCs are higher, which increases the probability of deep fades in the system. The severe deep fades due to WRF channel conditions can lead to a devastatingly poor bit error rate, BER performances, and significantly reduced link capacity.

### 1.5.6 Bussgang Decomposition

The Bussgang decomposition is well known as a general modelling approach for non-linear systems: it expresses the output of a memoryless non-linear system as the sum of a term which is proportional to the input plus an uncorrelated distortion term. It is derived from the Bussgang theorem [Bus52], which states that given two jointly Gaussian random variables $x$ and $z$, and a memoryless non-linear function $y = f(x)$, the product moment $\mathbb{E}[xz]$ of $z$ with the input $x$ and with the output $y$ are related by a constant $\alpha_x$ which depends only on the non-linearity and the standard deviation $\sigma_x$ of the input. It follows that the output $y$ of the non-linearity if the input $x$ is a Gaussian random variable can be expressed as the sum of a component which is proportional to the input (and thus fully correlated with it), and a component $\delta_x$ which is perfectly uncorrelated. Then:

$$y = \alpha_x x + \delta_x \tag{10}$$

where:

$$\alpha_x = \frac{\mathbb{E}[yx]}{\mathbb{E}[x^2]} = \frac{1}{\sigma_x^2} \int_{-\infty}^{\infty} x f(x) p_x(x) \, dx \tag{11}$$

and $p_x(x)$ denotes the PDF of $x$, assumed to have zero mean and standard deviation $\sigma$. We also define a second parameter $\gamma_x$ which relates the output power of the non-linearity to its input power:

$$\gamma_x = \frac{\mathbb{E}[y^2]}{\mathbb{E}[x^2]} = \frac{1}{\sigma_x^2} \int_{-\infty}^{\infty} f^2(x) p_x(x) \, dx \tag{12}$$

[Bus52] assumes that $x$ has the Gaussian distribution, and for this reason it is sometimes assumed that the decomposition is only applicable to Gaussian variables. Moreover [Zil10] derives a formula for the SDNR:

$$\text{SDNR} = \frac{1}{\frac{\gamma_x}{\alpha_x^2} - 1} \tag{13}$$

However it is easy to show using the proof of this formula given in [Zil10] that it does not depend on the Gaussian assumption, even though this is not explicitly stated and the assumption is used in general in that paper.

We next consider the case where the input to the non-linearity includes unwanted noise or interference $n$ as well as a wanted signal $s$ – we refer to this as the noisy case. Here we are usually interested in decomposing the output into a component which is correlated with $s$ and a combined distortion and noise component which has zero correlation with it. We may write:

$$y = \alpha_s(s + n) + \delta_s \tag{14}$$

We may show that if the overall output distortion plus noise $y - \alpha_s s$ is to be uncorrelated with the signal $s$ then:

$$\alpha_s = \frac{\mathbb{E}[ys]}{\mathbb{E}[s^2]} = \frac{1}{\sigma_s^2} \iint_{-\infty}^{\infty} y s p_{y,s}(y,s) \, dy \, ds = \frac{1}{\sigma_s^2} \int_{-\infty}^{\infty} s p_s(s) \mu_y(s) \, ds \tag{15}$$

where $\mu_y(s)$, the conditional mean of $y$ given $s$:

$$\mu_y(s) = \int_{-\infty}^{\infty} y p_{y|s}(y|s) \, dy \tag{16}$$

It is possible to show, using the Bussgang theorem, that provided both $s$ and $n$ have the Gaussian distribution (which we will call the noisy Gaussian case) then $\alpha_s = \alpha_x$ – but this only applies if the Gaussian assumption is valid. As before, we may also define:

$$\gamma_s = \frac{\mathbb{E}[y^2]}{\mathbb{E}[s^2]} = \frac{1}{\sigma_s^2} \iint_{-\infty}^{\infty} y^2 p_{y,s}(y,s) \, dy \, ds = \frac{1}{\sigma_s^2} \int_{-\infty}^{\infty} p_s(s) \mu_{y2}(s) \, ds \tag{17}$$

where $\mu_{y2}(s)$, the conditional second moment of $y$ given $s$:

$$\mu_{y2}(s) = \int_{-\infty}^{\infty} y^2 p_{y|s}(y|s) \, dy \tag{18}$$

It is easy to show that, regardless of the distribution of $s$ and $n$, $\gamma_s = \gamma_x(1 + \frac{\sigma_n^2}{\sigma_s^2})$. These expressions can be used to obtain expressions for SDNR – though we should note that in this case this quantity should be understood as "Signal to distortion-plus-noise ratio", while in the noiseless case it is "Signal to distortion noise ratio".

Thus in the noisy case we can always use the formula:

$$\text{SDNR} = \frac{1}{\frac{\gamma_s}{\alpha_s^2} - 1} = \frac{1}{\frac{\gamma_x}{\alpha_s^2}(1 + \frac{\sigma_n^2}{\sigma_s^2}) - 1} \tag{19}$$

but in the noisy Gaussian case we can express this as:

$$\text{SDNR} = \frac{1}{\frac{\gamma_x}{\alpha_x^2}(1 + \frac{\sigma_n^2}{\sigma_s^2}) - 1} \tag{20}$$

which has the advantage that $\alpha_x$ depends only on the non-linearity and the standard deviation of the signal, while in the more general case $\alpha_s$ depends also on the distribution of the signal and of the noise. (Note also that in the general case $\gamma_x$ would also have to be calculated using the actual distribution of $x$, which is the sum of two variables with potentially different distributions). However this result does show that a version of the Bussgang decomposition applies in general, regardless of the distribution of signal and noise.

For the general noisy case it is generally more convenient to use the formulae (15 – 18) to obtain the expression for SDNR. These allow expressions for $\mu_y(s)$ and $\mu_{y2}(s)$ to be calculated first, depending only on the conditional distribution of $y$ given $s$; once these have been determined they can be used with the distribution of $s$ to find $\alpha_s$ and $\gamma_s$, and hence SDNR.

It is important to note that even in the Gaussian case the distribution of the noise plus distortion is not necessarily (indeed not usually) Gaussian – thus results for quantities like BER which assume Gaussian noise may not be valid. This may limit the usefulness of the Bussgang decomposition in some practical cases.

## 1.6 O-RAN and disaggregation

### 1.6.1 O-RAN Key Architecture Principle

Traditional RANs have historically used proprietary hardware at base stations with tightly integrated network functions, making it challenging to adapt to changing demands without manual on-site interventions. The introduction of O-RAN has transformed this approach by providing a more flexible and open architecture that addresses these limitations [38.17]. To advance this, the O-RAN Alliance was established in 2018 by mobile network operators to create an open, intelligent, and cost-effective aRAN architecture. The concept of O-RAN is currently attracting a lot of attention, in the context both of the roll-out of 5G networks and of their evolution to 6G. Part of the motivation for this, from a commercial viewpoint, is the need to diversify the supply chain for RAN equipment, which is now regarded in many countries as vital national infrastructure [MH23]. The O-RAN vision builds on years of research into open, programmable networks, a concept central to the software-defined networking (SDN) transformation in wired networks over the past 15 years. Recently, these principles have extended into wireless networks. The xRAN Forum, led by operators, introduced standardized fronthaul interfaces and open interfaces for external RAN controllers. Simultaneously, the cloud-RAN (C-RAN) architecture centralized baseband processing in cloud data centers, improving signal processing and load balancing while reducing costs by sharing computational resources [PZK24].

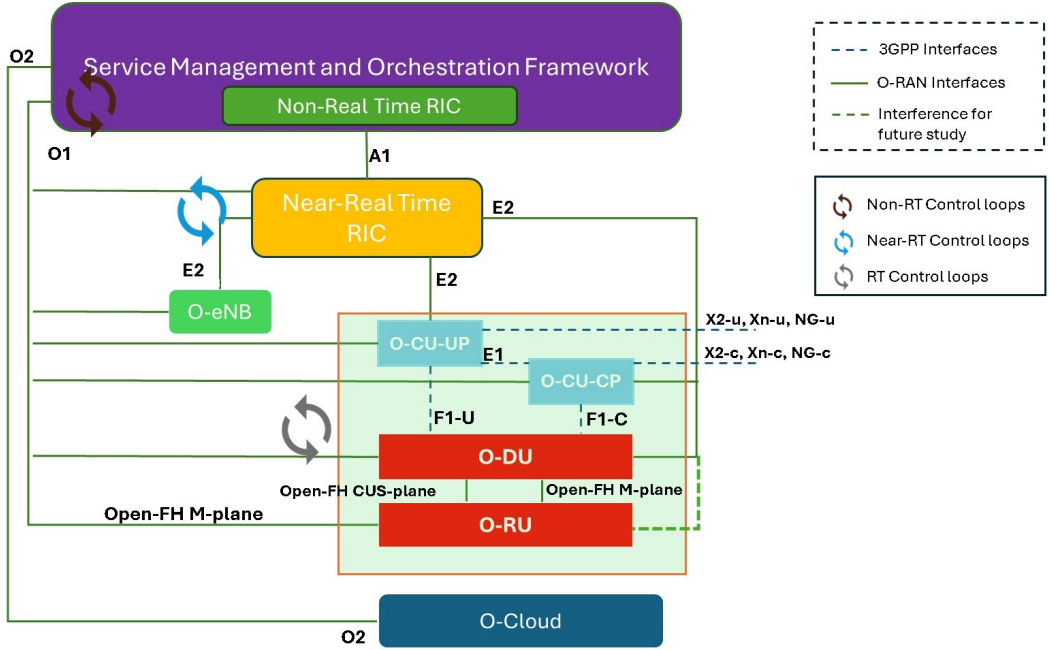

Figure 4: O-RAN Architecture [MH23]

The O-RAN architecture, as shown in Figure 4, is built on the principles of disaggregation, intelligent data-driven control through RAN intelligent controllers (RICs), virtualization, and open interfaces. Disaggregation separates base station functions into three main components: the open-radio unit (O-RU), open-distributed unit (O-DU), and open-central unit (O-CU). The O-CU further splits into the open-central unit control plane (O-CU-CP), responsible for tasks such as radio resource management and handover decisions, and the open-central unit user plane (O-CU-UP), which handles user data traffic, including encryption, decryption, and routing. These components communicate via open interfaces, promoting vendor interoperability and network flexibility. The RIC plays a central role in enabling intelligent network management by virtualizing traditional RAN functions as virtual network functionss (VNFs). The RIC allows the network to dynamically respond to changes, optimizing performance and resource allocation in real time [PBD+23]. Overall, the O-RAN architecture increases agility, flexibility, and interoperability, allowing operators to deploy new services faster, scale efficiently, and foster vendor competition. This open, disaggregated approach drives innovation, cost reduction, and accelerated technological advancement in the RAN domain.

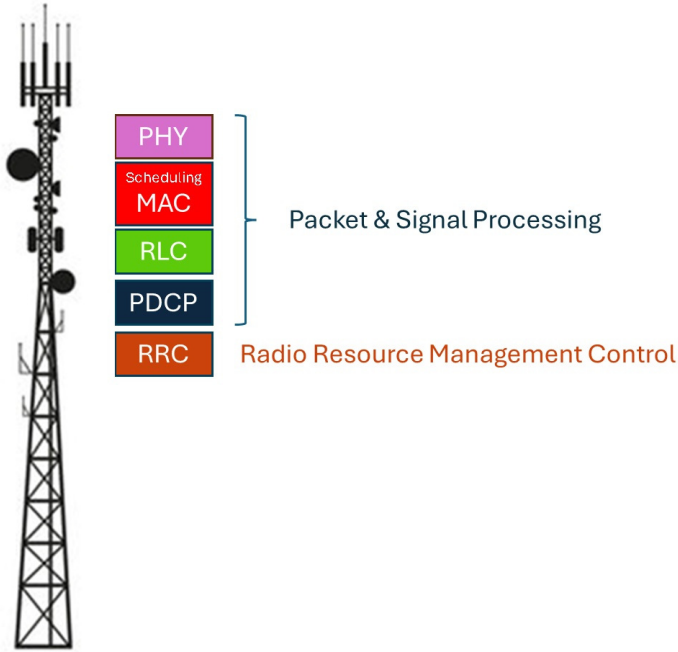

Figure 5: RAN Protocol and RRM control

**RAN Disaggregation:** In cellular networks, the RAN is responsible for providing wide-area wireless connectivity to mobile devices. To achieve this, it performs two essential functions [BPD+20]:

- It converts internet protocole (IP) packets into Physical Layer packets for transmission over the dynamic mobile channel, utilizing advanced packet and signal processing techniques.

- It manages radio resources to optimally allocate and control the limited radio spectrum, ensuring connectivity for active mobile devices.

As shown in Figure 5, the 3GPP has structured the RAN with a layered protocol stack, which has been disaggregated into three tiers across two dimensions [3gp22a]. The first tier is horizontal disaggregation, which divides the RAN protocol stack into independent components, enabling flexibility in implementation. This approach addresses challenges such as high costs, energy consumption, and the need for intelligent, dynamic resource management while promoting rapid innovation and multi-vendor interoperability. The 3GPP has established several options for horizontal disaggregation within the RAN architecture. These options provide flexibility in how different functional components of the RAN are separated and implemented, allowing for enhanced adaptability in network deployments [3gp18a]. A comprehensive summary of these disaggregation options is provided in Figure 6, outlining the various approaches to splitting the RAN protocol stack for optimal performance, scalability, and interoperability across different vendor solutions.

Several horizontal disaggregation options have already been defined by 3GPP. Following the O-RAN architecture, as shown in Figure 7, the disaggregation enables distributed deployment of RAN

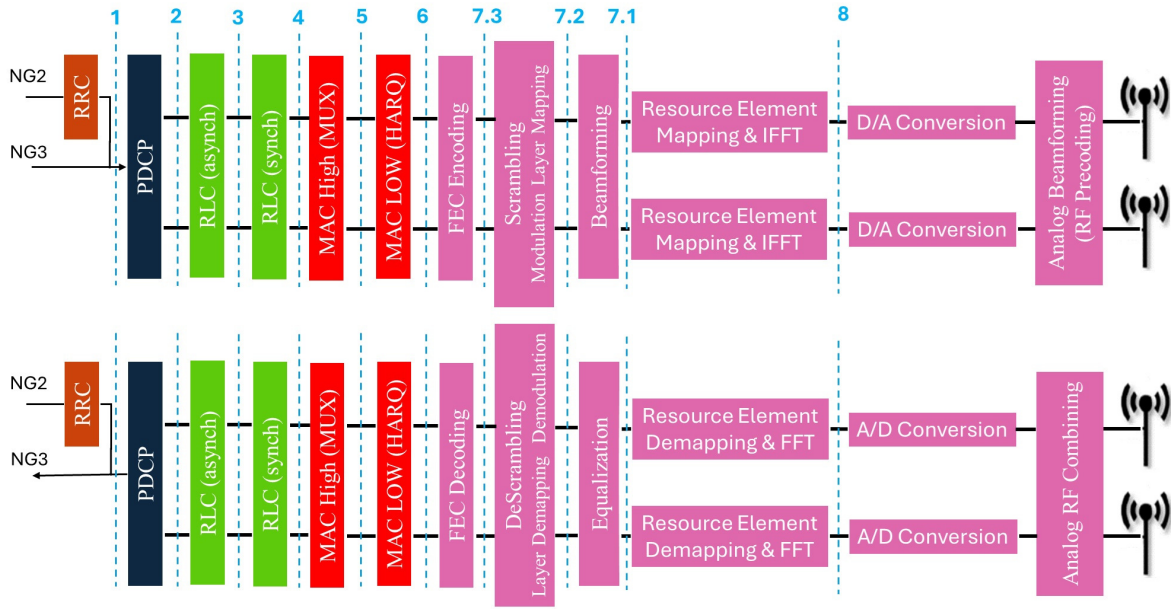

Figure 6: 3GPP Specific RAN Disaggregation Options

functions over the coverage area, involving the following key components:

O-CU: Centralizes packet processing functions, implemented as VNF on commodity hardware in telco edge cloud locations.

O-DU: Handles baseband processing functions at cell sites, also implemented as VNFs, with potential for hardware acceleration using technologies like field-programmable gate arrays (FPGAs).

O-RU: Manages radio functions and provides geographical coverage through specialized hardware at antenna sites. This horizontal disaggregation approach is designed to be flexible, allowing operators to implement various configurations based on use case, geography, and operational preferences. These configurations include:

- Splitting components as O-CU and O-DU+O-RU or O-CU+O-DU and O-RU.

- Combining all components (O-CU+O-DU+O-RU) into a single unit.

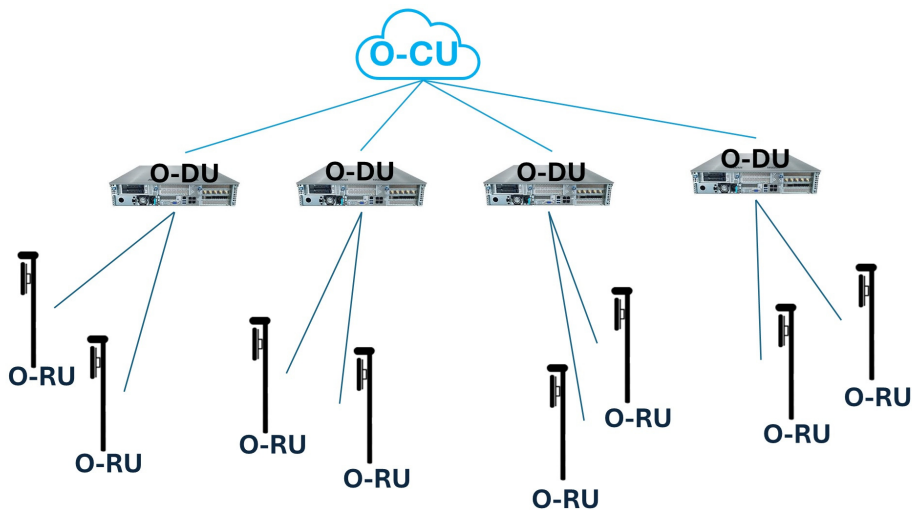

Figure 7: RAN Disaggregation and Distributed Deployment

The O-RAN Alliance is developing open specifications for interfaces between these disaggregated components. The Alliance has focused on a subset of 3GPP RAN split options (as shown in Figure 8); (1) The O-RU hosts Low-PHY and RF processing (split option 7.2). (2) The O-DU manages the radio link control (RLC), MAC, and High-PHY layers (split option 2). (3) The O-CU handles radio resource control (RRC), packet data convergence protocol (PDCP), and service data adaption protocol (SDAP) layers.

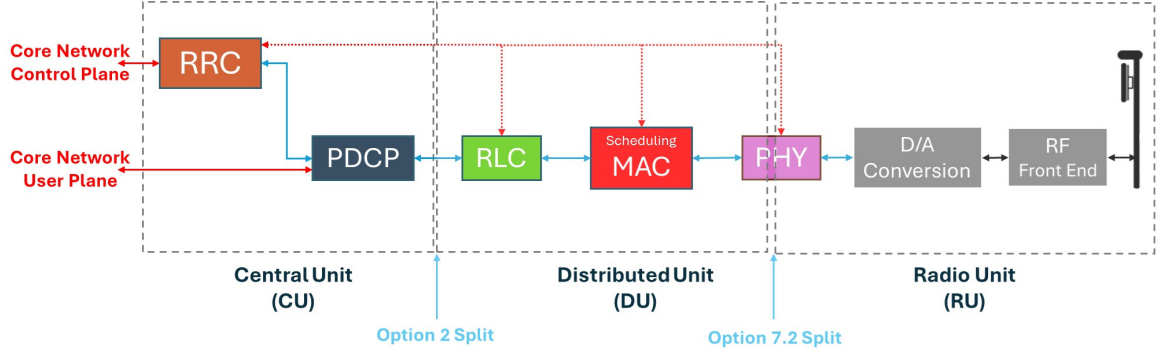

Figure 8: CU-DU-RU Disaggregation

The second tier of disaggregation is vertical and focuses on separating the control and user planes within the O-CU. This disaggregates the O-CU into:

- O-CU-UP: Manages the user-plane part of PDCP and SDAP protocols.

- O-CU-CP: Handles the control-plane part of PDCP and RRC protocols.

The third tier of disaggregation follows the SDN model by carrying vertical disaggregation one step further. It does this by separating RAN control functions from disaggregated components, primarily from O-CU-CP. These control functions are centralized into applications running on an SDN controller, called the near real-time RIC (nRT RIC) within the O-RAN architecture. The O-RAN Alliance has carefully analysed the various O-RU/O-DU split options proposed by 3GPP, with a focus on finding the optimal physical layer division between the O-RU and O-DU [38.17].

After consideration, the 7.2x split was selected as it offers a balance between the simplicity of the O-RU and the data rate and latency requirements on the interface between the O-RU and O-DU. In the 7.2x split, the O-RU handles time-domain tasks such as precoding, fast Fourier transform (FFT), CP addition/removal, and RF operations. This configuration keeps the O-RU cost-effective and easy to deploy. On the other hand, the O-DU manages the remaining physical layer functions along with the MAC and RLC layers. These include tasks like scrambling, modulation, layer mapping, partial precoding, and mapping data into physical resource blocks. The operations of the MAC, RLC, and physical layers are tightly integrated, as the MAC layer generates TBS from data buffered at the RLC layer for use in the physical layer. At the higher level, the O-CU is split into O-CU-CP and O-CU-UP functions. The O-CU implements the upper layers of the 3GPP stack, including the RRC layer, which manages connection life cycles, the SDAP layer, which ensures quality of service (QoS) for traffic flows (bearers), and the PDCP layer, which handles reordering, packet duplication, and encryption for the air interface [MARGS+20].

**RAN Intelligent Controllers:** The second key innovation is the introduction of RICs, which bring programmability to the network, enabling automated, closed-loop control of the RAN. Central to the O-RAN framework, RICs gather and analyze Key key performance metricss (KPMs) from the network in real-time, providing insights such as active users, network load, and resource utilization. By leveraging AI and ML, RICs can optimize functions like load balancing, user handovers, and scheduling without manual intervention [BDP+21].

The O-RAN Alliance outlines two types of RICs: the nRT RIC, which manages critical tasks like network slicing and handovers within 10 milliseconds to 1 second, and the non-real-time RIC (non-RT RIC), which focuses on long-term optimization using predictive analytics. xApps and rApps further enhance network efficiency, offering specialized services and localized control, reducing latency.

Real-time control loops for tasks like scheduling and beam management, operating within 10 milliseconds, are noted for future study in O-RAN specifications [ORA21b].

**Virtualization:** The third principle of the O-RAN architecture introduces new components to manage and optimize network infrastructure, from edge systems to virtualization platforms. All O-RAN elements can be deployed on the O-Cloud, a hybrid cloud platform that pools resources across multiple datacenters, enabling hardware-software decoupling, resource sharing, and automated RAN function deployment. O-RAN virtualization optimizes power consumption by dynamically scaling compute resources and enhancing sleep cycles for base stations, improving energy efficiency [ORA21a].

**Interfaces:** The O-RAN Alliance has introduced technical specifications defining open interfaces that connect various components within the O-RAN architecture. Figure 4 highlights these open interfaces, along with 3GPP intra-RAN interfaces, which partially enable the disaggregated next-Generation node B (gNodeB) architecture. The Open fronthaul between the O-DU and O-RU complements these interfaces. These open interfaces move away from the traditional "black box" RAN approach by providing data analytics and telemetry to RICs, facilitating control, automation, virtualization, and deployment optimization. Without O-RAN, radio resource management and network function optimization would remain closed and inflexible, limiting operators' access to their RAN equipment. The standardization of these interfaces is crucial for breaking vendor lock-in and enabling interoperability between components from different manufacturers, fostering innovation, competition, and faster upgrades. The E2 interface connects the nRT RIC to RAN nodes, enabling near-real-time control loops via telemetry and feedback. The A1 interface links the nRT RIC to the non-RT RIC, facilitating non-real-time control and policy deployment. The non-RT RIC also manages the O1 interface for network orchestration, while the O2 interface connects the non-RT RIC and service management and orchestration (SMO) to the O-RAN O-Cloud. Additionally, the O-RAN Fronthaul interface links O-DUs and O-RUs. To ensure interoperability, the O-RAN Alliance has established testing procedures, focusing initially on the fronthaul and E2 interfaces. These open interfaces allow flexible deployment across different network locations, such as cloud, edge, or cell sites, with multiple configurations available.

Table 1: MCS and TBS Configuration

| MCS | $Q_m$ | R | [Index]-TBS | m | MCS | $Q_m$ | R | [Index]-TBS | m |
| --- | --- | --- | --- | --- | --- | --- | --- | --- | --- |
| 0 | 2 | 30 | [7]-2408 | 4 | 12 | 6 | 517 | [3]-24072 | 8 |
| 1 | 2 | 64 | [6]-4872 | 4 | 13 | 6 | 616 | [3]-28880 | 8 |
| 2 | 2 | 120 | [5]-4872 | 4 | 14 | 6 | 772 | [2]-38556 | 8 |
| 3 | 2 | 193 | [4]-6280 | 4 | 15 | 6 | 910 | [2]-31752 | 8 |
| 4 | 2 | 308 | [3]-6144 | 4 | 16 | 6 | 682.5 | [3]-31752 | 10 |
| 5 | 2 | 526 | [2]-6016 | 4 | 17 | 8 | 754 | [2]-34816 | 10 |
| 6 | 2 | 679 | [2]-7808 | 4 | 18 | 8 | 841 | [2]-38936 | 10 |
| 7 | 4 | 340 | [4]-22032 | 6 | 19 | 8 | 948 | [1]-27656 | 10 |
| 8 | 4 | 434 | [4]-28168 | 6 | 20 | 10 | 803.5 | [0]-31240 | 10 |
| 9 | 4 | 553 | [3]-22032 | 6 | 21 | 10 | 853 | [0]-40440 | 10 |
| 10 | 4 | 658 | [3]-26632 | 6 | 22 | 10 | 900.5 | [1]-26120 | 10 |
| 11 | 6 | 438 | [3]-31752 | 8 | 23 | 10 | 948 | [0]-23040 | 10 |

### 1.6.2 Fronthaul Compression

Since the signals carried on the fronthaul are digitised, they are inherently quantized, and the quantization process may result in quantization distortion which will affect the performance of the physical layer. Clearly also the more finely the signals are quantized the smaller the quantization distortion but the higher the load on the fronthaul network. The objective of this part is to examine the trade-off between fronthaul load and the performance of the disaggregated physical layer, as determined by the parameters of the quantization process.

In this section we consider the uplink specifically, since this requires quantization of received signals at the O-RU and disaggregates the reception process, directly affecting its performance. O-RAN defines three compression methods applicable to user signals on the uplink, which all operate jointly on the samples of one PRB using floating point representation. The simplest (Block Floating Point Compression) calculates a joint exponent based on the sample (in-phase and quadrature (IQ)) with the largest magnitude from which the mantissae of all IQ samples are calculated. Compression then consists simply in dropping the least significant bits, leaving either 8 or 9 bits (these are two options for compression level). The next, Block Scaling Compression, further calculates a scale factor from the largest-magnitude sample and multiplies all samples by that, again dropping all but the most significant 8 or 9 bits. In this case the quantization interval is simply twice the maximum sample magnitude divided by 256 or 512. The third option is $\mu$-law compression, which applies a simple non-linear transformation prior to an effective quantization.

The fronthaul compression/quantization algorithm can be combined with the functional split 7.2. In this model, the O-RU performs initial physical baseband processing in the uplink, such as CP removal and FFT. The output of the FFT is then compressed before being transmitted over the fronthaul. Compression/quantization is applied on a per-PRB basis. Each PRB consists of 12 complex IQ samples, corresponding to 12 subcarriers in an OFDM symbol. The real and imaginary components of each IQ sample are represented by 16-bit signed integers. After compression, each complex sample is represented by a bit depth ranging from 2 to 16 bits. The term iqWidth is used to denote the number of bits used for the real or imaginary part of each sample, meaning a complex sample is represented by 2×iqWidth bits.

**Optimum quantization vs O-RAN Fronthaul Compression Algorithms :** The quantizer $\mathbf{Q}(\mathbf{x})$ is simulated using the expression

$$\mathbf{Q}(\mathbf{x}) = \Delta \cdot min\left(max\left(\left(\left\lfloor\frac{\mathbf{x}}{\Delta}\right\rfloor + \frac{1}{2}\right), -\frac{2^m - 1}{2}\right), \frac{2^m - 1}{2}\right), \qquad (21)$$

where $\Delta$ is the quantization interval and $m$ is the number of quantization bits. The quantization range is limited to $\pm\frac{\Delta}{2}(2^m - 1)$. This quantization process can generally lead to two forms of distortion, known as granular distortion due to quantization error within the range, and overload distortion due to truncation of the signal outside the quantization range. In practice $m$ and optimum $\Delta$ are chosen for optimal performance. Details on this can be found in [BBM18]. On the other side, the O-RAN fronthaul compression algorithms determine the quantization interval in a different way to the optimization quantization. The algorithm compresses one PRB at a time, determining the maximum absolute value $|x|_{\max}$ of the I/Q samples within the PRB and choosing $\Delta$ as $\frac{2|x|_{\max}}{2^m - 1}$. This avoids any overload distortion but may result in larger granular distortion than necessary. If the effect of phase shift is neglected it is easy to see that for QPSK the cumulative distribution function (CDF) of $|x|_{\max}$ is given by:

$$P(|x|_{\max} \leq z) = \left(1 - Q\left(\frac{z - A}{\sigma}\right) - Q\left(\frac{z + A}{\sigma}\right)\right)^M \qquad (22)$$

where $A$ is the amplitude of the IQ signals, $\sigma$ is the standard deviation of the I/Q sample noise, $Q(.)$ denotes the Gaussian Q-function, and $M$ is the total number of I and Q samples. Since the number of subcarriers in a PRB is 12, $M = 24$.

**Information theoretic analysis** Next, a theoretical approach to the effect of quantization is considered, focusing on its impact on the mutual information after quantization. Specifically, the mutual information between the source data and the quantized signal is analysed in comparison to that of the unquantized received signal. This approach is motivated by the effectiveness of LDPC codes in approaching the channel capacity, or the maximum mutual information. It is suggested that the degradation caused by quantization may primarily be influenced by the extent to which the channel capacity is reduced by the quantization process. The following formula can be used to estimate capacity of unquantized QPSK

$$\begin{aligned}I_{uq} = 1 - \frac{1}{2}\int_{-\infty}^{\infty}\bigg(&P_{\mathcal{N}}\Big(\frac{y-A}{\sigma}\Big)log_2\bigg(\frac{P_{\mathcal{N}}\big(\frac{y-A}{\sigma}\big)}{P_{\mathcal{N}}\big(\frac{y-A}{\sigma}\big) + P_{\mathcal{N}}\big(\frac{y+A}{\sigma}\big)}\bigg)\\&+P_{\mathcal{N}}\Big(\frac{y+A}{\sigma}\Big)log_2\bigg(\frac{P_{\mathcal{N}}\big(\frac{y+A}{\sigma}\big)}{P_{\mathcal{N}}\big(\frac{y-A}{\sigma}\big) + P_{\mathcal{N}}\big(\frac{y+A}{\sigma}\big)}\bigg)\bigg).\end{aligned} \qquad (23)$$

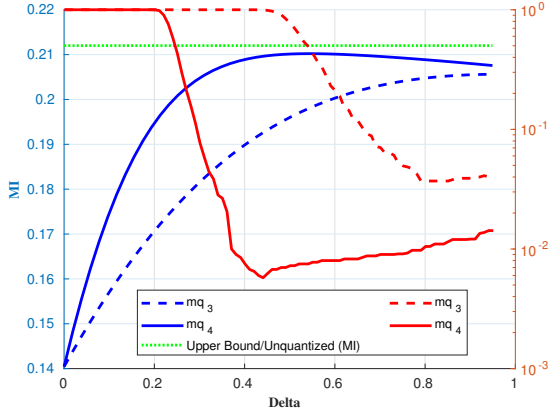
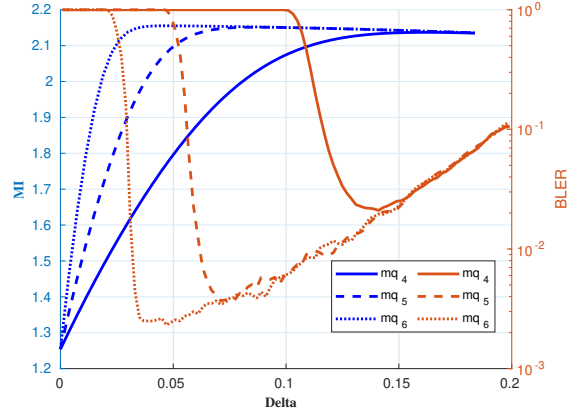

Figure 9: Finding optimum $\Delta$ (SNR -8dB).     Figure 10: Finding optimum $\Delta$ (SNR 5.8dB).

In the quantized case the channel can be treated as a discrete memoryless channel with input $x$ and $2^m$-ary output $y_q \in Y_q$. In the case of QPSK, $x$ is binary: $x \in -A, A$, while in the case of $M$-QAM it forms a $\sqrt{M}$-PAM constellation with the same average power. Then the quantized mutual information (MI) is

$$I_q = 1 - H(X|Y_q), \qquad (24)$$

where $H(X|Y_q)$ denotes the conditional entropy of the input X given the quantized output $Y_q$:

$$H(X|Y_q) = \sum_{x=-A}^{A} \sum_{y_q=-\frac{(\Delta 2^m - 1)}{2}, \Delta}^{\frac{\Delta(2^m-1)}{2}} P(x)P(y_q|x)log_2(\frac{1}{P(x|y_q)}), \qquad (25)$$

where

$$P(x|y_q) = \frac{P(x)P(y_q|x)}{P(y_q)} = \frac{P(y_q|x)}{\sum_{x=-A}^{A} P(y_q|x')}, \qquad (26)$$

and

$$P(y_q|x) = \begin{cases} Q\big(\frac{2^{m-1}\Delta - x}{\sigma}\big), & y_q = \frac{\Delta(2^m-1)}{2}, \\ Q\big(\frac{y_q - \frac{\Delta}{2} - x}{\sigma}\big) - Q\big(\frac{y_q + \frac{\Delta}{2} - x}{\sigma}\big), & \frac{\Delta(2^m-3)}{2} \geq y_q \geq -\frac{\Delta(2^m-3)}{2}, \\ 1 - Q\big(\frac{-2^{m-1}\Delta - x}{\sigma}\big), & y_q = \frac{-\Delta(2^m-1)}{2}. \end{cases} \qquad (27)$$

Our methodology uses simulation to select the optimum quantization interval $\Delta$, and the required number of bits $m$ to represent the quantized signals. This can be compared with the MI values for the quantized and unquantized cases. We include plots shown in Fig. 9 and 10 to illustrate this methodology. Fig. 9 and 10 show the results in two cases, MCS 1 and 8 as detailed in Table 1 for various given $m$, with MI and BLER (from simulation) plotted against quantization interval $\Delta$. The latter plots are shown in red and refer to the right-hand vertical axis, while the former are shown in blue and refer to the left hand vertical axis. Note that since we assume a fixed number of quantization bits, there is an optimum value of $\Delta$ in each case, since too small a value truncates the signal, while too large quantizes it too coarsely. We note that the maxima of MI and minima of BLER match quite well, given the limitations of the simulation, indicating that the information theoretic approach may be useful in selecting the optimum $\Delta$. In Fig. 9 we also add in green a line showing the maximum MI in the unquantized case, showing that quantization with $m = 4$ is extremely close to the maximum MI available. It is also possible to estimate the capacity gap (and hence the difference in required SNR) between the quantized and unquantized cases.

**Effects of Compression/Quantization on the PHY Performance:** To investigate the effect of quantization, a range of 5G-NR MCS options and TBS were considered, along with the impact of a frequency selective fading channel. Figure 11 shows the BLER-SNR curves for 24 different MCS from

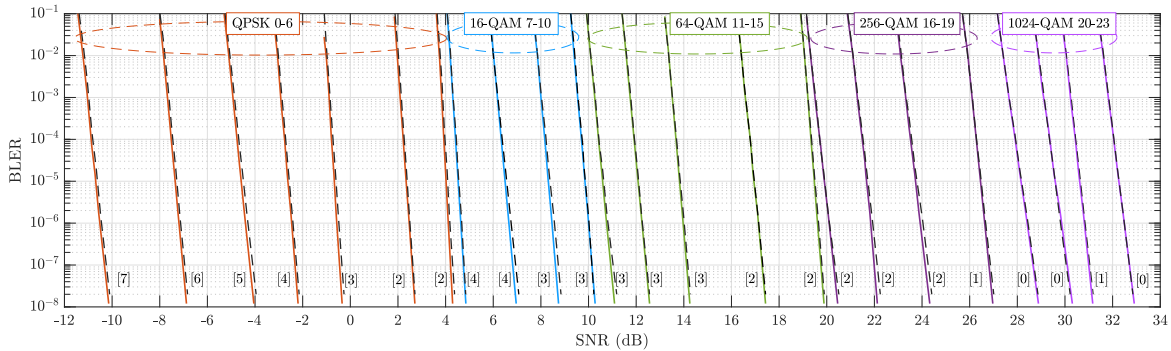

Figure 11: BLER-SNR curves for MCS 0-23 (left to right). Optimum quantized results shown as black dashed lines. TBS index as $[X_2]$.

Table 1. The optimum quantization results (black dashed line) match that of the unquantized case. The investigation reveals that the optimum quantization results (black dashed line) match that of the unquantized case. Interestingly, the required number of bits can be much smaller than assumed in the O-RAN specifications, such as 4 bits for coded QPSK, though it varies with modulation schemes, increasing to 6 bits for 16-QAM. This suggests that an adaptive quantization scheme could significantly reduce fronthaul load. The analysis also highlights that the optimal quantization intervals depend primarily on the modulation scheme and the number of quantization bits, rather than the code rate. The BLER decreases rapidly as the quantization interval increases from excessively small values, reaching a relatively flat minimum where performance is stable before gradually increasing again as the interval becomes too large. It is noted that the current O-RAN fronthaul compression algorithm leads to significant variation in the effective quantization interval, which can degrade performance. However, since the default number of quantization bits is higher than assumed in the study, the performance impact is not severe. Finally, information-theoretic analysis shows that the reduction in mutual information caused by quantization can only partially explain the performance loss observed with practical 5GNR LDPC codes, as the loss is somewhat greater than predicted by mutual information reduction alone. The analysis also effectively determines the optimal quantization interval, though this finding is currently limited to binary cases, corresponding to QPSK modulation.

## 1.7 Security/Vulnerability

### 1.7.1 Overview of Wireless Security

**The Role of Security in Wireless Communications** Security in wireless communication is paramount due to the inherent vulnerabilities associated with radio frequency transmission. Unlike wired networks, wireless signals can be intercepted, eavesdropped, or tampered by malicious nodes without physical access to the network infrastructure. Therefore, implementing robust security methods is essential to safeguard sensitive data and protect wireless systems against various threats. In the realm of wireless security, safeguarding sensitive information and maintaining the integrity of communication channels are paramount objectives. To achieve these goals, security techniques often rely on the foundational principles of the CIAA quartet [SFA23]: Confidentiality, Integrity, Authenticity, and Availability, as shown in Figure 12. These pillars form the cornerstone of robust wireless security techniques, ensuring that data remains secure, accurate, accessible, and trustworthy.

Confidentiality stands as the first line of defense in wireless security, encompassing measures designed to prevent unauthorized access to sensitive information. By safeguarding data from interception or eavesdropping, confidentiality shields sensitive information from malicious attackers seeking to exploit vulnerabilities in wireless networks. On the other hand, integrity ensures the accuracy and trustworthiness of data transmitted over wireless channels. Tampering with data, whether through malicious manipulation or inadvertent corruption, can have far-reaching consequences, undermining the reliability and validity of information exchanges. To mitigate this risk, security functions and digital signatures are employed to verify the integrity of data packets, detecting any unauthorized alterations and preserving the integrity of the communication stream.

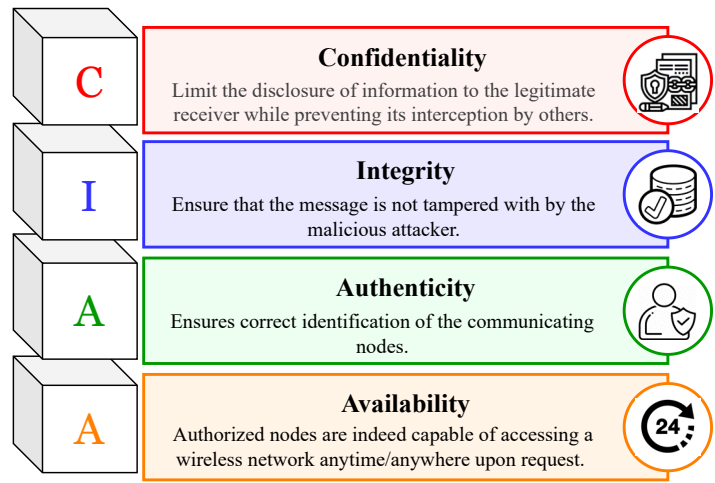

Figure 12: An illustration depicting the CIAA quartet as the foundational pillars of wireless security.

Authenticity, the third part of the CIAA quartet, plays a crucial role in wireless security by verifying the identities of users and devices seeking access to network resources. Strong authentication mechanisms validate user credentials and establish trust relationships, preventing unauthorized access and fortifying the perimeter of wireless networks. By ensuring that only authenticated entities are granted access to network resources, authentication enhances the overall security of wireless environments. On the contrary, availability emphasizes the importance of ensuring timely and uninterrupted access to network resources and services. Denial of service (DoS) attacks and network disruptions pose significant threats to the availability of wireless networks, potentially disrupting vital operations and impeding communication.

**Security in the context of 6G Networks** As the world eagerly anticipates the advent of 6G networks, the next generation of wireless communication technology promises unprecedented advancements in speed, capacity, and connectivity [ZXM+19]. A summary of 6G requirements is illustrated in Figure 13. However, amidst the excitement surrounding the potential benefits of 6G, it is essential to consider the paramount importance of security in shaping the future of wireless networks. The landscape of 6G networks is poised to revolutionize various industries, enabling ultra-fast data transmission, low-latency communication, and seamless connectivity for an array of emerging technologies, including AR, virtual reality (VR), and the IoT. Yet, with these transformative capabilities comes an expanded attack surface and heightened security risks, necessitating a proactive and robust approach to safeguarding network infrastructure and sensitive data.

One of the key security challenges in the context of 6G networks is the proliferation of connected devices and the exponential growth of data traffic. With billions of IoT devices expected to be interconnected within 6G ecosystems, the potential for security breaches and cyberattacks increases exponentially. Malicious attackers may exploit vulnerabilities in IoT devices to launch large-scale distributed DoS attacks, compromise network integrity, or exfiltrate sensitive information. To mitigate these risks, 6G networks must prioritize security by implementing advanced encryption protocols, authentication mechanisms, and intrusion detection systems.

Moreover, the integration of AI and ML technologies in 6G networks offers promising opportunities for enhancing security capabilities. AI-driven security solutions can analyze vast amounts of network data in real-time, identifying anomalous behavior patterns and potential security threats with greater accuracy and efficiency than conventional methods. ML algorithms can adapt and evolve in response to emerging cyber threats, continuously improving the resilience of 6G networks against evolving attack vectors. Furthermore, privacy considerations loom large in the era of 6G networks, as the proliferation of connected devices and ubiquitous data collection raise concerns about data privacy and user consent. Implementing privacy-enhancing technologies, such as differential privacy and homomorphic encryption, can enable data anonymization and secure computation while preserving individual privacy rights. Additionally, robust data governance frameworks and regulatory mechanisms are essential to ensure compliance with privacy regulations and protect user data from unauthorized access or misuse.

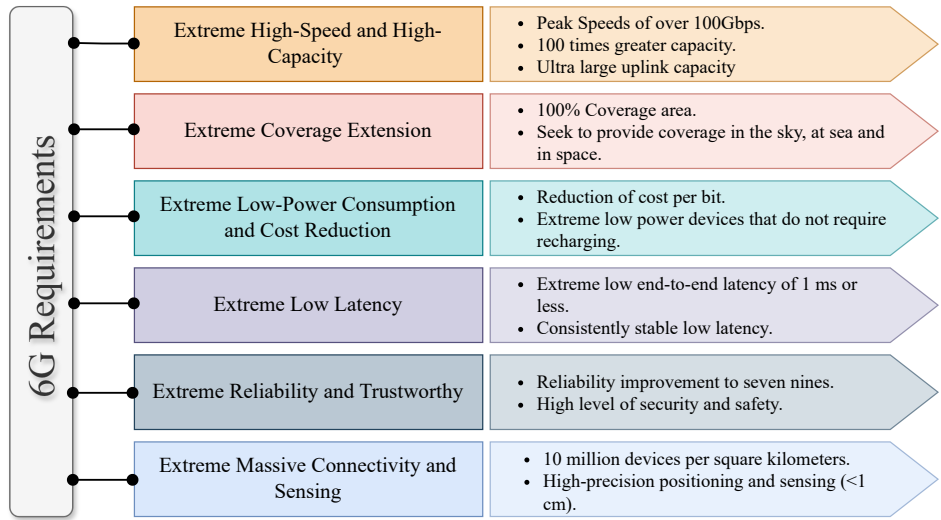

Figure 13: An overview of key requirements for 6G networks.

In conclusion, security is a critical enabler of the transformative potential of 6G networks, underpinning trust, reliability, and resilience in wireless communication ecosystems. By prioritizing security measures such as advanced encryption, authentication, intrusion detection, and privacy-enhancing technologies, stakeholders can navigate the evolving landscape of 6G networks with confidence, unlocking the full promise of next-generation wireless connectivity while safeguarding against emerging cyber threats and vulnerabilities.

**Overview of Wireless Threats** In the dynamic landscape of wireless communication, where data traverses through the airwaves, various threats lurk, poised to compromise the integrity, confidentiality, authenticity, and availability of wireless networks. Among the most prevalent and disruptive adversaries are eavesdroppers, spoofers, and jammers (as illustrated in Figure 14), each presenting unique challenges and security concerns to network operators and users alike.

**Eavesdropping** represents a pervasive threat in wireless communication, wherein malicious attackers intercept and surreptitiously monitor data transmissions between legitimate parties. By passively listening to wireless signals, eavesdroppers can glean sensitive information, such as login credentials, financial transactions, or proprietary business data, without the knowledge or consent of the communicating parties. This clandestine activity poses a significant risk to data confidentiality, potentially leading to unauthorized access, identity theft, or intellectual property theft. Conventionally, data encryption is the most commonly used technique for masking important and sensitive content, rendering them unintelligible to unauthorized interceptors.

**Spoofing**, on the other hand, involves the manipulation of wireless signals to deceive or impersonate legitimate entities within a wireless network. By spoofing MAC addresses, IP addresses, or APs, attackers can masquerade as trusted devices or networks, leading to unauthorized access, data interception, or intruder-in-the-middle attacks. Spoofing attacks can compromise network integrity and undermine the trustworthiness of communication channels, posing significant risks to data integrity and user privacy. To mitigate the threat of spoofing, robust authentication mechanisms are employed to verify the identities of communicating parties and prevent unauthorized access to network resources.

**Jamming** represents a disruptive threat to wireless communication, wherein malicious attackers transmit interference signals to disrupt or block legitimate communication channels. By flooding the wireless spectrum with noise or jamming signals, attackers can prevent authorized users from accessing network resources, rendering wireless networks unavailable or unusable. Jamming attacks can have far-reaching consequences, particularly in critical infrastructure sectors, such as transportation, healthcare, and public safety, where uninterrupted communication is essential for operational continuity and public safety. To mitigate the impact of jamming attacks, frequency hopping techniques, spread spectrum modulation, and adaptive modulation schemes can be employed to dynamically adapt to changing environmental conditions and mitigate the effects of interference caused by jamming.

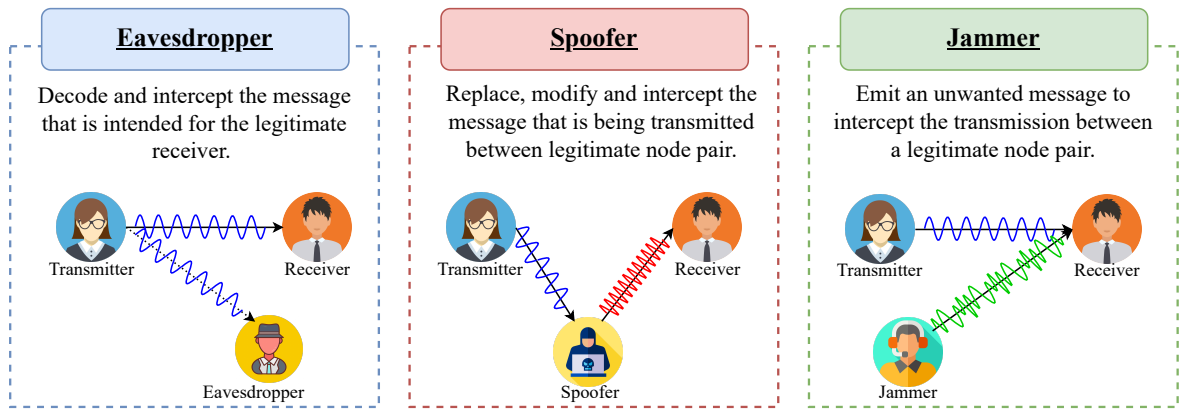

Figure 14: Schematic representation of the primary functions of wireless threats.

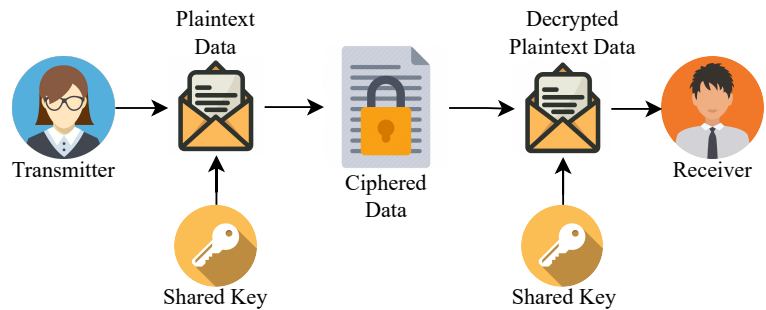

Figure 15: A schematic representation of symmetric cryptography.

In conclusion, eavesdroppers, spoofers, and jammers represent formidable adversaries in the realm of wireless communication, posing significant risks to data confidentiality, integrity, authenticity, and availability. By understanding the nature of these threats and implementing appropriate security measures, such as encryption, authentication, and interference mitigation techniques, organizations can protect their wireless networks against emerging threats and safeguard sensitive information from unauthorized access or misuse. Additionally, ongoing research and development efforts are essential to stay abreast of evolving threats and vulnerabilities, ensuring the resilience and reliability of wireless communication in an increasingly interconnected world.

### 1.7.2 Cryptographic Methods in Communication Security

**Fundamentals of Cryptography**   Cryptography serves as the cornerstone of communication security, employing various techniques to safeguard sensitive data from unauthorized access and manipulation. Symmetric cryptography, characterized by the use of a single shared key for both encryption and decryption, offers efficiency and speed in data processing. In symmetric encryption (as shown in Figure 15), algorithms like the advanced encryption system (AES) are commonly employed to encrypt plaintext data into ciphertext, ensuring confidentiality during transmission. Conversely, asymmetric cryptography utilizes a pair of public and private keys for encryption and decryption as illustrated in Figure 16, providing enhanced security through separate keys for encryption and decryption. Algorithms such as Rivest–Shamir–Adleman (RSA) and elliptic-curve cryptography (EcC) are prominent examples of asymmetric encryption techniques, facilitating secure key exchange and digital signatures to authenticate communication parties. Understanding the principles and applications of both symmetric and asymmetric cryptography is essential for designing robust communication systems that protect against eavesdropping and unauthorized access.

**Key Management and Distribution Strategies**   Key management and distribution are critical components of cryptographic systems, enabling secure communication between parties while mitigat-ing the risk of key compromise or interception. Effective key distribution mechanisms, such as key exchange protocols like Diffie-Hellman and public key infrastructure (PKI), facilitate secure sharing of cryptographic keys over insecure communication channels. Furthermore, key derivation functions and key agreement protocols play a vital role in generating session keys for temporary encryption purposes. Robust key storage mechanisms, including hardware security modules (HSMs), physical unclonable functions (PUFs), and secure key vaults safeguard cryptographic keys from unauthorized access or theft.

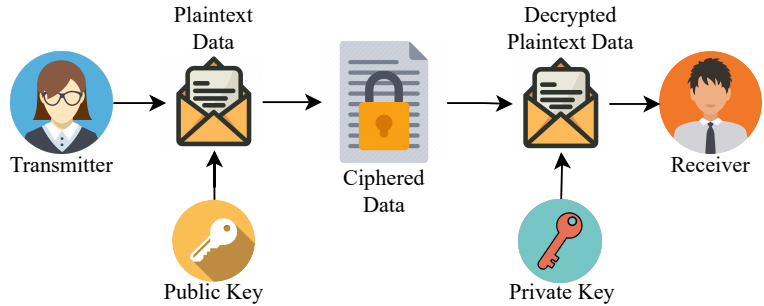

Figure 16: A schematic representation of asymmetric cryptography.

Moreover, key lifecycle management practices are essential for maintaining the security and integrity of cryptographic systems. Key rotation ensures that cryptographic keys are periodically updated to mitigate the risk of cryptographic attacks, while key revocation mechanisms allow compromised keys to be promptly invalidated. Key escrow solutions provide a means for securely storing cryptographic keys with trusted third parties, enabling key recovery in the event of key loss or unavailability. By employing robust key management and distribution strategies, organizations can enhance the security posture of their communication systems and mitigate the risk of unauthorized access or data breaches.

Additionally, advancements in cryptographic technologies, such as homomorphic encryption and post-quantum (i.e., quantum-resistant) cryptography, address key management and distribution challenges in modern communication systems. Homomorphic encryption enables computation on encrypted data without decrypting it, allowing for secure processing of sensitive information in untrusted environments. Quantum-resistant cryptography algorithms, designed to withstand attacks from quantum computers, offer enhanced security against emerging threats to traditional cryptographic systems. By leveraging these innovative cryptographic techniques, organizations can bolster the resilience of their communication systems against evolving cyber threats and attacks.

**Challenges and Limitations of Cryptography**  In the context of next-generation networks, several factors challenge the efficacy of traditional cryptographic security measures, which are summarized as follows:

- **Quantum Computing:** The computational complexity of breaking cryptographic systems based on integer factorization and discrete logarithms is undermined by the advent of quantum computing [MVZJ18].

- **Resource Constraints in IoT/mMTC:** Terminal devices in IoT and mMTC are often constrained in terms of power and computational resources, necessitating lightweight security mechanisms [QCZZ20, GS20].

- **High-Mobility Applications:** Applications like high-speed trains, V2X communications, and non-terrestrial networks (NTN) introduce continuously changing network topologies, requiring renewed key management and authentication procedures [FSHA19].

- **Latency Concerns in uRLLC:** For URLLC and eMBB applications, latency is a critical issue, potentially rendering conventional cryptographic methods impractical [LLH$^+$18].

In response to these challenges, there is a growing need for alternative approaches to traditional cryptography. physical layer security (PLS) emerges as a compelling candidate to address the evolving security requirements of next-generation networks.

### 1.7.3 Physical Layer Security in Wireless Communications

**Physical Layer Security: Definition and Domains** PLS is founded on information-theoretic principles and aims to establish provable security by leveraging inherent wireless channel characteristics, such as fading, interference, and multipath propagation [HFA18, B$^+$21]. These features are exploited to verify user identities and safeguard transmissions, ensuring that legitimate receivers obtain signals of higher quality than potential attackers. Furthermore, PLS techniques may utilize hardware and RF features of transceivers, as well as environmental factors like device proximity, to authenticate connections.

In the literature, a wide array of PLS approaches are extensively covered, with some focusing on confidentiality mechanisms while others explore anti-jamming strategies. These approaches include the categories: secret key extraction for data encryption and decryption, alignment of physical signals or transmissions using shared keys, and modification of physical signals or transmissions based on extracted keys. However, despite the breadth of existing methodologies, a unified framework that integrates these approaches and accommodates future PLS advancements is yet to be established. For instance, to address this gap, the authors in [SSA22] present a generalized PLS framework, expanding PLS scope from protecting data and communication to safeguarding the whole radio environment map. In order to comprehensively delineate the scope of PLS, it is pertinent to examine various domains within its purview. Each of these domains represents a distinct subset of PLS methods, which will be expounded upon in subsequent subsections. These domains can be succinctly summarized as encompassing the wireless channel, RF front-end, radio environment, data bits, wireless signal, and network. The following sections offer detailed discussions pertaining to each domain individually, thereby elucidating their significance and implications within the framework of PLS.

**Wireless Channel** Wireless signals, through propagation, interact with environmental elements, undergoing absorption, reflection, refraction, and diffraction. These phenomena, inherently stochastic and temporally variant, pose communication challenges in environments abundant with scattering due to limited coherence distance, bandwidth, and temporal characteristics. Nonetheless, from a PLS perspective, independent channel observations by legitimate and illegitimate nodes, maintaining a minimum separation of half-wavelengths, prove advantageous. The discussion in [KFSA23, KFA23] encompasses a range of wireless channel characteristics applicable to PLS. Despite the complexity, various mathematical models elucidate environmental impacts on wireless signals. Measures such as received signal strength indicator (RSSI), channel state information (CSI), channel impulse response (CIR), and channel frequency response (CFR) quantify channels, often represented as FIR filters for modeling. Reciprocal channels, with constant parameters, motivate the adoption of time-division duplexing (TDD) systems, enhancing reciprocity. Multipath propagation, resulting from variations in reflective and absorptive properties of environmental objects, contributes to distinct channel characteristics, with multipath components treated as random. Channel estimation, a primary advantage of wireless channels in PLS, enables nodes to mitigate environmental effects, facilitating tasks like key generation, link adaptation, and interference signal injection. Effective channel estimation reduces overhead in PLS approaches. Consequently, wireless channels find extensive usage in PLS applications, including channel-based key generation, link adaptation, and interference signal injection.

**RF Front-End** Apart from the wireless medium itself, the RF front-end is susceptible to various imperfections, including clock jitter, phase noise, carrier frequency offset (CFO), in-phase/quadrature imbalance (IQI), power amplifier non-linearity, and antenna imperfections [Ars21]. These impairments, varying among devices, serve as distinctive "fingerprints" for device differentiation [WSP$^+$16]. RF fingerprinting serves as a physical layer authentication mechanism, aimed at thwarting or detecting attacks on node identity or message integrity. Additionally, RF fingerprints complement channel-based authentication methods, as elucidated in [ZRS$^+$19], where devices are authenticated by their fingerprints while secure communication is enabled through channel-based key generation.

RF-based approaches exhibit stability in mobile environments, whereas channel-based techniques are more suited for indoor and relatively stationary environments where frequent authentication is unnecessary. In practical scenarios, the reliability of RF-based PLS is challenged by the variations in the fingerprints. For instance, a single impairment may fail to distinguish between devices due to a limited dynamic range. Solutions proposed in [HW15] advocate for combining multiple device characteristics in a weighted manner, while [WHH16] proposes a collaborative approach aggregating observations from multiple nodes. Both strategies aim to enhance reliability by assessing the significance of identified impairments. However, challenges arise when channel-related phenomena mimic hardware impairments, such as Doppler spread/shift induced by environmental mobility, akin to CFO caused by local oscillator imperfections. Distinguishing between these effects is crucial. One approach, as outlined in [HWCR14], involves recognizing that channel effects vary at a significantly slower rate than device impairments. Alternatively, [KPMA21] proposes integrating channel and RF impairments into a time-varying device fingerprint for authentication. The former utilizes imperfect or chaotic antenna geometries and activation sequences, while the latter employs beamspace representations of mutual coupling in a MIMO system.

**Radio Environment** The transmission of wireless signals through the air is influenced by surrounding objects and their properties. In rich scattering environments, independent determination of both the channel and environmental characteristics provides insight into the surrounding environment. Parameters such as distance, speed, angle, size of objects, and constituent materials serve as observable parameters utilized for authentication or security of wireless links [BEK+16]. Among the prominent environmental measurements for PLS is the distance or angle between communicating nodes. For instance, [BKEF+15] employs AoA-based key generation, utilizing azimuth, elevation, or both angles. This method is particularly advantageous in low SNR scenarios due to its lower mismatch rate compared to channel-based key generation. Another proposal, as outlined in [GCK14], involves generating keys based on nodes' relative locations, eliminating the need to share an entropy source between devices since relative location is a reciprocal quantity. Various approaches to distance calculation exist, including time difference of arrival (TDoA)-based and received signal strength (RSS)-based methods [PLK+11]. As demonstrated in earlier RF-based approaches, parameters obtained from the radio environment or sensing can be integrated with channel knowledge, as exemplified in [BKE+14]. It is noteworthy that sensing is not exclusively confined to the RF domain within the generic PLS framework. Moreover, external sensors such as cameras, LiDar, humidity sensors, and temperature sensors may be integrated into the system for enhanced environmental sensing capabilities.

**Data Bits** Conventional wireless systems commonly employ a bit-level security mechanism, where data security is achieved by encrypting plaintext or messages into ciphertext using cryptographic algorithms, as discussed in [Riv90, Section 2]. It is essential to differentiate between cryptography and key-based PLS, as both involve bit-level transformations. In cryptography, symmetric and asymmetric encryption utilize known or shared keys. Since this process, including key management, typically occurs at higher layers, it falls outside the scope of the PLS paradigm. PLS mechanisms develop key-generation techniques based on observable parameters associated with the wireless channel and radio environment surrounding communicating nodes [JWW+19]. Consequently, key exchange is unnecessary in PLS, as both transceivers observe the same channel and environment. Additionally, channel coding is utilized to enhance bit-level security in PLS. Although coding-based PLS mechanisms such as polar and LDPC codes exist [SV13], a significant limitation is that the eavesdropper's channel quality must be lower than the quality of the legitimate link to achieve non-zero rate secure communication [BBRM08, FHA20a]. Note that bit-level modifications primarily aim to thwart eavesdropping, yet they do not address issues such as jamming and spoofing, which represent notable limitations of standard cryptographic solutions.

**Wireless Signal** The primary focus of PLS research centers around the wireless signal domain. This domain encompasses all components from the transceivers' antennas to the coded bit stream. The overarching objective of security solutions within this domain is to enhance data decoding capabilities for legitimate receivers while deterring malicious attackers. This may involve intentionally degrading the performance of eavesdroppers or improving the QoS for legitimate recipients. To counter eavesdropping, numerous PLS signal modification techniques have been developed, which will be elaborated below. Signal parameter modification can also prove more effective in mitigating jamming compared to alternative strategies. A subset of anti-jamming techniques will be examined below. However, it is worth noting that while signal modification is a viable approach to counter eavesdropping and jamming, it may not always be the optimal or most widely adopted method for protecting against spoofing.

**Network** In the context of network-based PLS, the term "network" denotes a collection of nodes within the environment. Examples include relays employed in cooperative communication, BS integrated into coordinated multipoint architectures [SSK+21], and RISs utilized in smart radio systems. Within network-based PLS, macro-diversity strategies are employed to bolster user communication reliability while simultaneously diminishing the attacker's effectiveness. Despite requiring minimal implementation effort, these techniques serve to mitigate eavesdropping and jamming attacks.

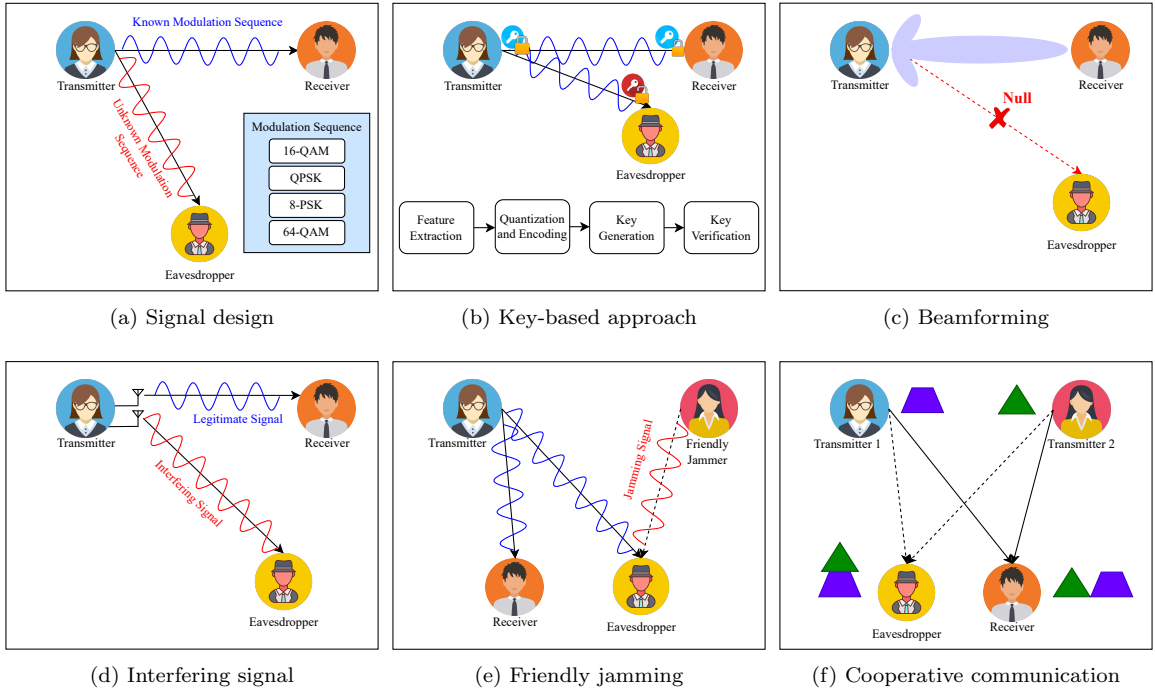

Figure 17: Basic illustration for anti-jamming based PLS approaches: (a) signal design, (b) key-based approach, (c) beamforming, (d) interfering signal, (e) friendly jamming, and (f) cooperative communication.

**Anti-Eavesdropping Strategies** The primary objective of anti-eavesdropping schemes is to achieve perfect secrecy, ensuring that legitimate transmitters convey sensitive information with positive secure communication rates to intended receivers while preventing eavesdroppers from accessing any transmitted information [Wyn75]. Note that the term "perfect secrecy" can have different meanings in the literature, and here we consider the most common definition for which the secrecy leakage rate tends to zero when the blocklength tends to infinity. These approaches encompass various strategies, including signal design, key generation, beamforming, interfering signal manipulation, and cooperative communication. Figure 17 presents a basic representation of the aforementioned techniques. Signal design methodologies entail modifications to the physical signal structure, encompassing waveform design, modulation and coding schemes, and frame structures. For instance, the approach proposed in [XRD+17] introduces a constellation selection method to suppress eavesdropping, employing diverse modulation modes and constellation structures to enhance PLS. To securely share constellation information between legitimate entities, QR decomposition of the channel is applied. Furthermore, secret keys extracted from observable parameters address key management challenges inherent in cryptography-based methods, thereby ensuring bit-level data security [FHA20b].

Beamforming represents another avenue for securing communication channels. For instance, the method proposed in [MS10] introduces a beamforming-based PLS approach that amplifies signal power at the legitimate receiver while suppressing it in other directions. Additionally, incorporating artificially generated noise signals aims to degrade eavesdropper performance. In the depicted scenario, the legitimate transmitter with multiple antennas utilizes the beam's null space to introduce interference signals, commonly known as artificial noise [GN08]. Furthermore, friendly jamming involves employing external nodes to generate interference signals, termed jamming signals [DHPP09]. Cooperative communication strategies have also been leveraged for PLS. For example, the approach proposed in [WW16] introduces CoMP transmission to ensure that signal components transmitted from different sources collide at the eavesdropper while remaining collision-free at the intended receiver, as depicted in Figure 17f.

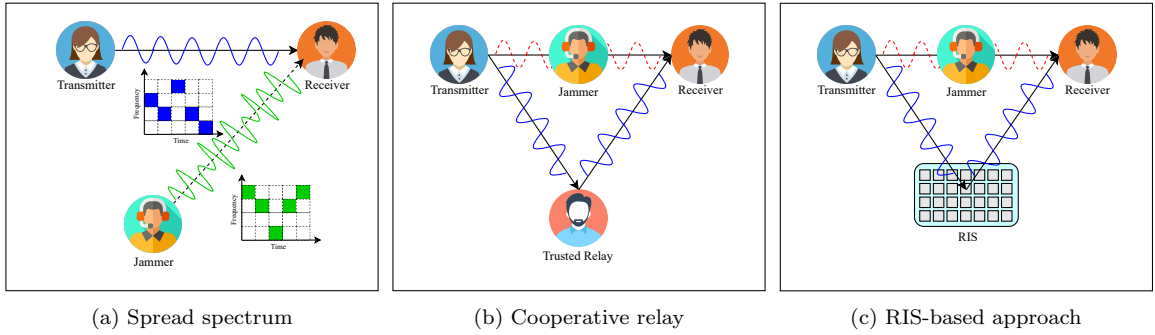

(a) Spread spectrum  (b) Cooperative relay  (c) RIS-based approach

Figure 18: Illustration of anti-jamming techniques: (a) spread spectrum, (b) cooperative relay, and (c) RIS-based approach.

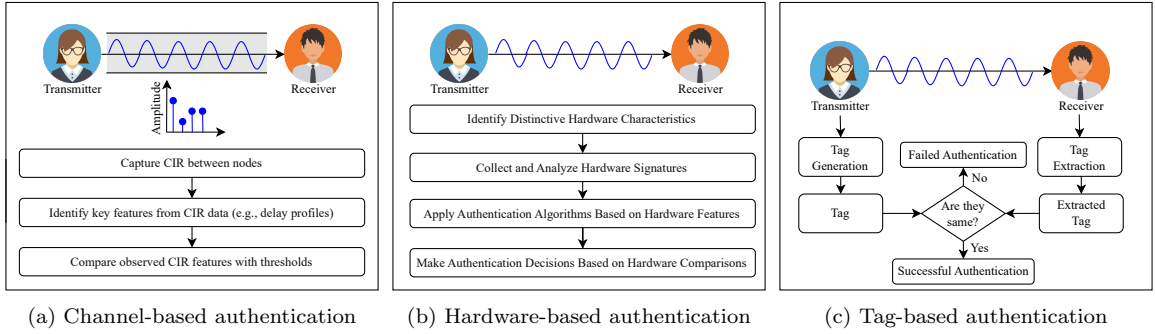

(a) Channel-based authentication  (b) Hardware-based authentication  (c) Tag-based authentication

Figure 19: Illustration of existing physical layer authentication techniques: (a) channel-based authentication, (b) hardware-based authentication, and (c) tag-based authentication.

**Anti-Jamming Techniques**  Various approaches have been proposed to mitigate jamming attacks, including spread spectrum techniques, cooperative relaying, and multi-antenna schemes. Spread spectrum techniques are widely used to counter jamming attacks by dispersing the signal's energy across specific domains such as time and frequency. In frequency-hopping spread spectrum (FHSS), the carrier frequency is periodically changed to evade jamming attacks, as depicted in Figure 18a. These frequency changes are based on identical frequency hopping sequences, which can be pre-shared [HLL07] or channel-based [WZL+18]. However, the effectiveness of FHSS may diminish if smart jammers target multiple channels simultaneously, necessitating additional spectrum resources to evade jamming.

Cooperative communication involving trusted relays represents a straightforward anti-jamming solution, providing alternative signal propagation paths between legitimate nodes to mitigate disruptor effects, as depicted in Figure 18b. Relay selection and beamforming design are jointly optimized to ensure secure and reliable transmission [GHX+20]. More recently, RISs have emerged as anti-jamming solutions due to their capability to adjust channel parameters [ATAK+20]. RIS enhances PLS by manipulating jamming signal reflections to weaken their impact at legitimate nodes, as illustrated in Figure 18c.

**Anti-Spoofing Strategies**  Authentication serves as the mechanism to verify the legitimacy of a communication request originating from a particular node. Common types of PLS-based authentication encompass channel-based authentication, hardware-based authentication, and tag-based authentication. Channel-based authentication involves utilizing RSS and CSI to detect identity-based attacks in wireless networks [CYTM10, MMV+11]. Figure 19a depicts the fundamental procedures involved in channel-based authentication. In [LWL+17], a ML-based authentication technique was proposed to discern between nodes, even when they possess similar signal fingerprints, using CSI measurements.

Hardware-based authentication relies on extracting physical layer characteristics unique to the legitimate transmitter. In Figure 19b, the schematic diagram delineates the foundational steps inherent to channel-based authentication. Hardware-based authentication leverages inherent variations introduced during the fabrication of analog components, such as IQI and CFO [HWB14, GS20, HWCR14]. However, the subtle differences in selected hardware attributes are often obscured by noise and interference, diminishing the accuracy of attribute estimation for authentication purposes [WHH16]. Tag-based authentication methods entail embedding a secret modulated signal into the transmitted signal, acting as a signal watermark, as depicted in Figure 19c. In [XZ18], a blind tag-based authentication scheme is introduced, employing blind known-interference cancellation and differential processing techniques for authentication purposes.

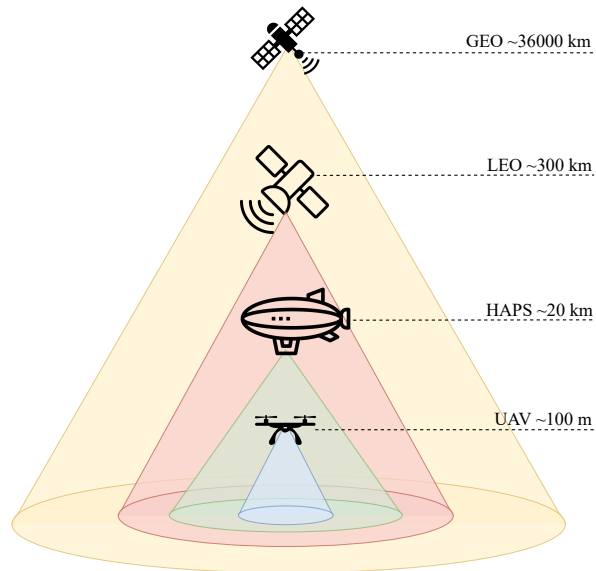

Figure 20: An illustration of different NTN deployments.

#### 1.7.4 Emerging Trends in Security for Communication Systems

**Security for Non-Terrestrial Network (NTN)** In response to the escalating demand for ubiquitous connectivity, academia, industry, and standardization bodies have intensified their focus on NTNs as a potential solution. Figure 20 illustrates NTN deployments such as satellites, high-altitude platform systems (HAPS), and unmanned aerial vehicles (UAVs), highlighting the differing attributes of height and coverage among these deployment options. NTNs aim to extend coverage to areas beyond the reach of conventional terrestrial networks. This expansion includes remote regions, marine vessels, and airplanes, addressing the imperative for connectivity regardless of location [RMT+20]. However, ensuring secure communication within NTNs presents unique challenges distinct from terrestrial networks. NTNs, particularly satellites, operate predominantly in LOS scenarios, leveraging reduced propagation losses to extend coverage [SDBA19]. Nonetheless, this expanded coverage also amplifies vulnerability to eavesdropping attacks. Addressing this concern, researchers propose leveraging the polarization domain to obfuscate communication signals from eavesdroppers. For instance, a dual-polarized antenna system has been developed for fixed downlink satellite communication, enabling legitimate receivers to extract polarization information while thwarting eavesdroppers.

Yet, devising robust security mechanisms for NTNs becomes more complex in hybrid networks, which integrate both terrestrial and non-terrestrial components. Studies have explored this scenario, focusing on relay selection and user scheduling strategies to bolster communication confidentiality. Additionally, analyses have been conducted to evaluate secrecy performance in multi-antenna NTNs to terrestrial recipient links, employing multiple cooperative relays to counteract eavesdropping threats effectively. In essence, as NTNs emerge as key enablers of extended connectivity, addressing the intricacies of securing these networks remains imperative. Researchers continue to innovate and devise sophisticated solutions to safeguard communication integrity and confidentiality, paving the way for a seamless and secure wireless future.

**Security for Vehicle-to-Everything (V2X) Communication**    V2X communication encapsulates a spectrum of interactions within the vehicular domain, ranging from vehicle-to-vehicle (V2V), vehicle-to-infrastructure (V2I), vehicle-to-pedestrians (V2P), to vehicle-to-network (V2N) communication. The overarching objective of V2X communication is to enhance road safety through the seamless exchange of data among vehicles and end devices. Central to this endeavor is the imperative for reliable and timely connectivity, underscored by the criticality of message confidentiality and security. However, the nature of V2X communication renders it susceptible to interception, potentially compromising the privacy of sensitive information such as user identity, position, and trajectory. PLS has emerged as a prospective solution to address these vulnerabilities, offering mechanisms to safeguard data exchange and protect against unauthorized access [EEBW19].

Nevertheless, a significant challenge lies in accommodating the diverse security requirements inherent to various V2X applications, which are contingent upon factors such as application type, geographic location, utility, environmental conditions, and contextual user information. As highlighted in [LLCG20], a one-size-fits-all security approach is impractical, necessitating the implementation of different PLS techniques tailored to specific V2X applications and scenarios. In pursuit of an effective security framework for V2X communication, collaborative efforts have been proposed to integrate multiple PLS techniques in a synergistic manner. While conceptual frameworks, such as those delineated in existing literature, offer initial insights into potential methodologies for cooperative PLS deployment in V2X environments, further empirical studies and feasibility analyses are required to ascertain the practical viability and efficacy of such approaches.

**Security for Internet of Things (IoT)**    The advent of IoT technology has revolutionized the capabilities of physical objects, endowing them with the ability to sense, communicate, and execute specific tasks on demand. This transformative potential has spurred the development of a diverse array of applications spanning domains such as smart homes, smart cities, and intelligent transportation systems (ITSs). As IoT technology becomes increasingly integrated into our daily lives, ensuring the security of IoT networks emerges as a critical imperative.

In addressing the security challenges inherent to IoT networks, PLS techniques offer a promising avenue for enhancement, focusing on three primary facets. Firstly, the dynamic nature of IoT devices, characterized by rapid mobility and frequent handoffs between APs or BSs, can lead to latency issues that exceed the tolerance of next-generation scenarios and applications [Zen15]. PLS techniques, such as RF fingerprinting, streamline the authentication process by providing a direct method of identification, thereby mitigating authentication delays. Secondly, the pervasive deployment of IoT devices across diverse sectors presents challenges in efficiently distributing and managing secret keys for secure communication [YWY+17]. PLS methodologies present an innovative solution wherein communicating nodes extract keys directly from their environment or channel, eliminating the need for centralized key distribution and management mechanisms. Thirdly, the resource-constrained nature of IoT devices necessitates lightweight security solutions that minimize processing overhead [WWAF+19, GK18]. While certain PLS methods, such as beamforming [CNGC16], noise aggregation [SD18], cooperative jamming [MFHS14], and artificial noise injection [GN08], may require additional hardware and energy resources, asymmetric PLS mechanisms (e.g., [SA23]) alleviate the processing burden on IoT nodes. In these asymmetric approaches, the responsibility for designing secure transmissions is shifted to the BS or AP side, eliminating the need for additional processing at the IoT node itself.

However, it is pertinent to note that many existing PLS mechanisms do not explicitly consider energy efficiency or attempt to optimize it alongside security performance. Addressing this concern, recent research endeavors have proposed user association strategies aimed at maximizing secure throughput while minimizing energy consumption in ultra-dense network environments. Moving forward, it is anticipated that asymmetric, lightweight, and low-power PLS mechanisms will witness increased adoption in next-generation networks, contingent upon their validation and performance assessment in real-world scenarios.

**Security for Integrated Sensing and Communication (ISAC)**    Sensing (radar) and communication have evolved as distinct wireless technologies, each with its own trajectory of development spanning several decades [LMP+20]. However, the proliferation of devices and applications, coupled with their growing interdependence, has prompted a shift towards ISAC in recent years. This transition is primarily motivated by challenges such as spectrum scarcity, power constraints, and shared hardware resources. Nonetheless, the amalgamation of sensing and communication functionalities in a joint design introduces certain trade-offs, potentially leading to performance degradation compared to conventional standalone systems [FSTA21]. Of particular concern in the context of sensing is the inherent reliance on wireless transmissions and reflections to acquire information about targets, raising significant apprehensions regarding user privacy and susceptibility to malicious nodes within the vicinity [GBSY23, WLM+22, WGY24, GBSY24]. To address these challenges, scholars have proposed frameworks for enhancing the security of ISAC and radio environment awareness. Additionally, investigations have explored the applicability of existing PLS methods to fortify sensing processes against potential threats.

Attacks on sensing systems can be in various forms, targeting the sensing process itself, individual nodes involved in radio environment awareness, or the physical-radio environment. Process-oriented attacks aim to manipulate the wireless sensing process, with defensive measures such as low probability of intercept (LPI) techniques [Law10] and randomized probing [SMY+15] employed to counter spoofing attempts. Node-oriented attacks focus on compromising nodes participating in radio environment awareness, potentially exposing sensitive information such as node identity, data, velocity, location, and RF characteristics [FTA12]. Environment-oriented attacks encompass alterations to the physical-radio environment, including changes to LOS and non-line-of-sight (NLOS) characteristics, channel properties, urban/rural categorization, mobility patterns, and interference levels. For instance, RIS can be exploited by attackers to manipulate multipath channels or distort coverage areas [FSTA21]. Despite the considerable attention devoted to the design and optimization of ISAC systems, there remains a conspicuous gap in the literature concerning security provisioning. Addressing this gap necessitates comprehensive research efforts aimed at developing robust security mechanisms tailored to the unique challenges posed by ISAC environments.

**Applications of Artificial Intelligence/Machine Learning (AI/ML) in Security** The escalating intricacy of wireless systems, characterized by diverse waveforms, propagation environments, and resource allocation mechanisms has compelled network operators and architects to embrace AI and ML techniques to enhance network performance. Instances of this integration include the utilization of DL algorithms within the citizens broadband radio service (CBRS) band for incumbent user detection and the optimization of network slicing methodologies to augment network resource utilization efficiency. However, these applications have unveiled vulnerabilities to adversarial ML, wherein malicious entities exploit the opaque nature of AI models to manipulate the learning process, resulting in decisions detrimental to legitimate users or network operations [SES21]. Beyond preemptive training strategies aimed at mitigating potential adversarial attacks [PMW+16, CG17], an alternative approach in [Guo20] involves the adoption of explainable artificial intelligence (XAI) techniques to bolster user trust in AI-driven systems. XAI endeavors to elucidate the decision-making rationale of AI models by furnishing interpretable explanations, thereby enhancing users' comprehension and confidence in system outputs [DSB17].

Nonetheless, a fundamental trade-off emerges between explainability and performance, particularly in scenarios devoid of explicit mathematical models. Furthermore, as articulated by scholarly discourse, the interpretability of AI/ML outputs remains contingent upon users' domain-specific knowledge, necessitating tailored approaches to XAI implementation. Consequently, it becomes imperative to design and tailor AI/ML models with comprehensive insights into the wireless communication domain, enabling the development of transparent and trustworthy systems that align with user expectations and operational requirements.

### 1.7.5 Open Issues and Challenges in Physical Layer Security

While substantial progress has been made in the domain of PLS, numerous challenges persist, hindering its practical implementation. This section outlines several open issues and challenges warranting attention in future research endeavors:

- **Assumptions Regarding Attacking Nodes:** A significant obstacle to realizing PLS in practice stems from the oversimplified assumptions made about attacking nodes. These assumptions often overlook factors such as the attackers' processing capabilities, number of antennas, and attack types (e.g., active/passive, individual/collaborative). This simplistic characterization impedes the development of proof-of-concept prototypes and the widespread adoption of PLS approaches in real-world security applications. It is crucial to consider the sophistication of malicious attackers, who can intelligently switch between different attack types based on link quality [VNJK18].

- **Reciprocity Challenges:** Reciprocity, a fundamental principle in PLS, poses challenges in practical scenarios. While reciprocity is assumed in theoretical frameworks, achieving it can be challenging due to factors such as noise, interference, and inconsistencies in channel estimation [ZZCZ20]. In scenarios where observable parameters differ at communicating ends, reciprocal compensation mechanisms are necessary, raising concerns about potential information disclosure.

- **Mobility Challenges in 6G and Beyond:** The advent of 6G and beyond networks promises ubiquitous connectivity, including scenarios with high-speed mobility. However, mobility introduces challenges such as Doppler spreading, channel selectivity, and low coherence time, exacerbating issues related to PLS. For instance, shorter coherence durations pose authentication challenges, necessitating the re-evaluation or redesign of authentication procedures to accommodate high-speed mobility scenarios effectively [WF16].

- **Need for New Security Metrics:** With the emergence of cognitive/adaptive PLS and PLS for ISAC, there is a pressing need to develop new metrics to quantify security in next-generation wireless networks comprehensively. While link-level metrics for communication security are well-studied, there is a dearth of research on quantifying location-based security. Novel metrics, such as a "secrecy map" as extensions of early works such as [SGE$^+$21], could provide insights into average secrecy capacity across the entire environment, facilitating a holistic assessment of security in sensing and communication applications.

- **Practical Implementations of PLS Approaches:** Despite theoretical advancements, practical implementations of PLS approaches remain limited. Few works have ventured into the practical validation of PLS techniques, with complex testbeds and replication challenges hindering widespread adoption. Efforts to bridge this gap include proof-of-concept implementations leveraging off-the-shelf hardware and innovative concepts such as spectrum programming. However, further research is needed to develop practical and scalable implementations of PLS techniques to protect legitimate communication against evolving threats effectively.

## 1.8 Underwater communications

As 5G is supposed to be an ubiquitous radiocommunication system that will enable to connected to everyone and everywhere the underwater environments seem to be forgotten. The reason for this might come from the knowledge that electromagnetic waves are very strongly attenuated in the water. The penetration depth of see water is the lowest among those presented in the ITU-R Recommendation P.527-6 [P.521]. However, it is much higher for pure water in frequency range lower than 400 MHz. Despite the fact that the most needed is the possibility of communication in the salt water (see and oceans), the other disadvantage in pure water is the size of antenna which should correspond to the wavelength. The lower the frequency, the longer the wavelength and the bigger antenna is needed. Another possible transmission medium under the water is an acoustic wave. It enables transmission also in the salt water. However, physical phenomena occurring during acoustic signal underwater propagation differ to those appearing during electromagnetic wave propagation in the air. Some of them must be taken into account when adapting the radiocommunication solutions to the underwater hydroacoustic communication systems.

In the first place the system user must be aware of the stratification of the environment. The hydroacoustic wave does not pass through water-air boarder. This is also a huge problem when the water is aerated. But the stratification leads also to the deflection of the wave. The stratification comes from the different pressure on different depths, change in water temperature and also different chemical composition. It leads to change the propagation speed in distance and depth. In Fig. 21 is presented a sound speed profile measured in Baltic See near Klaipeda [BD21]. In Fig. 22 is presented plot of rays generated by Bellhop tool [bel]. The red dot represents the place where the transmitter is situated and the blue dot is the receiver position. It can be seen that the direct path will never reach the receiver - only the reflected paths. This leads to the occurrence of multipath effect in most scenarios in shallow waters (Baltic See may be considered as shallow). It must be also noted that the speed profile significantly changes during the year.

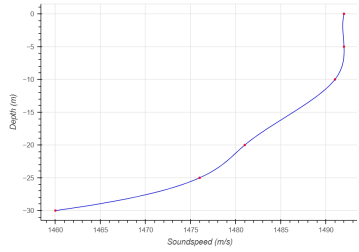

Figure 21: The sound speed profile plotted with Bellhop [bel] tool according to data from Baltic Sea measured in 2019.08.21 given in [BD21].

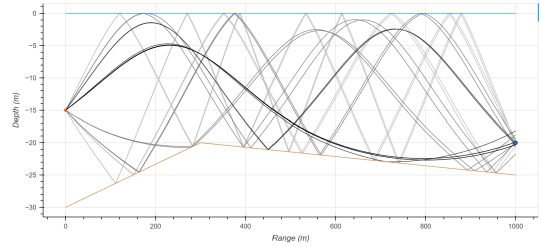

Figure 22: Propagation paths generated by Bellhop tool for the sound speed profile from Fig. 21

Another important phenomenon is the Doppler effect which is much more severe for acoustic wave than millimeter wave as the speed of sound in the water usually does not exceed 1500 m/s, what is 105 times less than speed of electromagnetic wave in the air. The severity of the Doppler effect reaches the level of the one observed for THz waves. The aspects of acoustic wave propagation under the water will also be addressed in COST Action CA20120. The contributions to date have been about realization of underwater data transmission in NLOS conditions TD(22)01082 [MJ21] and the analysis of impulse response measured in motion in a towing tank TD(23)04010 [A.22].

## 2 Signal Processing for Localization

This section focuses on signal processing techniques used for localization that are not directly aiming at their joint combination with communications. Techniques specific to ISAC are discussed in the next section.

Establishing highly accurate, robust, and reliable position determination is a topic that is attaining broad interest by the research community. There are several scenarios and applications where current satellite-based positioning systems are not available or suitable, e.g. assisted living, robot navigation, and asset tracking.

An example of a challenging application is outdoor localization in IoT asset tracking or search and rescue scenarios. Due to energy constraints or unavailability, global positioning system (GPS) cannot be used and an approximate localization (in the order of 50 meters) must be obtained with existing IoT networks or other low power solutions.

Another important application is gaming and virtual reality, where a very accurate localization (few cms) is required. GPS cannot offer the indoor coverage, so the service must rely on a particular infrastructure that can be deployed in a determined and controlled scenario. Here techniques such as UWB can be used and a map of the environment (fingerprinting) can be obtained at the service deployment stage.

Therefore, improving performance and resource-efficiency for reliable and highly accurate indoor radio positioning in challenging environments is a problem of great interest. This applies especially if the number of anchors is scaled up in a large-scale deployment, and if harsh radio channel conditions apply, e.g., in a factory with dense multipath propagation due to the presence of metallic objects.

In this section we start by presenting localization algorithms based on signal strength, direction/time of arrival, fingerprinting, among others. We then explain ways to achieve localization using several types of communication signals, including UWB or long range (LoRa), and also the use of RIS. Advances in waveform design for ranging and localization are discussed and finally some recent testbeds and demonstrations.

### 2.1 Localization techniques

This sub-section will discuss several localization algorithms, such as those based on AoA and fingerprinting, also using ML.

The implementation of round trip time (RTT) estimation methods is mainly motivated for localization purposes. Similar to conventional narrowband systems, UWB localization can be based on geometric methods where the localization of a target node from a network is derived from a channel parameter such as direction of arrival (DoA), RSS, time of arrival (ToA) and TDoA. In the first case, the position solution is obtained by means of triangulation of DoA estimates from different anchor nodes or stations. RSS and ToA provide range information that enables the application of trilateration techniques like in global navigation satellite system (GNSS), while TDoA leads to hyperbolic localization. In practice, DoA and RSS are not the preferred options for UWB localization. In the first case, DoA requires the use of antenna arrays that can be unaffordable to UWB-based wireless sensor networks. In the second case, an RSS-based approach does not benefit from the huge bandwidth capabilities of UWB systems. Therefore, time-based localization is the most widely adopted solution and the measurement employed (ToA/TDoA) is the base for RTT estimation. An alternative to geometric localization is known as fingerprinting. This approach consists in estimating the position of the target by finding the best matched pattern or fingerprint for a given measurement within a map that has been previously generated. This method is a good option in harsh environments, where the geometric relationship between the estimates and the position of the target cannot be established, such as under strong multipath or NLOS conditions. However, the price paid is a large amount of computation time and storage for generating the fingerprint map, which in practice limits the applicability of this technique to small-size indoor environments.

### 2.1.1 AoA localization

Here we discuss AoA localization with a single node combined with RTT/ToA/TDoA, and the importance of antenna error and RF error calibration in AoA estimation.

**Non-parametric methods**

Conventional methods for estimating the direction of arrival are based on beamforming and null-steering concepts and do not utilize the received signal vector model or the statistical model of the signals and noise. Rather than focus on a given set of parameters, the estimation is based on the computation of the full spatial spectrum, and for this reason, they are usually categorized as non-parametric methods.

Classical beamforming consists of electrically steering beams in all directions to look for peaks in the output power, which is done by scanning over the product of the beamforming vector and the covariance matrix of the array samples to find the maximum values [Bar50]. It is still widely used in practice as a very simple and robust approach. However, it suffers from the Rayleigh resolution limit, where improving the SNR has no impact and the only option is to increase the number of antenna elements in the array.

A more advanced alternative is the minimum variance distortionless response (MVDR) or Capon's method [Cap69], which consists of minimizing the variance due to noise while keeping the gain in the steering direction equal to the unity. By doing so, the resolution is not constrained by the number of sensors, and a reduced ripple in the sidelobes is obtained. However, the price paid is an increase of the computational cost (an estimation of the inverse of the covariance matrix is required) and certain limitations to perform when multiple coherent signals are present properly. Nevertheless, when compared against the more sophisticated approaches described next, its moderate complexity and its robustness against model errors, make this method a common choice in practical applications.

**Maximum-Likelihood methods**

Maximum-likelihood methods come from statistical estimation theory, where the problem consists of optimizing the likelihood function of the received signal model from the perspective of a set of parameters to be estimated, which include, at least, the direction of arrival of all signal components. This kind of approach provides the best asymptotic performance and remains stable in scenarios with limited numbers of snapshots, low SNR, and multiple coherent signals. The standard method involves a K-dimensional search, where K is the number of received signals, thus notably increasing the computational complexity of the whole process. In addition, it can be rather sensitive to the selection of the initialization parameters, given that the solution can converge to a local maximum instead of a global one.

In order to achieve a faster convergence in the optimization problem, different alternatives have been developed. The expectation-maximization (EM) algorithm replaces the original likelihood function with simpler maximization steps based on an augmentation scheme [FW88]. The space alternating generalized EM (SAGE) algorithm is a variation of the EM where the subsets of parameters are sequentially updated in one iteration [FH94b], which has been proven to achieve even faster convergence in the context of angle of arrival estimation [CB01]. In both cases, the complicated multidimensional search involved in maximizing likelihood functions can be simplified to a one-dimensional search. Further implementations can take advantage of this alleviation in the overall complexity by including array perturbation parameters estimation in the presence of mutual coupling and sensor location error [XLJ$^+$14].

**High-resolution subspace methods**

Subspace methods consist of separating the eigenspace of the covariance matrix of the array into signal and noise components to exploit this information for angle of arrival estimation. After the singular value decomposition of the covariance matrix, the multiple signal classification (MUSIC) method [Sch82] builds the spatial spectrum by combining the singular vectors associated with the noise subspace with the beamforming vectors, thus limiting the maximum number of unknown directions to the number of elements in the array minus one. The size of the array also determines the computational complexity of the algorithm. The spectrum obtained has sharp peaks at the estimated source locations but has no direct relation to actual power levels. In the case of uniform linear array (ULA), this scanning process can be transformed into solving the roots of a corresponding polynomial (method referred to as root-MUSIC [RH89]). Despite providing high-resolution consistent estimations, MUSIC is rather sensitive to array calibration errors and to the presence of coherent signals (although there are some studies that propose solutions in this respect [GCPW14, DK20]).

Estimation of signal parameters via rotational invariant techniques (ESPRIT) is an algebraic subspace method that does not require a search procedure [RK89b]. It reduces computational complexity by imposing a structural constraint on the array. The ESPRIT algorithm assumes that the antenna array consists of two identical subarrays, which may overlap. The individual elements of each subarray can have arbitrary polarization, directional gain, and phase response, as long as each element has an identical counterpart in the other subarray. The pairs of identical sensors, or doublets, are assumed to be separated by a fixed displacement vector. Consequently, the array possesses translational invariance, meaning array elements occur in matched pairs with identical displacement vectors. This property leads to the rotational invariance of the signal subspaces spanned by the data vectors associated with the spatially displaced subarrays, which ESPRIT then utilizes to determine the direction of arrival of the impinging signals.

Finally, both MUSIC and ESPRIT can be expanded in the 2-dimensional (2-D) space to estimate the multipath propagation delay as well as the angle of arrival by exploiting known characteristics of the received signals [VVdVP98, Ozi05, LCX$^+$20].

**Sparse signal reconstruction methods**

The concept of sparse signal reconstruction comes from the compressive sensing theoretical framework, where a discrete signal is considered sparse if it has a small number of nonzero components in certain representation domains. Therefore, the problem of recovering the entire signal can be seen as the problem of finding the sparsest representation in certain transform bases from the available set of measurements. This concept suits well in the context of angle of arrival estimation, where an overcomplete matrix of steering vectors can be defined and evaluated for a given scenario. This methodology maintains a spatial sparsity that remains unchanged, regardless of signal correlations.

Three main categories can be defined in this field [SSS$^+$19]. The first one belongs to the family of greedy algorithms, such as the matching pursuit [MZ93] and its different versions [PRK93, NT10], where the transform basis function is iteratively selected by evaluating the total signal energy in each round of the process. The second category relies on convex optimization, where linear programming and regression methods can be used to find a solution. Examples of this category are the $\ell$1-singular value matrix decomposition [MCW05], the basis pursuit method [CDS01], and iterative methods such as focal under-determined system solution (FOCUSS) [GR97], and the iterative adaptive approach (IAA) based on weighted least squares (WLS) [YLS$^+$10]. Finally, the last category is Bayesian methods based on probabilistic modeling [WR07, GMXN16]. A cost function containing the sparse vector is then derived to optimize the posterior information of the sparse vector. The critical factor of these methods lies in the incorporation of empirical priors, which depend on a set of hyperparameters.

**Calibration for Accurate AoA Estimation**

Additionally, the accuracy of AoA estimation is closely related to hardware impairments. Accurate AoA estimation in 5G systems necessitates phase coherence among multiple antennas [XFLS+23]. However, in practical RF front-end hardware, phase coherence is compromised due to factors such as clock drift, manufacturing inaccuracies, temperature variations, and radio conditions. Additionally, antenna-specific issues like gain-phase inconsistencies, mutual coupling, and manufacturing imperfections further exacerbate phase errors. Signal processing algorithms for AoA estimation typically assume an ideal antenna array, but deviations from this ideal model degrade performance [PZS+22, EEB16]. To mitigate phase and gain uncertainties, a calibration process is essential. Calibration involves determining the antenna response, typically conducted in controlled environments such as dedicated measurement chambers, which eliminate multipath propagation and interference. However, this method is costly. An alternative is In-Situ calibration, which uses transmitters with known directions to calibrate antennas under actual operating conditions. Although more practical and cost-effective, this method is limited to specific error models and does not account for propagation conditions [GBE+03, WWX+19].

Experiments conducted to evaluate calibration methods [XFLSSG23] showed significant improvement in AoA estimation accuracy. Theoretical and experimental array responses were compared, and phase errors were estimated and corrected. Despite multipath and interference conditions, the computed steering vector closely followed the theoretical model. Interference was mitigated by excluding affected subcarriers. In severe multipath conditions, the propagation delay of reference signals at each subcarrier introduces a phase shift, complicating AoA estimation. To address this, interpolation within the array modeling error data is performed. Results indicated a substantial improvement in AoA accuracy, with a maximum improvement of 46.74% after correcting RF channel errors. Further corrections for antenna errors yielded an additional 25.9% improvement, especially in end-fire array scenarios. The maximum error estimation deviation reduced from $\pm 17.32$ degrees to less than $\pm 0.58$ degrees after full calibration corrections.

These findings underscore the importance of a comprehensive calibration process, including In-Situ measurements, to achieve stable and accurate AoA estimation. Such calibration ensures the measured antenna array manifold aligns closely with the nominal manifold, significantly enhancing AoA estimation performance.

### 2.1.2 Fingerprinting

In indoor environments, localization algorithms such as AoA localization may face challenges due to signal obstruction and reflections, resulting in significant inaccuracies [ÁMLCKB21]. Time of flight (ToF) based ranging estimation offers a potential solution to mitigate these inaccuracies, albeit at the expense of requiring sophisticated hardware. On the other hand, indoor environments usually have multiple radio signals that can be detected and reported without requiring any hardware modifications. Within cellular networks, UEs are required to report information on all detectable base stations, allowing the serving base station to determine the optimal cell [WYH+20]. In stable indoor environments, received power levels tend to remain constant, making radio map techniques particularly advantageous. Here, we provide a comprehensive overview of fingerprinting techniques that are relevant to these scenarios.

**Classical Fingerprinting**

Classical fingerprinting is a localization methodology that generates a distinct fingerprint of wireless signal strength and other relevant characteristics of a specific location. This fingerprint is then used to pinpoint the whereabouts of a device. To create this fingerprint, wireless signal attributes are measured at multiple locations within a given area, such as a building or campus. In situations where conditions remain static and changes are minimal, the power received at a specific spatial point tends to remain relatively constant over time. Consequently, each spatial point $\boldsymbol{y_i}$ can be cataloged within a database comprising $T$ entries through the utilization of a vector encapsulating the RSSI measurements from $N$ APs received by the user equipment (UE) with known positions denoted as $\boldsymbol{R} = (RSSI_1, RSSI_2, ..., RSSI_N)$. This process, depicted in Figure 23, renders each spatial point uniquely identifiable through its distinctive RSSI vector, thereby facilitating precise location tracking.

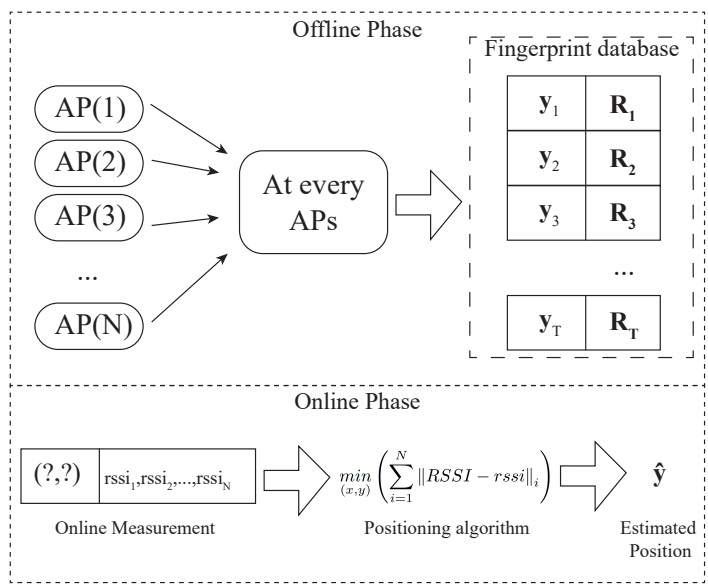

Figure 23: Fingerprinting method

Fingerprinting consists of two distinct phases, as depicted in Figure 23. The initial phase, commonly referred to as the offline or training phase, involves creating a radio map by assigning a unique fingerprint to each grid point within a predefined area. Subsequently, in the online or exploitation phase, the device requests signal strength information from nearby APs and generates a new vector $(rssi_1, rssi_2, ..., rssi_N)$. The vector is compared to different points on the map to determine the most similar fingerprint. To determine the most probable location $\hat{\boldsymbol{y}}$ on the grid, the algorithm aims to identify the point that minimizes the Euclidean distance between the new RSSI vector and the fingerprint repository, as described in the positioning algorithm equation in Figure 23.

The precision of the fingerprinting technique depends on several variables, such as the grid dimensions used during the training phase, the input vector magnitude, the variability of measured power across individual components, and the precision of UE measurements. If the input vector contains fewer than $N$ elements, the resulting estimation will degrade. In situations where multiple estimates are produced, the system may randomly choose a final location. Additionally, there is a trade-off between complexity and precision. This is important because selecting a smaller grid results in a shorter training phase but lower accuracy, while a finer grid requires a longer training phase but yields higher accuracy.

**Rank-based fingerprinting**

One of the problems of the classical fingerprinting approach is the use of RSSI values to estimate position. Since a localization system may be implemented on heterogeneous devices, i.e., smartphones, differences in hardware and software can lead to variations in reported RSSI values under the same conditions. To address this issue, rank-based fingerprinting, a modification of classical fingerprinting, was proposed [MBP11]. In this modified Algorithm 1, the RSSI values are used to sort APs in the localization request, and ranks are assigned to each AP based on their position in the sorted vector. Subsequently, the same ranks are assigned to APs in the radio map to create matched vectors. The rule is simple: if an AP in the radio map has the same MAC as an AP in the rank vector, the assigned value in the matched vector must be the same. These generated matched vectors are then used to estimate the position similarly to classical fingerprinting. However, in this case, Euclidean distance is not optimal, and distances of rank vectors should be used. Possible metrics include Spearman distance, Canberra distance, as well as Hamming distance. Comprehensive testing of rank-based fingerprinting (RBF) has been performed, e.g. in [MBA24] an impact of radio map interpolation on performance of Rank-Based Fingerprinting algorithm is presented.

**Algorithm 1:** RBF Algorithm

**Input:** LocalizationRequest, RadioMap
**Input  :** LocalizationRequest, RadioMap
**Output:** Estimated position
Sort APs based on RSSI and assign ranks based on position in the sorted vector;
SortedVector ← Sort(LocalizationRequest.RSSI, descend);
RankVector ← 1:length(SortedVector);
**for** *each vector in RadioMap* **do**
    **if** *SortedVector.MAC == RadioMapVector.MAC* **then**
        MatchedVector ← RankVector(SortedVector.MAC);
    **end**
**end**
MatchedRadioMap ← append MatchedVector;
**for** *each MatchedVector in MatchedRadioMap* **do**
    Distance ← CalculateSimilarity(RankVector, MatchedVector);
**end**
Estimate position based on the lowest Distance;

### 2.1.3 Simultaneous Localization and Mapping (SLAM)

Most of the classical radio-based localization schemes either detect the existence of line-of-sight (LoS) path and avoid non-LoS measurements, or try to mitigate the influence on the estimation of LoS path induced by multipath propagation. In contrast, many recent works [LMH+19, LVTM23, MMBW19, GKK+22, NSGJW23] leverage the environment geometry-related information conveyed in specular MPCs for localization. Specifically, signals are considered originated from map features (MFs) representing surroundings, which are time-synchronized to the physical transmitter. Thus, localization become possible in cases of LoS obstruction or insufficient infrastructure, even single base station localization becomes viable. By jointly estimating the mobile agent's position together with the environment information represented by MFs, simultaneous localization and mapping (SLAM) algorithm enables localization even in unknown environments. Furthermore, 5G and beyond networks exploiting mmWave spectrum and massive MIMO technique greatly enhance the localization and sensing services with the increased signal bandwidth and array aperture providing superior spatial resolution.

Here, we aim to provide an brief overview of the fundamentals and recent developments of multipath-based SLAM. First, the geometry-based environment model is introduced, which represents position-related information conveyed in specular MPC with MFs. The estimation of position-related information of MPCs relies on the geometry-based stochastic signal model which is introduced in the following section. Furthermore, probabilistic geometric formulations of multipath-based SLAM problems are provided.

**Geometrical Model of the Environment**

At each discrete time $n$, a physical anchor transmits radio signals and a mobile agent at an unknown time-varying position acts as a receiver. The signals propagate via different mechanisms, i.e., LoS propagation, specular reflections on smooth surfaces, scattering from point scatterers (PS) and rough surfaces, diffraction, etc. Geometric modeling of MFs constitutes an important component of multipath-based SLAM solution design. MFs for radio-based SLAM mostly refer to virtual anchors (VAs) denoting the mirror images of physical anchor (e.g., base stations) w.r.t., flat surfaces and modeling specular reflection. Different types of propagation mechanisms (represented by MFs such as VA, PSs or diffuse scatterers) typically coexist in complex propagation environments [KCT24] and are gradually starting to be considered in multipath-based SLAM approaches [GJW+16b, GKK+22, LCLT24]. More recent radio-based SLAM works also propose to model reflective surfaces as master VAs fusing correlated multipaths [LVTM23]. A more comprehensive modeling of the environment can help to improve the utilization of channel information.

It may happen that the visibility from the mobile agent's position $\check{p}_n$ to the MFs at position $\check{a}_{l,n}$ with $l \in \{1, \cdots, L_n\}$ is blocked at some time steps, i.e., the corresponding interaction are geometrically impossible. The time-varying visibility conditions in dynamic scenarios lead to death and birth process of MPCs. The number of visible MFs $L_n$ equals the number of existing MPC (i.e., model order), and model order is typically unknown and time-varying in dynamic scenarios. MPC parameters are geometrically related to the mobile agent and MFs positions, but the model can be different given different types of MFs.

**Geometry-Based Stochastic Signal Model**

The time-varying MPC parameters are usually estimated from multi-dimensional measurements using antenna arrays and multiple frequencies using super-resolution (SR) algorithms where accurate signal models serve as the basis. Geometry-based stochastic signal models (GSSM) are widely used given that they comprise geometry dependent specular MPC, stochastically modeled dense multipath and measurement noise. It is straightforward to account for the temporal behavior of MPCs using the GSSM, thus it is well suited for non-static environments.

Given each received signal vector , parametric channel estimation methods are applied to estimate time-invariant MPC parameters, for instance, the subspace methods like MUSIC [Sch86] and ESPRIT [RK89a] or ML methods [OVSN93]. ML methods are in general computationally prohibitive, where EM-based methods are often used as viable approximations. Moreover, there exist many SR methods providing MPC parameter estimates with high quality for example SAGE [FTH+99] and the SR-SR-sparse Bayesian learning (SBL) based methods [GLWF24, LLVT22] which inherently incorporate model-order estimation.

**SLAM Technical Challenges**

Multipath-based SLAM problems are often complicated by the following issues:

- The channels are in general observed within a finite aperture providing limited resolvability of MPC closely spaced in the dispersion domains. This further leads to inaccurate estimation of model-order (false alarms and miss-detection) and MPC parameters.

- data association (DA): To find the association between estimated MPC and MFs is another critical issue, and it becomes more challenging in case of MPC overlapping or in the presence of false alarm measurements.

- The dynamic nature of the propagation channel raises scalability requirements for SLAM algorithms.

Multipath-based SLAM can be interpreted as a framework that jointly performs detection of active MFs, sequential estimation of geometric states of mobile agent and MFs and system related states over time, which presents a very challenging inference problem.

**State-of-the-Art Methods**

For the probabilistic formulations of multipath-based SLAM problem, the system models are mostly non-linear and the posterior PDF is non-Gaussian and multimodal, the straightforward maximum-likelihood estimation or the sequential Bayesian estimation cannot be solved analytically. In this case, approximate sub-optimal Bayesian methods such as the extended Kalman filter (EKF) [LLO+19, GJW+16a] or particle filter (PF) are needed. It is noted that most EKF-based and PF-based methods solve the DA, MFs number estimation, and the sequential estimation of mobile agent and MFs states in separate blocks.

Bayesian factor graph-based inference techniques together with message passing algorithms constitute a well-known distributed computational scheme, which have been proven to be efficient in solving a wide variety of optimization and inference problems including multipath-based SLAM problems [LMH+19, LVTM23, KGFG+24]. Furthermore, they also exhibit strong scalability for dynamic scenarios and data/sensor fusion. The work in [LMH+19, LVTM23, LCLT24] presented a BP-based joint probabilistic DA method for joint localization and mapping. Due to nonlinear and non-Gaussian system models, the joint posterior PDF is complex and direct marginalization is infeasible. By following common assumptions as for example statistically independent measurements and MPC states, the joint posterior PDF can be factorised as a product of lower-dimensional local functions and represented with a factor graph (FG), and BP by means of sum-product algorithm can be performed on the factor graph for efficient calculation of marginal posterior PDFs with reduced complexity. It has been proposed that the statistics of MPC amplitudes can be exploited to determine the unknown and time-varying detection probabilities, which improves the detectability and maintenance of low SNR MFs [LGL+18].

Radio networks evolve towards using wider signal bandwidths and larger antenna arrays providing greatly enhanced spatial resolution for radio-based SLAM, which however also poses challenges for new channel models and more efficient parametric channel estimation methods for processing large-scale radio measurements. With the number of connected devices and accessible data further scaling up in beyond 5G networks, cooperative radio-based SLAM becomes prominent which fuses information over different observations by different devices.

### 2.1.4 Machine learning for localization

ML is a dynamic and versatile framework that can interpret complex patterns in data, making it very useful for improving localization accuracy. Advanced DNN can model complicated and non-linear relationships within the data collected from wireless environments, further enhancing location estimation. Some of the state of the art techniques are as follows.

- *Supervised learning:* This technique trains an algorithm using labeled data to predict outputs for new inputs.

- *Unsupervised learning:* This technique, unlike supervised learning, does not rely on labeled datasets. Instead, it aims to understand and identify patterns directly from input data without predefined labels.

- *Reinforcement learning:* This technique enables agents to adapt to changing conditions by learning robust policies through interaction with the environment. It integrates contextual information and long-term planning, allowing agents to make informed decisions even in complex settings.

- *Transfer Learning:* This technique involves reusing or adapting a model that has been trained for one task to solve a different, but related task. Instead of starting from scratch, transfer learning (TL) uses the knowledge gained from solving one problem to help solve another problem that has some similarities to the original one.

- *Federated Learning:* It provides a decentralized ML approach that involves training models on data distributed across multiple devices or edge nodes. It presents a promising approach for addressing the challenges inherent in localization and sensing systems.

- *Foundation Model and Fine Tuning:* Foundation models are large-scale pre-trained ML systems with extensive training on massive datasets leveraging DL techniques. This rigorous training process enables these models to discern intricate patterns and make highly informed predictions.

The main challenges that are found when applying ML techniques for localization are related to the difficulty of data collection, changes in the environment, and computational complexity. We describe here some of the ML methods that we propose to be used for localization.

**Decision tree regression**  Decision tree regression (DTR) methods are known for their simplicity and computational efficiency [ZLW19]. They simulate the functionality of the localization system by establishing a series of hierarchical comparison principles that are implemented sequentially.

The DTR learning process comprises two main stages: training and testing. In the training phase, 80% of the total available samples are randomly selected to construct the training dataset ($D_{train}$), while the remaining 20% constitute the testing dataset ($D_{test}$). The main objective of the training phase is to construct a tree that minimizes the regression error observed within the training set. Next we explore various algorithms based on DTR, selected for their notable attributes of high precision and computational efficiency. The algorithms include Random Forest, as well as two Adaboost-based training methodologies known as decision tree adaboost (DTA) and linear tree adaboost (LTA) described below.

- *Random Forests:* It is an ML technique that uses an assembly of individual models, called *base models* to make a final prediction. This technique is versatile and can be applied to various ML tasks, including classification, regression, and localization. Random Forest are particularly useful in localization tasks because they can effectively combine predictions from multiple decision trees to determine the device's location [MSR+19, GALL18].

  To generate decision trees, Random Forest use the bootstrapping technique, which involves selecting a random subset of the training data for each tree. The iterative process is repeated multiple times, resulting in an extensive array of decision trees trained on various data subsets. To derive the final prediction in the localization process, the predictions from all decision trees within the forest are averaged, as shown in Figure 24. Implementing Random Forest is straight-

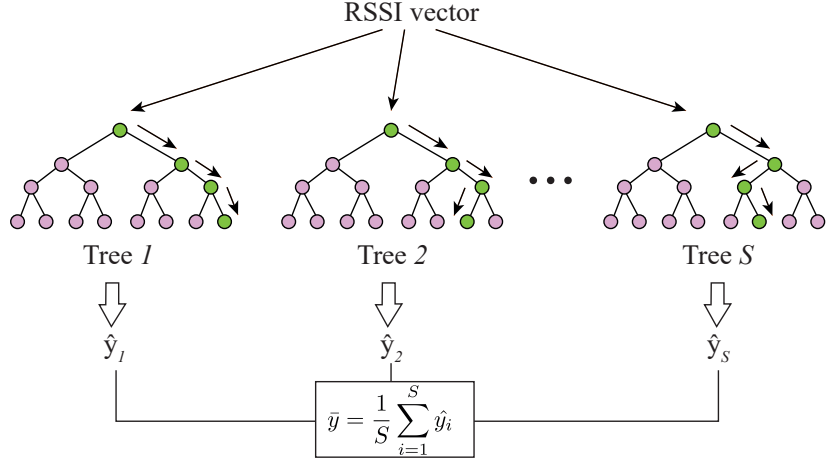

Figure 24: Random Forest scheme

  forward due to their use of decision trees, resulting in computational efficiency. Additionally, Random Forest are robust to data noise, as the averaging of outputs from multiple decision trees helps mitigate the impact of any single inaccurate estimation.

- *Adapting Boosting (Adaboost):* This technique uses predictions from multiple weak learners (WL) to make a final prediction. Boosting is the technique used to produce these WLs [LCR+23]. The model is trained iteratively on successive subsets of the data, with each iteration focusing on the misclassified data points from the preceding round. Here we analyze two Adaboost-based training algorithms: DTA and LTA. The DTA approach averages positions associated with a decision rule from diverse WLs for the final prediction [ZWC21], while the LTA method establishes an interpolation function among the various outputs within a set of decision rules [ZNZ19, IGCB21].

  Adaboost is capable of adapting and learning from changes in data over time, making it essential in dynamic settings with fluctuating wireless attributes. Although Adaboost is highly accurate, especially in indoor environments, it has a significant drawback in that it relies on intensive computational processing for the final estimation.

  The performance of fingerprinting, DTA, LTA, and Rain Forest is evaluated through an experiment using real 5G data as described in [AMKLC+24]. RSSI data from different points in the scenario is utilized to construct a radio map for fingerprinting. For Adaboost and RF, the training data is employed to build the trees or WLs. The testing data is employed to assess the accuracy of the different localization techniques. The results are visualized using the CDF of horizontal errors, as depicted in Figure 25.

**DNN-based wireless localization** DNN-based models have demonstrated outstanding accuracy among various ML approaches [WGMP15, VLS+17, BGP20, WCSW24]. Unlike traditional methods that rely on derived features like signal ToA, AoA, or RSS [RS19, STK05], DNN-based techniques primarily leverage raw CSI [BGP20, WCSW24], particularly crucial for massive MIMO systems. Massive MIMO, a cornerstone of 5G and future mobile network generations, bolsters localization accuracy through its extensive antenna arrays. Implementation strategies encompass centralized massive MIMO and spatially distributed antennas such as distributed antenna systems and cell-free massive MIMO, which enhance the effectiveness of ML-based positioning methods by leveraging spatial diversity [SL15].

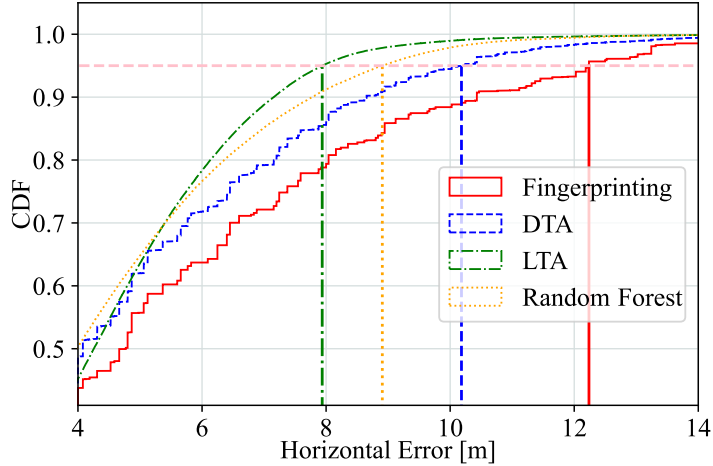

Figure 25: CDF of the error of different methods with 5G data

A basic approach for DNN-based fingerprinting involves splitting the CSI into its real and imaginary parts to obtain two vectors. These vectors, corresponding to the pilot subcarriers, are then fed into a multi-layer perceptron (MLP). The output layer directly represents the three-dimensional user position (regression). Alternatively, for spot-based localization (classification), the output layer can be a soft-max layer.

For massive MIMO scenarios with a large number of antennas, considering many subcarriers leads to high computational complexity of the MLP. To mitigate this, the CSI can be interpreted as a two-dimensional "image" and fed into a CNN. However, the choice of the convolution kernel introduces a strong bias, which may not be suitable for CSI. This approach is used in existing literature such as [VLS+17, BGP20] is referenced.

DNNs typically provide output without information about the uncertainty in their estimate. To address this, approaches supporting inherent location uncertainty estimation for CSI-based wireless localization have been proposed. In [SSR21], a two-pronged approach is presented: 1) Augmenting the base DNN with an additional fully connected output layer implementing a parametric GMM to estimate both UE localization and position uncertainty. 2) Estimating model uncertainty through Monte Carlo sampling, either from an ensemble of independently trained networks or using Monte Carlo dropout.

To demonstrate the performance, this approach is applied to ray-tracing data of DeepMIMO [Alk19]. In Figure 26, the nnormalized root mean squared positioning error (nRMSE) is compared to the uncertainty estimate provided by the DNN, showing qualitative information about the reliability of the location estimate.

**Wireless Transformer**  While CNNs have been widely adopted for DNN-based wireless localization, they may underperform in dynamic environments due to their local spatial bias. Inspired by attention mechanisms, transformer-based models offer advantages in modeling large-scale dependencies efficiently. The wireless transformer (WiT) presented in [SSR22] uses subcarriers as input representations for a transformer-based DNN. The attended features at the output of the transformer block are averaged and fed into an MLP to produce the final location estimate. In Figure 27, WiT is shown to significantly outperform the direct use of raw CSI for localization in outdoor ray-tracing scenarios.

**Self-Supervised Channel Features**  Supervised learning and transformer-based models are limited to scenarios with large amounts of labeled training data. To address this, a self-supervised learning method for wireless channel feature extraction called self-supervised wireless transformer (SWiT) is proposed in [SRS24]. SWiT learns features from unlabeled CSI using pretext tasks, exploiting redundant and complementary information across subcarriers. These self-supervised features support effective learning and are transferable across different scenarios/environments with little fine-tuning.

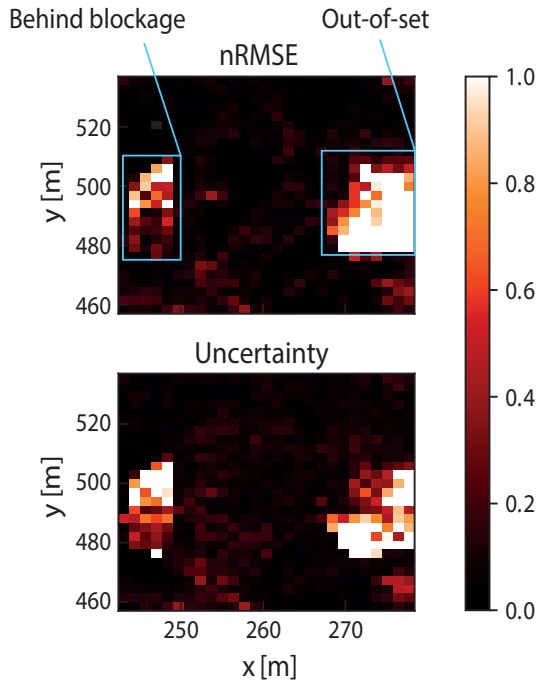

Figure 26: Comparison of simulated normalized root mean squared positioning error and uncertainty estimate obtained through a Gaussian mixture model and Monte Carlo methods [SSR21]. The uncertainty estimate provided by the DNN allows at least a qualitative assessment of the reliability of the location estimate.

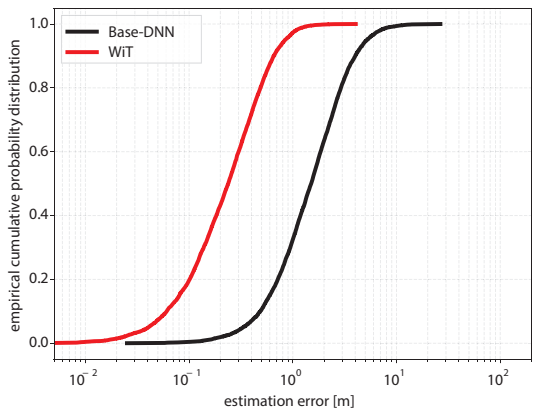

Figure 27: Performance comparison of base DNN and wireless transformer (WiT) [SSR22]. The attended features provided by the transformer enable more accurate localization than directly using raw CSI.

In Figure 28, SWiT is compared to WiT in terms of localization accuracy on ultra dense indoor massive MIMO CSI data. SWiT achieves significantly better localization accuracy, especially in scenarios with limited labeled training data.

Through these approaches, DNN-based methods for wireless localization are enhanced with location uncertainty estimation, improved feature extraction, and increased adaptability to varying scenarios and data availability.

**Multipath Assisted Positioning** Multipath-assisted positioning leverages spatial information from MPCs to enhance localization accuracy. Each MPC arriving at the receiver is treated as a LoS signal from a virtual transmitter. This concept is depicted in Fig. 29, where different propagation paths

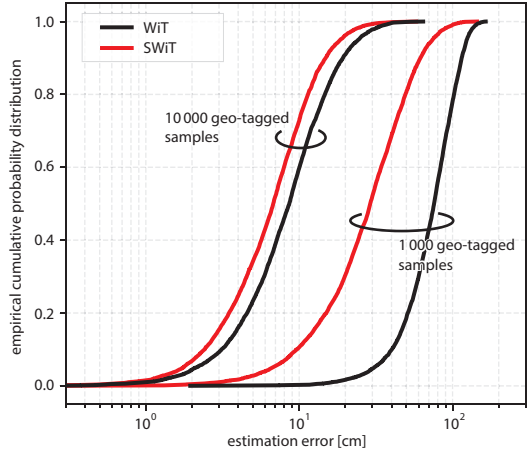

Figure 28: Performance comparison of base wireless transformer (WiT) and self-supervised wireless transformer (SWiT) [SRS24]. Through self-supervision, SWiT is able to extract higher level features from unlabeled (non-geo-tagged) training data. These features can then be fine-tuned for wireless localization with relatively little labeled (geo-tagged) training data.

from the physical transmitter to the receiver are illustrated. The red MPC, reflected at the wall,

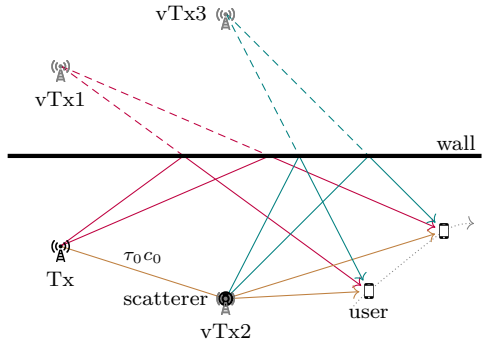

Figure 29: The idea of multipath assisted positioning is to treat every MPC as a LoS signal from a virtual transmitter.

represents a LoS signal from the virtual transmitter vTx1, located at the mirrored position of the physical transmitter. The brown MPC, scattered at a point scatterer, is regarded as a LoS signal from the virtual transmitter vTx2 at the scatterer's location. In cases of scattering, the virtual transmitter exhibits a propagation delay towards the physical transmitter, akin to a clock offset.

While idealized assumptions are made here regarding perfectly straight walls and point scatterers, this framework reasonably models real-world scenarios, including diffraction and multiple interactions of signal components with various objects. Fig. 29 exemplifies this with a propagation path first scattered at the scatterer and then reflected at the wall, corresponding to a virtual transmitter vTx3.

Typically, multipath-assisted positioning involves two steps. First, a channel estimator detects and tracks signal components, estimating parameters like ToA or AoA. These parameters serve as measurements for the transmitter positions in the second step, where receiver localization is jointly estimated with the physical and virtual transmitter states. Robust DA methods are crucial for mapping between signal components and transmitters.

As the locations of virtual and physical transmitters are generally unknown, they can be simultaneously estimated with receiver positions using SLAM approaches. Various methods exist in the literature, such as Channel-SLAM [GJW+16c, Ulm21].

**Multipath Assisted Positioning-based Fingerprinting** While multipath-assisted positioning is computationally complex for real-time applications, it facilitates offline fingerprinting database creation, reducing physical effort significantly. By estimating fingerprint locations with multipath-assisted positioning, channel features such as CIR or CSI can be used as fingerprints. A novel approach intro-duced in [UG23] involves utilizing multipath-assisted positioning in the offline phase of fingerprinting schemes relying on channel features. The scheme, depicted in Fig. 30, employs Channel-SLAM to estimate fingerprint positions $\boldsymbol{\tau}_i$, which are then used to train a DNN. In the online phase, fingerprints are measured and passed to the DNN for inference, followed by position tracking using a Kalman filter.

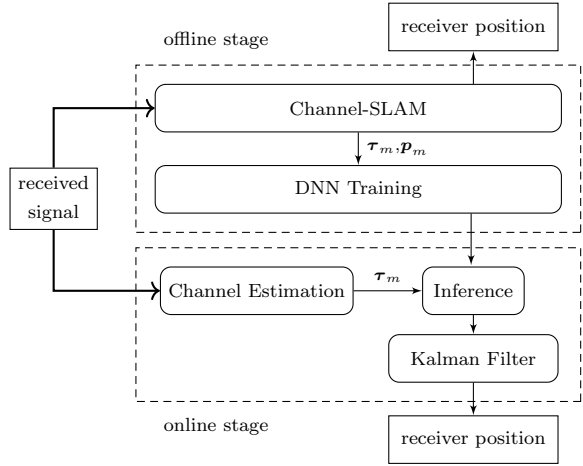

Figure 30: The flow chart of a fingerprinting scheme using Channel-SLAM for collecting fingerprints $\boldsymbol{\tau}_i$ and a DNN learning the receiver positions $\boldsymbol{p}_i$.

Fig. 31 illustrates a simulation scenario with a physical transmitter, reflective walls, and point scatterers. A mixture density network (MDN) is trained with Channel-SLAM data from multiple receivers traversing the scenario. Fig. 32 presents positioning performance for different numbers of GMM components in the MDN, indicating that a number of two yields optimal performance.

This fingerprinting approach utilizing multipath-assisted positioning offers low physical effort in offline fingerprint collection and reduced computational complexity in the online phase. MDNs effectively handle fingerprinting ambiguities, enhancing positioning performance.

### 2.1.5 Cooperative localization

To enable localization in wireless networks that cover large areas with a small number of anchor nodes, cooperative localization algorithms are used. Cooperative localization algorithms also use agent-to-agent proximity measures to determine the locations of the agents.

The wireless network can be represented by a graph $\mathcal{G} = (\mathcal{V}, \mathcal{E})$, where each vertex is assigned to one node. $N$ denotes the number of all nodes (agents and anchors). $\mathcal{E}$ are edges. We assume the first $N_a$ nodes are agents. The set of vertices $\mathcal{V}$ is a union of agents $\mathcal{A}$ and anchors $\mathcal{S}$, $\mathcal{V} = \mathcal{A} \cup \mathcal{S}$. Vertices are adjacent if there exists direct communication between those nodes. The set of all neighbors of node $i$ is denoted by $\mathcal{N}_i$, its cardinality by $N_i = |\mathcal{N}_i|$, and the number of all links in the graph is denoted by $M = |\mathcal{E}|$. A node location is $\mathbf{x}_i$. A range $r_{ij}$, corresponds to the estimated distance between two adjacent nodes $i$ and $j$. The range is assigned to an edge in the graph. The range includes measurement error $\epsilon$. We assume $r_{ij} = r_{ji}$. For a more compact notation, we fold all the measurements into a vector $\mathbf{r}$ of length $M$. In addition, we introduce the vector of node locations $\mathbf{X} = [\mathbf{x}_1, \ldots, \mathbf{x}_N]^T \mathbb{R}^{lN}$. The mapping $D : \mathbb{R}^{lN} \to \mathbb{R}^M$ computes the distances between nodes such that

$$\mathbf{r} = D(\mathbf{X}) + \boldsymbol{\varepsilon}. \tag{28}$$

Solving eq. (28) is the main task of the localization algorithm and there are several different approaches for estimating the parameters $\hat{\mathbf{X}}$ [WLW09]. The algorithms can be classified into deterministic and probabilistic. The deterministic algorithms minimize the mean squared error (MSE) between measured ranges $r_{ij}$ and distance calculated from estimated node locations $d_{ij}$. The deterministic algorithm exhibits relative computation simplicity but lacks statistical interpretation. The probabilistic algorithms consider the statistics of measurement error, which results in a probability of the estimated location correctness. Mathematically the problem can be tackled by LS, or probabilistic, i.e. maximum-likelihood estimator, approach. Both estimators fall into the class of Non-Bayesian estimators. An alternative approach uses Bayesian estimators, which takes into account the uncertainty of proximity estimation, for example, the likelihood of measured distance, and prior probability density function of the positions of unknown nodes.

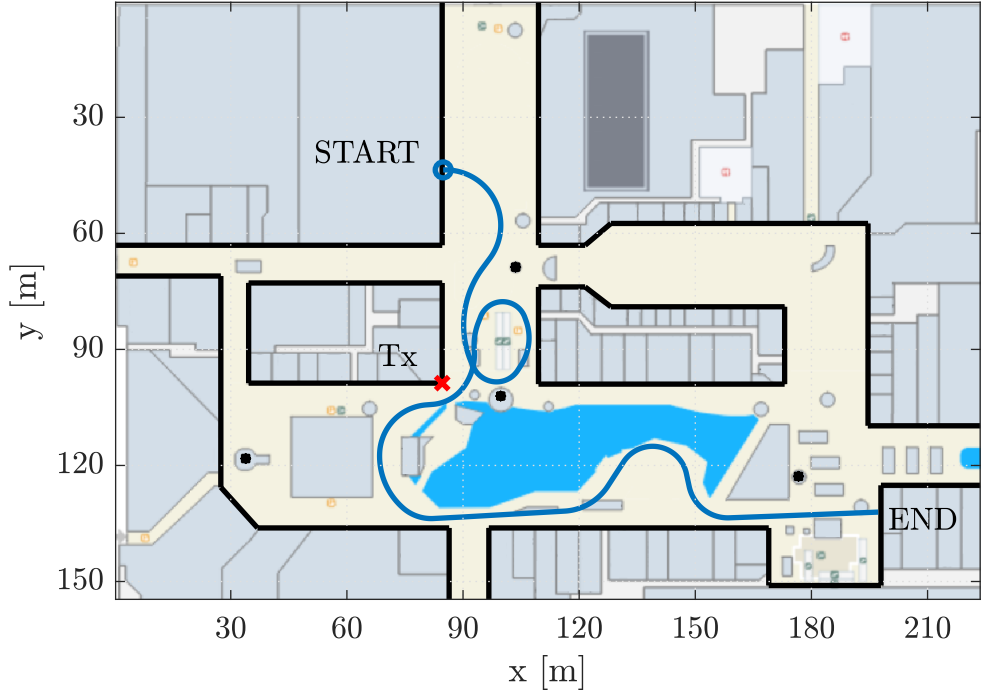

Figure 31: A top view of the simulation scenario with one physical transmitter labeled transmitter (Tx) and one receiver track.

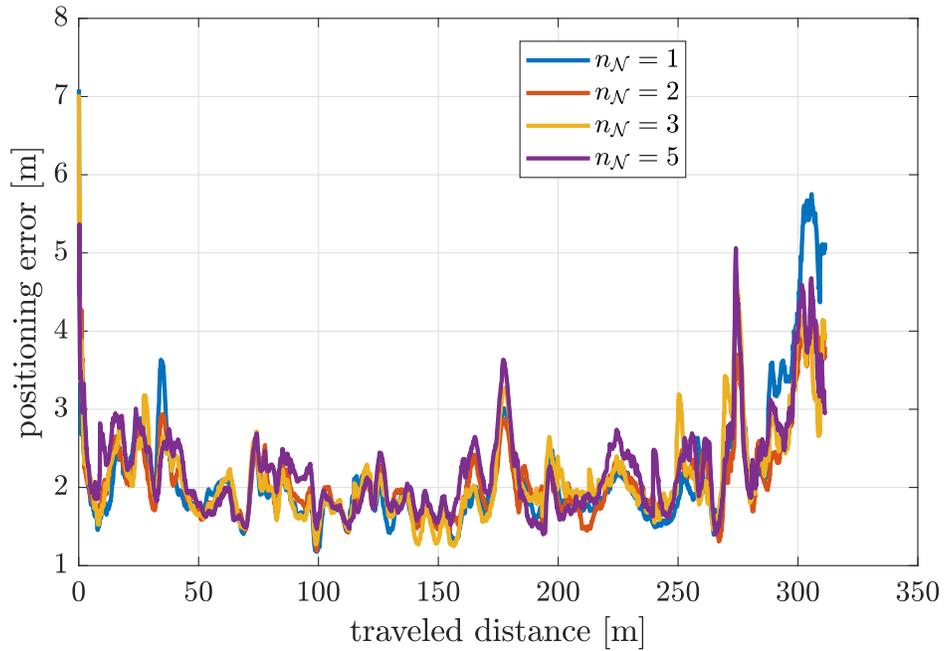

Figure 32: Results for different numbers of components in the MDN GMM.

**Non-Bayesian cooperative localization algorithms**

The LS estimator is a standard non-Bayesian method to find solutions by minimizing the square error between the measured distances and calculated distances from the location of nodes as

$$F(\mathbf{X}) = \sum_{i \in \mathcal{V}, j \in \mathcal{N}_i} (r_{ij} - \|\mathbf{x}_i - \mathbf{x}_j\|)^2$$

$$F(\mathbf{X}) = \|\mathbf{r} - D(\mathbf{X})\|^2 = \|\boldsymbol{\varepsilon}(\mathbf{r}, \mathbf{X})\|^2 .$$

$$\hat{\mathbf{X}} = \arg\min_{\mathbf{X}} \|\boldsymbol{\varepsilon}(\mathbf{r}, \mathbf{X})\|^2 . \tag{29}$$

The relation $F(\mathbf{X})$ is in general non-linear and non-convex, and finding the global minimum is not a straightforward task, so some approximate methods are often in use. A minimum is often found by an iterative approach, which can in the case of non-convex problems finishes in a local minimum.

The LS method does not give us the likelihood of the obtained solution, but if the probability density function of $r_{ij}$ given $\mathbf{X}$ $p(r_{ij}|\mathbf{X})$ is known the maximum-likelihood estimator can be applied. If the observations $p(r_{ij}|\mathbf{X})$ are statistically independent, we can write

$$p(\mathbf{r}|\mathbf{X}) = \prod_{i \in \mathcal{V}, j \in \mathcal{N}_i} p(r_{ij}|\mathbf{X}). \tag{30}$$

The optimal maximum-likelihood solution can be found by maximizing

$$\hat{\mathbf{X}} = \arg\max_{\mathbf{X}} \prod_{i \in \mathcal{V}, j \in \mathcal{N}_i} p(r_{ij}|\mathbf{X}) \tag{31}$$

If we assume a normal distribution of errors $\varepsilon_{ij} \sim \mathcal{N}(0, \sigma)$, the maximum-likelihood method (31) converges to the LS method (29). Finding the global minimum of (29) and (31) is not straightforward, so some approximation methods are developed. Among them the most popular are, sequential cooperative localization, multidimensional scaling (MDS), convex relaxation, and alternative direction method of multipliers (ADMM).

**Sequential cooperative localization:** Sequential cooperative localization is the common sense approach to cooperative localization [ZB19]. The basic idea is to find agents in a network with a list of three connections to the anchors in the first step and estimate their locations using the non-cooperative algorithm multi-lateration algorithm. In the next step, the agents with the estimated locations are assumed virtual anchors, and step one is repeated using the set of initial anchors and virtual anchors till no agent with at least $l+1$ connections to initial anchors and virtual anchors is found in the network.

The algorithm is straightforward but fails in the beginning if no agents have at least three connections to anchors and the location of some agents may not be found if the network is not sufficiently dense. Furthermore, the error propagation is another serious problem of this algorithm. The erroneous location of the virtual sensors is propagated to all estimated agent locations in the next procedure steps.

**Convex relaxation:** The problem described in (29) is not convex, thus the solver may end in the local minimum. To solve this problem several methods are developed, to relax the minimization problem in the convex form and use known solving methods on the relaxed optimization problem [BY04b, BLT+06, Tse07, SXG15]. The LS optimization problem can be slightly reformulated (29) to transform it into the semidefinite programming (SDP) or second-order cone programming (SOCP) [Tse07]. The SDP problems can be solved by the SDP solvers [MOS24, TTT17, GB14].

The authors in [SXG15, HB05, GTSC13] solved the problem relaxation by the vector projection on the convex set, disc, or sphere. The Nesterov's accelerated gradient descent method was proposed for achieving faster convergence in [SXG15]. The algorithm requires global synchronization of wireless networks, which is in practice hard to achieve, thus also an asynchronous algorithm is proposed in [SXG15], which applies the Nesterov's gradient asynchronously to a particular node.

**Multidimensional scaling:** MDS is a set of data-analytic methods that attempt to arrange nodes in low dimensional space by processing the estimated distance between nodes. In its basic formulation, the algorithm is used to analyze higher-dimensional data in social sciences, it can be adapted for cooperative localizations [Li07, CPH06, YR04].

The MDS algorithm tries to find the minimum of (29) and it applies the LS approach. The function $F(\hat{\mathbf{x}})$ is called the stress function in MDS. One method to solve the minimization in (29) is majorization [dLM09], which suggest to find a simpler more manageable function $g(x,y)$, which majorizes function to be minimized $f$

$$g(x,y) \leq f(x), \text{ for each } x,$$

where $y$ is some fixed value called the supporting point. The surrogate function $g(x,y)$ should touch function $f(x)$ at the point $y$, meaning $f(y) = g(y,y)$

$$f(\hat{x}) \leq g(\hat{x},y) \leq g(y,y) = f(y).$$

The MDS method estimates the relative locations of sensor nodes, thus we need to translate these estimates into a physical coordinate system. This alignment typically involves translation, rotation, calling, and mirroring of coordinates by applying three anchors for two-dimensional space.

**Alternating direction method of multipliers:** The ADMM is a robust method for solving optimization problems of arbitrary size, which is used in machine learning, statistics with huge datasets, and dynamic optimization in large networks. There exists also a distributed version, where a set of distributed nodes cooperatively solves a large problem by sending relatively small messages to nodes in their neighborhood. Accordingly, the ADMM method has been suggested for solving cooperative localization problems [Ers15, PE18].

The ADMM approach is used to solve problems of the form

$$\arg\min_{\mathbf{x},\mathbf{z}} \quad f(\mathbf{x}) + g(\mathbf{z}),$$
$$\text{subject to:} \quad \mathbf{x} \in \mathcal{X}, \mathbf{z} \in \mathcal{Z},$$
$$\mathbf{A}\mathbf{x} + \mathbf{B}\mathbf{z} = \boldsymbol{\gamma}.$$

where $f$ and $g$ are convex functions, $\mathbf{A}$ and $\mathbf{B}$ are matrices, $\boldsymbol{\gamma}$ is vector, while $\mathbf{x}$ and $\mathbf{z}$ are optimization variables in limited spaces $\mathcal{X}$ and $\mathcal{Z}$. The problem can be formed with the Lagrangian function with the penalty term

$$\begin{aligned} L_c(\mathbf{x},\mathbf{z},\boldsymbol{\lambda}) &= f(\mathbf{x}) + g(\mathbf{z}) + \\ &+ \boldsymbol{\lambda}^T(\mathbf{A}\mathbf{x} + \mathbf{B}\mathbf{z} - \boldsymbol{\gamma}) + \\ &+ \frac{1}{2} c \, \|\mathbf{A}\mathbf{x} + \mathbf{B}\mathbf{z} - \boldsymbol{\gamma}\|_2^2. \end{aligned}$$

The main disadvantage of the ADMM algorithm is that it can get trapped in the local minimum, which may not even be close to the true solution. Convergence to the global minimum is more likely if the initial values of $\hat{\mathbf{x}}_i^0$ are close to the true solution, which can be realized by setting the initial values to the result of some other method that is not sensitive to this type of problem. In the [PE18] a hybrid method is presented that uses the convex relaxation method to compute an approximation of the solution, which is then used as the initial value of the algorithm.

**Bayesian cooperative localization algorithms** The Bayesian methods are probabilistic methods that take into account the uncertainty of proximity estimation, for example, the likelihood of measured distance, and prior probability density function of the positions of unknown nodes. The methods look at finding the maximum a posteriori (MAP) of the posteriori probability density function or minimum mean square error (MMSE) of posteriori probability. The main drawback of the probabilistic methods is the high computational complexity.

**Minimum mean square error:** The MMSE estimator is a Bayesian probabilistic method. MMSE is defined as

$$\hat{\mathbf{X}} = \arg\min_{\hat{\mathbf{X}}} \int_{\hat{\mathbf{X}}} p(\mathbf{X}|\mathbf{r})(\hat{\mathbf{X}} - \mathbf{X})^T(\hat{\mathbf{X}} - \mathbf{X}) \, d\mathbf{X}. \tag{32}$$

Minimum of (32) by its derivation and setting the derivation to zero

$$\frac{d \int_{\hat{\mathbf{X}}} p(\mathbf{X}|\mathbf{r})(\hat{\mathbf{X}} - \mathbf{X})^T(\hat{\mathbf{X}} - \mathbf{X}) d\mathbf{X}}{d\hat{\mathbf{X}}} = 0$$

which yields to

$$\hat{\mathbf{X}} = \int_{\hat{\mathbf{X}}} \mathbf{X} \, p(\mathbf{r}|\mathbf{X}) \, d\mathbf{X}. \tag{33}$$

In the case of linear system $\mathbf{r} = \mathbf{AX} + \mathbf{1}\varepsilon$ and measurement errors have normal distribution with mean value 0 ($\varepsilon \sim \mathcal{N}(0, \sigma^2)$), (33) leads to

$$\hat{\mathbf{X}} = (\mathbf{A}^T\mathbf{A} + \sigma^2\mathbf{1})^{-1}\mathbf{A}^T\mathbf{r}.$$

**Maximum a posteriori MAP estimation:** The location estimated using MAP estimation can be expressed as

$$\hat{\mathbf{X}} = \arg\max_{\mathbf{X}} p(\mathbf{X}|\mathbf{r}) = \arg\max_{\mathbf{X}} \frac{p(\mathbf{r}|\mathbf{X})p(\mathbf{X})}{\int p(\mathbf{X}|\theta p(\theta)d\theta}. \tag{34}$$

The expression in the denominator is constant and thus has no impact on the maximum, thus the expression (34) yields to

$$\hat{\mathbf{X}} = \arg\max_{\mathbf{X}} p(\mathbf{X}|\mathbf{r}) = \arg\max_{\mathbf{X}} p(\mathbf{r}|\mathbf{X})p(\mathbf{X}). \tag{35}$$

The a posteriori distribution $p(\mathbf{r}|\mathbf{X})$ is difficult to describe, and obtaining its mean, mode, or marginals is a very hard problem. The problem can tackled by using the pairwise pairwise Markov random fields (MRF) and the Hammersley-Clifford theorem [Bis06, ZB19]:

$$\hat{\mathbf{X}} = \arg\max_{\mathbf{X}} \prod_{i \in \mathcal{V}, j \in \mathcal{N}_i} p(r_{ij}|\mathbf{x}_i, \mathbf{x}_j) \prod_{i \in \mathcal{V}} p(\mathbf{x}_i). \tag{36}$$

The BP is commonly applied on the pairwise MRF [Bis06]. The BF splits the computation of the marginal beliefs into smaller computational tasks and distributes the computation to several nodes. It is an efficient method for distributed inference, also applied in cooperative localization.

The standard BP algorithm finds the posterior marginal probability density function $M_i(\mathbf{x}_i)$ at the node $i$ using the product of local evidence $\psi_i(x_i)$, which is in the case of cooperative localization presented as the probability the node is located at $\mathbf{x}_i$ and messages coming from neighboring nodes $m_{ji}(\mathbf{x}_i)$

$$P_i(\mathbf{x}_i) \propto \psi_i(x_i) \prod_{i \in \mathcal{V}, j \in \mathcal{N}_i} m_{ji}(\mathbf{x}_i), \tag{37}$$

where $\mathbf{x}_i$ is the location of node $i$. The product includes neighboring nodes of node $i$. The message from node $j$ to node $i$ $m_{ji}(\mathbf{x}_i)$ is calculated by:

$$m_{ji}(\mathbf{x}_i) \propto \int_{\mathbf{x}_j} \psi_k(\mathbf{x}_j)\psi_{i,j}(\mathbf{x}_i, \mathbf{x}_j) \prod_{j \in \mathcal{V}, k \in \mathcal{N}_j \setminus i} m_{k,j}(\mathbf{x}_j) d\mathbf{x}_j, \tag{38}$$

where $\psi_{i,j}(\mathbf{x}_i, \mathbf{x}_j)$ is the pairwise potential between node $i$ and $j$. In the case of cooperative localization expressed as distance error measurement probability function $\psi_{i,j}(\mathbf{x}_i, \mathbf{x}_j) = p_n(r_ij - \|\mathbf{x}_i - \mathbf{x}_j\|)$. The product $\prod$ is over all messages going in node $i$ except the message from node $i$ ($\mathcal{V}, k \in \mathcal{N}_j \setminus i$). Replacing (37) into (38) yield to an iterative solution for BP algorithm (39) into (40).

$$P_i(\mathbf{x}_i)^t \propto \psi_i(x_i) \prod_{i \in \mathcal{V}, j \in \mathcal{N}_i} m_{ji}^t(\mathbf{x}_i), \tag{39}$$

$$m_{ji}^t(\mathbf{x}_i) \propto \int_{\mathbf{x}_j} \psi_{i,j}(\mathbf{x}_i, \mathbf{x}_j) \frac{M_j^{t-1}(\mathbf{x}_j)}{m_{ij}^{t-1}(\mathbf{x}_j)} d\mathbf{x}_j. \tag{40}$$

At the first iteration, an initialization is necessary, by setting $m_{ij}^0 = 1$ and $M_i^0$ to delta function for anchors, and $M_i^0$ to uniform distribution for anchors. The nonlinear relationships and potentially non-Gaussian probability density functions in sensor localization of the standard BP messages may overload the wireless sensor network. The particle-based representation via nonparametric belief propagation (NBP) [9] enables the application of BP to localization in WSN [IFMW05, SZ13].

**Algorithm comparison**

We generate a representative wireless network to compare cooperative algorithms with 12 anchors and 50 agents. The agents are randomly distributed within a square with the size of 1 m by 1 m, while anchors are randomly placed along the sides of the square. The node locations and their connections are plotted in Figure 33. The node range is set to 0.3 m. If the distance between nodes is below the specified range there is no connection, i.e. information exchange between nodes. In Figure 33 the anchors are presented with squares, while agents with circles. The additive noise with random distribution is added to the ground true distance between nodes to introduce measurement errors

$$r_{ij} = d_{ij}(1 + \mathcal{N}(0, \sigma)),$$

where $\mathcal{N}(0, \sigma)$ is normal distribution with mean value equals to 0 standard deviation $\sigma$. The addition of measurement errors corresponds to the estimation of the distance from the received signal strength.

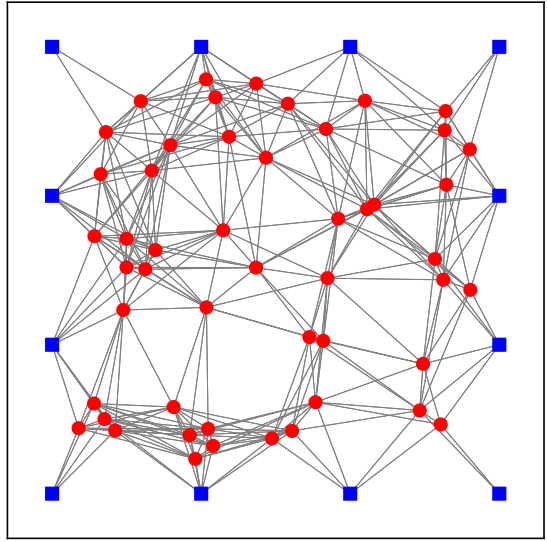

Figure 33: The node locations and their connection estimation: ● agent, ■ anchor, — connection.

The RMSE of several algorithms, namely LS, cooperative LS, Convex relaxation, MDS and ADMM is presented in Figure 34 for the network shown in Figure 33. When testing, the fully connected network is assumed, but in the standard LS algorithm only the connections to agent-to-anchors are considered, while the cooperative LS algorithm takes into account also agent-to-agent connections. It is obvious the cooperation between agents significantly decreases the RMSE of the estimated location. The MDS algorithm is a non-optimal algorithm, which performs majorization, and this leads to an increase of RMSE compared to LS algorithm fully connected wireless network. Convex relaxation and ADMM algorithm show performance compared to the fully connected cooperative LS algorithm, despite only the connection being limited to neighboring nodes. The error of short estimated ranges is lower compared to longer estimated ranges thus the lower average error is introduced in the algorithm.

## 2.2 RSSI-based localization

The active localization of RF emitters is widely researched in both terrestrial and non-terrestrial wireless communication systems due to the widespread availability of instruments that can measure the RSSI. Localization techniques like ToA, TDoA, and AoA, just to cite a few familiar techniques, offer precision but demand complex hardware and synchronization. Conversely, RSSI localization is simpler, ideal for weight-sensitive UAV and less complex devices, despite its susceptibility to noise and signal distortion.

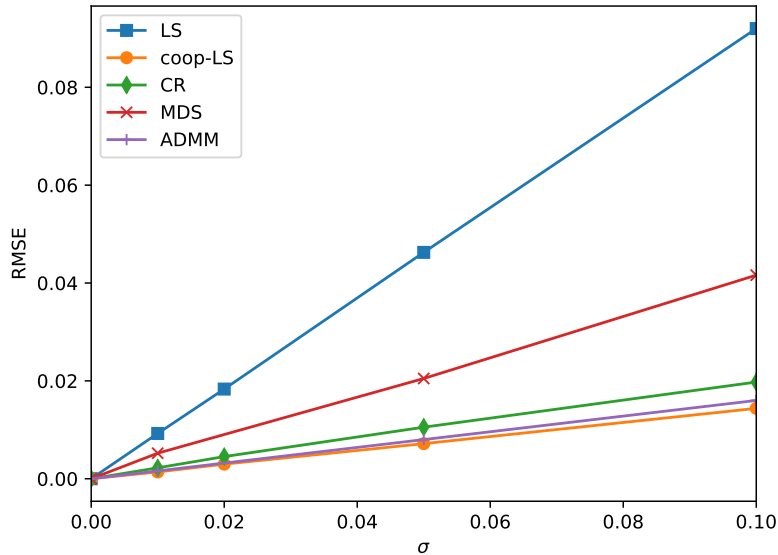

Figure 34: root mean square error (RMSE) of compared cooperative localization algorithms.

While RSSI localization is widely adopted for its speed and cost-effectiveness, it also presents serious challenges. Some of these are: sensitivity to jamming or interference and multipath; the need of many measurements to improve the precision; and the need of calibration and/or knowledge of transmitted signal power.

In the sub-sections that follow we will discuss some recent works on RSSI-based indoor localization using different technologies and RSSI-based outdoor localization from UAV.

### 2.2.1 RSSI-based localization in indoor scenarios

Because of the instability of the surroundings, radio localization in indoor environments is still a difficult problem. In [WNTC22] the localization accuracy achieved by the RSSI-fingerprinting approach using BLE is compared for two distinct environments: a wide, empty hall and a small, confined hallway. Six distinct smartphones from three different manufacturers were used to perform the measurements, which makes them interesting because the accuracy of the various devices can be compared.

In [ACQ22] virtual antenna arrays are used to locate RF transmitters using a mobile receiver. Accurately locating antennas and controlling frequency offsets that impact signal phase are challenging. Software-defined radio (SDR) tests demonstrate that the approach of dual-antenna receivers improves azimuth and elevation accuracy even with low-quality oscillators, and that results improve with more antennas.

The work in [KSB$^+$22] tackles the problem of optimizing anchor numbers for large-scale deployments under challenging situations including multipath propagation, with a focus on improving indoor radio placement in complicated surroundings. It is shown that fewer anchors are required to achieve performance targets, hence increasing localization accuracy and system scalability. This is achieved by utilizing a geodetic network optimization technique and developing signal-to-interference-plus-noise ratio (SINR)-based REMs.

The article in [SP22] looks at the increasing need for precise indoor tracking systems and investigates the potential of LoRa technology in this field. It describes how trilateration and RSSI measurements allow LoRa, operating in the 2.4 GHz band, to efficiently estimate item positions indoors. Experiments conducted in a range of environments demonstrated LoRa's potential accuracy, albeit reliant on node placement and signal configuration; data was made accessible for more research.

The study in [ACQ23] addresses issues such as antenna location and frequency drifts in order to improve RF transmitter localization using virtual antenna arrays. In addition to accounting for low-quality oscillator drifts and introducing an enhanced algorithm for dual-antenna receivers, experimental results validate the theoretical advantages of a dual-antenna arrangement over a single antenna.

### 2.2.2 Hybrid RSSI/AoA estimation

A different method of using RSSI to estimate position, which aims at estimating the AoA by using directive antennas, is gaining attraction in the literature [HYA18, TSH+18, DK18, WH20]. The AoA is given by simply rotating the antenna [TSH+18, WH20], or using multiple antennas pointing in different directions[DK18]. In [TSH+18], authors provide another example of this with a hybrid RSSI/AoA algorithm that combines triangulation from AoA with RSSI information to enhance estimation accuracy. Positions are determined via angulation, with measures derived from relative RSSI between antennas. AoA measures are assessed relative to actual RSSI, prioritizing measures with higher RSSI values and larger differences between maximum and minimum across all antennas.

### 2.2.3 RSSI-based localization from UAVs

UAVs are set to revolutionize the telecommunication industry with their mobility, primarily for search and rescue operations and deploying temporary base stations. Researchers have focused a lot of attention on using UAV to find a ground-based radio transmitter in the past few years. In some approaches, a UAV is used as an anchor point to help localizing one or several ground RF emitters, and its measurements are merged with those from the ground anchor nodes. It is noteworthy that very few experimental works use UAV as anchor points, and even fewer of them make use of RSSI-based localization. Some methods described in the literature use a UAV to gather data and learn how to maneuver it. By comparing the RSSI from two antennas installed at the front and rear, it is demonstrated how to identify and maneuver a UAV toward a ground RF emitter.

The work in [MTS+22] proposes a technique that gathers RSSI data at different sites using UAV as low-altitude platforms to locate ground RF emitters. The main innovation is an experimental arrangement that uses an Adalm Pluto SDR and GPS/inertial measurement unit (IMU) sensors to record RSSI and UAV positions simultaneously. Two unlicensed frequency bands are used for data collection of which the 865 MHz range has less interference. A maximum likelihood approach is used to estimate the transmitter's location, yielding an average localization error of 4 meters in the absence of interference and 5 meters in its presence. The work also proposes a threshold-based technique to improve the accuracy in the presence of interference. One of the main novel aspects of the work is to evaluate the impact of the interference on the performance of RSSI-based localization.

The study in [JK23] introduces a method for positioning a drone base station (DBS) to enhance signal quality and fairness among users, using an adaptive algorithm that relies on channel measurement history instead of exact user locations. It highlights that strategic DBS movement can lower energy use, and presents a multi-agent Q-learning approach that's computationally efficient, demonstrating through simulations its superiority in fairness and reliability over existing models.

Future advancements in RSSI-based localization could delve into more complex environments. The data collection campaign could be broadened to facilitate the training of sophisticated machine learning models. Additionally, the framework could be enhanced by incorporating user feedback based on real-time measures or autonomous UAV path planning in order to maximize localization accuracy.

## 2.3 Localization by radio systems with Ultra-Wide Band, LoRa and Signals of Opportunity

Satellite-based navigation systems are often the preferred technology for positioning and navigation in outdoor settings. However, in densely populated urban areas or urban canyons where there are strong multipath echoes or complete blockage of GNSS signals, the accuracy of positioning can be reduced. In indoor environments or tunnels, positioning may not be available at all. Moreover, satellite-based navigation systems are susceptible to jamming [NO16] and spoofing [BPP20]. In recent years, various approaches have been proposed in the literature to address these issues. Combining hybrid approaches with multiple sensors and technologies in data fusion can help mitigate the vulnerabilities of satellite-based systems in such environments [GBTM+16, ANM+18]. Various sensor and technology combinations are appropriate for positioning and its applications, based on factors such as the environment, dynamics, budget, accuracy, requirements, and level of robustness or integrity needed [Gro13]. An alternative for outdoor and indoor positioning is based on signals of opportunity (SOP)-based localization systems, that depend on the signals present in the vicinity [MK21, KKAL22, SKE21]. A contribution of INTERACT addresses this topic in section 2.3.4.

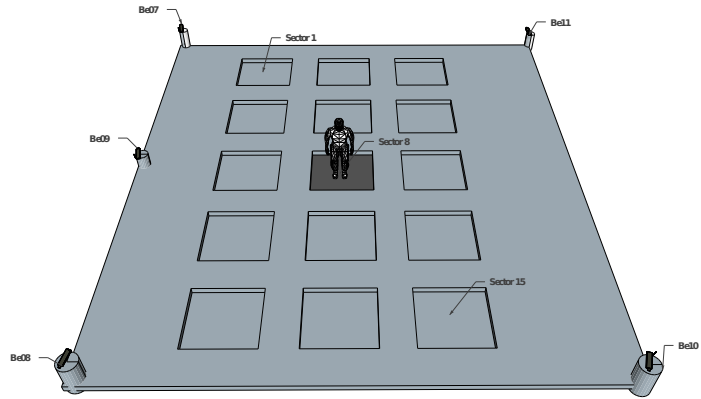

Figure 35: **Beacon indoor experimental area setup.**

With respect to complex indoor environments, they could be challenging scenarios due to multipath and NLOS effects that are known to cause errors in ranging and positioning [DCD15]. In these setups, technologies such as WiFi stand out as one of the most widely employed wireless technology, although it was not initially developed for the purpose of localization. In the future, the list of these technologies may expand to include LoRa [HHDG+23], where an INTERACT contribution is presented in Section 2.3.3. LoRaWAN is a network protocol that uses LoRa technology for low power wide area network (LPWAN) for IoT applications. The performance study shows results of RSSI-based indoor localization using LoRa at 2.4 GHz and trilateration algorithm.

UWB technology, originally used for wireless communication, has evolved to be utilized in localization systems. According to the Federal Communication Commission (FCC) in 2002, UWB is defined as a radio wave with a fractional bandwidth greater than 20 or at least 500 MHz. UWB with ultra-short impulses (nanosecond durations) is robust to multipath effects and offers high time resolution, enhancing ranging and positioning accuracy in centimeter scale with ToA estimation and ToA based methods. The IEEE 802.15.4 standard [std] and its amendments specify different Physical layers for low-rate wireless personal area network (LR-WPAN). The formation of the FiRa Consortium (Fine Ranging) is working on precise localization with UWB and its use cases, applications, standardization and more. An INTERACT contribution presents data-driven FG models for anchor-based positioning with UWB radio systems in Section 2.3.2. A FG model is explained for improved anchor-based positioning that takes into account both, i) the effects of the propagation conditions (e.g., multipath and NLOS) and ii) the effect of geometry, involved in this type of positioning.

In order to improve performance in challenging scenarios, other works focused on other types of solutions that do not require additional hardware or very complex algorithms, although they may require coordination between devices. These solutions aim to decrease the effects due to multipath on measurements for ranging, improving the positioning results. Examples include the channel hopping mechanism in the standard IEEE802.15.4e and the Asynchronous Power Transmission (Asynchronous Transmission Power (TxPower)) technique [SGW+23, CCLMBR+17]. Another INTERACT contribution addresses this topic in section 2.3.1.

The following subsections present the commented contributions.

### 2.3.1 Combating Multipath Fading in Indoor Positioning with Asynchronous Power Transmission Techniques

Asynchronous TxPower is a novel approach to mitigate multipath fading in IPS by transmitting signals at different power levels [BHL22]. The idea is to exploit the power differences between the multipath components to enhance the signal quality and improve the positioning accuracy [LMCCHC+19]. By transmitting signals at different TxPower levels, the receiver can distinguish the direct path from the reflected paths and reduce the impact of multipath fading [SGW+23, CCLMBR+17]. The proposed technique does not require any additional hardware or modifications to the existing infrastructure, making it a cost-effective and practical solution for IPS applications [TCCCOBGC23].

Once the signals are received at different TxPower levels, signal processing techniques are applied to extract the relevant information for position estimation. This may involve filtering, correlation, and other signal processing methods to separate the direct path signal from the reflected paths. The position of the target object can then be estimated using range-based, e.g., RSSI, ToA, TDoA or AoA, or range-free techniques, e.g. proximity, Fingerprinting or centroid [WXDZ20].

Table 2: Accuracy (%) and Mean Error (m) for each Classification Model. Best Values for Each TxPower Are Highlighted in Bold.

| Model | 0x01 | | 0x02 | | 0x03 | | 0x04 | | 0x05 | | 0x06 | |
|---|---|---|---|---|---|---|---|---|---|---|---|---|
| | **Acc** | **ME** | **Acc** | **ME** | **Acc** | **ME** | **Acc** | **ME** | **Acc** | **ME** | **Acc** | **ME** |
| LoR | 60.50 | 0.933 | 49.30 | 1.270 | 50.80 | 1.433 | 57.20 | 1.111 | 48.90 | 1.605 | 60.30 | 1.104 |
| LDA | 63.80 | 0.869 | 52.00 | 1.162 | 57.90 | 1.309 | 62.60 | 0.937 | 54.30 | 1.387 | 66.70 | 0.912 |
| GNB | 73.90 | 0.644 | 64.80 | 0.835 | 72.20 | 0.885 | 73.60 | 0.674 | 65.00 | 0.991 | 77.70 | 0.649 |
| MLP | 56.00 | 1.078 | 45.30 | 1.190 | 44.00 | 1.650 | 53.20 | 1.191 | 43.80 | 1.765 | 55.40 | 1.131 |
| SVM | 72.20 | 0.772 | 68.70 | 0.766 | 74.90 | 0.783 | 74.10 | 0.753 | 67.20 | 1.056 | 79.20 | 0.636 |
| DT | 73.80 | 0.623 | 69.70 | 0.706 | 72.70 | 0.882 | 75.00 | 0.640 | 64.20 | 1.024 | 78.20 | 0.583 |
| k-NN | 77.50 | 0.555 | 72.70 | 0.702 | 77.00 | 0.660 | 75.70 | 0.594 | 70.00 | 0.958 | 81.10 | 0.525 |
| RF | 79.10 | 0.526 | 76.50 | 0.570 | 79.10 | 0.660 | 79.80 | 0.529 | 71.30 | 0.892 | 83.60 | 0.499 |
| ET | 78.50 | 0.516 | 74.90 | 0.585 | 79.20 | 0.647 | 78.70 | 0.525 | 71.30 | 0.927 | 84.20 | 0.490 |
| **GBM** | **81.70** | **0.426** | **78.80** | **0.536** | **82.00** | **0.561** | **82.30** | **0.463** | **76.50** | **0.714** | **86.10** | **0.384** |
| AB | 80.00 | 0.522 | 75.30 | 0.603 | 80.00 | 0.628 | 78.70 | 0.505 | 72.10 | 0.832 | 83.70 | 0.532 |
| 0.933 | 79.40 | 0.502 | 76.20 | 0.592 | 80.00 | 0.597 | 80.10 | 0.491 | 73.80 | 0.852 | 84.70 | 0.490 |

**Illustrative performance by simulations**

The performance of the asynchronous TxPower technique can be evaluated through simulations and experiments in real-world indoor environments. Key performance metrics include positioning accuracy, robustness against multipath fading, and computational complexity. Comparisons with traditional multipath mitigation techniques can also be performed to assess the effectiveness of the proposed approach [ZCA+21, CCLMBR+17]. To evaluate the performance of the asynchronous TxPower technique in mitigating multipath fading and improving indoor positioning accuracy, both simulations and experiments in real-world indoor environments can be conducted [TCCCOBGC23, LMCCHC+19, CCLMBR+17].

The experimental area, depicted in Figure 35, shows the five beacons labelled as Be07, Be08, Be09, Be10, and Be11 of BLE technology. The beacons are Beacon IB0004-N Plus modules from Jaalee Inc. Four beacons were positioned at each corner of a 9.3m × 6.3m rectangular area, with a fifth beacon placed in the centre of one of the longest edges of the room. The area was divided into 15 sectors, each of $1m^2$, separated by a guard distance of 0.5m. A 1.5m-wide strip was maintained around the experimental area, allowing for better differentiation of the RSSI level of joint sectors when presenting the results. Measurements were taken by positioning the mobile device at the centre of each of the 15 sectors, as detailed below. The minimum distance between a beacon and a receiver was set to 1.5m [LMCCHC+19, CCLMBR+17].

For the purposes of this experiment, beacon devices with varying TxPower were employed. In accordance with the specifications of the five beacons utilised in the experiments, operation at one of eight distinct TxPower levels is possible. Adhering to the specifications, the TxPower levels are designated in sequential order from the highest to the lowest level as TxPower=0x01, TxPower=0x02, ..., TxPower=0x06 (ultra wide range transmission power: 4dBm to -40dBm). Throughout the experiments, multiple measurement campaigns were conducted by establishing the TxPower level of all beacons at the onset of each campaign. Moreover, all measurements were executed under LoS conditions.

**Symmetric vs. Asymmetric Transmission Power Setups**

In the context of indoor location fingerprinting, Supervised Learning Algorithms (SLAs), particularly classification models, are useful for solving the problem. SLAs involve knowing the input parameters and the output, which are used to iteratively train the dataset. Once the model is trained, predictions are made and compared with ground truth values to estimate the model's performance. The classification algorithms can be categorised into a taxonomy of three models:

- **Linear models**: In these models, the target value is expected to be expressed as a linear

Table 3: Accuracy (%) and Mean Error (m) Results for Each Classification Model Using Asymmetric TxPower Configuration. The Values of the Best Classification Model Are Highlighted in Bold.

| Model | TxPower | Acc | ME |
|---|---|---|---|
| LoR | [4–1–2–3–1] | 70.85 | 0.647 |
| LDA | [6–1–3–3–5] | 77.18 | 0.486 |
| GNB | [4–1–2–6–1] | 86.01 | 0.289 |
| MLP | [4–1–4–6–1] | 68.18 | 0.692 |
| SVM | [6–1–6–6–1] | 83.49 | 0.460 |
| DT | [6–1–3–6–5] | 85.95 | 0.289 |
| $k$-NN | [6–1–3–6–1] | 87.78 | 0.264 |
| RF | [6–1–3–3–5] | 89.84 | 0.228 |
| ET | [6–1–3–3–5] | 89.25 | 0.197 |
| **GBM** | **[6–1–3–3–5]** | **91.45** | **0.185** |
| AB | [6–1–3–6–5] | 89.40 | 0.234 |
| VC | [6–1–3–3–5] | 90.43 | 0.200 |

combination of constant values or the product between a parameter and a predicting variable. Algorithms evaluated: Logistic Regression (LoR) and linear discriminant analysis (LDA)

- **Non-linear models**: These models do not make strong assumptions about the relationship between the input attributes and the output attribute being predicted. Algorithms evaluated: Gaussian Naive Bayes (GNB), MLP, support vector machine (SVM), $k$-Nearest Neighbors ($k$-NN) and decision tree (DT).

- **Ensemble models**: These models combine prediction models to improve the strength and performance of the classification model. Algorithms evaluated: Bootstrap Aggregation (Bagging), extremely randomized trees (ET), adaptive boosting (AB) gradient boosting machine (GBM) and Voting (VC).

Therefore, the evaluation is based on various classification models and TxPower configurations. For synchronous TxPower setups, Table 2 presents the localization accuracy and mean error results, respectively. Ensemble models, such as GBM and VC, outperform other models in terms of accuracy and mean error, while linear models exhibit the poorest performance. The GBM model achieves the highest accuracy of 86.10% and the lowest mean error among all models.

In contrast, asynchronous TxPower setups yield significant improvements in both accuracy and mean error. Table 3 summarizes the results of all evaluated Asymmetric TxPower level combinations. Comparing these results with those obtained from synchronous TxPower levels, ensemble algorithms, particularly GBM, continue to demonstrate superior performance. The GBM model achieves an accuracy of 91.45% and a mean error of 0.185m with asynchronous TxPower, specifically, Be07=0x06, Be08=0x01, Be09=0x03, Be010=0x03 and Be11=0x05.

The improvement in accuracy and mean error for asynchronous TxPower setups can be summarized as follows:

- Linear models: 10% improvement in accuracy and 0.500 m reduction in mean error.
- Non-linear models: 9% improvement in accuracy and 0.340 m reduction in mean error.
- Ensemble models: 5% improvement in accuracy and 0.260 m reduction in mean error.

In conclusion, asynchronous TxPower setups outperform synchronous setups in terms of localization accuracy and mean error. Ensemble algorithms, especially GBM, consistently demonstrate superior performance in both scenarios. However, it is important to note that searching for the optimal asynchronous Tx Power level requires the use of metaheuristic search techniques, as a brute-force search is computationally unfeasible due to the exponential growth of the search space. Therefore, employing asynchronous transmission power in an indoor localization environment is a highly powerful technique to mitigate multipath fading and consequently enhance location accuracy.

### 2.3.2 Localization with Data-driven Factor Graph Models Using UWB Radio Systems

Recently, there have been proposals for algorithms utilizing FG to enhance positioning and navigation. As an example, in [WXW+22], an overview of the state of the art in recent years of FG for navigation and localization was presented. In [WPBH21], a FG was proposed for GNSS/inertial navigation system (INS) integration. In [PP19], a FG for positioning was introduced with multimodal Gaussian mixture model to the error distribution.

A FG is a bipartite graph that allows the graphical representation of the dependencies between variables and factors. A brief introduction to FG is presented in Appendix A of [MFP23].

The data-driven FG model presented in [MFP23] was designed to perform anchor-based positioning. The system makes use of the FG to compute the target position, given the distance measurements to the anchor node that know its own position. The aim was to design a hybrid structure (that involves data and modeling approaches) to address positioning models from a Bayesian point of view, customizing them for each technology and scenario. The FG with a BP (or sum-product) algorithm converges to an optimal solution as it is shown in results. The convergence is ensured, since the FG was modeled as a tree (without loops). The FG is a linear model with Gaussian BP (Gaussian aleatory variables), which simplified the calculation of the integrals of the messages between the factors and variables that formed the FG.

The FG model for anchor-based positioning takes into account both the effect of the propagation conditions and the effect of geometry involved in this type of positioning. Therefore, the weighted geometric dilution of precision (WGDOP) metric, which measures the effect on the positioning solution of distance error to the corresponding anchor node and network geometry of anchor nodes, was taken into account. In the presented case, anchor-based positioning techniques that used radio devices, such as UWB, IEEE 802.15.4, were considered.

### Discussion of performance results of Factor Graph Model for Anchor-Based Positioning

The results with collected data were obtained with measurements of the published datasets which were collected from commercial UWB devices. The "Datasets of Indoor UWB Measurements for Ranging and Positioning in Good and Challenging Scenarios" [Mor22] are available at the Zenodo repository and INTERACT database of datasets. The datasets were collected at CTTC's Indoor Navigation Laboratory [Nav]. Measurements related to range and positioning were collected using the MDEK1001 system, which is comprised of DWM1001 UWB modules (DW) [MDE] from Decawave (Qorvo). The nodes in the system are equipped with the DW1000 chip, which is compliant with the IEEE 802.15.4-2011 [std] standard with UWB physical layer. This real-time localization system (RTLS) based on UWB is located in the indoor navigation lab. The lab is a static indoor laboratory environment, as shown in Figure 36.

Anchor nodes (represented by red triangles) were positioned on the walls, while the tag node (represented by a black dot) was placed on a tripod. Both the anchor and tag nodes were positioned at specific reference locations, and the coordinates of these reference positions were determined using a total station. This static indoor lab environment allowed setting up scenarios and changing the main conditions that affect the anchor-based positioning performance: different propagation conditions and different geometrical situation of the anchor nodes. Figure 37 shows the scenarios from [Mor22] and the table with propagation and geometry conditions of the nodes. Thus, the UWB DW nodes were mounted (*i*) at different positions in the laboratory with different geometries, and (*ii*) with different propagation conditions (*e.g.*, multipath and NLOS conditions) that affected the distance error. LoS conditions were considered with no obstacles between the anchor and tag nodes. *softNLOS* and *hardNLOS* conditions were configured with moving obstacles (*e.g.*, obstacles made of cardboard and metal, respectively) between the corresponding anchor node and the tag node in a controlled manner. Some results obtained for some challenging, benign and intermediate scenarios are showed next.

- Challenging Scenarios (A2): the RMSE value of the target position is high (iteration 1), but the algorithms did not diverge and iterations reduced the error (Figure 38). The RMSE of the presented FG ('WLS with FG' in figure legend) was 0.1 m (iteration 4), which was lower than the RMSE of 0.9815 m provided by the DW nodes. It is similar to LS techniques.

Challenging Scenarios (A3): this scenario involves histograms of ranges with bias, an increased standard deviations, and there is a histogram that does not follow the Gaussian distribution curve. In this scenario the tag position was estimated with high error (Figure 38 and 39), but algorithms still did not diverge. It can be seen that the algorithm iterations reduced the error, but the FG algorithm with WLS ('WLS with FG') needed more iterations than the iterative WLS ('WLS' in figure legend) to converge. However, it converged to a solution with the minimum positioning error.

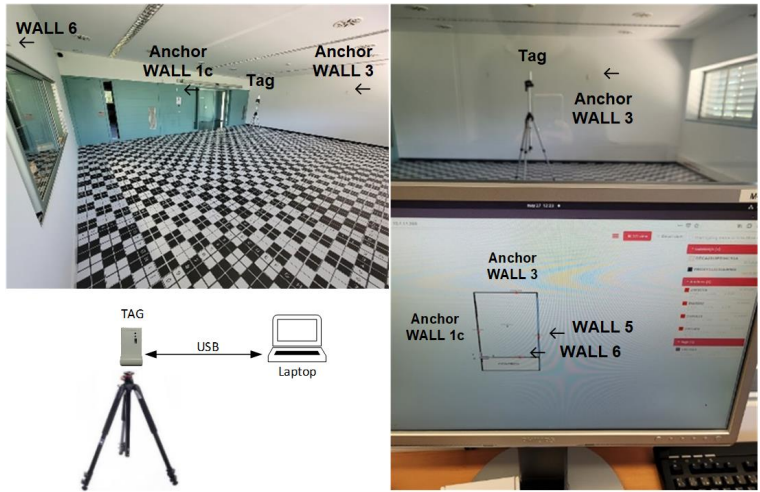

Figure 36: Dataset was collected at the Indoor Navigation Laboratory of the CTTC [Nav]. The pictures show the lab, the tag UWB node placed on a tripod, anchor nodes placed on the walls, and a map of the lab (Scenario B).

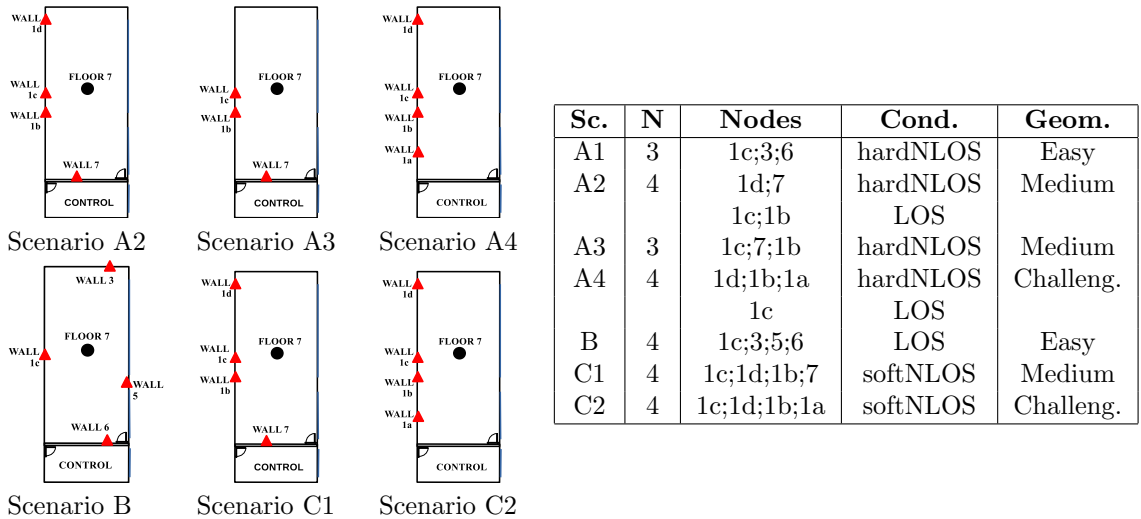

| Sc. | N | Nodes | Cond. | Geom. |
|---|---|---|---|---|
| A1 | 3 | 1c;3;6 | hardNLOS | Easy |
| A2 | 4 | 1d;7 | hardNLOS | Medium |
|    |   | 1c;1b | LOS |  |
| A3 | 3 | 1c;7;1b | hardNLOS | Medium |
| A4 | 4 | 1d;1b;1a | hardNLOS | Challeng. |
|    |   | 1c | LOS |  |
| B  | 4 | 1c;3;5;6 | LOS | Easy |
| C1 | 4 | 1c;1d;1b;7 | softNLOS | Medium |
| C2 | 4 | 1c;1d;1b;1a | softNLOS | Challeng. |

Figure 37: Scenarios (left) from [Mor22] with node placements in the map of the lab. The red triangles are the anchor UWB nodes and the black dot is the target UWB node. The table (right) shows the propagation and geometry conditions of each anchor node in the scenarios. Scenarios A are challenging, B is a benign scenario and Scenarios C are intermediate.

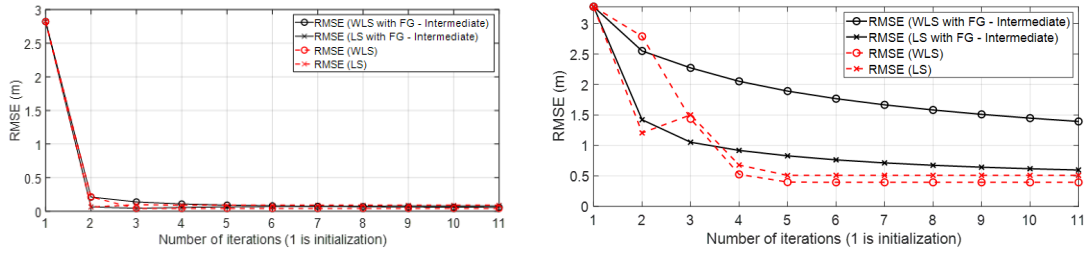

Figure 38: RMSE of the target node position with Scenario A2 (left) and A3 zoom (right).

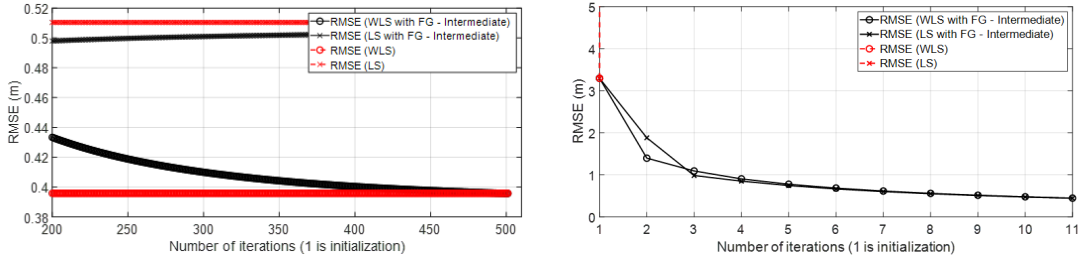

Figure 39: RMSE of the target node position with Scenario A3 (left) and A4 (right).

Challenging Scenarios (A4): the RMSE error of the target node position is high (Figure 39), and the classical algorithms diverged (iterative LS and WLS, 'WLS' and 'LS' in figure legend) but FG-based ones with LS and WLS did not. Iterations reduced the error. The RMSE of the position X-Y with DW system was 0.9730 m, and the RMSE of the position for FG with WLS was around 0.9 m (iteration 4) and 0.55 m (iteration 8).

- Benign Scenario: the results (Figure 40) of the algorithms without FG did not diverge, they were similar to the results of the WLS algorithm with FG. In LoS conditions, the histogram of the distances between each anchor node and the tag followed a normal distribution curve. If the errors follow a normal distribution and the model is linear, the LS based estimators are also the maximum likelihood estimators. This is shown in the RMSE results. The RMSE results were similar: the RMSE values of algorithms with FG ('FG with WLS' in figure legend), RMSE of algorithms with iterative WLS ('WLS' in figure legend), and DW RMSE Position X-Y (0.0172 m). Note that DW solution is based on a maximum likelihood algorithm for positioning.

- Intermediate Scenarios (C1): the results of the algorithms with FG were similar to the results of iterative WLS algorithm and the commercial solution DW.

Intermediate Scenarios (C2): iterative classical LS based solutions diverged, however the FG-based algorithms converged to a solution with a value similar to the DW RMSE value.

As the results show, the belief propagation algorithm converges to an optimal solution. The convergence was ensured, since the data-driven FG model for anchor based positioning was modeled as a tree, a connected and acyclic graph. Moreover, the performance results of anchor-based positioning depend on the propagation conditions between nodes and also the anchor nodes' position. The presented algorithm based on factor graphs took advantage of weighted techniques. It allowed distances with better covariance results to contribute more than the others to the final position estimation. Moreover, the proposed algorithm grouped distances (random variables) to the anchor nodes, on the one hand, to avoid loops in the FG and, on the other hand, so that the position solution of each group was weighted based on its covariance. The covariance results of each group were related with the metric WGDOP. Therefore, the presented factor graph solution considered position estimations depending on the covariance results and taking advantage of Bayesian inference. The results showed that it is an iterative algorithm that converges to a solution, outperforming the classical algorithms and a UWB commercial system. The presented factor graph algorithm for anchor-based positioning takes into account both, the effect on the positioning solution of distance error on the corresponding anchor node and the network geometry of the anchor nodes.

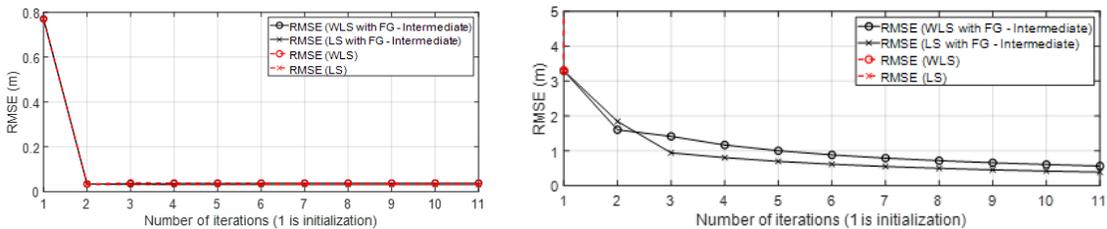

Figure 40: RMSE of the target node position with Scenario B (left) and C2 (right).

The system can be improved in several aspects. One of them is the convergence of the BP algorithm in challenging scenarios such as A3, since the number of iterations is greater for the FG with WLS than for the classical algorithms.

The FG modeling could be be explored using techniques derived from alternative ranging and positioning methods, as well as other technologies and its data fusion. This exploration could consider factors such as propagation conditions, geometry, and other variables that may impact the accuracy of positioning results. Other points of interest to be studied are related to the complexity and resource consumption versus the positioning accuracy of the system.

### 2.3.3 LoRa-based Indoor Localization in the 2.4 GHz Band

LoRa technology is a wireless communication protocol that enables long-range, low-power communication for IoT applications. Developed by Semtech Corporation, LoRa is designed to address the challenges of long-range communication with low power consumption, making it, among others, ideal for applications such as smart cities, industrial automation, agriculture, and environmental monitoring [AYCT16]. LoRa operates in unlicensed RF bands, especially in the sub-1 GHz band (in Europe, 868 MHz). Since 2017, LoRa-based wireless links can also be established in the 2.4 GHz Industrial, Scientific, and Medical (ISM) band [PM20].

In recent years, researchers have turned their attention to studying the performance of LoRa-based communication links in the 2.4 GHz band [JBA$^+$20, PM20, HHDG$^+$23]. Simultaneously, there has been a growing number of studies focusing on LoRa-based localization. Particularly, investigations into indoor localization based on RSSI have garnered significant interest [IIKN19, SS18, KAR$^+$21, PPT$^+$18, KLH$^+$21a].

The study presented in [IIKN19] compared the performance of BLE, WiFi, and LoRa technologies for indoor localization. Measurements were conducted in a long corridor (23 m) and a large open room (25 m × 23 m), encompassing both LoS and NLOS conditions. Localization accuracy for all technologies was evaluated based on packet drop and RSSI. The study found that LoRa achieved the best results. However, unlike BLE and WiFi, LoRa utilized the sub-1 GHz band, which may be one of the main reasons for its excellent performance. Unfortunately, the paper did not provide details about the system used or signal configurations.

Sadowski et al. [SS18] conducted a series of RSSI-based measurements to analyze and compare BLE, WiFi, ZigBee, and LoRa wireless technologies for indoor localization. Once again, LoRa utilized the sub-1 GHz band (915 MHz). The measurements took place in two indoor office environments on a university campus, each with distinct characteristics: a research lab measuring 10.8 m × 7.3 m and a meeting room spanning an area of 33 m$^2$, furnished with tables and chairs. Location estimation results, covering distances up to 5 m, were obtained using the trilateration technique. The study revealed that LoRa exhibited an extended transmission range and remarkably low power consumption. However, its localization accuracy was slightly lower compared to other wireless technologies, with an estimation error of approximately 1.19 m, particularly in environments with significant reflections. Although details about the signal configurations were not provided, the dataset obtained from the measurements is publicly available.

A study similar to that in [SS18] was presented in [KAR$^+$21]. However, it employed only one hardware device, Pycom's Lopy v1.0, which supports WiFi, BLE, and LoRa communication modes. The measurements were conducted in three distinct indoor environments—a graduate lab, a classroom, and a corridor—at a university when only a few students were present in these spaces. The results indicated strong performance by LoRa, which ranked as the second best among the technologies tested, utilizing the sub-1 GHz band. However, the measurements considered only one signal configuration for LoRa, with a signal level of 14 dBm and spreading factor $SF = 7$.

In [PPT+18], a comparison of different RSS algorithms for BLE (2.4 GHz) and LoRa (868 MHz) indoor localization was presented based on simulation and measurement data. The measurements were conducted in a large open hall characterized by line-of-sight conditions, with dimensions of 69 m × 69 m. A LoRa signal with high data rate and a power of 18 dBm was utilized, while both BLE and LoRa anchors were positioned at a height of 2 m. Simulation results indicated superior performance for LoRa compared to the BLE-based localization system, although minimal differences were observed in the measurements. The feasibility of using LoRa for smart home indoor localization at 915 MHz was explored in [KLH+21a]. RSSI values were obtained using the Adafruit Feather 32u4 RFM95 LoRa module. The measurements were conducted in a furnished two-bedroom apartment covering an area of 114.4 m$^2$. LoRa was configured with a signal bandwidth of 125 kHz, $SF = 8$, and a coding rate (CR) of 4/5. The transmit power ranged from 5 to 23 dB, and five tags and three anchors were utilized. A total of 2,000 RSSI values were collected for each of the five tags from each of the three anchors. Offline processing of the RSSI data revealed a localization accuracy of approximately 1.6 m and 3.1 m for line-of-sight and non-line-of-sight conditions, respectively.

As can be concluded, an in-depth performance study of RSSI-based indoor localization using LoRa at 2.4 GHz has yet to be conducted extensively. Next, in the experiments, the focus has primarily been on a single system or signal configuration of LoRa. In many instances, detailed system parameters of LoRa have not been presented. In the next sub-section, results of the RSSI-based indoor localization employing LoRa in the 2.4 GHz ISM Band are presented. The results presented below were published in [SP22]. Link to the obtained dataset is also available in this work.

**Experimental Setup and Measurement**

In this study, we utilized the WiMOD iM282A starter kit to establish a LoRa-based wireless communication link operating at 2.4 GHz. The kit comprises two separate modules serving as the transmitter and receiver, both equipped with the iM828A radio module. The iM828A is a bidirectional radio module that incorporates the SX1280 RF module, enabling the establishment of a LoRa-based wireless link in the 2.4 GHz band. Additionally, it offers support for three different modulation options: LoRa (main), fast long range communication (FLRC), and Gaussian frequency shift keying (GFSK).

Measurement campaigns were conducted in the building of the grammar school in Zdar nad Sazavou, Czech Republic. We selected three indoor environments (see Fig. 41) with varying characteristics. RSSI values were recorded under conditions where both the transmitter and receiver modules were mounted on tripods at a height of 1 meter above the floor level. This height approximately simulates scenarios where a person holds a terminal device in their hands or places it in their pocket. Minimal movement of people was observed during the measurements in all indoor environments.

The first set of measurements took place in the Hall of the grammar school. The hall spans an approximate area of 200 m$^2$. It is constructed primarily of concrete, with brick walls supplemented by supporting concrete columns. The ceiling is made of mineral wool with a paint finish. In the left corner of the hall, there are tall wooden cabinets accompanied by a glass display case. The second indoor environment where the measurement campaigns were conducted is the Locker Room of the grammar school. It spans an approximate area of 300 m$^2$. The walls of the room are constructed of bricks, while the floor is covered with tiles, and the ceiling, along with the columns, are made of concrete. Throughout the room, there are several metal wardrobes arranged in rows. The larger size of the room, numerous obstacles, and favorable conditions for NLOS-based transmission were the primary reasons for including this indoor environment in our study. The Corridor of the grammar school was the third indoor environment. The floor and ceiling partitions are constructed of concrete with mineral wool, while the walls are made of bricks. Additionally, there are wooden doors present, and minimal obstacles obstructing the path. Consequently, there are favorable conditions for LoS transmission.

**Discussion of performance results**

In all measurement scenarios, the carrier frequency was set to 2.45 GHz, and $CR = 4/5$ was employed. The selected values of parameters predominantly represent settings for the receiver with the highest sensitivity -low bandwidth (BW) and high spreading factor (SF) - and lowest sensitivity - high BW and low SF -. These configurations directly impact localization accuracy.

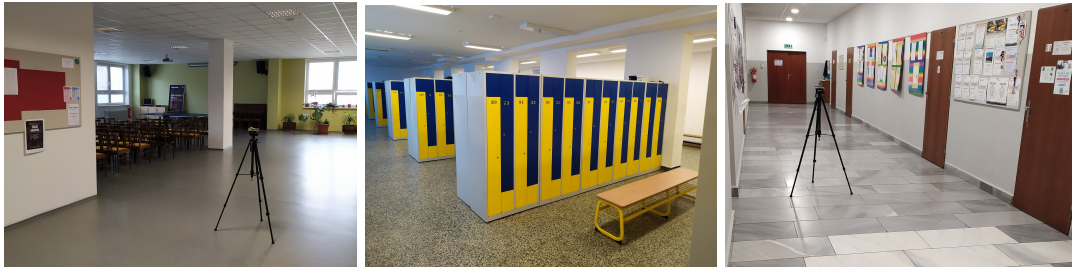

Figure 41: Indoor measurements environments: [left] Hall, [middle] Locker room, and [right] Corridor.

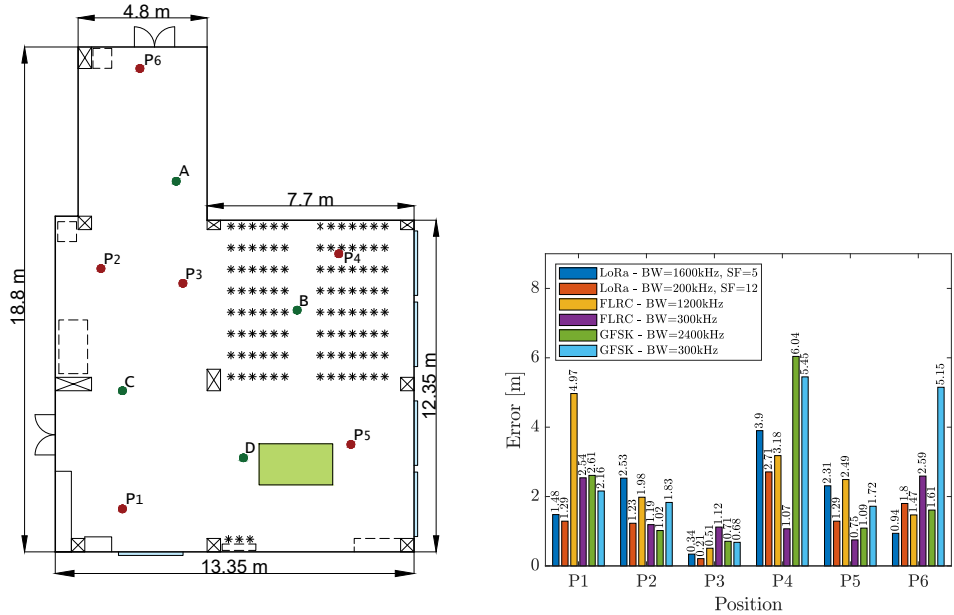

Figure 42: Floor plan of the environment *Hall* (green and red dots mark the positions of Tx and receiver (Rx)) and localization errors in the environment *Hall* [SP22].

Results from the measurements conducted in the environment **Hall** are shown in Fig. 42. These results were derived using reference nodes labeled A,[4.5; 13.8], C,[2.5; 6.0], and D,[7.0; 3.5] (see floor plan of *Hall* in Fig. 42). The horizontal axis represents the positions (P1, P2, etc.) where the receiver was installed. The vertical axis indicates the error between the real and estimated positions of the receiver. The MSE [SS18] was utilized to compute this error.

The analysis demonstrates varied performances of LoRa in localization accuracy across different signal configurations. In the basic LoRa mode, differences between configurations representing the lowest and highest sensitivity levels are generally insignificant. However, increased errors occur at positions P4 and P5 with the $BW = 1600\,\text{kHz}$ and $SF = 5$ configuration, attributed to nearby furniture reflections. The FLRC mode utilizes coherent GMSK demodulation with higher FEC protection, including interleaving [dat]. Two signal bandwidths were tested, with performance comparable to the basic LoRa mode. Lower localization errors are noted at positions P2, P4, and P5 with $BW = 300\,\text{kHz}$. In the frequency shift keying (FSK) mode, utilizing conventional GFSK modulation, the LoRa signal shows lower resistance to multipath reflections, evident in errors at P4 and P6. Nevertheless, its performance is akin to the FLRC mode in other cases. On average, localization errors for LoRa, FLRC, and GFSK modes are 1.67 m, 1.98 m, and 2.51 m, respectively.

The results from measurements in the Locker Room are depicted in Fig. 43. Due to the room's large size and numerous obstacles, the transmitter was placed at five reference positions. The most accurate results were obtained using reference nodes 'ABD,' with coordinates A $[10.7\,;14.2]$, B $[8.0\,;20.3]$ and D $[1.0\,;17.5]$. Localization accuracy is mainly influenced by the LoRa signal configuration and the receiver position. The highest errors occurred at positions P2 and P6, where obstacles are abundant. Low accuracy was observed with $BW = 1600\,\text{kHz}$. Generally, lower $BW$ and higher $SF$ values led to lower errors. We investigated the impact of transmission power on accuracy for $BW = 1600\,\text{kHz}$ and $SF = 12$. Theoretical assumptions held true: lower signal power corresponded to higher error. However, at position P2, accuracy was poorest at $P_{\text{Tx}} = 0\,\text{dBm}$, despite repeated measurements ruling out invalid RSSI data. At $P_{\text{Tx}} = 8\,\text{dBm}$, average errors were 1.45 m and 1.67 m for $BW = 200\,\text{kHz}/SF = 12$ and $BW = 1600,\text{kHz}/SF = 12$, respectively.

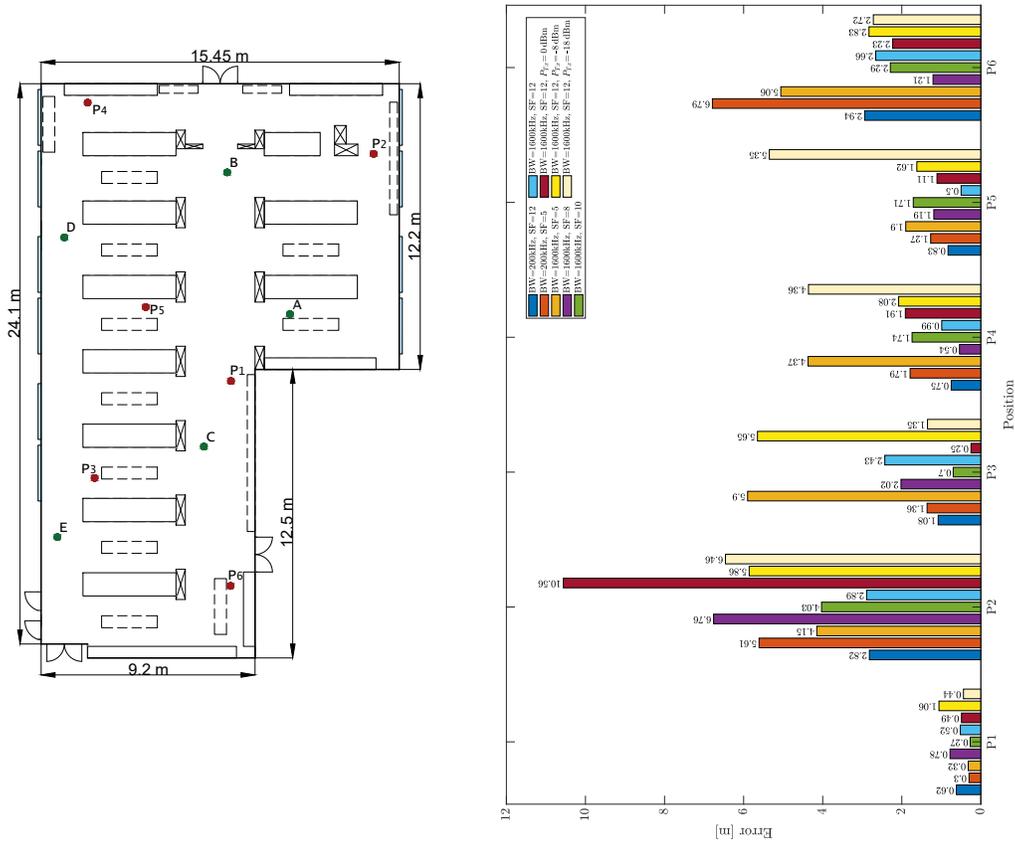

Figure 43: Floor plan of the environment *Locker room* (green and red dots mark the positions of Tx and Rx) and localization errors in the environment *Locker Room* [SP22].

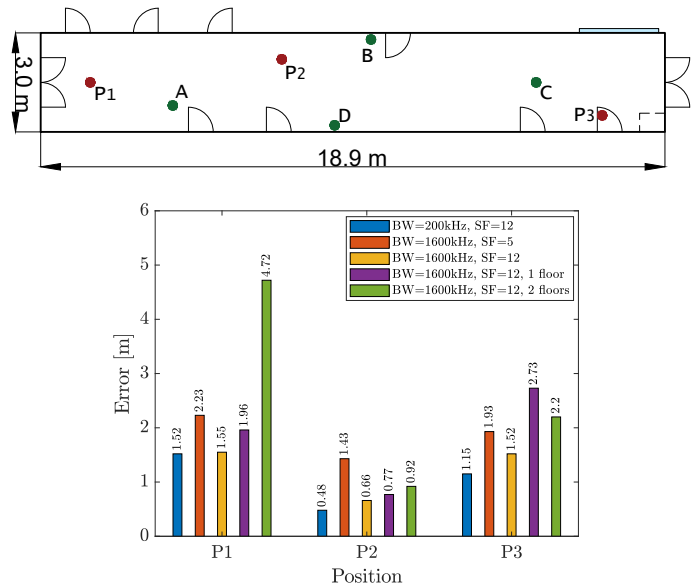

Figure 44: Floor plan of the environment *Corridor* (green and red dots mark the positions of Tx and Rx) and localization errors in the environment *Corridor* [SP22].

The results from measurements conducted in the Corridor are depicted in Fig.44. The transmitter was positioned at four different locations, and nodes A $[2.2\,;4.0]$, B $[0.2\,;10.5]$ and D $[2.8\,;8.9]$ were utilized to determine the receiver position. Two LoRa signal configurations corresponding to low and high sensitivity levels of reception were used. The measurements were split into two parts. In the first part, both transmitter and receiver were positioned in the corridor. The results indicate no significant difference between LoRa signals with low or robust configurations. The variance in localization error between the configurations is less than 1 m, primarily due to the prevalence of LoS conditions in this environment. The second part of the measurements employed LoRa signal with $BW = 1600,\text{kHz}$ and $SF = 12$. Transmitter was placed on the third (highest) floor corridor, while the receiver was placed successively on the second and first floors, maintaining the same transmission environment. Consequently, the lowest localization accuracy was observed at position P1. The average localization errors in the corridor, on the first and second floors, were 1.39 m, 1.82 m, and 2.6 m, respectively.

The evaluation of the results indicates that LoRa operating at 2.4 GHz holds promise as a potential radio technology for indoor localization or tracking applications in the future (e.g., relatively small localization errors for LoS and NLOS conditions [KLH$^+$21b]). However, its performance, particularly in terms of localization accuracy, is influenced by several factors. The signal configuration of LoRa and the chosen operation modes (such as FLRC) directly affect the level of localization error. In environments with abundant NLOS conditions, like locker rooms, it is advisable to utilize a robust signal configuration of LoRa, characterized by a combination of low $BW$ and high $SF$ values. However, it is evident that the LoRa-based indoor localization needs further study.

### 2.3.4 Outdoor Localization Using Signals of Opportunity

Localization based on SOP, which rely on the available signals in the surrounding, are nowadays gaining growing interest [MK21, KKAL22, SKE21]. SOPs are RF signals that are not intended for navigation, such as digital television, analog television (TV), amplitude modulation (AM) & frequency modulation (FM) radio, WiFi signals, global system for mobile communications (GSM), code division multiple access (CDMA), and LTE transmissions. There is potential for incredible signal diversity, in both direction and frequency, when using signals of opportunity.

There are several advantages for using SOP for navigation. SOP can be relatively high power and are able to penetrate buildings. Furthermore, no infrastructure is required to transmit the signals. SOP are already being transmitted for other purposes (by definition), so they are essentially "free" for the localization of a user for navigation purposes.

There are some challenges that faces navigation with SOP. Unlike GPS and other signals transmitted for the purposes of navigation, SOP are usually not designed with localization or navigation in mind. One of the most important factors is timing. In order to use the time of arrival to determine position, the transmission time must be known. However, most communication systems are not time-synchronized to an accuracy of several nanoseconds (like GPS), which would be required in order to navigate without an additional reference receiver. One of the system requirements is that the transmitter locations must be known. Also, multipath and NLOS problems are significant.

Multiple-element antennas can be used to determine the AoA of a signal, and this knowledge from multiple transmitters enables the user to use triangulation to determine their position.

TDoA measurements calculate the difference in arrival time between two different receivers.

### TV as an SOP

Different outdoor localization methods using TV signals as an SOP have been proposed in the literature [YHF08a, YNB14, CJT+15]. TDoA localization technique is often used with TV signals. A typical setup to self-localize a moving receiver using the TDoA technique consists of several transmitters, a reference receiver, and the moving node to be localized [MLD10]. TV transmission has the advantage of a wide coverage area, low-frequency carriers, and large bandwidth [WWC06].

Additionally, TDoA localization requires accurate time synchronization between either the transmitters [NAM12] or the receiving nodes. This requirement increases the computational cost and lowers the localization accuracy [XSYY13]. Thus, it is an advantage to use an alleviated time synchronization scheme such as in [LDRB14, KHLJ14], or to remove the need for it such as in [KC15].

For illustration, we present the measurement results of an outdoor TDoA SOP-based receiver-localization campaign. We employ a reference receiver to collect signals required for the TDoA algorithm. In the applied algorithm, no time synchronization scheme is employed or assumed between different nodes. The SOPs are assumed to be 4 spatially different analog phase alternating line (PAL) TV broadcasting sources. The positions of the reference receiver and the SOP sources are predefined. Only the position of the target receiver is to be determined.

We adopt the signal model of PAL systems B and G. These systems are characterized by a channel bandwidth of 7MHz and 8MHz, respectively. The video bandwidth is 5MHz. The sound is FM and separated by +5.5MHz from the luminance carrier [itu90]. In PAL systems B and G, the TV broadcasting signal consists of frames. A frame sweeps the whole screen display in 40ms. A frame consists of 2 fields, of 20ms each. PAL transmission has 625 lines/frame. Each field starts with a *vsync* signal and each line starts with a *hsync* signal [itu86].

### Setup and measurements

In our system, we have 2 receiving nodes: the target receiver that we localize and the reference receiver. Each receiving node consists of 4 National Instruments (NI) 2930 universal serial radio peripherals (USRPs) connected to a common Octoclock as shown in Fig. 45. The Octoclock allows multiple USRPs to use the same physical clock, thus, unifying their timing. The USRPs interface with a host computer through a switch. Unlike GPS-based localization where the transmitters are synchronized, in our scenario, every transmitter has its independent clock. Instead, we aim to synchronize the 2 receivers without using a time synchronization scheme. This requires unifying the sampling time of different received stations through the Octoclock so that the relative time offset between both receiving nodes is common for all channels. Additionally, employing the reference receiver allows applying TDoA as the PAL signal does not have any information about the transmitter's time or location.

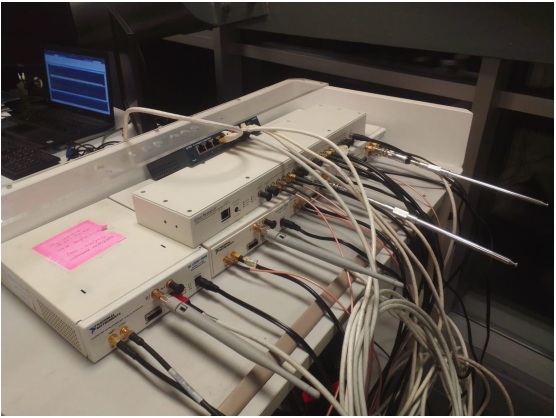

| Parameter | Value |
|---|---|
| $f_{c_1}$ | 479.25MHz |
| $f_{c_2}$ | 551.25MHz |
| $f_{c_3}$ | 679.25MHz |
| $f_{c_4}$ | 687.25MHz |
| 3-dB $BW$ | 4MHz |

Figure 46: The receiver parameters

Figure 45: The setup of a receiving node

To configure the USRPs, we use GNU Radio framework with its graphical interface known as GNU Radio Companion. In the receiving mode, the 2 types of blocks are designed: USRP source and file meta sink. USRP source block interfaces with the USRP hardware and allows to change its configuration. File meta sink blocks save the received signal samples, in addition to some extra information such as the received signal bandwidth and timestamp of the first received sample. The receiver parameters are given in Fig. 46.

The measurement campaign took place in Istanbul, Turkey. The scenario consists of 4 analog TV broadcasters and 2 receiving nodes distributed as shown in Fig. 47. The exact geographical coordinates of the SOP transmitters and receivers are given in Table 4. Due to the relatively low height of the transmitters and the territory shape between the transmitters and receivers, a LoS could be achieved between the receivers and Tx1 only. Thus, we receive the 4 channels from Tx1, then, we apply signal processing on received signals of channels 2 to 4 to have the same effect as if they are transmitted from Tx2, Tx3, and Tx4, respectively.

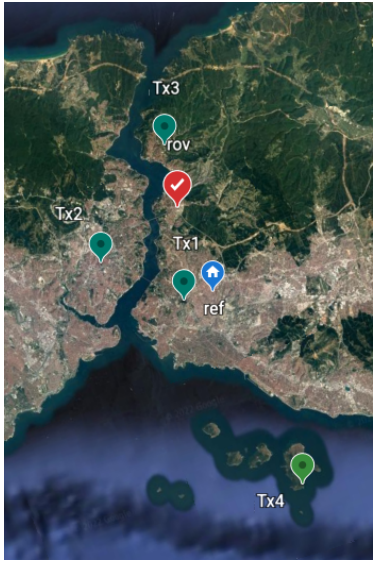

Figure 47: Map of the testing scenario.

Table 4: Geographic coordinates of nodes.

| Node | Coordinates |
|---|---|
| Tx1 | $41°00'59''N$ $29°03'57''E$ |
| Tx2 | $41°04'06''N$ $28°59'31''E$ |
| Tx3 | $41°08'39''N$ $29°06'03''E$ |
| Tx4 | $40°50'20''N$ $29°07'25''E$ |
| rov | $41°05'27''N$ $29°05'30''E$ |
| ref | $41°00'56''N$ $29°05'55''E$ |

The applied algorithm is similar to that in [YHF08b]. First, we detect the fields of received TV channels at each receiving node. Within a window of 1 field duration, we detect 1 field from each channel. Then, we find the local TDoAs between received fields of different TV stations, received at the same receiving node. The local TDoAs cancel out the receiver's common effects such as the time offset between the receiving nodes. After that, we find the difference between the local TDoAs of the 2 receiving nodes. The differential TDoAs remove the transmitter's nuisance parameters such as the transmission time. We apply the localization algorithm to the recorded signals using MATLAB.

For the aforementioned setup, the achieved localization error through detected fields and its histogram are as shown in Fig. 48 and Fig. 49, respectively. It can be seen that the localization error in the X and Y directions is on average below 500m. However, a spike in the localization error happens from time to time as the receiving hardware drops samples. This is mainly due to hitting the Ethernet maximum capacity in the connection between the USRPs and the host computer in one or both receiving nodes.

To improve the localization results, the detected X and Y coordinates can be smoothed using a moving average filter. The smoothing operation decreases the average error, so that the overall localization error can go below a 100m, as shown in the localization error through fields and its histogram after smoothing the detected coordinates in Fig. 50 and Fig. 51, respectively.

The minimum achievable error $E$ in detected TDoA depends on the bandwidth $BW$ as shown in equation (41).

$$E \geq \frac{C}{BW}, \tag{41}$$

where $C$ is the speed of light [KBR97]. Considering the 2 dimensions of X and Y, and the bandwidth, the minimum error is around 105 m. The error histogram show that the majority of the localization error falls within the same tab below 200 m. The error histogram for the smoothed coordinates shows a performance better than the theoretical minimum achievable error.

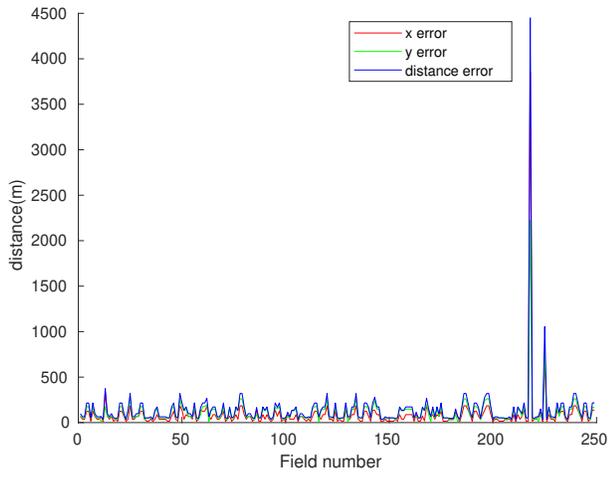

Figure 48: Localization error through fields.

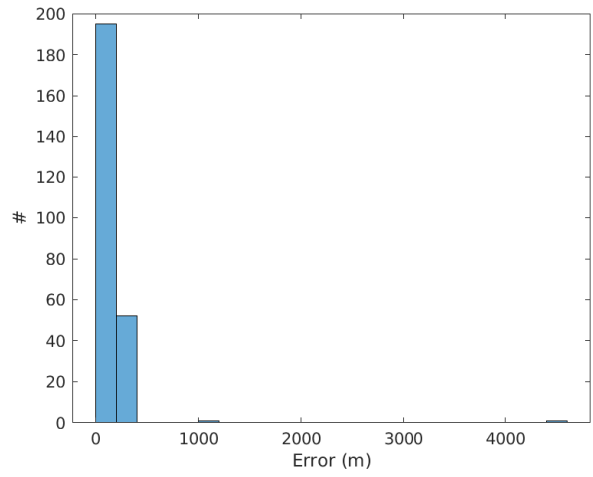

Figure 49: Localization error histogram

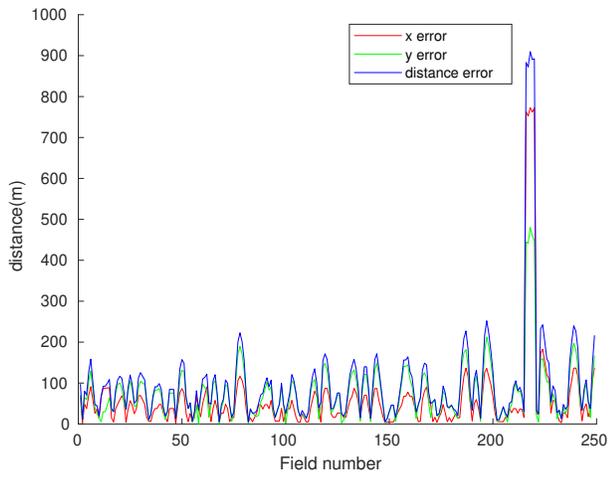

Figure 50: Localization error though fields for after smoothing the detected coordinates.

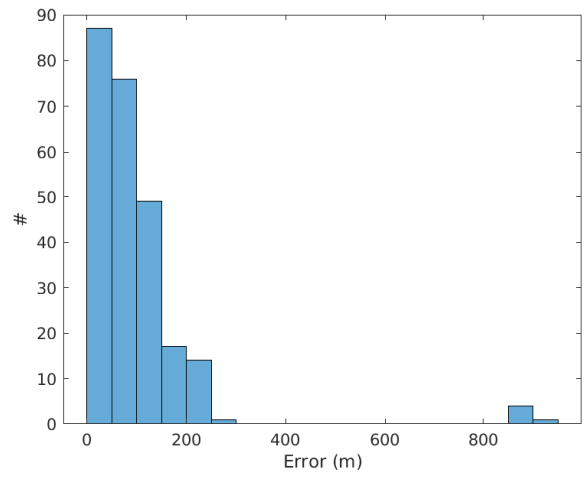

Figure 51: Histogram of localization error after smoothing the detected coordinates.

The ability to self-localize is an essential requirement for many systems. Mostly this is accomplished by using a GPS receiver. TV broadcasting can be utilized for SOP-based self-localization through proper mathematical models. For this work, we collected measurements of PAL analog TV broadcasting to aid the localization model. Specifically, we used these measurements to achieve a TDoA localization of one receiver with the knowledge of the coordinates of four transmitters and a reference receiver. The histogram of the localization error showed that the average error hit the theoretical limit of the employed bandwidth or the limit set by the time resolution of the samples. The accuracy of the localization exceeded this theoretical limit when further signal processing was applied. We claim that an average error of 20 m can be achieved if a bandwidth of 10 MHz similar to that of GPS L1 band is used. Collecting the measurements using a drone can further provide a solution for the drone localization problem in GPS-denied environments. Introducing a tracking algorithm such as Kalman filter shall enhance the achieved accuracy.

## 2.4 Localization utilizing low functional network elements like RIS for beyond 5G and 6G systems

The next generation of communication systems, extending beyond 5G and even toward the emerging concept of 6G, is expected to elevate fundamental performance metrics related to the QoS beyond the current capabilities of 5G system's ultra-reliable low-latency communication URLLC, eMBB, and mMTC [HHA$^+$20, SBC20]. To effectively support unprecedented applications and meet their stringent requirements, cutting-edge methodologies will be integrated with the emerging technologies of the *RIS*, which will pave the way for novel paradigms such as ISAC [WQW$^+$23a, CF23] and the utilization of mmWave/terahertz (THz) communications [JZH$^+$24]. RISs represent a revolutionary technology able to turn conventional structures such as building walls and ceilings into engineered objects outfitted with numerous closely arranged subwavelength-sized reflecting meta-surfaces. Each meta-surface element can be remotely controlled via software and can be dynamically re-configured to enact specific transformations on impinging electromagnetic waves [BDRDR$^+$19]. This capability grants a higher level of control over the radio propagation environment, potentially offering advantages such as mitigating potential LoS blockages in mmWave communications [LHZZ23]. Recent research findings suggest that with precise channel state information, the integration of RIS into cellular networks can lead to significant enhancements in both energy and bandwidth efficiency [WHA$^+$21, WHA$^+$22], especially when joint optimization of the transmit beamformer and the reflection coefficients at the RIS is pursued [GLCL20, WZ19]. On the other hand, it should be considered that RISs possess distinct characteristics that set them apart from conventional antenna arrays: i) they are usually composed of numerous reflective elements, resulting in significantly larger apertures and fostering near-field propagation conditions; ii) RISs are typically fabricated as passive objects (with no RF chains) with low-cost hardware components that face several limitations [DRZD$^+$20] to keep costs at a reasonable level. Nevertheless, recent advancements have witnessed the emergence of active and hybrid active-passive configurations, albeit with increased costs [ZDC$^+$23, HMGZ24].

RISs have been recognized as one of the key enabling technologies toward beyond 5G *localization* and *sensing* since they fundamentally offer two big benefits: they act as VA introducing an extra known location reference and provide additional measurements, independent of the passive, uncontrolled multipath [WHD$^+$20]. Consequently, they circumvent the necessity for prominent reflectors in the surroundings, which are essential for traditional multipath-assisted localization methods. Moreover, RIS-based approaches hold promise for simplified model-based solutions, contrasting the complexity of DL methods. Not least, the smart signal propagation enabled by RISs allow for solving localization problems that were unfeasible in previous mmWave setups [KDA$^+$23].

### 2.4.1 RIS capabilities for enhanced localization

Feasibility studies on the position error bound (PEB) and the orientation error bound (OEB) of a UE demonstrated that localization accuracy can be greatly enhanced thanks to the very high number of geometric channel parameters (including time, angle, phase and Doppler information) that can be leveraged in RIS-aided systems [KKD$^+$21, EGGA21, HRE18]. The relation between measurements and position parameters hinges on the foundational channel model, primarily geometric: each path aligns with a cluster of rays, dictated by the electromagnetic characteristics of the surrounding environment. Essentially, the spatial distribution and electromagnetic attributes of the environment establish a direct mapping between the positional and the measurement domains. Building upon this premise, recent research endeavors have delved into various methodologies. Some research on the topic focused on exploiting RISs to bolster LoS links through coherent combining techniques [WD20, KKSGW21]. This endeavor not only enhances three-dimensional localization accuracy but also augments synchronization efficiency. When leveraging multiple RISs for localization, an intricate aspect resides in separating received signals associated with specific RISs. In addressing this challenge, Hadamard matrices have been proposed as a potential solution, allowing the receiver to discern channel parameters linked with each RIS [KW21]. Furthermore, RIS-aided localization methodologies have been studied in contexts where a controlling base station is absent, as elucidated in [AVW22] and [KSGAW22]. In [AVW22], a two-step positioning strategy is achieved without a controlling base station, employing a single receive RF chain at each RIS. Similarly, in [KSGAW22], backscatter modulation techniques are employed to facilitate localization in scenarios devoid of a controlling base station. Another very interesting finding is that RISs allow for solving very challenging localization problems, such as single-antenna UE localization supported by a single-antenna base station in LoS and even NLOS conditions [RDK+21]. Similarly, in [FCWSG21b] it has been shown that thanks to the use of a RIS, joint *localization* and UE *synchronization* is possible in mmWave multiple-input single-output (MISO) systems supported by a single base station and using only downlink (broadcast) transmissions, whereas in the counterpart of the same mmWave system without RIS only accurate UE localization [FCWSG19, FCWSG20, GLN+20], low-complexity channel estimation [FDMC+22, KFJ+23], and simultaneous localization and mapping [FCWSG21a] are feasible. Further enhanced performance can be obtained when RISs are configured with suitably designed beamforming strategies for optimized localization and synchronization [FKC+22].

### 2.4.2 RISs in near-field localization

When RISs are equipped with a large number of reflective elements, UEs likely experience *near-field* propagation conditions and, in turn, additional location information associated with the spherical electromagnetic wavefronts can be leveraged for positioning [EDB24]. Within a multiple RIS setup, continuous positioning capability can be achieved while simultaneously improving accuracy in near-field NLOS estimation [ASKK+21]. Two low-complexity algorithms based on OFDM and suitably designed time-varying RIS reflection coefficients are presented in [DDGG22], allowing individual elements of a single RIS to act as virtual anchors for UE localization. In [PPJW23], a high-frequency THz communication system supported by a RIS is considered and a suitable strategy based on a down-sampled Toeplitz covariance matrix is adopted to decouple estimation of range and angle information, subsequently used for UE localization. A novel framework operating at two different time scales is presented in [PGAD23] to design optimized multi-RIS reflection coefficients and track the position and velocity of a multi-antenna UE in either far-field or near-field, with a lowered complexity achieved by optimizing RIS configurations less frequently than performing localization tasks.

### 2.4.3 RISs in monostatic sensing applications

RISs have significant applications in monostatic sensing, which is used for target detection and tracking similar to radar systems. RISs have been incorporated into MIMO radar systems, as described in [CASW21], to determine the positions of multiple passive targets. In [KFC+23], a new framework for monostatic sensing performed by UE is proposed, which is suitable for environments characterized by single- and double-bounce signal propagation. The approach involves designing UE-side precoding and combining techniques to separate signals and estimate passive objects using both double-bounce signals via passive RIS (referred to as RIS-sensing) and single-bounce multipath signals directly to the objects (referred to as non-RIS-sensing), employing a mapping filter.

Another approach using an RIS for self-sensing is proposed in [SYM+22]. The RIS controller emits probing signals that undergo direct reflection by the target, called the direct echo link. These signals are then reflected successively by the RIS and then the target called the RIS-reflected echo link. Specialized sensors are deployed at the RIS to capture both the direct and RIS-reflected echo signals from the target. This configuration enables the RIS to perceive the direction of its proximate target.

## 2.5 Waveform design for ranging and localization

Signal propagation delay based positioning methods require precise ToA estimation to perform an accurate ranging. Here we examine the properties of the signal that have an impact on the achieved accuracy. As an example, let us consider a baseband multi-carrier signal

$$s(t) = \frac{1}{\sqrt{N}} \sum_{\ell=-\frac{N-1}{2}}^{\frac{N-1}{2}} S_\ell \, e^{\mathrm{j} 2\pi \ell \Delta f \, t} \tag{42}$$

with bandwidth $B = N \Delta f$, which we generate from $N$ complex frequency domain samples $S_\ell$ assigned to $N$ subcarriers with subcarrier spacing $\Delta f$. In (42) we have grouped the subcarrier frequencies symmetrically around zero.

We define the Cramér-Rao lower bound (CRLB) by

$$\sigma^2_{\mathrm{CRLB}} = \frac{c_0^2}{8\pi^2 \, \beta^2 \, \frac{E_\mathrm{s}}{N_0}}, \tag{43}$$

is often used to evaluate the accuracy of ToA estimation [Kay93]. The CRLB is inversely proportional to the squared equivalent signal bandwidth

$$\beta^2 = \frac{\Delta f^2 \sum_\ell \ell^2 \, |S_\ell|^2}{\sum_\ell |S_\ell|^2} \tag{44}$$

and the signal-to-noise power ratio $\frac{E_\mathrm{s}}{N_0}$ at the receiver. The squared equivalent bandwidth $\beta^2$ as well as the energy $E_\mathrm{s} = \frac{1}{B} \sum_\ell |S_\ell|^2$ of the signal are calculated from its discrete power spectrum $|S_\ell|^2$. $N_0$ denotes the noise power spectral density and $c_0$ is the speed of light.

Particularly at low SNR, an estimator might erroneously pick the delay of the sidelobe of the autocorrelation function of $s(t)$ instead of the mainlobe with non-negligible probability. Due to this behavior, the ToA estimation variance rapidly increases for lower SNR. This threshold effect is not accounted for by the CRLB, which is known to be tight for reasonably high SNR only. The Ziv-Zakai lower bound (ZZLB), however, takes this effect into account. There are several forms of the ZZLB in literature. For our purposes we require the ZZLB for scalar parameter estimation. According to [MO07], the ZZLB for range estimation calculates to

$$\sigma^2_{\mathrm{ZZLB}} = c_0^2 \int_0^{T_\mathrm{o}} \tau \left(1 - \frac{\tau}{T_\mathrm{o}}\right) Q\left(\sqrt{\frac{E_\mathrm{s}}{N_0} \left(1 - \varphi(\tau)\right)}\right) \mathrm{d}\tau, \tag{45}$$

where $Q(x)$ denotes the Gaussian Q-function. $T_\mathrm{o}$ describes the length of an observation interval. The signal propagation delay, as the parameter to be estimated, is assumed to be equally distributed within the interval $[-T_\mathrm{o}/2, +T_\mathrm{o}/2]$. The autocorrelation function in (45),

$$\varphi(\tau) = \frac{1}{E_\mathrm{s}} \sum_{\ell=-\frac{N-1}{2}}^{\frac{N-1}{2}} |S_\ell|^2 \, e^{\mathrm{j} 2\pi \ell \Delta f \tau}, \tag{46}$$

is calculated from the discrete power spectrum $|S_\ell|^2$, which itself is obtained as the absolute square of our frequency domain signal samples in (42).

For signal propagation delay based ranging, there is a trade-off between the estimation resolution obtained from the CRLB and the detection ambiguities between the mainlobe and the sidelobes in the autocorrelation function. For a given SNR, a dedicated PSD exists, which minimizes the mean-square error of ranging.

## 2.6 Testbeds and Demonstrations

Localization testbeds and demonstrations are crucial to unveil the limits of current signal processing technologies to support positioning capabilities, as well as to promote novel research directions to solve the problems discovered on those field tests. Real-world experimentation in representative field environments is indeed necessary to demonstrate the feasibility of the signal processing concepts and algorithms studied for positioning through theoretical frameworks, simulations or controlled laboratory testbenchs. These field experimentations are of special interest to observe unmodelled effects in the analytical, simulated or laboratory scenarios, which then motivate enhanced positioning algorithms. This sub-section presents three relevant examples based on a pioneering flying 5G network testbed for positioning, a self-organised UWB localization and sensing network for planetary missions, and an extensive outdoor and indoor sensor network testbed to validate RF localization techniques.

**HOP-5G Testbed**

The Hybrid Overlay Positioning with 5G and GNSS (HOP-5G) testbed is considered the world's first airborne 5G network dedicated for positioning purposes. The HOP-5G testbed is an experimental and flexible proof-of-concept (PoC) of advanced 5G and hybrid positioning features. This first-of-a-kind demonstrator is based on a dedicated 5G network for on-demand positioning services, by deploying multiple aerial and ground base stations in a target positioning area. The key building blocks of the HOP-5G testbed are the OpenAirInterface (OAI) open-source software, commercial-of-the-shelf (COTS) computing, SDR equipment and real-time software algorithms, as well as custom-built payload platforms for UAV or drones in aerial networks. One of the major developments of this testbed is the real-time experimentation of the 5G positioning reference signal (PRS) with up to 80-MHz of bandwidth from ground and aerial base stations for ToA measurements at the UE, resulting in a key feature of OAI positioning support. Furthermore, this advanced testbed is able to perform stand-alone and hybrid positioning with multiple technologies, i.e., 5G FR1 (at 3.7-3.8 GHz band), 5G FR2 (at 27 GHz), with GNSS and sensor technologies.

The HOP-5G is a predecessor of future 3D network testbeds within 5G evolution and 6G systems supporting native positioning capabilities. Specific network design considerations are discussed in [dPRNR+22a] and to be expanded to enable integrated communication and positioning capabilities in future 5G and 6G networks. Target positioning metrics already consider precise and reliability navigation as key requirements for innovative applications, such as in urban air mobility (UAM) or autonomous vehicles. Certainly, positioning integrity is expected to require advanced and extended testbed operations to assess the impact of the potential faults in the 5G and 6G network solution in the position estimation, as discussed in [dPRNR+22b, RdPR23]. Thanks also to extensive laboratory assessments, synchronization issues between base stations and UE can be characterized over long operation periods as in [dPRNR+22b], in order to propose feasible synchronization solutions for TDoA positioning methods. One of the solutions proposed in the HOP-5G testbed is the use of a positioning reference unit (PRU) to mitigate the base station clock offsets, which is demonstrated in [dPRYK+23, dPRYN+23] to enable sub-meter level positioning even with aerial base stations. In addition, the exploitation of mmWave technologies is also studied within HOP-5G testbed in [dPRYS+24] with a 16-element antenna array, in order to assess the maturity of this technology for AoA positioning. The flexible deployment of flying base stations for positioning purposes, as proven in the HOP-5G testbed and shown in Figure 52, is considered an example of enhanced positioning features within future 3D networks in 6G.

**Self-Organised UWB Localization and Sensing Network for Planetary Missions**

Robotic swarm with portable sensor nodes is an emerging technology for sensing dynamic physical processes both on Earth and in future space exploration missions. These nodes are equipped with radio transceivers, providing precise time and position references without additional infrastructures, such as GNSS. Each node is additionally equipped with environmental sensors, for example a photonic sensor to sense the illumination in caves beneath the lunar surface, a hydrogen sulfide sensor to explore the volcanic activities, or a methane sensor to track organic traces.

At the German Aerospace Center (DLR), compact and robust sensor "eggs" with UWB technology are designed for decentralized position and environmental estimation. These eggs can be easily deployed by robots, and are thus suitable for technology demonstrations in space-analog missions. Details of the UWB sensor egg design can be found in [ZRB+23, BWZ+24].

The sensor eggs transmit UWB signal with exclusive channel access, utilizing a self-organized time division multiple access (SO-TDMA) scheme with a 200 ms cycle. Within this 200 ms, the clock offset and drift cannot be neglected. Therefore, a three-way ranging scheme is applied for obtaining precise distance measurements between the eggs. In addition, novel passive TDoA measurements can be obtained by eggs that only listen [ZRB+23]. These measurements are then fed into a decentralized positioning filter on each egg for position [ZSJ+20, BWZ+24], and optionally orientation [ZRB+23], estimation. The position solution is finally used for a sensing application according to the knowledge of the physical process of interest [BWZ+24].

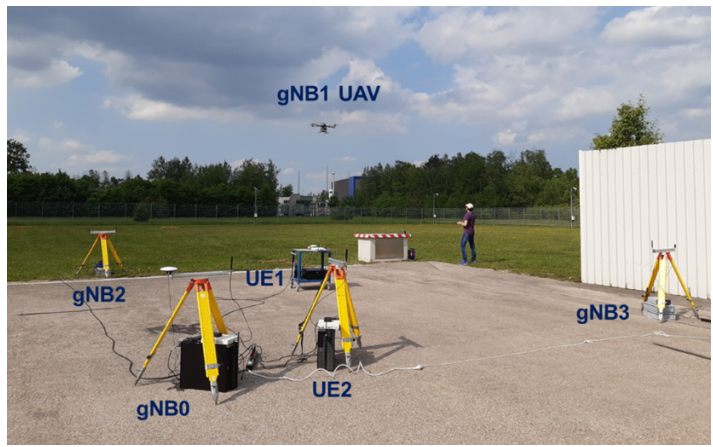

Figure 52: HOP-5G dedicated and aerial 5G positioning testbed [dPRYN+23].

A sequence of DLR space-analog missions have been conducted to demonstrate swarm localization and sensing. In 2022, multi-robot relative pose estimation has been performed on the volcano Mountain Etna, Italy. As shown in Figure 53, four UWB transceivers are mounted on each robot, so that both relative position and orientation of the robots can be estimated, which is shown in Figure 54. In 2023, two measurement campaigns have been conducted on the Vulcano Island, Italy, for volcanic gas source sensing Figure 55, and in a lava cave on Lanzarote, Spain, for cave structure sensing Figure 56. These eggs are also served as educational platforms, assist students to gain intuitions and in-depth knowledge in signal processing for communications, localization, decentralized estimation or sensor fusion, among other applications.

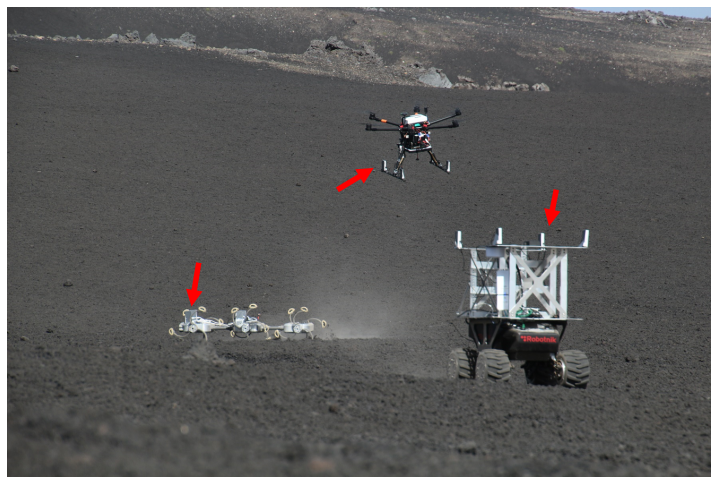

Figure 53: Mission scenario on Mt. Etna robots with attached UWB devices [ZRB+23].

**LOG-a-TEC Testbed**

The LOG-a-TEC testbed [Log, VSG+18] consists of several clusters of permanently mounted radio nodes that are dedicated to experimentation with spectrum sensing, radio communications and RF localization. In addition to permanently mounted nodes, several kinds of mobile nodes or instruments

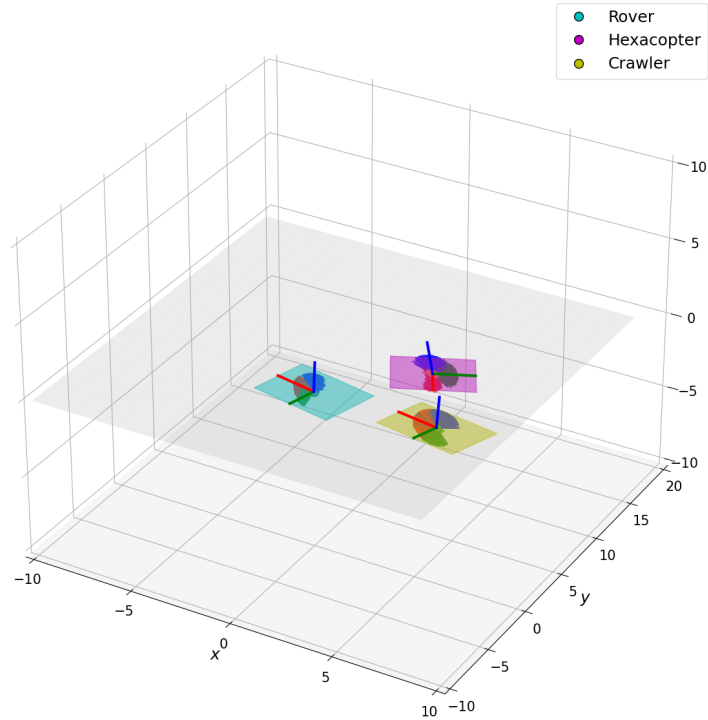

Figure 54: 3D position and orientation estimation [ZRB+23].

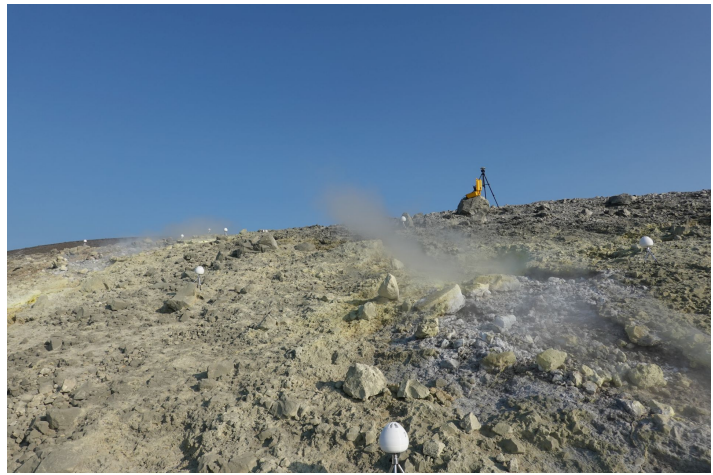

Figure 55: Sensor eggs deployed on the fumarole field of the volcano "La Fossa", Vulcano, Italy [BWZ+24].

can be added to the testbed.

The testbed is located at the offices and park of Jožef Stefan Institute, Ljubljana, and is comprised of 77 nodes, Figure 57, providing the combined outdoor and indoor environment. Each node is equipped with multiple reconfigurable radio interfaces that can be used in various modes. A especially developed embedded wireless node [MHV+21] can host four different wireless technologies and seven types of wireless transceivers as presented in Figure 57. Furthermore, the testbed infrastructure provides a Sensor Management System [MHV+21] equipped with comprehensive tools for node management based on DevOps practices, providing the environment for an iterative research process and fast controlled experimentation.

The number of locations, their density, and the possibility that each location can host multiple radio interfaces enable deployment of the very dense and heterogeneous wireless experiments. In addition, the versatile location of the testbed and modern wireless technologies provide an experimentation area

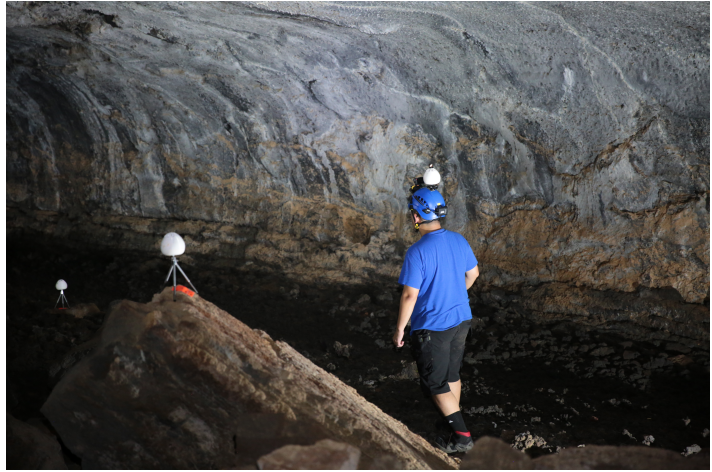

Figure 56: Sensor eggs deployed in a lava cave on Lanzarote, Spain.

suitable for real-world validation of RF localization techniques and algorithms [BMC22, MGHJa].

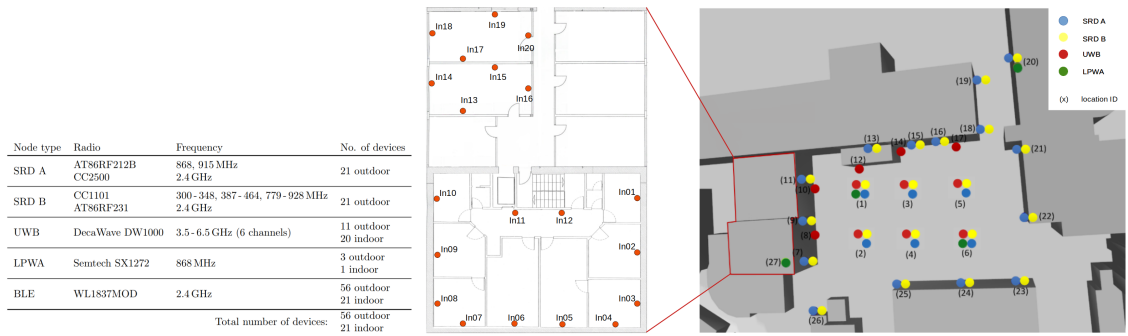

Figure 57: The LOG-a-TEC testbed, with the position of the nodes.

# 3 Integrated Sensing and Communication

This section focuses on the ISAC activities at INTERACT. ISAC systems provide communication and sensing functionalities, either through a fully integrated or partially shared system.

Given the increasing prominence of ISAC as a research topic, particularly in the context of mobile communication and WiFi, it has been designated as a subgroup of working group 2. ISAC covers a wide range of applications, research topics and disciplines. This leads to a rich interest in today's research topics, including signal processing, localization techniques, channel measurements and new modeling methods. As a subgroup of WG2, ISAC also shares channel modeling and measurement topics with WG1. Consequently, this section includes topics that have been discussed in joint sessions between WG1 and WG2.

To enhance clarity, we will first categorize and define the term "sensing," as it encompasses a wide range of applications and research disciplines. This will help avoid misunderstandings in the subsequent technical sections. In section 3.2, we explore waveforms and radio resource allocation methods specifically designed for ISAC applications. Section 3.3 focuses on channel measurements related to ISAC. In Section 3.4, we present new modeling and estimation techniques. Finally, Section 3.5 examines ISAC applications within WiFi and cellular networks.

## 3.1 Introduction/defining ISAC terminology

Communication and radar technology, historically developed separately, exhibit notable differences in waveforms and hardware design to meet their distinct objectives. Recent trends have driven the concept of integrating sensing and communication into a joint system [TAG$^+$19a, WQW$^+$23b, PTHL$^+$21]. One driving factor is the constraint of limited radio bands, coupled with the increasing bandwidth of communications and radar systems. An integrated communication and sensing (ISAC) system combines both functionalities within a joint framework, enhancing radio spectrum and infrastructure efficiency. In addition, the mutual assistance of communication and radar functionalities can support new digital applications, which underlines the importance of ISAC systems.

However, integrating both functionalities into a joint system introduces new technical challenges, requiring additional efforts in signaling and resource allocation, signal processing, and hardware design. The novelty of ISAC, its wide range of applications, and the rapidly evolving research landscape in academia and industry contribute to the ambiguity of its terminology. Terms such as C&S, joint communication sensing (JCS), joint communication and sensing (JCAS), integrated communication and sensing (ICAS), and others were used with slightly different meanings, creating potential confusion within the ISAC community.

Currently, the term ISAC has gained prominence but is often used without specific demarcation, leading to ambiguity and misinterpretation. In that context, we define the term ISAC as follows: The fundamental concept of ISAC describes a system using fully or partly shared infrastructure/hardware and radio signals for communication and Radio sensing purposes. Fig. 58 illustrates both functionalities, the data transmission for communication and the radio sensing to gather information from the environment. In the following, we further categorize and elaborate this term.

**Radio Sensing**

The term *sensing* is employed solely in the context of radio sensing, which entails the use of radio signals to perceive the surrounding environment. Although the term sensing can also be applied to sensors of other types, such as sonar and lidar, these are excluded from this definition. Numerous sensing applications differ in terms of the specific requirements and the methodologies employed. These include detection and tracking, environmental sensing, smart human interaction, imaging and recognition, as illustrated in Fig. 59.

The field of detection and tracking is concerned with the estimation of the presence and localization of a target of interest. In addition, further information, such as the geometrical state vector (including position, velocity, and orientation) or the classification of a target, may be of interest. For this reason, accurate channel models are required. A pure localization process can furthermore be distinguished according to the type of sensing technology employed. In radar sensing, the target of interest does not actively transmit any signals and requires illumination by a transmitter. In contrast, in emitter localization, the target actively transmits a signal for detection and localization purposes. Typical applications of detection and tracking include surveillance, automotive driving, and robotics.

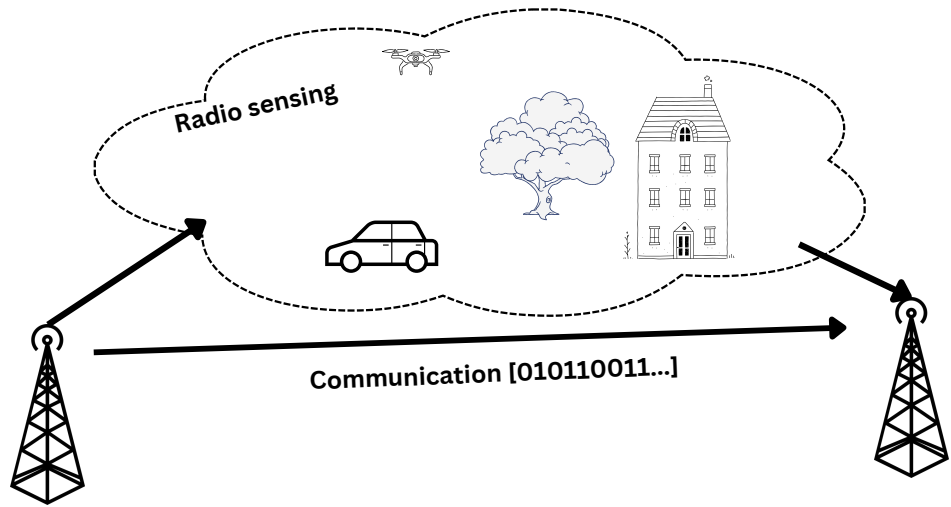

Figure 58: Communication and radio sensing on .

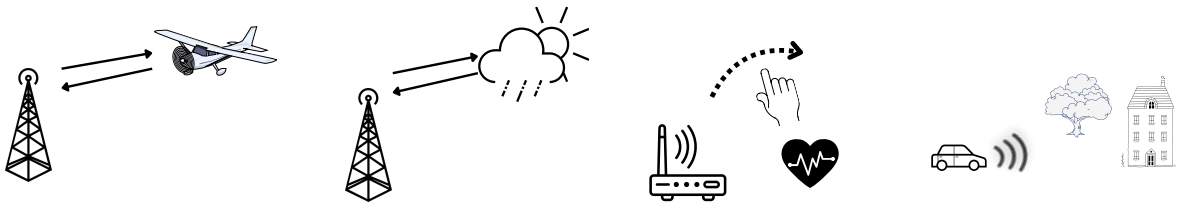

Figure 59: ISAC Applications: detection and tracking, environmental sensing, smart human interaction, and imaging. as illustrated

Environmental sensing employs radio signals to determine the condition of the air, including the presence of dust, as well as to predict weather patterns. Smart human-supporting systems utilize radio signals, e.g., to detect human motions such as gestures, vital signs, or falls. Imaging makes use radio signals to sense the surrounding geometrical structure and subsequently create a digital reconstruction. This technique is employed in the context of digital twins in industry 4.0. Our research has a particular focus on the detection and tracking of objects.

**Integration level**

The integration level describes the extent to which hardware, information, or radio signals are shared between the sensing and communication functionalities within a system. Given the parallel development of radar and communication systems, the degree of integration varies from co-existent to cooperating and collaborative systems, shown in Fig. 60. In *co-existent* systems, separate RF-frontends and distinct waveforms are employed for each functionality, with little to no information sharing. While this represents the simplest form of an ISAC system, interference between the two functionalities raise a significant challenge. *Cooperating* systems also utilize separate RF-frontends and individualized waveforms for sensing and communication. However, in these systems, information is shared between the two functionalities to mitigate interference. *Collaborative* systems are intentionally engineered to accommodate both functionalities. It is characterized by the use of joint waveforms and a unified RF front-end, serving both functionalities jointly. It is important to mention that the integration level is more of a continuum, spanning from minimal to maximal integration, with various degrees in between.

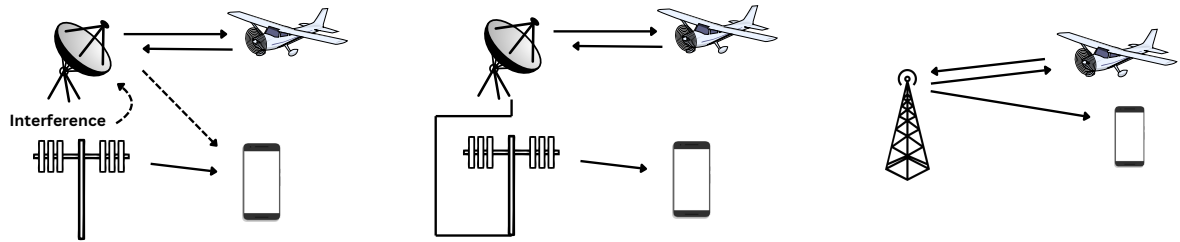

Figure 60: Integration levels: co-existent, cooperating, and collaborative systems

**Underlying system**

The underlying system refers to the original purpose for which a system was designed and pri-marily optimized to fulfill its specific function, either radar sensing or communication, shown in Fig. 61. The underlying system typically exhibits differences in the fundamental transceiver structures and the waveforms utilized, which are dependent on the underlying system. The functionality that the system was not initially designed to accommodate is integrated into the system afterwards. A *Communication-centric* system, initially designed for communication, incorporates radar sensing as an additional feature. Examples include mobile communication technologies like 5G/6G and WiFi. These systems employ waveforms developed for reliable data transmission, at high data rates and to multiple users simultaneously. Common waveforms include OFDM. A *radar-centric* system is initially designed for radar purposes, with communications functionality added later. Since data transmission is not the primary focus, the transceivers of the system and waveforms are optimized to provide a narrow ambiguity function that enables the detections of multiple closely spaced targets. This optimization often includes keeping some basic signal processing components in the analog domain, which reduces transceiver costs. In contrast, communications signals require demodulation of received signals which requires more complex transceiver hardware. Common waveforms used in radar-centric systems include FMCW, pseudo-noise, and pulse-shaped signals. These systems are used, e.g., in industries such as automotive and robotics for tasks such as autonomous driving and surveillance. The challenge for such systems is to utilize these specialized waveforms for data transmission as well. A third category, *co-design*, involves designing a system from the beginning to incorporate both communication and sensing functionalities. The system is designed to optimize performance for both communication and sensing requirements simultaneously, with careful consideration of trade-offs.

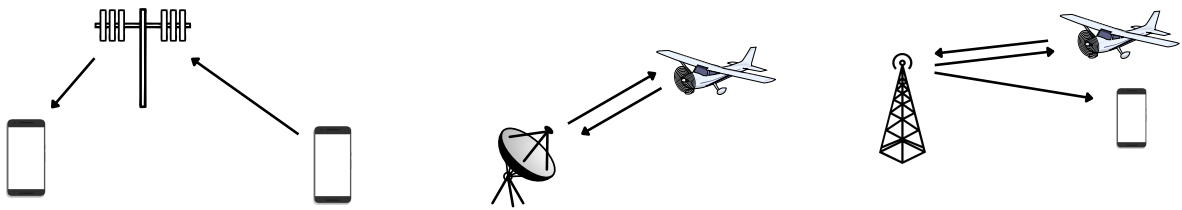

Figure 61: Underlying systems: Communication-centric, radar-centric, and co-design

**Link constellation**

The link constellation for sensing refers to the spatial distribution of the sensing nodes, illustrated in Fig. 62. In monostatic sensing, a single node is employed for both transmission and reception. In bi-static sensing, the transmitting and receiving nodes are spatially separated. In multi-static or distributed sensing, multiple transmitting and multiple receiving nodes are deployed. This different link constellations have advantages and disadvantages. A monostatic system can be mounted in a single location, which means less installation work. However, a monostatic system requires a full-duplex transceiver, whereas a bi- or multistatic system does not. Without any further pre-knowledge, monostatic and bi-static localization require DoA estimation and therefore the use of antenna arrays. In a multi-static constellation, the use of antenna arrays is not necessary for target localization. Instead, localization can be achieved through multilateration. However, synchronization is a challenge due to the spatial separation of the nodes. A distributed detection network can utilize multiple connections and achieve reliable detections even if one link has a low probability of detection, e.g. if the target is moving at a Doppler blind spot. When considering a mobile communication network like 5G, the sensing links can be employed by the downlink, uplink, or sidelink. Moreover, in radar sensing, the target can be illuminated intentionally or unintentionally. This means that for radar sensing, either dedicated sensing signals or reused signals initially intended for communication purposes can be employed. This is referred to as active radar sensing and passive radar sensing, respectively.

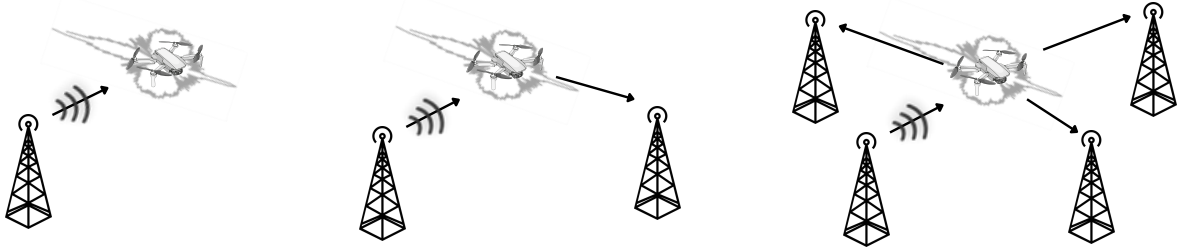

Figure 62: Link constellations: Monostatic, bistatic, and mulit-static

**Potential ISAC System Architecture**

To enable ISAC, from BS point of view, there are a few possibilities for system architecture, with a different level of integration in the air-interface. First, as is shown in Fig. 63a, an ISAC system with TDD communication can be realized by having a JCAS aperture and an auxiliary aperture as the air interface. The JCAS aperture is a shared aperture that could produce multi-beam, some beams for communicating with UEs and others for illuminating sensing targets (or receiving radar echos). The necessity of an auxiliary aperture is due to the long period of packet frames in TDD communication during downlink, whereas the radar requires continuous illuminating of the sensing domain and the receiving of echos, therefore conventional single TDD aperture architecture is not proper for JCAS application [AKM23]. The auxiliary aperture cooperates with the JCAS aperture to achieve radar sensing functionality: e.g., when the JCAS aperture is transmitting, the auxiliary aperture will be in receive mode and receive the back-scattered signals from the target, and while the JCAS aperture is in receive mode (some beams receive uplink coms signal and other beams receive target echos), the auxiliary aperture will transmit and illuminate the target. Additionally, since the auxiliary aperture is only utilized for sensing, the number of antennas can be decreased by sparse or tiled solutions, guaranteeing the angular resolution of sensing [AKM24].

Second, as is shown in Fig. 63b, ISAC system could also be realized by a separated Tx and Rx antenna array, instead of the JCAS and auxiliary aperture. Both the transmit and receive arrays could produce multi-beams to do either coms or sens tasks. As communication often needs a large number of antennas to provide appropriate capacity for the users in the uplink and downlink periods, this solution needs a large number of antennas for both Tx and Rx apertures. This is advantageous even for sensing in comparison to the first solution due to its higher gain and consequently higher signal-to-noise ratio for the sensing, however with the cost of more hardware including antennas and RF chains.

In addition to the above two types of architecture, ISAC system could also be realized by fully shared array aperture, as is shown in Fig. 64, where each array element is supported by PA that also works well as low noise amplifier (LNA) (this is a concept proposed by the Dutch NWO OTP project 3D-ComS, kicked off in fall 2023, with partners from University of Twente and Technical University of Eindhoven). The antenna-LNA/PA is then connected to a switch-able up or down converter; then the IF band signal will be transmitted via optical domain for furthre processing and control.

## 3.2 Resource Allocation and Waveform Design

This section examines challenges and strategies in resource allocation and discusses different waveform types for ISAC systems. In these systems, it is crucial to consider the performance parameters of both communication and sensing. Although both services compete for radio resources, they can provide mutual benefits. Traditionally, waveforms have been designed for communication or sensing purposes.

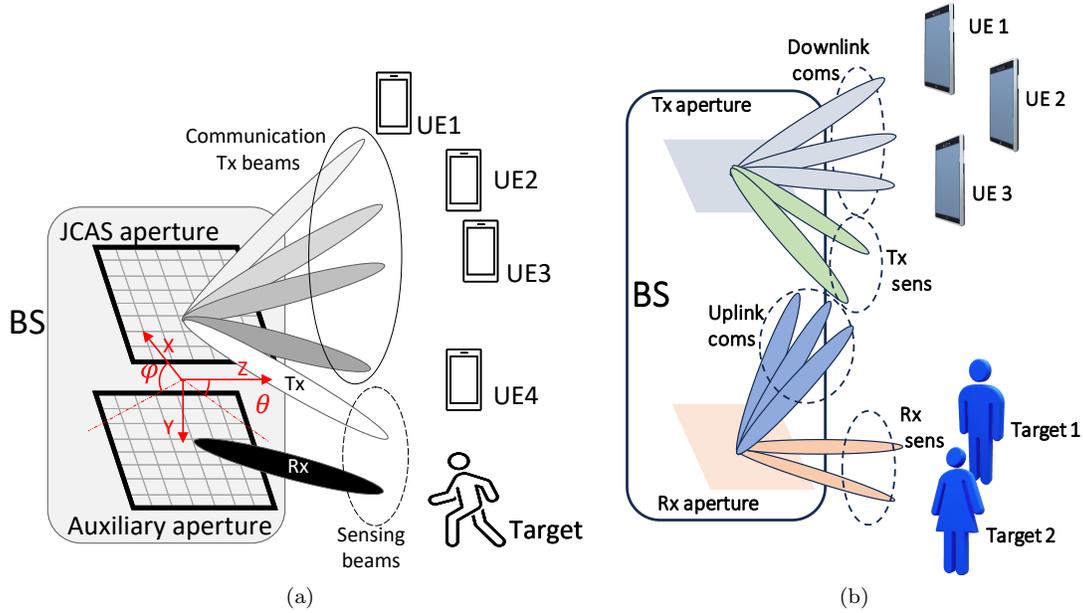

Figure 63: (a) An illustration of an ISAC system with TDD communication and 'duplex' monostatic sensing [AKM23], where the use of auxillary aperture is to assist the JCAS aperture to form 'duplex' monostatic sensing. For instance, when JCAS aperture forms multiple transmitting beams for downlink communication and transmitting sensing beam, the auxillary aperture receives radar echos. (b) An illustration of an ISAC system with separated Tx and Rx apertures, where each could generate multi-beam for simultaneous downlink and uplink communications and radar sensing.

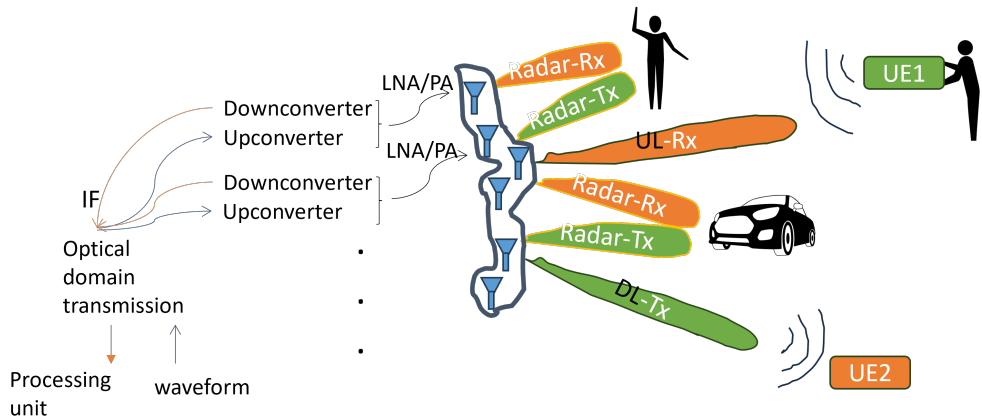

Figure 64: An illustration of an ISAC system with fully shared aperture: each antenna is connected to a unit working as both LNA or PA at the back plane of the array, and then to up/down converters that are switch-able. This architecture is with TDD assumption.

However, in ISAC systems, waveforms must be designed to serve both functions simultaneously. In section 3.2.1 different type of waveforms are categorized. Section 3.2.2 introduces an deep reinforcement learning (DRL) approach to allocate the radio resources in time-frequency and Tx-power. Another DRL approach is discussed in 3.2.3 to reduce the signaling overhead in downlink using radar sensing. In 3.2.4 a sensing approach for blockage prediction is introduced. In 3.2.5 an extension to the 5G-NR resource allocation algorithm is given to also consider radar sensing as a service. The contribution in 3.2.6 introduces an ML approach for resource management in terrestrial networks.

### 3.2.1 Waveform Design

This section offers an overview of the current waveform design for ISAC, categorizing the existing waveforms based on the level of design coupling. Detailed descriptions of these waveforms can be found in Subsection 1.1 of this document. Depending on the integration level, the waveform design for ISAC differs. The design can be categorized into two main groups: loosely coupled technology and tightly coupled technology [XYZT22], which can be described as follows:

**Loosely coupled technology**

Loosely coupled technology can be understood as multiplexing sensing and communications. This involves the independent operation of sensing and communication waveforms, which can be multiplexed based on the orthogonality of one or more dimensions. The typical options are time-division multiplexing (TDM), frequency division multiplex (FDM) and spatial-division multiplexing (SDM).

In TDM, different waveforms are emitted at different times to serve radar detection and data communication functions independently. Some examples of ISAC by TDM can be found in [KK97, MW16]. The advantage of this approach lies in the low complexity of implementation, although radar may experience detection blind spots due to separate time resource allocation for communication.

In the case of FDM, radar and communication signals operate in different frequency bands, allowing simultaneous implementation of both functions in time, as proposed in the literature in [WD07, HEAS17]. However, blank frequency bands between subchannels are required to prevent interference, reducing spectrum efficiency, and nonlinear distortion can lead to intermodulation problems.

Finally, for SDM, radar and communication functions are realized in different beam directions by dividing the radar active phased antenna array into subarrays. This allows simultaneous target detection and data communication without interferences [KEK21]. However, dividing the antenna array reduces the energy available for radar detection, impacting radar performance despite concurrent operation.

**Tightly coupled technology**

The tightly coupled technology refers to a deeper level of communications and sensing integration, and can be further classified into the following three categories according to the design priority and signal formats: communication centric, radar centric and dual-function (joint design). It is important to note that, for every system, this can be further classified into monostatic/bistatic or active/passive radars, which involve different waveform designs.

**Communication waveform-based technology**: This refers to the direct use of communication signals as radar waveforms to achieve target detection. The channel estimation process provides the sensing information, such as delay and Doppler, which can be directly related to the range and velocity of targets. Typically, this involves modifying existing systems originally designed for communications to obtain the sensing parameters. In this category, we include OFDM-based, Chirp-based and Delay-Doppler-based systems, which were also introduced in Subsection 1.1 of this document.

An OFDM-based system consists in obtaining the sensing information from the OFDM waveform used for communications. For example, the work in [LWJJ22] performs sensing with a OFDM-linear frequency modulation (LFM) (OFDM-LFM). The work in [SZW09] analyzes the OFDM-based radar and proposes a novel technique. Most of the existing literature is focused on improving the OFDM waveform characteristics, such as the autocorrelation function (in active radar applications), PAPR, etc.

Chirp-based systems involve waveforms as OCDM [BMAM22], AFDM [RdAC$^+$] and chirp spread spectrum (CSS). Particularly, CSS has been used in radar for some time and is currently adopted for communications by means of LoRa modulation in long range wide area network (LoRaWAN). We highlight the INTERACT contribution in [SGEX], in which the authors present the waveform structure and receiver processing of CSS for sensing, communications and localization. The use of CSS for radar/sensing is based on the transmission of the following waveform

$$s_{\text{CSS}}(t) = e^{(j\pi(\mu[t]_{T_c}) - B)[t]_{T_c})} \tag{47}$$

which stands for a signal that linearly varies its frequency in the range $[-B/2, B/2]$ with a slope given by $\mu = B/T_c$, where $B$ is the signal bandwidth and $T_c$ the chirp period. Furthermore, the operator $[\cdot]_x$ is the modulus-$x$ operator. The key point of CSS and its extended use in radar is the simplicity of the receiver given by the so-called de-chirp process. This process consists of mixing the received signal with a local replica of the CSS waveform $s_{\text{CSS}}(t)$, i.e., $r_{dc}(t) = r(t) s^*_{\text{CSS}}(t)$. The time-delay estimate can be obtained taking the maximum peak of the Fourier transform of $r_{dc}(t)$ [REF]. Similar results are obtained when considering velocity radars in which the received signal is modeled as $r(t) = s_{\text{CSS}}(t) e^{(j2\pi f_D t)}$. In this case is the frequency Doppler that can be estimated with the de-chirp process.

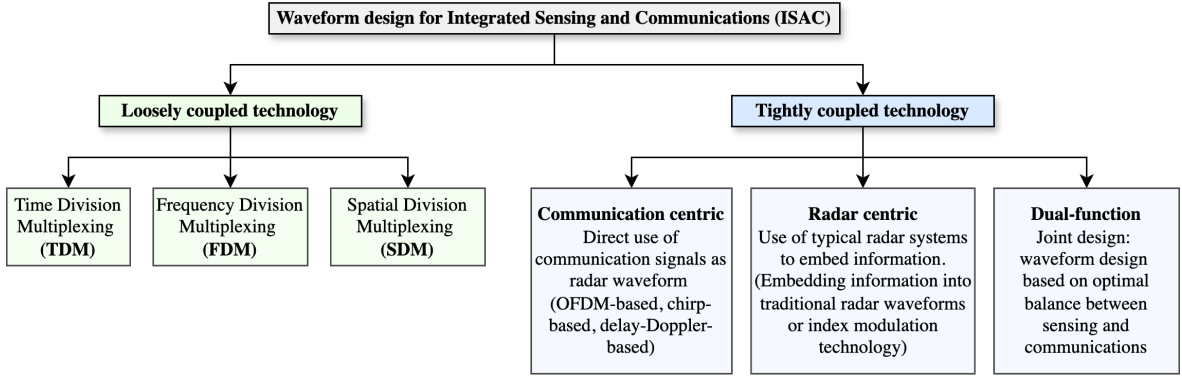

Figure 65: Classification of waveform design for ISAC [WQW+23a].

In the case of CSS for communications, we give a summary of the advantages and drawbacks of different CSS modulations used in the literature for communications:

- BOK-CSS: A digital modulation scheme sometimes considered in practice for chirp signals is the so-called binary orthogonal keying (BOK) for communications [16-ENC]. Note that although considering up- and down-chirp signals, the BOK chirp scheme considered here is different for the one used in the pilot-component [4-ENC]. In this case, the transmitted signal is written as

$$s_{\text{CBOK}}(t) = b(t) s_{\text{CSS}}(t) + (1 - b(t)) s^*_{\text{CSS}}(t), \tag{48}$$

with $b(t) = \{0, 1\}$ the data bit. Note that only one bit per Tc can be sent, which limits the data-rate capability.

- PSK-CSS: A phase shift keying (PSK)-CSS modulation is given by [17-ENC]:

$$s_{\text{PSK}}(t) = s_{\text{CSS}}(t) e^{j\theta(t)}, \tag{49}$$

where $\theta(t)$ stands for the phase corresponding to the transmitted symbol at time $t$. This modulation could be useful to increase the data-rate capabilities w.r.t. the BOK-CSS when considering a higher order modulation than a binary phase shift keying (BPSK).

- FSK-CSS: As long as the initial frequency of the transmitted chirp is not used for other purposes (e.g., user identification), it can be used for data delivery as done in LoRa modulation [13-ENC]. In that way, each symbol in an M-ary constellation is assigned to a given initial frequency, $s_k$. Then, the transmitted signal is written as

$$s_{\text{FSK}}(t) = s_{\text{CSS}}(t) e^{(j2\pi s_k[t]_{T_c})}, \tag{50}$$

for $(k-1)T_c \leq t \leq kT_c$, and the instantaneous frequency given by

$$f_k(t; s_k) = [s_k + \mu t]_B. \tag{51}$$

This option can be useful to increase the data-rate w.r.t. a BOK-CSS and it can provide a better alternative to the PSK-CSS in terms of hardware (HW) impact.

Finally, CSS can be used for localization. For a localization signal, we have to consider the presence of data information, time-delay ($\tau$) and frequency Doppler ($f_D$). The challenges of a CSS-based localization signal come from the traditional de-chirp process used in radar and communications [4-ENC]. Traditionally, the frequency of the de-chirped signal has been used to estimate either time delay, frequency Doppler, or data information. In a localization scenario, all the parameters of interest are embedded in the de-chirped frequency [4-ENC]. This scenario leads us to transmit two components: the pilot and data components. The pilot is used to synchronize the signal. Once the signal is synchronized, the data component is used for the reception of the useful information needed for localization. Second, in the act of synchronizing the signal, we will extract the ranging information needed to localize the user. Finally, it is worth noting that the design of a localization signal must include a multi-satellite access (MSA) scheme.

In Delay-Doppler based systems, OTFS is the most popular waveform for communications-based ISAC, due to the direct mapping of targets in the DD domain. However, OTFS systems suffer from a large pilot overhead to effectively estimate the channel and sensing parameters. To avoid this, overhead-free systems based on superimposing the pilots to the data have been proposed in the literature. However, these works rely on computationally intensive self-induced data interference cancellation techniques. In the INTERACT contribution in [MMCHFGGGA22], authors propose a robust overhead free channel estimation technique based on superimposed training for OTFS for ISAC, which improves the channel estimates and, at the same time, it reduces computational complexity. The work proposes a superimposed pilot-sequence design which allows strategic averaging techniques at the receiver, which enable to filter out the data and noise. The design first obtains the delay and Doppler shifts based on a bank of correlators, and later the channel gains based on a simple two-step process. All the operations can be performed as additions, avoiding multiplications and thus significantly reducing complexity.

**Radar waveform-based technology**: this refers to the use of typical radar systems to embed information. We describe the following approaches: embedding information into traditional radar waveforms and index modulation technology.

When embedding information into traditional radar waveforms, one approach to embedding communication information into radar waveforms involves introducing differences in the radar waveforms and utilizing these differences for communication modulation. Typically based on classical radar waveform LFM signals, communication information can be characterized using parameters such as pulse starting frequency, step frequency, phase, pulse width, and others. For pulsed LFM signals, waveform parameters can be categorized into inter-pulse coding and intra-pulse coding. Inter-pulse coding, achieved through initial frequency, frequency adjustment, and pulse repetition interval, exhibits low coding efficiency and may not meet communication data rate requirements. Intra-pulse encoding, which divides a pulse into subpulses and modulates amplitude, phase, or frequency, is the predominant method with high coding efficiency and resource utilization. However, communication modulation can impact radar detection performance, necessitating modulation schemes designed to mitigate this influence.

Index modulation divides information bits into index bits and modulation bits, unlike traditional systems that map all information bits directly to amplitude or phase symbols. Index bits determine which portion of radio resources, such as antennas, subcarriers, or spread spectrum codes, are activated, while modulation bits are mapped via traditional modulation methods. This technique offers several advantages. Firstly, information transmission via the index requires no additional transmission power, resulting in higher energy efficiency compared to traditional systems. Secondly, activating only part of the resources leads to sparse signals, facilitating low-complexity detection algorithms. Lastly, sparse signals reduce adjacent interference, enhancing index modulation's anti-interference capabilities. In summary, index modulation technology offers low power consumption, low complexity, high energy efficiency, and is well-suited for designing ISAC waveforms.

**Design of dual-function waveform technology**: Apart from modifying the existing communication systems to obtain sensing parameters, or embed information into the radar ones, a further category can be added here: the design of a dual-function waveform. This is typically based on solving an optimization problem to design a waveform with optimal balance between sensing and communications. We highlight the INTERACT contribution in [MYA22]. In this paper, the authors emphasize that the design of the time-frequency waveform for OFDM in dual-function radar communications is essential to meet future communications and sensing needs. This work introduces a novel approach to minimize the Cramér-Rao bound (CRB) for delay and Doppler estimation, thereby enhancing the radar capabilities of an OFDM dual-functional system. Existing methods in the literature aim to improve the CRB, but they often rely on feedforward signaling or subcarrier reservation. However, by leveraging the constellation extension of QAM, it is possible to achieve lower CRB without these prerequisites. Consequently, the proposed method enables seamless communication while minimizing the CRB in conventional OFDM systems. To evaluate the proposed method, simulations consider both CRB and symbol error rate (SER). Additionally, a theoretical SER analysis is conducted to understand the impact of CRB minimization on communication performance.

### 3.2.2 Deep Reinforcement Learning for Radio Resource Allocation in ISAC System

ISAC enables monitoring in smart cities and autonomous driving by leveraging existing mobile communication infrastructure. However, because the ISAC system makes use of shared radio resources, it is necessary to develop new radio resource management (RRM) strategies to guarantee that the performance requirements for radar sensing and communication services are met [TDJ+21b]. The ability to flexibly adapt radio resources for efficient use is a critical requirement for novel RRM strategies. This contribution introduces a schema to optimize the radio resources for a radar service integrated into a communication system by means of DRL, described in [CZC+24] with more details. Fig. 66 illustrates the concept of the DRL mechanism incorporation with the ISAC simulation framework.

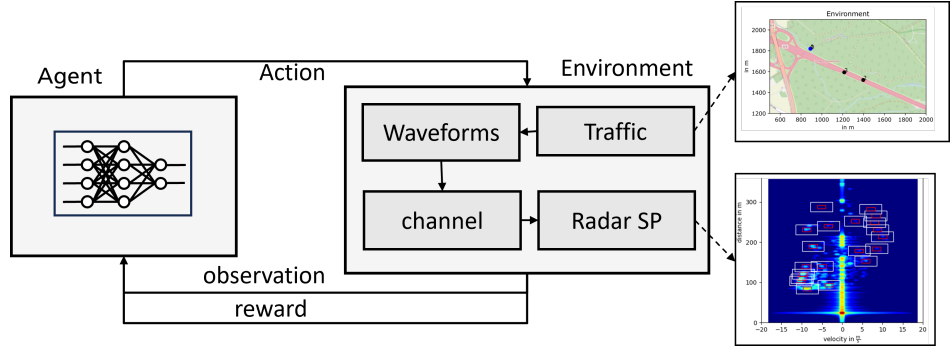

Figure 66: DRL concept with the agent and the environment including the simulation models.

DRL is a suitable approach to solve complex problems such as resource allocation. The DRL algorithm solves problems by a decision policy represented by a neural network that has been trained on its own experience. This approach promises the advantages of real-time capable methods that do not require the collection of labels and are robust against model mismatches.

In literature, many DRL approaches exist for resource allocation. In contrast to existing literature on DRL methods to adapt radio resources, such as [XZS22, TKB+20, LCN+22], our approach emphasizes minimizing the radio resource usage in a communication-centric for a radar sensing service. This efficient radio resource allocation not only enables eco-friendly services, but also conserves resources for other communication purposes.

A potential application for the proposed approach is analyzing the traffic situation to support autonomous and intelligent driving. However, the concept of the proposed algorithm is more expansive than this use case. It can also be adapted to air space monitoring for UAVs, or monitoring fields in smart farming.

**Resource Allocation**

We consider a mobile communication system, including a transmitting and a receiving antenna collocated at the same base station, and OFDM waveforms. The monitored area is a highway and the Radar targets are modeled with bi-static radar cross section (RCS). In addition, we consider scattered path components from static objects as clutter.

We distinguish between two radar sensing modes, illustrated in Fig. 67: the search and tracking modes, which are well-established techniques for radar systems [Sko02]. During the search mode, a predefined pattern of radio resources is transmitted periodically. This mode ensures that all targets entering the monitored area are detected for specific requirements. The requirements depend on the use cases and could include target acquisition time, minimum SNR of the target echoes, or minimum resolution. In contrast, the tracking mode employs an optimized radio resource pattern to efficiently track targets initially detected in the search mode. To track the targets we are using a classic OFDM radar-processing-chain, including the channel estimation by a point-wise division of the received OFDM resource grid with the reference resource grid, the object detection by a constant false alarm rate (CFAR) detector, and the multi-target tracking unit based on a Kalman filter. The signals and the data from the radar signal processing is generated by the ISAC simulation framework described in [SST23]

A comparison can be made between the radar modes and the connection process in mobile communication. In this context, the search mode can be considered analogous to "always-on" signals used for the initial connection of a new UE to the network. In 5G, this is comparable to the synchronization Signal Block (SSB). Subsequently, the tracking mode can be compared to the data transmission, where dedicated signals are used to provide the actual service.

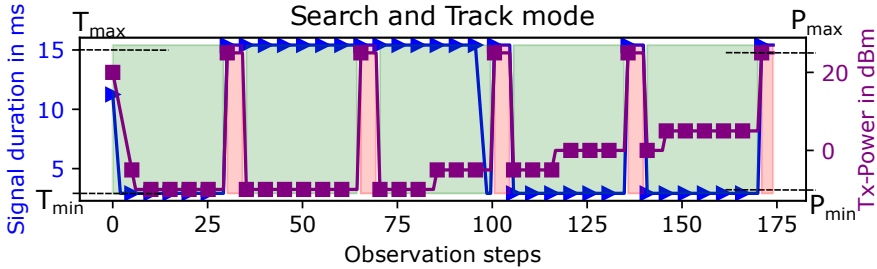

Figure 67: Radar sensing modes: The search mode in green and the track mode in red. Allocated radio resources Tx-power and signal duration.

reinforcement learning (RL) is a branch of machine learning where agents learn to make decisions based on their experience. As defined [SB18a], the agent interacts with an environment by observing its state and acting based on a specified policy. Subsequently, the environment evolves one step forward and provides a new state and feedback called reward. The agent collects the reward and refines its policy. The period during which the agent begins to act within the environment until the end is reached is referred to as an episode. Typically, the episode concludes when the agent either loses or wins the game, or when the environment reaches a specific iteration. For training, the agent expires numerous episodes.

In order to design an environment for resource allocation, we follow the previously described structure. The episode begins after the target search mode detects radar targets inside the monitored area. In the subsequent tracking modes, the agents optimize the radio resources, for a given reward function. The observations are given from the output of the radar signal processing chain. Depending on the use case the relevant parameters, such as SNR or target density, can be extracted from the radar signal processing. The actions in this context are the radio resources, which include bandwidth, signal duration, or transmit power. These actions can be defined as either continuous or discrete. The reward function is the crucial element to let the agent learn a "good" behaviour and needs to be carefully designed. The challenge is to consider all relevant aspects to quantify the value of the agent's action in a certain state. For designing the reward function in the context of radio resource management, We consider the SNR of the target echo, observation losses, and estimated positional parameters.

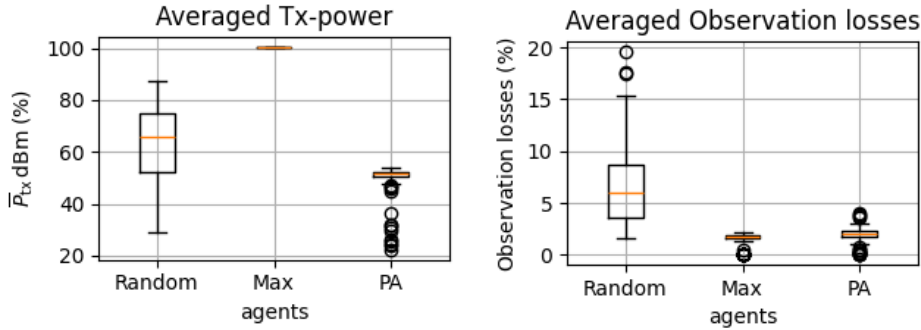

Figure 68: Averaged allocated power (left) and Averaged Observation losses by an agent selecting random actions, max actions, and the DRL agents for power allocation(PA).

To evaluate the trained agent's behavior, we evaluate the trained policy. In the evaluation mode, the agent interacts with the environment over multiple episodes while we collect all quality parameters that indicate the radar sensing service. Since the reward function represents how good a decision was only in one scalar value, it has to be validated that all performance criteria are fulfilled sufficiently. During the evaluation mode, the network weights are not updated. Furthermore, we compare the quality parameters from the DRL agent and further agents that are following a deterministic policy as benchmarks. We consider the deterministic policies of selecting random actions and selecting always the maximum available resources.

In Fig. 68 the performance quantified by the averaged transmission power ('Tx-power' in the figure) and the observation losses are compared of a random acting agent, an agent allocating the maximum available resources (max agent), and the DRL-agent. The DRL agent allocates less power than the random acting agent while having almost the same percentage of observation losses as the Max agent. This shows that the agent was able to allocate the radio resources to achieve an efficient resource usage while reliable detecting the radar targets.

This proposed approach demonstrates the proof of concept. However, in real scenarios the DRL method can reach its potential of performing better than model driven approaches. Further possible features to be investigated could be considering standard compliant radio frames, considering the beam dimension, and multiple-transmission and reception points (TRPs).

### 3.2.3 Beam management in ISAC system

Modern wireless communication taps into mmWave frequency, seeking more bandwidth. This brings a higher data rate for communication and enhances the sensing performance. At mmWave frequency, the high propagation loss can be compensated by leveraging massive antenna arrays and performing high-directional beams. However, narrower beams and blockage sensitivity make beam management (BM) a great challenge. The problem is not only about establishing and maintaining accurate illumination but also about minimizing the system overhead spent on BM [XJM+24].

In recent years, research on ISAC has become increasingly popular, with one of the most important solutions being the communication-centric ISAC. In this approach, sensing will be integrated into the communication systems, reusing communication hardware and signals, and sharing radio resources. The merge of two functions aims to reduce hardware costs and improve spectrum efficiency. However, different requirements and signal processing techniques may cause communication and radar to constrain each other [ZRW+22]. This conflict particularly occurs at mmWave frequencies, where BM becomes critical.

BM in modern communication systems is primarily designed based on communication protocols, focusing on minimizing overhead and latency. However, sensing has distinct demands, performance constraints, and signal processing techniques, making direct integration into a communication system challenging. Therefore, a unified beam management scheme is essential for harmonious integration.

In the next two sections, we will analyze the ISAC BM problem from two perspectives based on the type of objectives: cooperative targets and non-cooperative targets. Cooperative targets usually refer to communication UE and have the ability to receive signals and estimate channel states, while non-cooperative targets typically only act as reflectors. Because of their different capabilities, different channel estimate techniques are required, leading to different demands on radio resources and optimization approaches.

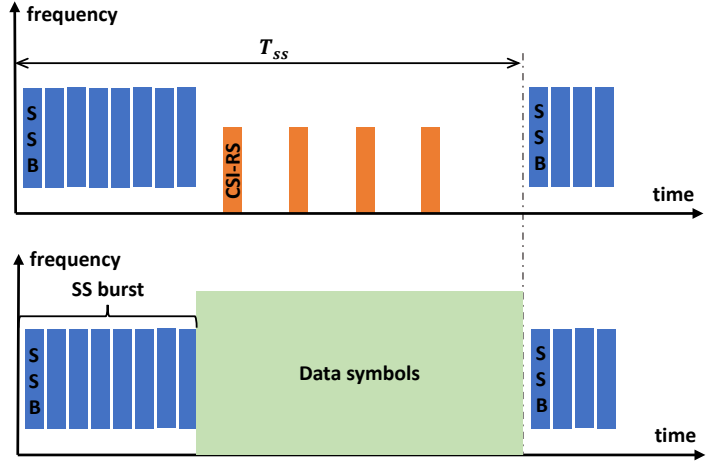

Figure 69: 5G NR frame structure vs. proposed downlink BM frame structure.

**Beam management for cooperative targets with DRL**

In a communication system, beam management is defined as the operation of establishing and maintaining the directional link between the base station and users with optimal beam pair. This operation is categorized into two phases: initial access (IA) and beam tracking [GPR+19]. IA aims to build up initial physical connection links between idle users and gNodeBs [GMZ16], while beam tracking not only includes adapting beam for mobile users but is also responsible for handover, path selection, and failure recovery.

In 5G NR, several reference signals, such as the synchronization signal (SS) and channel state information - reference signal (CSI-RS), are designed for downlink measurements. These reference signals can be transmitted on different beams from the base station, allowing UE to receive them and estimate beam strength. Once the UE identifies the optimal beam, it reports this information to the base station to align the beam [Da23]. This estimation and feedback mechanism enables the communication system to align beams with low resource costs.

Even though reference signal-based beam alignment is effective, it still requires optimization for several reasons. Firstly, 5G NR CSI-RS is user-specific and requires a dedicated set of resources to prevent interference, which increases overall costs in multi-user access scenarios. Additionally, a typical CSI-RS codebook consists of hundreds or thousands of beamforming vectors; searching for an optimal beam within such an extensive space incurs significant overhead [DHJ23]. Moreover, even after a connection is established, reference signals must be transmitted periodically to refresh channel estimates and adapt the beam to track the moving UE [Da23].

To reduce overhead, we propose a radar-sensing-assisted BM approach. During initial access, wide beam selection is still accomplished by transmitting SSBs and estimating beam strength on the UE side. After that, the CSI-RS feedback loop, typically used for beam refinement and tracking, is replaced by radar sensing. This step involves transmitting the user's data signals through the selected wide beam, then having the base station receive the echoes backscattered from the UE and perform radar signal processing. Channel parameters, such as Doppler shift and DoA, are estimated for beam refinement. Consequently, further data transmission is conducted on the refined beam.

In our proposed method, once the initial connection is established, the continuously transmitted data signals can be reused to update the channel states for beam tracking. Additionally, we employed a DRL agent to process channel estimates and adapt beams. DRL is particularly effective for sequential decision-making and, being data-driven can mitigate the effects of model mismatches compared to the Kalman filter. In our approach, the DRL agent observes the estimated channel parameters and analyzes them in conjunction with the channel history, subsequently selecting one of the predefined beams for transmitting the next data frame.

Table 5: Performance comparison of beam tracking

| Policy | DoA estimates | DRL agent | Random policy |
|---|---|---|---|
| Mean reward | 1 | 0.81 | 0.36 |
| Mean SNR (dB) | 36.9 | 29.5 | 20.5 |

Using the proposed method, we conducted simulations and trained a DRL agent for beam tracking. Our simulations demonstrate that the CSI-RS feedback loop can be replaced by radar sensing without compromising accurate beam alignment. This allows radio resources typically allocated for reference signals to be repurposed for data transmission, thereby improving spectrum efficiency. Additionally, as shown in Table 5, the trained DRL agent outperforms a random policy in terms of episode reward and mean SNR, highlighting its potential for effectively handling beam management tasks. However, compared to DoA estimate-based beam alignment, DRL tends to be less accurate, leading to lower SNR. This could be due to flaws in the design of the DRL algorithm or imperfect selection of hyperparameters.

A promising extension of this study would involve quantifying the reduction in overhead and comparing it with the reference signal-based method. Furthermore, conducting a comparative analysis between the DRL approach and the Kalman filter-based approach would provide valuable insights for future research.

**Beam training for non-cooperative targets**

As mentioned in the previous section, cooperative target can be perceived by base stations with the help of reference signals. On the contrary, non-cooperative target detection requires passive radar sensing, and in mmWave frequencies, beam sweeping is the most direct way to cover a spatial area and capture emerging targets.

Unlike communication channel estimation, passive radar sensing relies on coherent processing and deals with the echoes backscattered from the target [TAG+19b]. To guarantee a sufficient signal-to-noise ratio (SNR) and resolving capability, radar signal often occupies more bandwidth and time duration. However, since beamforming is normally employed in a digitally controlled analog or hybrid way in high frequencies, resulting in only one beam can be formed at a time [AKS+18]. Therefore, in beam sweeping, multiple beams and their specific data frames have to be transmitted in chronological order. These two facts lead to significant communication overhead and latency, opposing sensing to communication requirements.

To address the challenges outlined above, we introduce a dual-band sensing approach. The beam sweeping is replaced by a two-step sensing approach: target acquisition and fine detection, which combine high and low frequencies for different goals. As shown in Fig. 70, in target acquisition, a base station working at sub-6 GHz is employed, aiming for larger coverage but coarse estimation. Then, the estimates can be used to narrow down the sweeping space in mmWave frequency, where more bandwidth is available for high-resolution detection. The combination mitigates radio resources spent on beam sweeping, reducing communication overhead and latency. Ongoing research and simulations are actively being conducted, and we anticipate sharing more comprehensive results soon.

### 3.2.4 Guard Beam

Human blockage has impact on the highly directive mmWave communications. The human body attenuation is significant and can be as high as 28 dB for 26 GHz and 34 dB for 39 GHz [ZWL+17]. The interruption time due to human blockage could also last for more than 200 ms when they pass through the LOS path of a mmWave link [SSA+18]. Utilizing other sensors like Camera, Radar, Lidar, or even sub-6 GHz band radio system could help predict the human blockage; but these solutions need additional devices in addition to the mmWave communication infrastructure. Alternatively, utilizing existing communication-centric waveform and architecture gets feasible, as sensing as a service is envisioned in 6G. Allocating communication infrastructure resources for sensing can also preemptively detect potential mmWave link obstructions in the surrounding environment. With this sensing capability, we could assess the potential blocker and implement pre-cautions when necessary, e.g., handover, waiting, re-steering, etc.

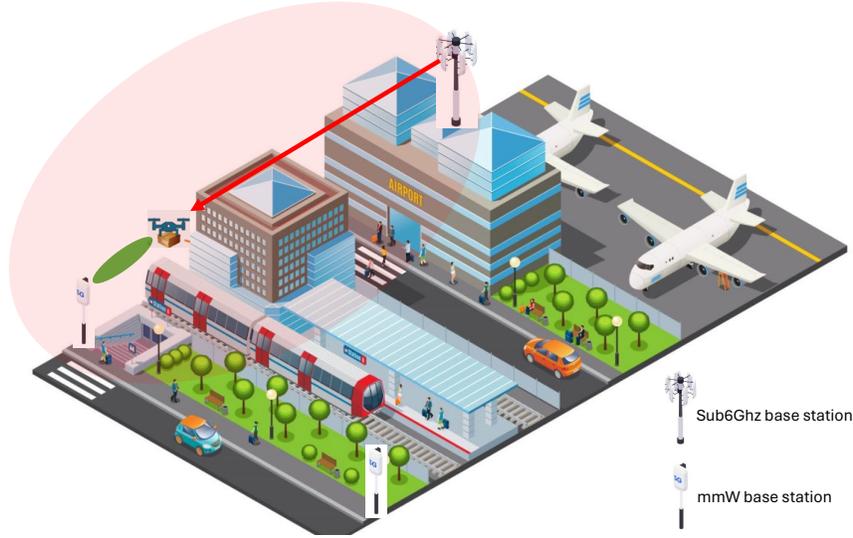

Figure 70: Dual-band sensing scenario. (JUST AS A PLACE HOLDER)

The key function of allocating resources for sensing in mmWave communication infrastructure is to achieve 'early detection' of potential human blockers, especially near UE. 'Detection' is to illuminate the target by radio waves, and then the target will influence the time-frequency-spatial domain radio signals. Conventionally, the illumination of target in surrounding environment could be done by using in-band communication beam depending on its beamwidth. 'Early' indicates the proactive prediction of occurrence and/or length of blockage; such prediction could be achieved by using power signatures of received signal at link ends, or could be based on machine learning, e.g., meta-learning using prior statistics [KSP21], or deep learning of power signature with limited extend to future [WACA22], or liquid time constant network using ordinary differential equation [NYSM23].

In [HMP22], and as is shown in Fig. 71, we propose to have a dedicated in-band guard beam for early detection of potential blockers to mmWave communication link. The aim of guard beam is to protect the area around the main communication beam. Aforementioned conventional in-band communication beam detection [WACA22] results in a limited prediction time; our proposed guard beam allocates additional Rx beam steered at a certain angle from the main beam; we can control the field of view of the guard beam to expand the detection range for early blockage indication. The guard beam could be implemented in only one side of the link, e.g., Rx side, at the cost of one additional RF chain and use of antennas.

We have both simulations and measurement for analyzing different design parameters of guardbeam and its resulting detection range. It's found that 1) widening the guard beam's half power beam width (HPBW) does not improve the detection range as effectively, compared to the enhancement of the separation angle between guardbeam and main communication beam. 2) we can not expand the separation angle as large as we want, due to the misalignment of the radiator and the guard beam; there is a limit on the largest separation angle.

Guard beam is a simple way of sensing beam allocation in ISAC. There are multiple other active approaches of beamforming utilized commonly in advanced radar/communication systems and could also be used for ISAC. For instance, steerable multi-beam, or hierarchical adaptive beam scanning. The hierarchical beam scanning, compared to the exhaustive beamscanning, could sense the target (or potential blockers) more efficiently. It's our future work to adaptive optimize our sensing beam resources for the ISAC scenarios.

### 3.2.5 Resource Allocation in ISAC-capable 5G-V2X Sidelink

The advancement of wireless mobile communication technologies in the ITS has led to the development of cellular vehicle-to-Everything (C-V2X) communication, which facilitates the exchange of information between vehicles, infrastructure, and other roadside units. This sharing of information paves the way for safe, efficient, and comfortable driving through standardized vehicular technologies (e.g., LTE-V2X and NR-V2X) [BBC+21]. C-V2X supports two modes of communication, namely direct or V2V communication and network-assisted or V2I communication [ZBW+22]. In direct communication, vehicles utilize the proximity-based services (ProSe) Communication at 5.9 GHz (PC5) interface to communicate with one another directly without relaying over the network. This mode of communication is called *Sidelink* mode. Sidelink communication in 5G-V2X has further two transmission modes: Mode 1 and Mode 2. In Mode 1, the serving base station (e.g., gNodeB) provides scheduling decisions that the vehicle-UE follows within the network coverage, whereas in Mode 2 the vehicle-UE relies on the scheduling decisions made on its own using the sensing-based semi-persistent scheduling (SB-SPS) algorithm outside the network coverage. The reader is referred to [GMGB+21, CMR+19] for in-depth details on sidelink communication including its physical layer structure, resource allocation mechanisms, and quality of service (QoS) management.

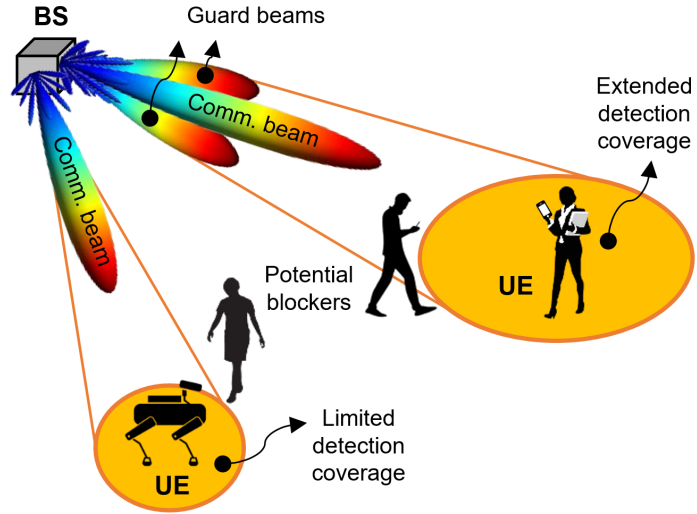

Figure 71: Illustration of the concept of guard beam

**ISAC-capable 5G-V2X Sidelink**

Till now, communication and radar sensing are operated separately in two physically distinct modules. The combination of sensing and communication into a single system is known as Integrated Sensing and Communication (ISAC) [TDJ+21a]. As 5G is becoming more and more common in the commercial automotive sectors, there are constant discussions about enhancements and improvements in the 5G-V2X sidelink communication. One of the new opportunities is the inclusion of radar sensing capabilities into sidelink communication where sidelink signals will not only be used for communication among vehicles but will also be exploited to detect other vehicles, acting as targets, in the vicinity to enable an ISAC-capable sidelink system. This concept is shown in Fig. 72.

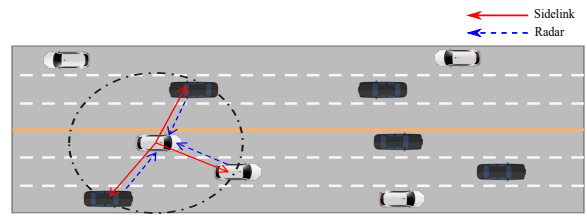

Figure 72: ISAC-capable Sidelink 5G-V2X Highway Scenario

The ISAC-capable sidelink system has distinct performance requirements for radar sensing and communication. The radar sensing requirements include high range and velocity resolution, low detection error, and large sensing range. For communication, low latency, high reliability, and high throughput are considered the required performance metrics. Since the radar and communication systems have different requirements, they need different radio resources to operate effectively. This means that in a shared radio resource pool, the ISAC system needs to allocate the radio resources in a way that balances the needs of communication and radar sensing.

There have been a few studies related to the resource allocation in ISAC-capable sidelink system in recent literature [BDM22, GDB+23, DBB+24]. Wherein the authors evaluated the impact of sidelink Mode 2 resource allocation on radar sensing performance by allocating the same radio resources for both sidelink communication and radar sensing. However, due to the distinct performance requirements of radar and communication, it may not be efficient to allocate the same set of resources for both tasks.

**Proposed Approach**

To address this issue, the authors in [SMSR+23] proposed an efficient radio resource allocation scheme based on SB-SPS keeping in mind the distinct performance requirements of radar sensing and communication in a high-density road traffic use case. The modified SB-SPS enables resource allocation for both communication and radar sensing and is shown in Fig. 73. The proposed scheme consists of three steps: 1) Resource Calculation, 2) Resource Sensing, and 3) Resource Selection.

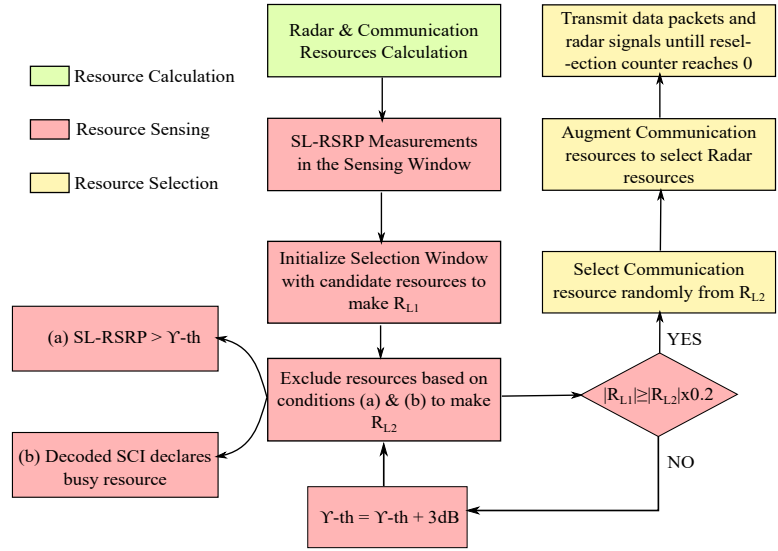

Figure 73: Workflow of Radar-enabled resource allocation algorithm

*Resource Calculation:* First, the vehicle computes the number of OFDM resources (i.e., time slots and PRBs) needed to fulfill the requirements of radar sensing and communication. The radar range resolution and velocity resolution decide the number of PRBs and time slots respectively. On the other hand, for communication, the time and frequency resources are calculated based on data packet latency and size (e.g., Transport Block) respectively.

*Resources Sensing:* Before selecting the resources for packet transmission and radar sensing, the vehicle first senses the past 1 s window for sidelink reference signal received power (SL-RSRP) measurements across the subchannels. Unavailable subchannels are excluded, if sidelink control information (SCI) declares them busy or if SL-RSRP is greater than a predefined threshold.

*Resource Selection:* Once the resource sensing task is done, the vehicle randomly selects the resources for communication and radar sensing operations from the available set of resources. To reduce channel occupancy, it reserves communication resources periodically, dynamically allocating radar resources only when needed. For further details of the proposed approach, the reader is referred to [SMSR+23].

### 3.2.6 Leveraging Machine Learning for Enhanced Radio Resource Management in Non-Terrestrial Networks

This section explores the integration of machine learning (ML) techniques for radio resource management (RRM) in NTN. With the increasing importance of NTNs in global connectivity, there is a pressing need for efficient resource allocation strategies. Traditional methods often fall short due to the unique challenges posed by NTNs, such as variable connectivity and limited resources. This paper proposes new ML-based approaches to optimize RRM in NTNs, demonstrating their potential to improve performance, reliability, and scalability.

Non-Terrestrial Networks (NTNs), comprising satellite systems, HAPS, and Unmanned Aerial Vehicles (UAVs), have emerged as pivotal components in the global communications landscape. Unlike traditional terrestrial networks, NTNs operate from elevated orbits and altitudes, offering unparalleled coverage and connectivity, particularly in remote, rural, and underserved areas. These networks are instrumental in bridging the digital divide, supporting disaster recovery operations, and enabling global Internet of Things (IoT) applications. Their ability to provide seamless connectivity across vast geographical areas makes them indispensable in an increasingly connected world [ASC$^+$22].

Machine Learning (ML) has revolutionized various industries by enabling systems to learn from data, adapt to changing conditions, and make informed decisions. In the context of NTNs, ML holds immense potential to transform operations and resource management. Traditional methods of managing radio resources in NTNs often rely on static rules and predefined algorithms, which can be inefficient and inflexible in dynamic and unpredictable environments. ML, with its capability to analyze large volumes of data, identify patterns, and predict future trends, offers a paradigm shift in managing NTNs. By integrating ML, NTNs can achieve higher levels of automation, efficiency, and adaptability, leading to improved network performance and user experience [KAR24, OGLL$^+$22].

The motivation for integrating ML into radio resource management for NTNs stems from the inherent limitations of current methods and the potential benefits ML can offer. Traditional resource allocation techniques often fall short in addressing the unique challenges posed by NTNs, such as highly variable connectivity, limited and shared resources, and the dynamic nature of the operational environment. These challenges necessitate advanced strategies that can dynamically adjust to real-time conditions and optimize resource usage [FOL$^+$23].

ML-driven approaches are poised to overcome these limitations by providing adaptive, data-driven solutions. For instance, reinforcement learning algorithms can continuously learn and adapt to changing network conditions, while supervised learning models can predict traffic patterns and optimize resource allocation accordingly. The adoption of ML in NTNs promises to enhance network efficiency, reduce latency, improve Quality of Service (QoS), and ensure more equitable distribution of resources among users. Moreover, ML can facilitate proactive maintenance and fault detection, further enhancing the reliability and resilience of NTNs [FOL$^+$23].

**Machine Learning as a Solution for RRM in NTN**

ML offers transformative solutions for the complex and dynamic challenges faced by NTNs. By leveraging advanced algorithms and data analytics, ML can significantly enhance RRM in NTNs.

*Adaptive Resource Allocation:* One of the primary advantages of ML in NTNs is its ability to provide adaptive resource allocation. Traditional RRM methods typically employ static or rule-based approaches, which may not be effective in the highly variable and dynamic environments of NTNs. ML, particularly reinforcement learning (RL), can continuously learn from the environment and adapt allocation strategies in real-time. RL algorithms can optimize resource distribution by dynamically adjusting to fluctuating network conditions, user demands, and mobility patterns. This leads to more efficient utilization of resources, reducing wastage and improving overall network performance [OGLL$^+$22, DRP$^+$24].

*Predictive Analysis:* ML models excel at predictive analysis, which is crucial for proactive resource management in NTNs. Supervised learning techniques can analyze historical data to predict future network states, such as traffic patterns, user behaviour, and potential congestion points. By forecasting these trends, network operators can make informed decisions about resource allocation, preemptively address potential issues, and optimize network performance. For example, predictive models can foresee periods of high demand and allocate additional resources to maintain QoS, thus preventing network degradation [VHP$^+$21, HVB22, FA21].

*Automation and Scalability:* The automation capabilities of ML significantly reduce the need for manual intervention in RRM, which is particularly beneficial for NTNs with large and complex infrastructures. ML algorithms can autonomously manage routine tasks, such as spectrum allocation, power control, and handover decisions. This not only increases operational efficiency but also allows human operators to focus on more strategic tasks. Additionally, ML models can scale efficiently to handle the growing complexity and size of NTNs, adapting to new technologies and expanding networks without requiring extensive reconfiguration[VHP+21].

*Fault Detection and Network Resilience:* ML techniques, particularly anomaly detection algorithms, are adept at identifying and mitigating network faults. By continuously monitoring network performance and analyzing patterns, ML models can detect anomalies that indicate potential failures or security threats. Early detection allows for prompt corrective actions, minimizing downtime and enhancing network resilience. Moreover, ML can assist in predictive maintenance by forecasting equipment failures and scheduling maintenance activities before issues escalate, thus ensuring uninterrupted service [CXP22, GLF21].

*Resource Optimization in Diverse Environments:* NTNs operate in a variety of environments, each with unique challenges and constraints. ML models can be trained to optimize resource management across different types of NTNs, such as satellite constellations, HAPS, and UAV networks. For instance, clustering algorithms can group users or devices based on similar characteristics or requirements, allowing for more efficient resource distribution. Similarly, ML can optimize the use of limited spectrum and power resources, balancing trade-offs between different performance metrics such as coverage, capacity, and energy consumption [FOL+23].

*Cross-Layer Optimization:* ML can facilitate cross-layer optimization, where decisions are made by considering multiple layers of the network stack simultaneously. This holistic approach ensures that resource management decisions are aligned across physical, MAC, and network layers, leading to more coherent and effective strategies. For example, ML can jointly optimize link scheduling, power control, and routing to maximize network performance and reliability[FOL+23].

*Advances in ML integration to Resource Allocation Methods in NTNs:* Resource allocation in NTNs is a complex and critical task due to the unique operational challenges these networks face. Current methods primarily involve heuristic algorithms, optimization techniques, and rule-based systems. Heuristic algorithms, such as greedy algorithms and genetic algorithms, offer quick and often effective solutions but can lack the precision needed for optimal resource allocation in highly dynamic environments. Optimization techniques, including linear programming and dynamic programming, provide more precise solutions but can be computationally intensive and less adaptable to real-time changes. Rule-based systems rely on predefined rules and thresholds, which, while straightforward, often fail to adapt to varying network conditions and demands [FA21].

In recent years, the application of Machine Learning in terrestrial networks and satellite communications has shown promising results, which can be adapted to NTNs. Key ML techniques include:

1. Supervised Learning: Used for traffic prediction, anomaly detection, and user behavior analysis. Algorithms like decision trees, support vector machines (SVM), and neural networks are employed to train models on historical data to predict future network conditions and optimize resource allocation.

2. Unsupervised Learning: Applied for clustering and anomaly detection without labelled data. Techniques such as k-means clustering and hierarchical clustering help group users or network segments with similar characteristics, enabling more efficient resource distribution.

3. Reinforcement Learning (RL): Particularly suited for dynamic resource management, RL algorithms like Q-learning and deep Q-Networks (DQN) learn optimal policies by interacting with the environment. These algorithms adaptively allocate resources by maximizing long-term rewards, making them ideal for the constantly changing conditions of NTNs.

4. Deep Learning: Leveraged for complex pattern recognition and high-dimensional data processing. Convolutional Neural Networks (CNNs) and recurrent neural network (RNN)s are used for advanced tasks like video traffic analysis and predictive maintenance.

5. Federated Learning: An emerging approach where models are trained across decentralized devices while keeping data localized. This technique enhances privacy and reduces latency, making it suitable for NTNs with distributed architectures.

**Gaps and Challenges**

Despite the advancements, there are notable gaps in the application of ML for RRM in NTNs:

1. Scalability: Many existing ML models are not designed to scale across the vast and heterogeneous environments of NTNs. Scalability issues arise from the diverse nature of NTN components and their operating conditions.

2. Real-Time Adaptability: While ML models offer adaptability, integrating them into real-time systems remains a challenge due to computational and latency constraints.

3. Data Availability and Quality: High-quality, comprehensive datasets for training ML models in NTNs are limited. The unique operational conditions and the need for continuous data collection pose significant hurdles.

4. Integration with Existing Systems: Seamlessly integrating ML models with existing network management systems requires addressing compatibility issues and ensuring minimal disruption to ongoing operations.

The gaps identified present opportunities for significant innovation in the field of NTNs. Future research can focus on developing scalable and real-time adaptable ML models tailored specifically for the unique characteristics of NTNs. Additionally, creating comprehensive datasets and standardized testing environments can aid in the development and benchmarking of ML algorithms. Addressing integration challenges and ensuring robust security and privacy measures will be critical to the successful deployment of ML in NTNs.

## 3.3 ISAC Channel Measurement

This section discusses radio measurements with ISAC application scenarios, including dynamic sensing targets. Measurements are important for deriving new models that accurately consider for the physical environment and communication channels. Different applications, such as automotive radar, indoor localization and wireless communication, require specific measurement techniques.

Section 3.3.1 is on tri-band measurements including 3.2 GHz, 34.3 GHz and 62.35 GHz for automotive scenarios. Section 3.3.2 is on V2X measurements and derived models at 28 GHz. Section 3.3.3 is on indoor distributed beamforming measurements at 24 GHz and 60 GHz. Section 3.3.4 discusses a dual channel model including monostatic and bi-static channels. section 3.3.5 introduces the bistatic radar (BiRa) Measurement Facility.

### 3.3.1 Tri-band Automotive ISAC

Cooperative connected automated mobility depends on sensing and wireless communication functions for realizing novel advanced driver assistance system (ADAS) features. With increasing carrier frequency both sensing and communication can be realized within the same radio frequency front end, being implemented in an ISAC framework. The attenuation of radio signals increases with increasing carrier frequency. Thus, beamforming by means of antenna arrays becomes key for data communication to achieve a signal to noise ratio (SNR) at the receiver that allows for a reliable data link and it allows for direction of arrival estimation for localization purposes. For communication and multi-static sensing, beamforming from the transmitter to moving receivers is required. This is a challenging task, especially at higher velocities, but can be improved by exploiting correlations of channel responses in different frequency bands.

The key idea in this work is the utilization of lower frequency bands below 6 GHz to first establish a coarse but fast direction estimate using virtual aperture algorithms with a single antenna. In a second step accurate beamforming at the mmWave frequency band with an antenna array can be obtained for direction of arrival estimation and reliable broadband link setup. This approach supports both communication and localization tasks and operates two frequency bands in a tandem fashion.

Therefore, M. Hofer et al. investigate the statistical properties of the vehicle-to-infrastructure (V2I) radio channel in three frequency bands with center frequencies of 3.2 GHz, 34.3 GHz and 62.35 GHz. The measurement results of a tripleband measurement campaign conducted in an urban V2I scenario [HLB$^+$21] have been utilized for the investigation. The measurement scenario is shown in Fig. 74. In this campaign a transmitter (Tx) car that is equipped with three omni-directional antennas passes by a static receiver (Rx) with three directive antennas mounted on a tripod and stops at a road intersection. The directive antennas have a HPBW of approximately $17-19°$. The channel impulse responses were collected simultaneously at all three carrier frequencies data with 155.5 MHz bandwidth and a sounding repetition rate of 31.25 µs.

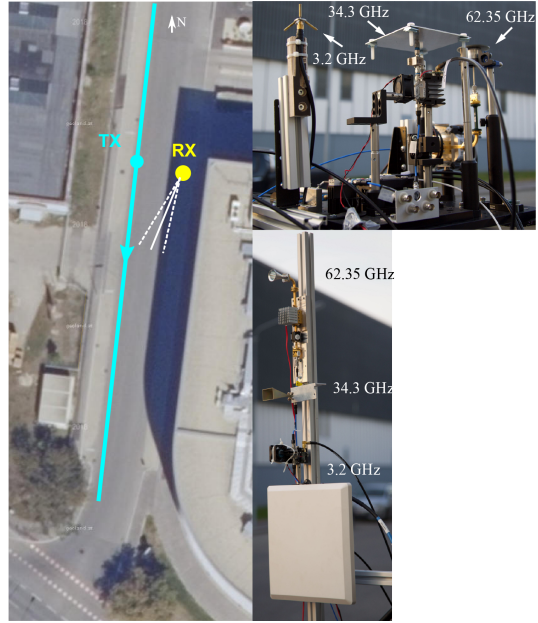

Figure 74: Scenario description and antenna setup: (left) A car approaches a "T"- intersection and stops. The directive horn antennas are pointed towards the road intersection. (top right) omni directional antennas mounted at vehicle (middle) directional antennas mounted at tripod

Using the high temporal sampling rate, the CLEAN algorithm was applied to estimate the path loss, delay and Doppler frequency of different multipath components. This information can be used to establish a broad band communication links.

It is found that the collinearity between the 3.2 GHz and 34.3 GHz band as well as between the 3.2 GHz and 62.53 GHz is smaller in the NLOS region but increases for the LOS region. Strong multipath reflections in lower frequency bands are also available at higher frequency bands. Analyzing the collinearity depending on the stationarity region lengths in terms of wavelength revealed that small and very large stationarity region lengths for investigated scenario lead to reduced collinearity results. For the investigated scenario a wavelength of $10\lambda$ gave the best result. Note that the non-stationary aspect has to be considered for the estimation of the statics on how many path we can find in one band and how many path we can find in another band.

### 3.3.2  28 GHz ISAC Vehicular Channel Measurements and Modeling

As one of the most potential ISAC application scenarios, vehicular ISAC networks have been expected to connect vehicles with surrounding environments, where communication transceivers and sensors are equipped on the vehicles to simultaneously conduct radio-based communication and echo-based perception. Different from conventional vehicular channel sounding activities, vehicular ISAC channels are measured using a monostatic system, typically capturing the propagation of echoes from transmitters to scatterers and from scatterers back to transmitters within the dynamic environment. Due to the sensitivity to the environment, novel features of vehicular ISAC channels, such as sensing multipath, clutter multipath, semantic information, etc., are expected to be captured in the measurement. Besides, the typical features such as multipath birth-death, sparsity, non-stationarity, etc., differ significantly from conventional vehicular channels, which is necessary to study.

ISAC aided vehicle-to-everything (V2X) is an integral component of ISAC application scenarios. It includes autonomous driving vehicles, which can enhance traffic efficiency and safety through inter-communication and synchronous perception. A comprehensive understanding of the vehicular ISAC channel is crucial for system design, performance evaluation, and standardization. However, most research on ISAC channel measurement and modeling has focused on static scenarios, with limited attention given to dynamic vehicular ISAC channels.

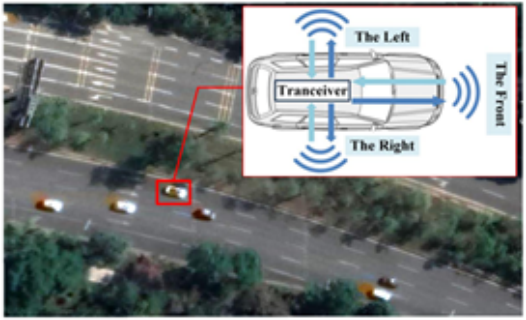

Figure 75: Different sensing direction for vehicular ISAC

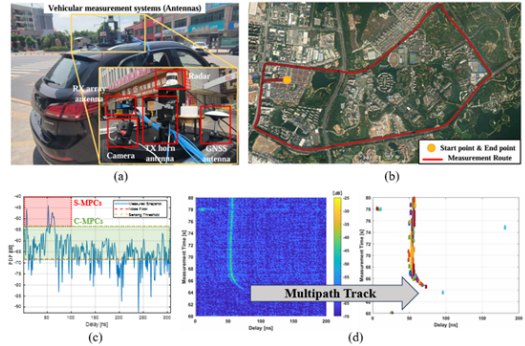

Figure 76: Illustration of vehicular ISAC channel measurements and features

In [ZHea24], the dynamic vehicular ISAC channel measurement was conducted in the Song- shanhu district of Dongguan, located in Guangdong, China, at a frequency band of 28 GHz. The mean vehicle speed during the measurements was 30 km/h, and the sensing directions encompass the front, left, and right directions. It is evident that each direction possesses its own primary scatterers, resulting in distinct environmental characteristics. This diversity is essential for conducting corresponding measurements. The vehicular ISAC channel measurement system based on vector signal transceiver (VST) is designed, including a signal generator based Tx, a signal digitizer based Rx, and a power supply based on the vehicular battery. The Tx and Rx used a directional horn antenna and a 4x8 array antenna respectively. The sounding signals are multicarrier signals with 1 GHz bandwidth, and are transmitted with the maximum power of 28 dBm. During the measurement, red, green, and blue (RGB) video data from the camera and point-cloud data from the radar are collected and stored simultaneously to aid in mapping between channel multipaths and environmental scatterers.

The sensing direction is the direction in which the vehicular ISAC perceives while driving. Due to different environmental features, channel also vary in different directions. Therefore, we conduct measurements separately from the front, left, and right directions. Fig. 75 shows different sensing direction for vehicular ISAC, which may help with understanding. As shown in the following Fig. 76, (a) is a diagram of vehicular ISAC measurement equipment setup, (b) is the RGB image of the measurement scene, (c) shows the sensing multipath componentss (S-MPCs) and clutter multipath componentss (C-MPCs) in a single channel snapshot, and (d) provides the tracks of dynamic multipath from raw measured data.

Based on actual measurements, a dynamic ISAC channel tapped-delay line model is presented, which is composed of S-MPCs and C-MPCs. For S-MPCs, they typically exhibit higher power and more pronounced clustering, attributed to the strong reflection from sensing targets, and occur continuously over time. In contrast, C-MPCs have lower power and are generally distributed across the entire delay domain due to widespread scatterers and noise, and occur randomly over time. A tracking algorithm is utilized to identify S-MPCs and C-MPCs, and a series of characterization, including number of new S-MPCs, lifetimes, initial power and delay positions, dynamic variations within their lifetimes, clustering, power decay and fading of C-MPCs, is statistically modeled to enable simulated channel with dynamic evolution process.

### 3.3.3 Human-Centric Indoor ISAC: mmWave Dual-band Distributed Beamforming Channel Measurement

The 5G NR and IEEE 802.11ay standards have already included the mmWave bands, aiming to provide wireless communications with ultra-high speed and ultra-low delay to serve for demanding use cases like virtual reality, industrial 5.0 and autonomous driving. The 5G NR mmWave bands start from 24 GHz and go up to 71 GHz. The wireless gigabit (WiGig) targets the 60 GHz band starting from 58 GHz to 74 GHz. These two standards are expected to co-exist in the complex and dynamic indoor environment, where the influences of human body and body dynamics are often unpredictable, making the indoor mmWave communications challenging in practice. Moving forward to 6G, researchers have been focusing on ISAC, where both high-speed communication and accurate sensing of the surrounding environment are important for the envisioned use cases like remote health monitoring, autonomous driving, and industry 5.0. Dual-functionality including communication and sensing will be integrated into future radio systems [AKM23, LDBM$^+$23]. Sensing is especially promising at the mmWave bands of the 5G NR and WiGig standards, where the smaller wavelength, the larger bandwidth and the array size/complexity are converging to radars that are conventionally used for sensing.

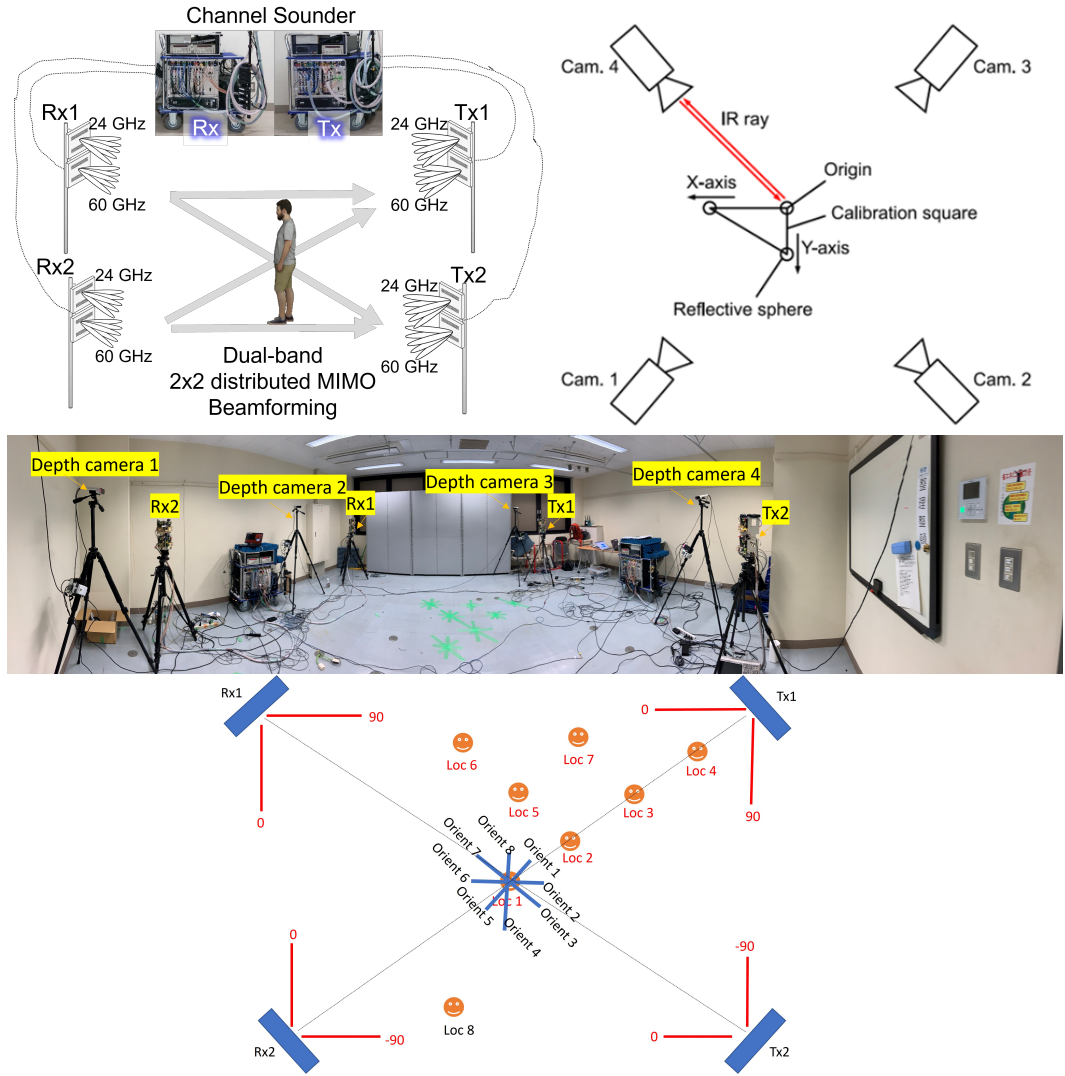

Figure 77: (Left Up) System configuration of dual-band distributed MIMO channel sounder; (Right Up) System configuration of distributed red, green, and blue- depth (RGBD) cameras; (Middle) Panoramic view of the measurement setup and the room; (Down) A sketch of the measurement scenario, human body locations and orientations.

The environment seen by 5G NR and WiGig systems is however much more dynamic and complex than the environment seen by outdoor radars. To enable ISAC in mmWave 5G NR and WiGig applications in dynamic indoor environments, an accurate modeling of human body interactions with the mmWaves is crucial. Radio waves are potentially reverberating [MPG$^+$19] between multiple indoor reflectors or human; such interactions, given a multi-link scenario, involve not only absorption but also scattering (bistatic scattering or backscattering). In particular, one of the ISAC use cases aims to prevent mmWave communications from human blockage, hence the proactive sensing (detection, localization and tracking) of humans before any blockage is the goal, wherein the human body serves as a scatterer/reflector in radar sensing. However, relevant works have mainly focused on the modeling of human blockage [ONYM16] and skin reflectivity [WRC15]; multi-link and multi-band characterizations in mmWave ISAC scenarios with the presence of humans are largely unexplored.

In paper [MKK+24] the authors present the dual-band and distributed MIMO beamforming measurement as well as the characterization of the channels with the presence of one person with controlled location and facing orientations. Novelties are:

1. As is shown in Fig. 77, the novel measurement setup includes (i) a dual-band mmWave channel sounder which can measure at 24 GHz and 60 GHz concurrently with a $2 \times 2$ distributed MIMO topology, and (ii) a distributed RGBD camera system which captures the depth image of groundtruth via a global registration procedure to obtain point clouds from distributed local to a global coordinate.

2. The novel indoor measurement scenarios have the presence of up-to-3 persons, located at eight different positions in relative to the arrays, separated by different distances, and facing eight directions (body orientation). From the characterizations of the distributed dual-band channels in both the delay and the azimuth-angular domains, insights for ISAC channel modeling and system design can be drawn. The data can be further exploited for scattering and blockage model parameterization, adaptive beamforming or channel prediction.

### 3.3.4 Bistatic Channel Estimation by Monostatic Sensing Channels

Channel estimation in the mmWave band is still challenging due to the hardware complexity and budget required to build a channel-sounding system, and it is even more demanding for dynamic scenarios. From a pure sensing point of view, some of the main challenges, for instance, in V2X scenarios, are related to the coexistence of many sensor signals that interfere with each other. However, some works exploit this problem to obtain valuable information [NCL23]. A two-port ISAC dual channel model is defined as the combination of two mainly differentiated signals, where 'T' refers to the multipath communication channel between the transmitter and the receiver, and 'R', which refers to the sensing channel, in the cases of $R_{11}$ and $R_{22}$ to the monostatic sensing channel, while $R_{12}$ and $R_{21}$ refer to the interference between the sensors or the bistatic sensing channel.

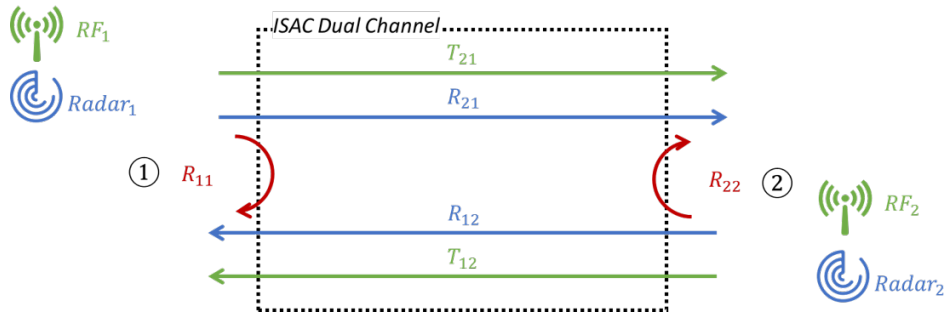

Figure 78: Definition of ISAC dual channel model

Some new and innovative approaches, such as [NC23], to the dual channel model in ISAC, show great potential for channel sounding. These approaches involve advanced signal processing techniques and sophisticated radar systems to extract and separate the monostatic and bistatic channels. In Fig. 78, it can be noted that if there is an integration of sensing and communications in one single system, the bistatic sensing channel $R_{21}$ will be equal to the communication channel $T_{21}$.

By using FMCW radar sensors in the mmWave band, the multipath communication channel can be obtained from monostatic-only measurement in outdoor and indoor scenarios. Processing beyond this technique is based on exploiting the mutual interference between two or more radar sensors. This interference, which occurs when the signals from different sensors overlap and mix on the receiver, is captured and processed by the system. The sensing signal captured by the system is composed of a mix of monostatic and bistatic channels; by separating both channels, as shown in Fig. 79, the monostatic sensing channel will provide information about the environment. Still, the most interesting part is the extraction of the bistatic channel, which is equal to the multipath communication channel, as demonstrated in the literature [AC24].

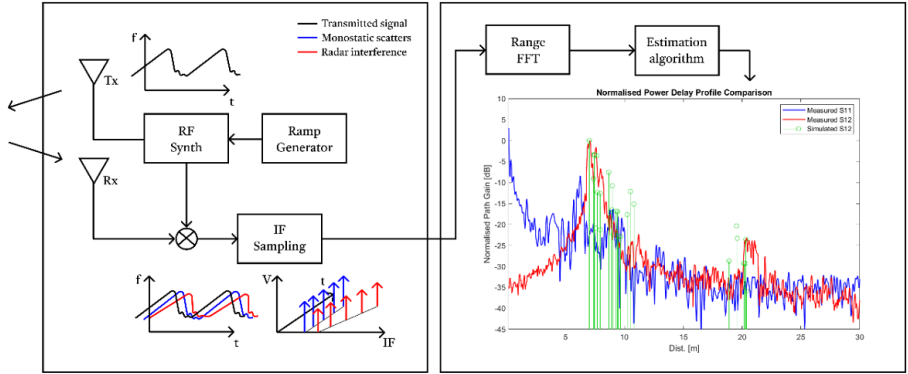

Figure 79: Flowchart of the channel estimation algorithm and comparison between measured S11, S12 and simulated S12

Using simpler off-the-shelf radar equipment instead of expensive channel sounder systems distinguishes this technique from other channel-sounding techniques, providing an accurate estimation of the multipath communication channel from monostatic measurements only. In Fig. 79, the graph on the right shows a measured result of this estimation technique compared with a Ray-Tracing simulator. Also, the monostatic sensing channel can be observed compared with the bistatic sensing channel; the correlation between those channels is proved to be low, which is significant as it indicates the independence of the two channels. This independence is crucial for accurate channel estimation and communication system design.

### 3.3.5 Bistatic Radar Measurement Facility, BiRa

Integrated Communication and Sensing (ICAS) is considered to be one of the key features of future 6G mobile radio. ISAC stands for the detection and analysis of non-cooperative objects ("targets") by exploiting the inherent resources of the mobile radio system on both the radio access and network level, using the signals transmitted simultaneously for both communication and target illumination purposes. Therefore, ISAC performs target detection in configurations typical for communication systems, which comprise the waveform — usually Orthogonal Frequency Division Multiplexing (OFDM) and derivatives, its numerology, multiuser access (OFDMA, time-division multiple access (TDMA)), pilot schemes, multi-link (multistatic) propagation, channel state estimation, and synchronization [TAB+23].

In order to detect and classify a target, its electromagnetic signature is important. Electromagnetic reflectivity is the signature of a target due to its geometrical shape and composed material. The term "reflectivity" is used here because of its general definition for both near and far-field conditions. The other signature is micro-Doppler, which refers to the small fluctuations in the Doppler shift caused by the target's local moving parts, e.g. rotating blades of drones, or the flapping wings of birds.

For a comprehensive experimental characterization of a target object we have to consider illumination from all possible directions in azimuth and elevation (practically only in the upper half sphere in case of vehicles and mostly in lower half sphere in case of flying objects) and observation also from all possible directions. This way, we end up with a four-dimensional data structure. Since Tx (illumination) and Rx (sensing/observation) angles can take any value, we have the specific cases of 0° and 180° aspect angles included. The former corresponds to the monostatic case, whereas the latter is called forward scattering case. Forward scattering includes shadowing or obstruction of the LOS between Tx and Rx but also diffraction around the target. So it is most distance and frequency dependent. However, it is also an interesting radar operation mode as diffraction can increase received power and, hence, enhance target visibility.

In general, the radar target has to be considered as an extended target. This means that it is wider than a resolution cell, which is given in radial distance by the range resolution. Therefore, we have to collect multiple range bins for depth information depending on the available bandwidth. The number of relevant range bins depends on the electrical size of the object, which may be larger than the mechanical one if the target is concave and has structural resonances, e.g., because of cavities.

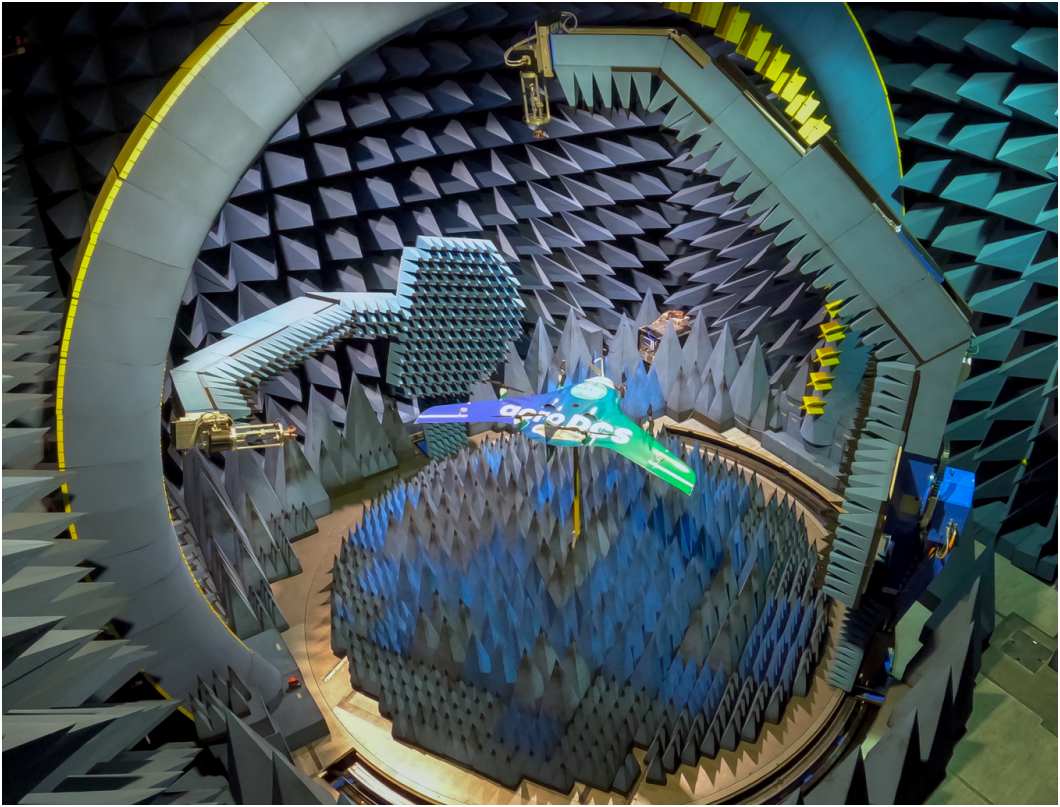

Figure 80: BiRa measurement range for bistatic radar reflectivity at TU Ilmenau (the Stargate antenna arch is not used in this case).

This way we effectively get another dimension besides the four-dimensional (4D) angles. Moreover, the radial distance of Tx and Rx antennas from the target matters.

Therefore, we introduce a new measurement range for the bistatic reflectivity of extended objects up to the size of a passenger car. This unique measurement range is capable of real-time wideband measurements of time-variant targets for the bistatic micro-Doppler response of extended targets [ZLaC17]. BiRa (Bistatic Radar Measurement Facility) is situated at the Thuringian Innovation Center for Mobility (ThIMo) of the Ilmenau University of Technology. This advanced system comprises two pivoting gantries, each equipped with a transmitting (Tx) and receiving (Rx) antenna, respectively [TAB+23]. The target object of interest is positioned at the center of the turntable as in Fig. 80. Therefore, the target can be illuminated and observed from any desired bistatic angle in azimuth, from any elevation bistatic angle in the upper hemisphere, and from 25° elevation in the lower hemisphere. Moreover, BiRa can perform long fully automated measurement runs without requiring any operator interaction.

The accessible frequency range covers all relevant frequency ranges from frequency range (FR)1 and FR2 up to 170 GHz. A remarkable feature is that besides of a standard vector network analyzer (VNA), we can use a wide-band $2 \times 2$ channel sounder for target illumination and reflectivity sensing with an instantaneous bandwidth of up to 4 GHz. The development of this RF measurement system can be found in [EGA+24]. This does not only accelerate the total measurement time. It allows also to investigate micro-Doppler of time-variant targets, e.g., multi-copters with rotating propellers. Analyzing, studying, and learning the micro-Doppler signature of these objects will help to classify their type and to identify and separate them in real-life multiple target scenarios. The fast and programmable availability of the BiRa measurement range allows the collection of training data for AI/ML-based target classification and modeling. In the subsection 3.4.1, present measurement examples that demonstrate some capabilities of BiRa.

## 3.4 ISAC Channel Parameter Estimation and Modeling

This section discusses channel modeling and estimation techniques for ISAC applications. Compared to traditional communication channel modeling and estimation techniques, ISAC requires more accurate models and techniques that consider the physical environment. For radar sensing applications in particular, the spatial and temporal consistency of the channel is required. These conditions must be satisfy with an appropriate compromise between accuracy and computational complexity.

In section 3.4.1 micro-Doppler modeling for UAV, pedestrians, and bicycles based on BiRa measurements are discussed. Section 3.4.2 introduces a hybrid channel model based on stochastic and raytracing techniques. Section 3.4.3 is on a quasi-deterministic channel model for gesture recognition. Section 3.4.4 introduces a physics-informed generative AI approach to simulate RF propagation for localization. Section 3.4.5 discusses a model informed machine learning approach for parameter estimations.

### 3.4.1 Reflectivity and Micro-Doppler Measurement and Processing

The integration of wireless communication and radar sensing is now getting a huge interest from researchers in both fields. The road map to the final goal and individual solutions to the challenges might differ in developing the ICAS system. However, since the detection, localization, and classification of the targets are involved, the electromagnetic signature of the targets will be still valid for all variants of the ICAS system. Therefore, the study on static reflectivity and micro-Doppler signatures of targets such as drones, pedestrians, and cyclists is of great importance to the ICAS community. We present some micro-Doppler and reflectivity measurements, performed using the state-of-the-art measurement system, BiRa 3.3.5, situated at the ThIMo of the Ilmenau University of Technology.

**Radar Targets**

**DJI Phantom II:** This is a very popular drone of intermediate size, shown in Figure 84a, which gives a good idea of how static reflectivity and micro-Doppler signatures of drones behave. For micro-Doppler measurements, it has the disadvantage that the measurement team had no control of the rotors speeds, which are set autonomously by the drone flight control system.

**Pedestrian target:** One of the most important applications for distributed ISAC is vehicular safety. In this context, bistatic micro-Doppler signature of vulnerable road users, such as pedestrians and cyclists, is of great importance. In order to collect data to understand how the micro-Doppler returns of arms and legs behave in pedestrian and cyclist cases, targets that can produce these movements with known and stable speed were used. For the pedestrian case, an Euro pedestrian target adult (EPTa) from the European- new car assessment programme (Euro-NCAP), as specified in [ISO18], was used. This model represents a 50 percentile male adult pedestrian, with realistic proportions, and arms and legs movements. It is built with material that emulates the reflectivity of an average human being. The EPTa moves arms and legs, emulating a walking person, at a constant period of 1.09 s. Figure 86c illustrates the dynamic reflectivity measurement of the EPTa.

**Cyclist target:** For the cyclist case, an official Euro-NCAP Euro bicyclist target adult (EBTa) was used, representing an 50 % adult male on standard average European utility bike, as specified in [ISO18]. As well as the EPTa, it is also built with material that emulates the reflectivity of an average cyclist.

**Static Reflectivity Modeling**

When the far-field condition is satisfied, the bistatic RCS can be described by the following equation.

$$\sigma_{\text{bi}} = \lim_{d_{\text{Tx,Rx}} \to \infty} 4\pi \cdot d_{\text{Tx}} \cdot d_{\text{Rx}} \cdot \frac{|E_{\text{scat}}|^2}{|E_{\text{inc}}|^2} \quad (1)$$

However, electromagnetic scattering behaves differently in far-field and near-field. Therefore, the term "reflectivity" is used as in this paper [SHSH20] to generalize for the far-field and near-field conditions.

$$\text{Reflectivity} = \frac{|E_{\text{scat}}|^2}{|E_{\text{inc}}|^2} \qquad (2)$$

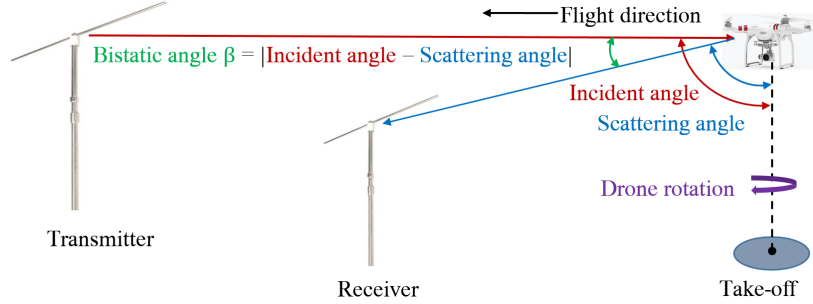

Figure 81: Analyzed scenario for drone detection

The analyzed scenario of drone detection in this section is illustrated in Fig. 81. The incident, scattering, and bistatic angles are also labeled in the figure. For simplicity, the take-off and landing part of the drone flight is omitted. The drone will fly over the receiver until the position between the transmitter and the receiver, which introduces the forward scattering case.

Therefore, the transmitter, the drone, and the receiver lie in a straight line. It perfectly forms the bistatic angles $\beta$ of 10° to 180°°, which can be translated into the angular sweep of the two gantries in the BiRa measurement system, as described in Table 6. In addition, the drone can rotate on its own axis, and it can be created in the BiRa Measurement system as the rotation of the turntable. Since the analyzed scenario is the flyover scenario, the reflection from the lower hemisphere of the drone is interesting, therefore, the drone is placed upside down on the support at the center of the turntable as in Fig. 82.

The number of frequency sweeps is 1601 from 2 GHz to 18 GHz. By using this ultra-wideband, the extended information of the target can be observed, and it is also an important signature of a target for classification. However, the analysis of the extended target signature is omitted in this section, and it will appear in upcoming publications.

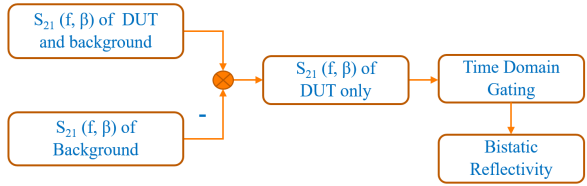

| Parameter | Setup |
|---|---|
| System | VNA |
| Frequency sweep | 2−18 GHz, 1601 steps |
| Static Gantry | 10°:5°:180° |
| Dynamic Gantry | 90° |
| Turn-Table | 0° and 45° |
| Polarizations | HH, HV, VH, VV |

Figure 82: The block diagram of signal processing  Table 6: Measurement set-up

The measurement is performed two times with the drone on the support or without the drone. Both measured data are calibrated to compensate for the propagation loss, the effect of the transmitting and receiving antenna, and the measurement system. Then the data are processed as in Fig. 82. Firstly, the background reflection is reduced by the background subtraction, which is the subtraction of the measurement data without the drone from the measurement data with the drone. However, all the reflections of the anechoic chamber can not be eliminated. By knowing the propagation distance, the location and the physical dimensions of the drone, it is possible to filter out the unwanted reflections by time domain gating. The post-processed data is the bistatic reflection of the drones over different bistatic angles and frequency bands.

Fig. 83 illustrates the agreement between the measurement reflectivity by BiRa and the simulated reflectivity by Feldberechnung bei Körpern mit beliebiger Oberfläche (FEKO). Therefore, not only measurement data but also the simulated data of each target can be utilized in training some machine learning algorithms for target reorganization and classification.

Normalized reflectivity of DJI P2 is illustrated in Fig. 84 for different bistatic viewing angles (10° to 180°) and polarizations over frequency of 2 GHz to 18 GHz. Since these three drones are very distant in their composed material and geometrical shape, their normalized reflectivity is also different over different bistatic angles and operating frequency bands. For the target classification, the joint estimation over the wide range of angles, operating frequency bands, and full polarization will be advantageous.

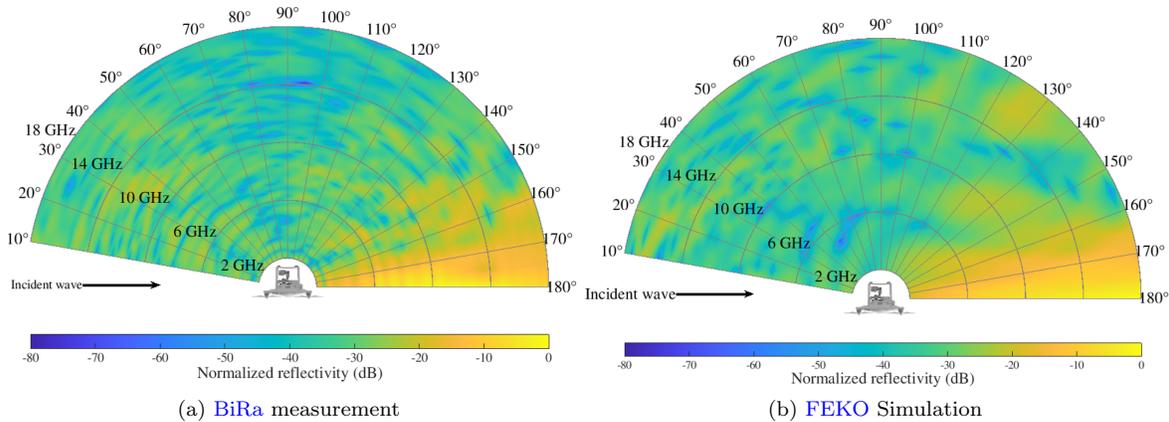

(a) BiRa measurement  (b) FEKO Simulation

Figure 83: Comparison of normalized reflectivity by BiRa measurement vs FEKO Simulation: DJI P2 for different bistatic viewing angles (10° to 180°) and polarizations over frequency of 2 GHz to 18 GHz.

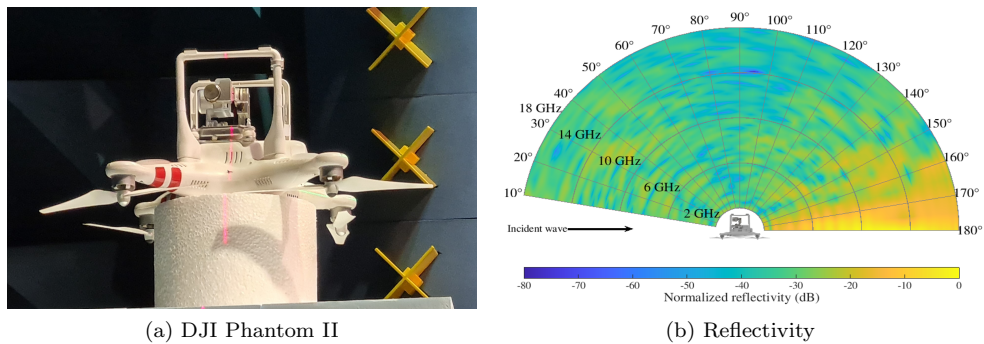

(a) DJI Phantom II  (b) Reflectivity

Figure 84: Normalized reflectivity of DJI P2 for different bistatic viewing angles (10° to 180°) and polarizations over frequency of 2 GHz to 18 GHz.

**Micro-Doppler Modeling**

In the dynamic reflectivity side, an important aspect of Multi-Link Propagation and Bistatic Target Reflectivity characterization is micro-Doppler. It refers to the small fluctuations in the Doppler shift caused by the motion of a target's internal parts, such as rotating blades or moving limbs. Micro-Doppler provides valuable information about the target's structure and motion, which can be used to identify the type of target and distinguish it from other objects in the environment. Due to these reasons, micro-Doppler is a phenomenon that is becoming increasingly important in the context of ICAS systems.

There is plenty of literature about micro-Doppler under monostatic continuous wave (CW) or LFM radars. Furthermore, in [SSZ11], the author analyzes Doppler estimation in the context of an OFDM ICAS, but there is no micro-Doppler analysis, and no considerations about the bistatic or multistatic cases. In [SICS20], micro-Doppler is employed in a multistatic Joint Radar Communication System, using chirp-like waveform, to perform target classification of people, cyclists, and dogs. However, in order to understand how this phenomenon appears under ICAS systems, it is important to analyze target responses both under different bistatic constellations and using OFDM-based waveforms, since these are the most common situations in ICAS.

Bistatic micro-Doppler measurement and analysis from various aspect angles are important tools for characterizing targets, since bistatic geometry provides diversity advantage over monostatic case, giving additional information about the target's motion and structure [Smi08]. Therefore, bistatic micro-Doppler is an important field to be researched. Bistatic Radar (BiRa) can play an interesting role in providing data for this kind of research, since it is able to perform, among many other applications, multiaspect OFDM-based bistatic micro-Doppler measurements in an automatized way.

About OFDM-based waveforms, they reflect the target's micro-movements in a way that differs from typical CW and LFM radars, since each frequency component of the OFDM signal produce a different Doppler response. This effect can be neglected in narrowband systems, but they are important in broadband environments, and even more important for ultra-wideband (UWB) and high frequency sensors. Besides, the use of multiple subcarriers increases system performance with respect to the SNR of the micro-Doppler signature [Sen14].

Two measurement setups were implemented, using a wideband OFDM-based transmit signal called Newman sequence with constant spectral magnitude and minimal crest-factor [Boy86]. Table 7 details the setups used with this measurement system.

Table 7: MICRO-DOPPLER MEASUREMENT SETUPS

| Parameter | Setup 1 | Setup 2 |
| --- | --- | --- |
| Waveform | OFDM-based | OFDM-based |
| Central frequency | 3.7 GHz | 7 GHz |
| Bandwidth | 200 MHz | 2.46 GHz |
| Total number of carriers | 1600 | 2500 |
| Carriers with energy | 1280 | 2048 |
| Symbol duration | 8 µs | 1 µs |

For micro-Doppler analysis, we process the signal in slow-time, i.e., a vector of the returns of the target in a single range bin, along different symbols. Some subsampling can also be applied in order to reduce the amount of data and the slow-time sampling frequency, and thus, to achieve a better visualization of the data, focusing the analysis on the bandwidth of interest. In this work, a subsampling factor of 8 means that we use only every eighth symbol available.

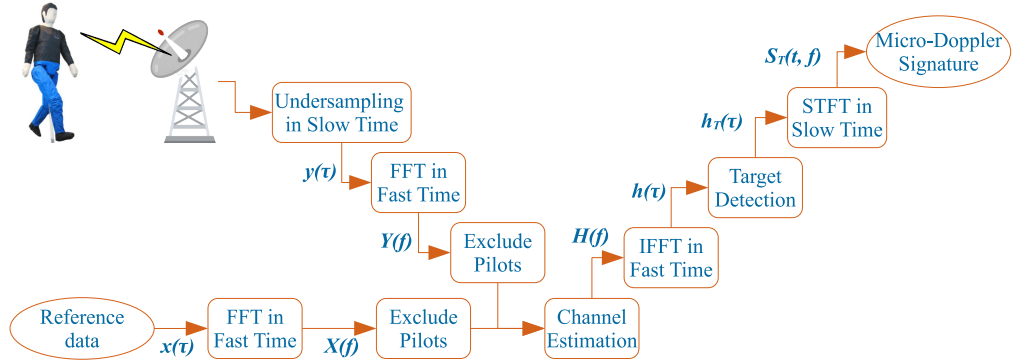

Figure 85: The block diagram of OFDM Micro-Doppler signal processing

To achieve this, we follow the signal processing scheme showed in Fig. 85. First, we convert both the received signal and the reference from time domain ($y(\tau)$ and $x(\tau)$) to frequency domain ($X(f)$ and $Y(f)$), respectively. Then, we exclude pilot frequencies, and perform channel estimation by doing $H(f) = \frac{Y(f)}{X(f)}$ [Bra14].

Finally, we perform target detection and calculate the Spectrogram of the slow-time profile of the target, in order to obtain a time-frequency representation. Especially for these measurement setups, where high symbol repetition frequencies are used in the measurement, some subsampling can also be applied in order to reduce the amount of data and the slow-time sampling frequency, and thus, achieving a lower processing cost, a better visualization of the data by focusing the analysis to the interest bandwidth, and allowing to check what happens with the system uses a different symbol repetition frequency.

**DJI Phantom II:** The first micro-Doppler measurement performed used the DJI Phantom II drone with 1 carbon propeller and 3 plastic propellers rotating. It was produced with measurement Setup 1 from Table 7. Fig. 84b shows a sample time-Doppler signature obtained from this measurement. The image shows lines approximately vertical (along the Doppler dimension), that repeat themselves periodically along the time dimension. These lines correspond to the micro-Doppler returns of the carbon propeller, since it has stronger reflectivity than the plastic propellers. Both the lines length and repetition rate can be used to calculate the rotor speed rotation. This measurement was very useful, since it allows the analysis of a single propeller return, which is an important step to understand the signature of a drone.

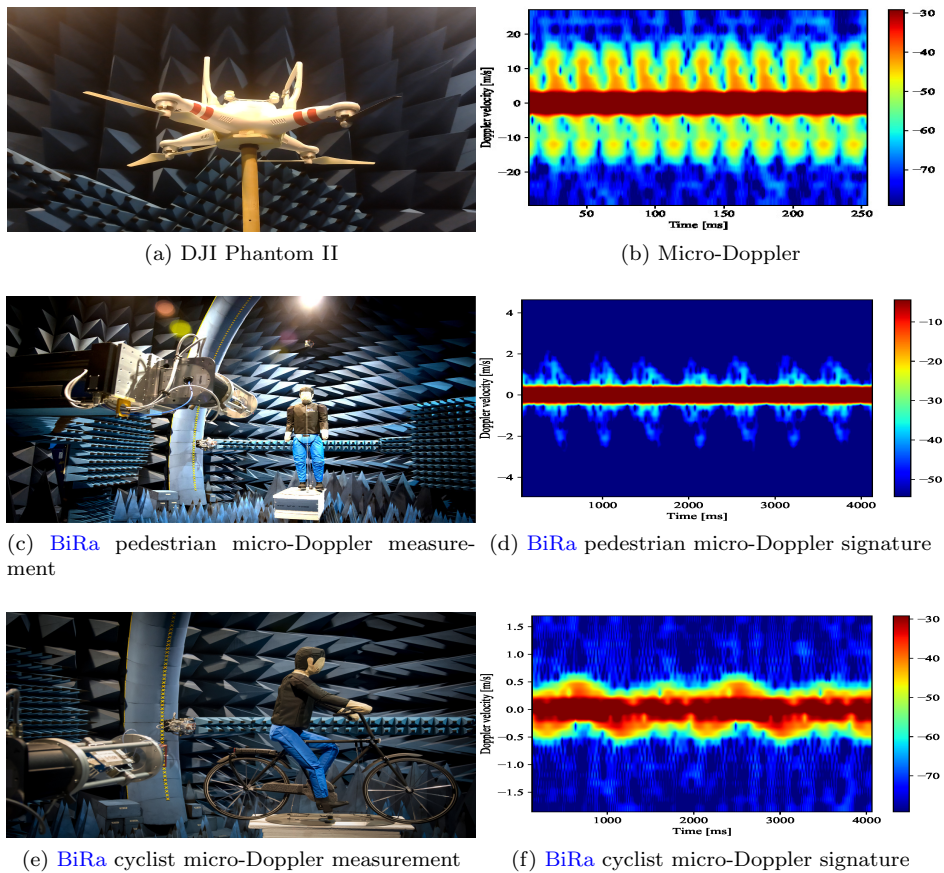

(a) DJI Phantom II  (b) Micro-Doppler

(c) BiRa pedestrian micro-Doppler measurement  (d) BiRa pedestrian micro-Doppler signature

(e) BiRa cyclist micro-Doppler measurement  (f) BiRa cyclist micro-Doppler signature

Figure 86: BiRa micro-Doppler measurement of DJI Phantom II drone, vulnerable road user (VRU) pedestrian model target, and VRU cyclist model target.

**Pedestrian target:** Fig. 86d shows a sample time-Doppler signature of the EPTa produced with Setup 1. The image shows the signal has oscillatory nature, alternating moments of maximum and minimum Doppler in a periodic way. This was expected since the target's arms and legs have maximum speed when hands and feet are next to the body, and minimum speed when they are distant from the body.

**Cyclist target:** Fig. 86f shows a sample time-Doppler signature of the EBTa produced with Setup 1. It is possible to see that the signal has an oscillatory nature, similar to that of the EPTa, but with less Doppler spread, since there is no arm movements, and the movements of the feet are narrower than that from the pedestrian target.

### 3.4.2 Hybrid ISAC Channel Modeling

ISAC systems are expected to combine traditional communications with sensing applications. Testing is an essential step in building a reliable and robust solution, as any system has to be tested before deployment. While taking real measurements is quite an expensive method, simulating on a computer is a common and cheap way of testing. Simulation on a computer implies representing or modeling a wireless channel in a realistic way, so the goal of channel modeling is to provide a realistic channel with the necessary application features.

Geometry-based stochastic channel model (GBSCM)s, for example, 3GPP TR38.901 [3GP22b], are commonly used for regular communication systems. However, this model cannot provide some features necessary for sensing applications. Sensing implies finding parameters like distance, position, velocity, etc, for objects or targets. These parameters are the ground truth information that should be modeled deterministically. The stochastic processes in GBSCM cannot adequately represent such deterministic parameters. The solution is to include deterministic modeling, for example, with the ray tracing (RT) tool [FVB+15].

An additional aspect crucial for sensing applications is that the channel model is expected to accurately present scattering from the targets, which includes parameters such as scattered power, Doppler, and delay. Scattered power is usually presented using the Radar Cross-Section (RCS). The RCS is known to be a far-field characteristic where the target is assumed to be a point scatterer. However, the point scatterer assumption might be too harsh when the target is located at a short distance and takes a significant angular size. For such cases with extended targets, another term should be exploited, for example, reflectivity function, which is not a single value because of the target's size. Also, the general bistatic scenario is crucial to be considered, meaning that the RCS becomes a bistatic parameter [TAM+22].

Another important aspect is to make the channel modeling algorithm as fast as possible. Computational performance may become problematic for a large scene or network with many simulated in parallel links. Stochastic channel models are widely used for communication applications because of their low computational burden, which allows the simulation of the whole network with a huge number of links. Ray tracing, in contrast, requires more computational resources in general.

This work's contribution is twofold. One is a hybrid channel model that inherits both stochastic and deterministic modeling in different proportions to address ISAC application requirements and reach a fast channel model. Another is enabling a ray tracing tool to accurately model the scattering from complex-shaped bodies like a vehicle.

**A scalable hybrid channel model**

Computational performance is an important factor for the applicability of the channel model. This chapter discusses a proposal aimed at constructing a channel model that is both appropriate for the ISAC application and computationally efficient.

The ISAC system comprises two key components: communication and sensing. Any object that has to be sensed needs to be included in the channel model deterministically. The stochastic methods can be used to include any other information in the channel to make the channel complete and realistic.

The concept of a hybrid channel model was inspired by the 3GPP TR38.901 hybrid channel model [3GP22b, Section 8.4], which integrates an RT tool to find some of the propagation paths. After that, additional paths are obtained by generating a certain number of stochastic clusters using the conventional stochastic 3GPP TR38.901 model.

Different ISAC applications have different requirements for the channel model. For example, an application of target positioning minimally requires the inclusion of targets deterministically in the channel model, whereas the influence of the remaining environment (which is not sensed) can be modeled stochastically. On the other side, imaging of the whole environment is a more complicated application that requires modeling the whole scene deterministically. To address these requirements, we introduce the channel model's scalability to achieve a tradeoff between model accuracy and complexity.

Figure 87 shows how scalability can be adjusted by the number of stochastic clusters and the propagation path types resolved by the RT tool. As the hybrid model shifts towards a deterministic approach, it becomes more accurate yet computationally demanding.

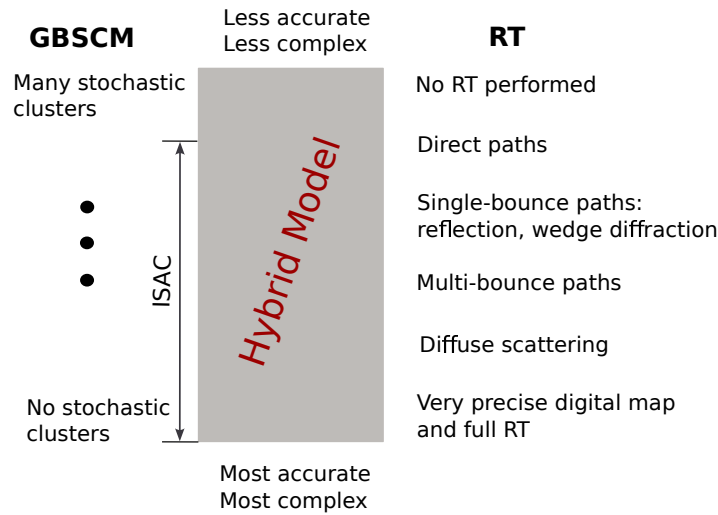

Figure 87: Scalability between stochastic and deterministic channel modeling components and accuracy vs computational complexity.

**Scattering from the extended target**

Vehicles are significant components affecting the wireless channel in vehicular applications, including ISAC, as their influence by scattering and blockage can be more noticeable than that of buildings and other objects. Therefore, scattering from these objects needs to be characterized, which is a challenging task as the strength of reflectivity depends on the target's size, curved shape, and material properties. We propose a solution in which we discretize the complex-shaped curved body into polygons, such as a triangle mesh, and employ conventional RT to compute the reflected and diffracted fields.

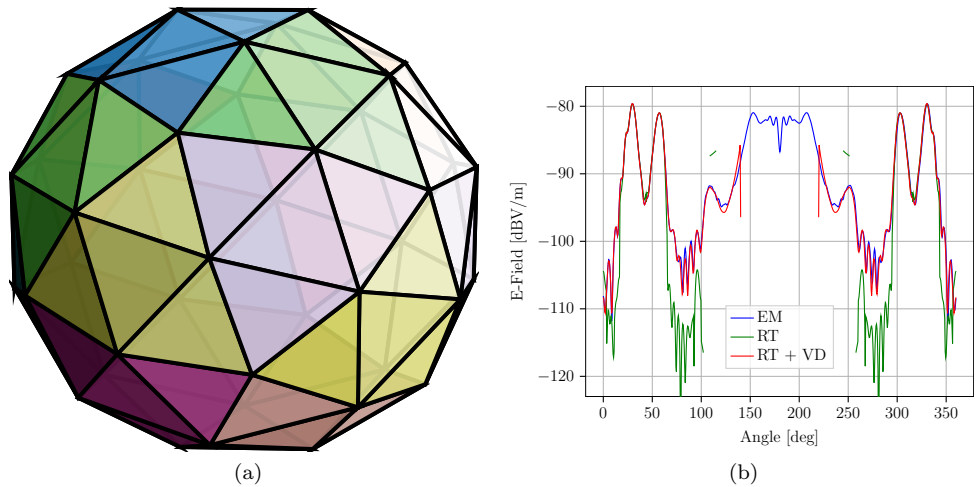

Figure 88: (a) Discretized sphere used for the simulation; Radius $12.8\lambda$, facet side $7\lambda$. (b) Scattered field from the sphere at 2 GHz computed by full-wave simulation (blue), conventional RT (green), and conventional RT augmented with vertex diffraction (red)

This method is applied to a discretized sphere depicted in Figure 88a. It was found that diffraction from vertices, which is not usually considered in RT tools, is essential to describe the scattered field accurately. A method for accounting vertex diffraction [ACCM09] is implemented in the RT tool.

The resulting scattered fields can be seen in Fig. 88b, where Electromagnetic (EM) simulation is compared against conventional RT and RT with vertex diffraction (VD). The conventional RT tool (green color) shows generally inaccurate results, while adding vertex diffraction (red color) improved the overall result. A slight improvement was also in the shadowed region behind the object because vertices are more likely to be exposed to both transmitter and receiver than wedges. However, advanced diffraction methods, such as slope diffraction [MPM90], are still required to characterize this region.

The results show that the RT tool with vertex diffraction can provide an opportunity to model the scattering from the discretized complex-shaped bodies accurately. However, the RT tool with single-bounce interaction often fails to predict the field in the shadow region of the object.

**Future prospects**

Incorporating vertex diffraction is essential for computing the scattering in the "reflected" region of the discretized body. However, more advanced diffraction methods, such as slope diffraction or creeping wave diffraction [MPM90], are still required to accurately predict the scattered field in the "shadow" region of the object.

Another important aspect involves the computational efficiency of channel modeling, which can be enhanced by utilizing faster hardware, such as compute unified device architecture (CUDA) graphics processing units (GPU) cores, for running the GBSCM and RT. Additionally, advancements in faster path tracing algorithms, commonly employed, for example, in the computer design and gaming industry, offer viable solutions for improving efficiency.

### 3.4.3 Quasi-Deterministic Channel Propagation Model for Human Sensing: Gesture Recognition Use Case

In [CCS$^+$24], we describe a quasi-deterministic channel propagation model for the human gesture recognition use case, reduced from real-time measurements with our context-aware channel sounder. The measurement campaign recreates a scenario in which a human is seated on a sofa facing a monitor equipped with a monostatic Radar sensor and moving their hands for sensing by an enhanced reality application. The campaign consists of a total of four human subjects with diverse body dimensions and 20 different body motions for each subject, for a total of 80 cases. Each motion lasts a period of 3.9 s, over which the channel is sampled every 2.6 ms, for a total of 1500 distinct times.

The sounder features a radio-frequency (RF) system with 28 GHz phased-array antennas to extract discrete multipaths backscattered from the body in path gain, delay, azimuth angle-of-arrival, and elevation angle-of-arrival domains per time. Thanks to our switched beamforming technique and super-resolution algorithms, the multipaths have an average error of only 0.1 ns in delay and 0.2° in angle. While viewing the 3-12 multipaths per time provides little feature information, clusters of multipaths clearly emerge when aggregating them over time: Fig. 89 (a) shows the extracted multipaths for an illustrative case aggregated over the whole collection period. Most notably, two distinct clusters around 15 ns can be associated with the left and right hands, the two clusters that arrive earliest as they are closest to the receiver. Once extracted, the multipaths are classified according to which body parts backscattered them into the receiver via our density-based algorithm. While most algorithms cluster multipaths over the path gain, delay, and angle domains only, our algorithm clusters over time as well. This enables multipaths backscattered from different body parts that are seemingly close per time to be discriminated via the tracks they form over time. Fig. 89 (b) shows the multipaths clustered into eight different body parts.

The sounder also features a camera / Lidar system to extract discrete keypoints that correspond to salient parts of the body in the same domains as the multipaths. During the same collection period, the data acquired from the camera/Lidar system was processed into three-dimensional (3D) keypoints: two-dimensional (2D) keypoints are first extracted from the images that the camera captures through the high-resolution network (HRNet) algorithm, an AI-based algorithm trained from a rich database of human images, to reliably classify body parts. The keypoints are then projected onto the 3D point cloud that is generated by the Lidar, as shown in Fig. 89 (c), yielding 3D keypoints that exist in same geometrical domain as the multipaths. In Fig. 89 (d), the keypoints extracted over the collection period are aggregated, and then clustered according to the body parts classified in the legend. Thanks to the precision of the RF system, we can reliably associate the multipaths to the keypoints geometrically, through visual inspection. The multipaths then inherit the classification of their counterpart keypoints; the eight counterpart clusters found in Fig. 89 (b) and Fig. 89 (d) are assigned the same colors.

The multipaths are subsequently reduced into channel model parameters, namely Radar cross-section (RCS) and channel coherence time. With the multipaths classified by body part, the channel model parameters can be rendered part specific. For example, the hands, which are moving, have notably different coherence times than the chest, which is static.

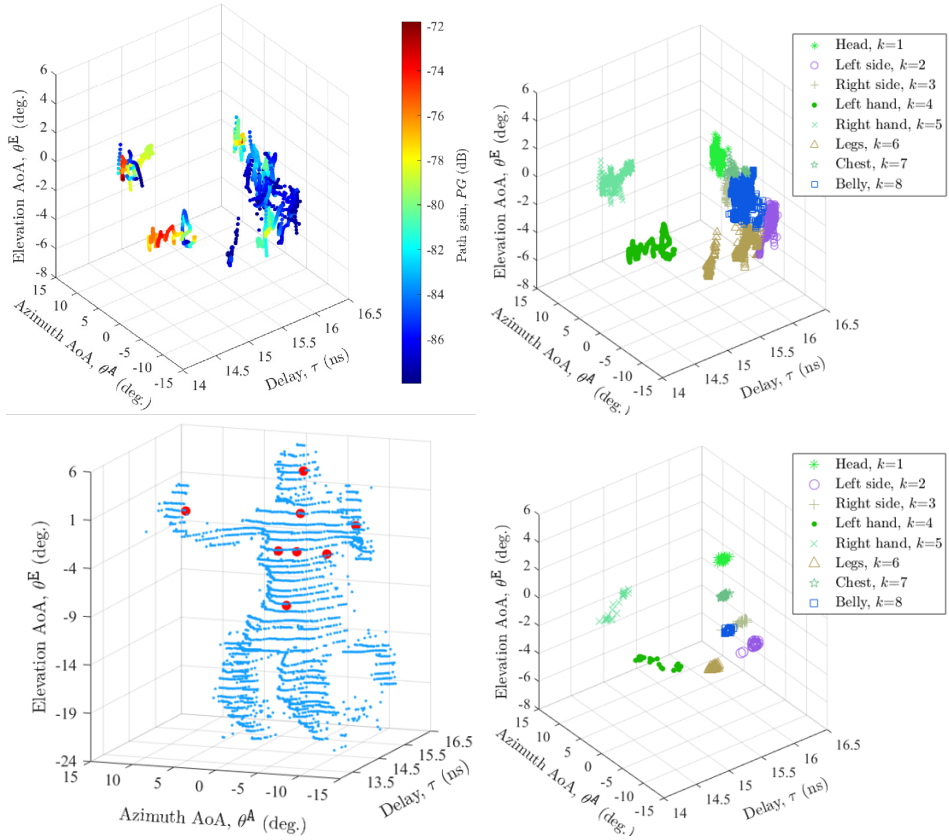

Figure 89: Illustrative case of a human moving their hands outwards (a) Multipaths extracted from the RF system per time, aggregated over the whole collection period (b) Aggregated multipaths clustered into eight dominant scatterers of the body (c) Keypoints extracted from the camera / Lidar system at one time (d) Keypoints clustered and classified into the same eight body parts as the multipaths, to enable association between them.

### 3.4.4 Physics-Informed Generative Neural Networks for RF Propagation Prediction with Application to Passive Radio Localization

Passive radio-frequency (RF) sensing enables communication while sensing paradigm as it employs ambient RF signals to sense, detect, locate, and track people that do not carry any electronic device (device-free or passive). RF sensing adopts electromagnetic body models to infer geometrical information about the presence and location of bodies inside the monitored area. These models are time consuming and prevent their adoption in strict real-time computational imaging problems such as tomography, holography, and Bayesian estimation. To address this key issue, generative neural network (GNN)s have recently attracted a lot of attention thanks to their potential to reproduce a process by learning relevant physical constraints. GNNs such as variational autoencoders (VAE) and generative adversarial network (GAN) families, can be used to build a surrogate prior model for RF sensing which is helpful in several localization algorithms.

The contribution explores emerging opportunities of GNN tools targeting real-time passive RF sensing, particularly in multiple-input multiple-output (MIMO) communication systems. The proposed GNN tool comprises of a generator of electromagnetic (EM) field samples which can be used for EM full-wave propagation analysis. GNNs are shown to reproduce human body blockage effects under configurations which might be unseen during the training phase, or rather difficult to predict through traditional EM field computing methods. The proposed contribution demonstrates the effectiveness of the proposed GNN tools to support real-time passive Bayesian localization with WiFi devices and reproduce the Bayesian prior.

**Synthetic generation of RF signals**

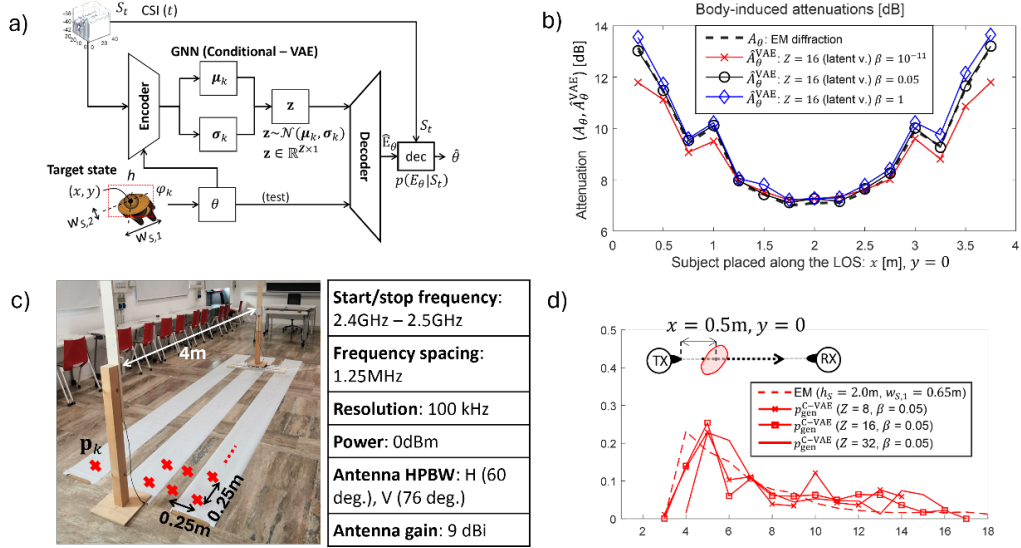

Figure 90: (a) generative neural network (GNN) for reconstruction of RF data - conditional variational autoencoders (C-VAE) implementation. (b) and (d) Effects of GNN reconstruction on dataset accuracy (comparison between measurement of body-induced attenuations and synthetic RF data reconstruction). (c) Experimental setup for validation.

The GNN tool with schematic reported in Fig. 90 (a) is based on a C-VAE implementation [ZYB23]. It has been evaluated to reproduce RF tensor structures considering different configurations and size of the 3D human target(s), varying positions orientation and size (for 2D/3D reconstruction), varying antenna array structures. Fig. 90 (b) and (d) consider a comparison between measurement of body-induced attenuations for the setup in Fig. 90 (c) and reconstructed synthetic RF data. We compare the ability of the generative model to reproduce the distribution of the RSSI considering different positions of the subject along the LOS path (b) and motions (d). Goal is to reproduce the randomness and distribution of samples. This is also necessary to optimize the hyperparameters of the generative model and achieve a good compromise between accuracy and randomness of samples. Validation is done with measurements taken in a hall with parameters detailed here (size 6 m x 14 m. Tx and Tx nodes are spaced 4 m apart, while the LOS is parallel to the lateral walls and horizontally placed at 1 m from the floor. Both Tx and Rx are equipped with directional antennas with parameters summarized in the Table.

**Passive localization example**

In the following example, the goal is to detect the distance of the target from the multi-antenna Rx device (Fig. 90 (c)) in real-time, as in typical radar applications. The proposed use case is critical in industrial scenarios where human workers operate in areas featuring increasing levels of safety or privacy. Using a Bayesian estimation method [SFK+24], the Table of Fig. 91 analyzes precision and recall for varying distance from the Rx and compares three cases:

1. Estimated prior: assumes the Bayesian prior probability [SFK+24] being estimated from large calibration measurements (ideal case, human subjects involved during training data collection);

2. C-VAE prior: adopts to generate synthetic RF samples using the C-VAE generator tool (Fig. 90 (a)) while no human subjects are involved during training;

3. Uniform prior: the prior is replaced with a uniform probability function.

| Distance | Estimated prior | C-VAE prior | Uniform prior |
|---|---|---|---|
| 0.5m | 0.95/0.98 | 0.92/0.85 | 0.67/0.48 |
| 1m | 0.92/0.91 | 0.76/0.66 | 0.35/0.20 |
| 1.5m | 0.92/0.90 | 0.87/0.67 | 0.41/0.55 |
| 2m | 0.89/0.88 | 0.70/0.64 | 0.47/0.30 |
| 2.5m | 0.78/0.85 | 0.47/0.69 | 0.19/0.18 |
| 3m | 0.86/0.89 | 0.73/0.77 | 0.18/0.19 |
| 3.5m | 0.76/0.74 | 0.62/0.83 | 0.19/0.26 |

Figure 91: Precision/recall metrics for distance estimation. Estimated prior from calibration data, C-VAE prior and uniform prior cases compared.

Note that scenario (1) gives the best performance, as expected; on the other hand, it requires time-consuming data collection and a calibration stage which might not be feasible in practice. Case (3) corresponds to the worst-case scenario since no prior information on body-induced attenuations is available. Finally, case (2) does not need any calibration as it uses the C-VAE tool to real-time generate synthetic samples from the prior. From the results in the table, the performance of C-VAE prior scenario approaches the estimated prior case, with an average drop of about 10

### 3.4.5 Parameter estimation for ISAC propagation environments with model-informed machine learning

Due to its applicability in urban or indoor areas, ISAC systems must be able to perform target sensing in multipath-rich scenarios. In such scenarios, detecting and localizing sensing targets presents a significant algorithmic challenge. To localize a target, the signal processing chain must estimate target properties, which are geometrical related to a target's location. The most widely used propagation model describes the electromagnetic propagation using rays (usually referred to as specular propagation paths), which assumes planar electromagnetic waves interacting with the sensing environment by reflection, refraction, diffraction, and scattering. The properties of these path (e.g., length, Direction-of-Arrival) are linked to objects the path interacts with, and hence, also link to sensing target location. Localization is enabled by estimating these path parameters and then using geometric relationships to perform localization, e.g. via triangulation or multilateration. Hence, the estimation of propagation paths is the first step in an ISAC system, serving as the input for the localization and tracking processing. Consequently, the achievable localization accuracy for any target depends on the accuracy of the propagation path parameter estimates.

Different approaches exist to solve the parameter estimation step, differing mainly by the assumptions contributing to the imposed signal model, which serves as an algebraic model for the measured signal at the receiver. Conventional RADAR sensing applications typically use non-coherent approaches like CFAR to detect and estimate the target parameters from a range-Doppler map. To achieve real-time requirements, the algorithms sacrifice estimation accuracy by utilizing a relatively simple signal model allowing peak detection in the range-Doppler maps. These grid-based approaches are limited by the available sensing signals spectral resources. Since ISAC is also intended for dense multi-user scenarios, these resources might be more limited compared to dedicated radar systems. For example, the signal bandwidth impacts the sensing range resolution. The varying resource allocations of the signal (e.g., from OFDMA), additionally degrade the performance. Other signal models known from channel sounding, e.g. as used in algorithms such as radio channel parameter estimation from channel sounding measurements (RIMAX) [Ric05] or SAGE [FH94a] can achieve better estimation accuracy. With algorithmic techniques such as successive interference cancellation they can even achieve improved target separation beyond the Rayleigh limit. This allows to cope with strong and weak paths that are close to each other. This more refined signal model allows the algorithms to compensate the effects of possibly unstructured OFDMA signals. However, they have a high computational complexity that currently makes them ill-suited to perform under the real-time constraints of sensing systems. Yet, if it were possible to optimize for them for the real-time requirements of ISAC, it would render

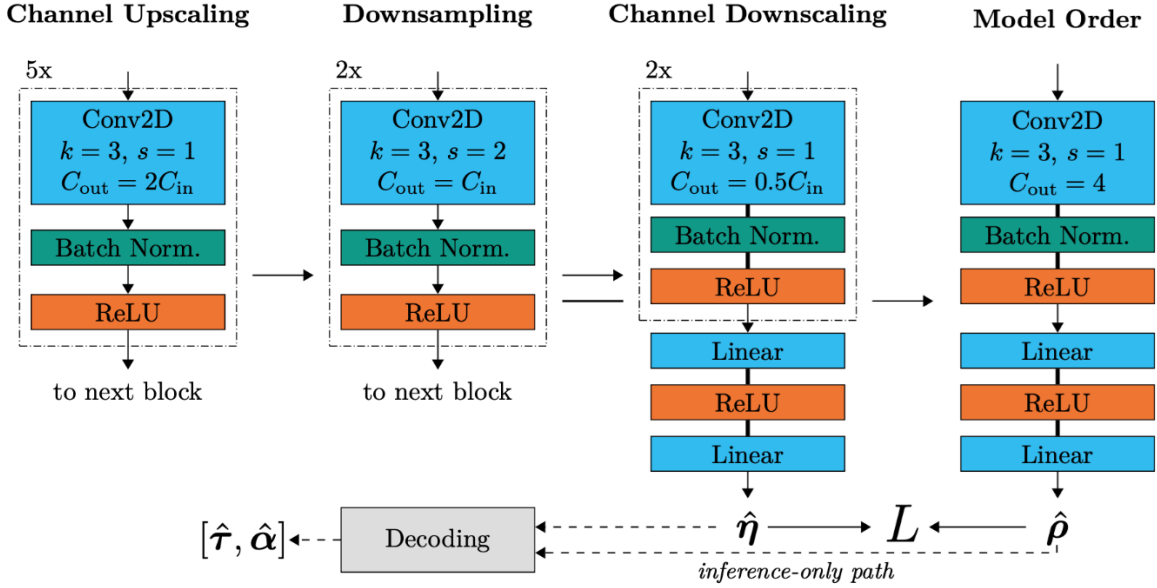

Figure 92: The architecture of the CNN used in [SSF+]. The network uses mainly convolutional layers, which the number of trainable parameters, enabling multiple dimensions. The predictions contain both the parameters and the model-order of the data.

them prime candidates for the parameter estimation task.

Optimizing the algorithms for runtime requires an understanding of their individual steps. In [Ric05] the estimation process is split into multiple iterations. In each iteration, the algorithm estimates and adds a single path to the model. Each iteration is divided into an initial global search followed by gradient-based optimization (for all paths currently in the model). During the whole procedure the algorithm uses the imposed signal model to achieve coherent processing both in amplitude and phase. This process is then repeated iteratively until a stop criterion is reached (e.g., by checking the "whiteness" of the residual). As each iteration depends on the result of the previous one, the process is not easy to parallelize. However, it is possible to reduce the overall number of iterations, if the global-search algorithm can initialize multiple paths at once. When we aim searching for multiple new paths simultaneously, the accuracy of the initial estimates is of importance. The following gradient-based optimization requires initializations sufficiently close to the global minimum of the highly non-convex likelihood function for the subsequent gradient steps to converge. Recently works [SSF+, BU21] utilize Deep Learning to learn an approximation of the inverse mapping, which is then used to obtain multiple initializations in the global-search procedure. While [BU21] shows results for a single parameter domain (e.g., only delay) with a fully connected network, [SSF+] presents an approach for multiple dimensions (i.e., delay and Doppler) with a convolutional neural network (CNN), shown in Fig. 92. Due to the use of convolutional layers, the approach can also be extended to more dimensions, as shown in [SSF+23] for joint estimation of delay, Doppler-shift, direction of arrival (DoA) in azimuth and elevation. The approaches in [SSF+, SSF+23] also require no prior knowledge of the model-order and work directly on the sampled, complex-baseband channel transfer functions. For the training, the approaches[SSF+, BU21, SMR+] utilize the available algebraic structure provided by the signal models. They can therefore be classified as model-informed machine learning approaches, similar to model-based deep learning.

This model-informed machine learning combines model- and data-based approaches into hybrid versions which exhibit reduced algorithmic complexity, fixed computation time, and physical explainability of the Machine Learning model. As an intersectional research field it attempts to overcome shortcoming of either purely model- or data- based approaches. Fig. 93 show the groundtruth (circle) and parameter estimates (crosses) with the data $|Y_1|^2$ (rectangular window) in the background. We observe the detection probability and estimate accuracy improves with increasing SNR. The results highlight that it is possible to obtain parallel initializations for the paths which are sufficiently close to the global minimum of the non-convex likelihood function, such that the gradient-based optimiza-tion can converge. In addition, [SSF+] also highlights the superior model order selection performance of the Machine learning approach compared to information-theoretic solutions, such as the efficient detection criterion (EDC). A comparison of algorithm runtimes in [SSF+] also shows, that the new hybrid algorithm exhibits a fixed runtime, which is of great interest in applications with real-time requirements, such as ISAC. Simultaneously, the hybrid algorithm achieves on par accuracy with the iterative Maximum Likelihood in the simulative study. For a realistic performance estimate and as-sessment of real-world performance, more investigations are required. Particularly, training strategies must be found to reduce the requirement for large amounts of site-specific measurement data. This could be done in the form of pretraining on synthetic data generated from the signal model and subsequent fine-tuning on very few site-specific samples. Another possibility would be to perform on-site finetuning of the models, using available information about the propagation channel, such as accurately known reflectors in the channel with known positions. Also, it is not yet clear to what degree the usage of Machine Learning can alleviate the impacts of model mismatch. The work in [SMR+] demonstrates that is possible to reduce the impact of imperfect array calibration. However, it is still unclear to which degree other mismatch sources, such as measurement equipment, polarization, or coloured noise impact these new approaches. On the other hand, this is of great interest in ISAC applications, as mobile communication deployments foreseen for the sensing task are not calibrated sufficiently for sensing, particularly localization.

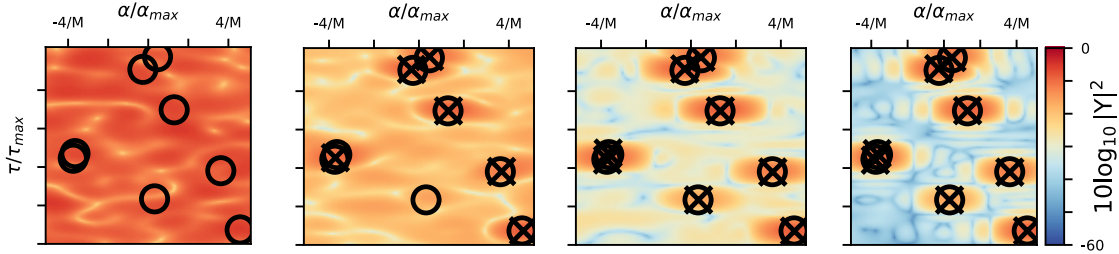

Figure 93: Delay $\tau$ and Doppler $\alpha$ estimations with -40, -20, 0, 20 SNR

## 3.5 Sensing in WiFi and Cellular Networks

This section focuses on WiFi and cellular network technologies for ISAC, exploring their potential for sensing and communication. WiFi networks provide coverage and flexibility in primary indoor environments, while cellular networks provide wide-area connectivity and robust infrastructure. Both technologies promise to enable ISAC applications across diverse scenarios.

Section 3.5.1 covers the possible errors and limitations of CSI, highlights its applications, and provides an example of localization using CSI. Section 3.5.2 focuses on sensing calibration techniques. In Section 3.5.3, we discuss how WiFi signals can be used for human activity recognition. Section 3.5.4 delves into WiGig multi-person positioning and respiration sensing. Section 3.5.5 addresses radio localization using cellular systems. Section 3.5.6 examines sensing and localization in IoT networks. Finally, Section 3.5.7 introduces integrated localization and communication (ILAC), enabled by ambient backscatters.

### 3.5.1 CSI-Based Sensing in WiFi Networks

channel state information (CSI) refers to the information about the state of the wireless channel between the transmitter and receiver. In general, it describes the impact of factors such as scattering and fading over distance. This information can then be utilized by communication systems to make decisions about efficient data transmission over the wireless channel. CSI is calculated on the receiver. In the beginning of the packet the transmitter sends long training field (LTF) with predetermined symbols in the preamble of the packet. Upon receiving the LTFs, the receiver estimates the CSI $H(f,t)$ using equation

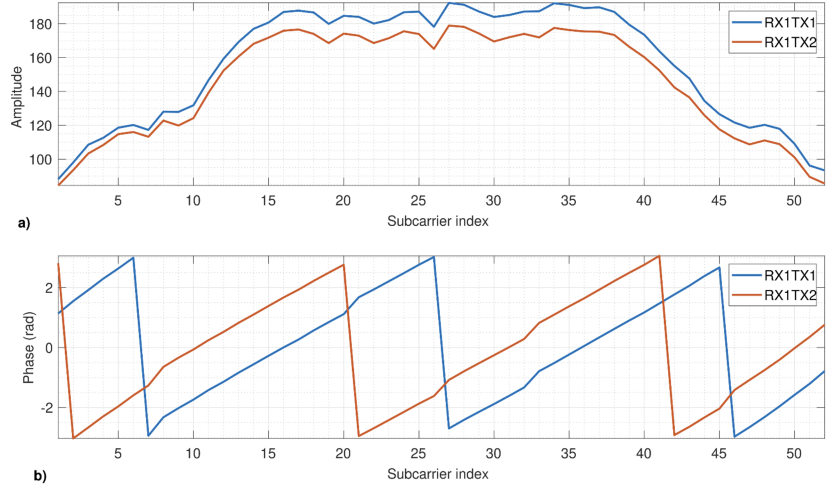

Figure 94: Example of CSI amplitude (a) and Example of CSI phase (b).

$$H(f,t) = \frac{Y(f,t)}{X(f,t)}, \tag{52}$$

where $Y(f,t)$ and $X(f,t)$ are received and transmitted signals in the frequency domain [DZ22]. In the IEEE 802.11 networks, OFDM is a key signal forming technique employed for efficient data transmission. OFDM divides the available frequency spectrum into multiple orthogonal subcarriers. CSI is extracted for each subcarrier to provide detailed data about the channel state of communication between a transmitter and receiver. By obtaining CSI data for individual subcarriers, the system can adapt its modulation and coding strategies dynamically based on the varying channel conditions. CSI is represented by the following equation

$$\mathbf{y} = \mathbf{H} \cdot \mathbf{x} + \mathbf{n}, \tag{53}$$

where $\mathbf{y}$ is the received signal, $\mathbf{x}$ is the transmitted signal, $\mathbf{n}$ is the noise vector and $\mathbf{H}$ is the CSI matrix. CSI matrix is defined by

$$\mathbf{H}_i = \left\{ \begin{array}{cccc} h_i^{11} & h_i^{12} & \cdots & h_i^{1N} \\ h_i^{21} & h_i^{22} & \cdots & h_i^{2N} \\ \vdots & \vdots & \ddots & \vdots \\ h_i^{M1} & h_i^{M2} & \cdots & h_i^{MN} \end{array} \right\} \tag{54}$$

where $h_i^{mn}$ represents the CSI of the $i$-th subcarrier for $m$-th receiving antenna and $n$-th transmitting antenna. The CSI value $h$ is a complex number

$$h_i^{mn} = |h_i^{mn}|e^{i\angle h_i^{mn}}, \tag{55}$$

consisting of amplitude $|h_i^{mn}|$ and phase $\angle h_i^{mn}$ [KW13]. An example of the CSI is shown in Fig. 94.

Raw CSI data may be corrupted by hardware or interference from nearby devices. The main sources of errors are [LNK19]:

- Random initial RF oscillator phase offset - until the network card is reset or recalibrated on Rx and Tx chains are added constant random phase shift. It is because the network card has random phase shifts on Tx and Rx chains during initialization.

- CFO – occurs when the carrier frequency of the transmitter and receiver are not well synchronised. This offset is constant for all subcarrier frequencies across the antenna array.

- sampling frequency offset (SFO) - the digital-to-analogue converter in the transmitter and the analogue-to-digital converter in the receiver sample the signal at different times and this causes a time offset. This offset is constant for all subcarriers and antennas.

- Packet Boundary Detection Delay - is the time delay incurred in accurately detecting the start points of the preamble of packets within a received signal. The packet detection delay in the time domain causes phase rotation. It is variable across packets.

- automatic gain control (AGC) - the power of the signal varies according to the distance between the transmitter and the receiver. Automatic gain control therefore tries to maintain a suitable signal amplitude. However, it cannot accurately compensate for the signal attenuation.

- antenna coupling errors (ACE) - the small distance between the antennas affects each other and this causes the antennas to be coupled. This violates the requirement for independent MIMO channel characteristics.

CSI is implemented on receivers as a standard feature but access to this information is not easy and common on off-the-shelf devices [FG19]. CSI extraction on commercial network cards is quite limited. It requires special software and the specific hardware for which the software is designed. The main CSI extraction tools for commercial off-the-shelf network interface cards are:

- Nexmon CSI - supports devices with BCM4365,66 and BCM4339,58,455 chipsets. It is 802.11ac compatible and uses a maximum bandwidth of 80 MHz [FG19].

- 802.11n CSI Tool – a tool that is designed to extract CSI from the network card WiFi Link 5300 with three antennas. It supports Institute of Electrical and Electronics Engineers (IEEE) 802.11n standard. Supported bandwidths are 20 MHz and 40 MHz [LNK19].

- Atheros CSI Tool - a tool that allows you to measure CSI on certain types of Atheros devices. Supports 802.11n standard and 20 MHz and 40 MHz bandwidths [YX15].

- PicoScenes – a robust platform which supports different types of network cards and software-defined radio devices. Based on the used hardware it is possible to extract CSI up to IEEE 802.11ax standard [ZJ22].

- FeitCSI - an open-source tool that enables CSI extraction up to IEEE 802.11ax standard and bandwidth up to 160 MHz on certain types of intel devices [MHM23].

CSI extracted from IEEE 802.11 networks can offer valuable insights and applications across various sectors and industries. In healthcare applications, CSI can be utilized for monitoring patient movements and vital signs without the need for invasive sensors. By analyzing the changes in CSI patterns, healthcare professionals can track patient activities, detect falls, and ensure timely assistance in case of emergencies [BT18]. CSI can be leveraged for real-time monitoring of the operational state of machines in industrial environments [Dem23]. Human presence and activity recognition using CSI can be used in smart home systems for adjusting lighting, temperature, and other settings based on the occupants' preferences and activities, enhancing comfort and energy efficiency [JY18]. CSI can also be used for enhancing security and surveillance systems. CSI patterns are used for detecting when a door/window is open or closed and for detecting movements inside the house. This enables prompt responses and ensures the safety of the premises [BT18]. CSI can be employed for indoor localization and tracking applications in various settings. The first localization system using CSI can be considered PinLoc [SRCM12a] which achieves localization accuracy of 1 m. There are also localization systems [XW17], [WW21], which from phase and amplitude create radio images and then use various techniques to estimate the position of the target. We also realized basic experiments in standard 802.11ac using radio images and fingerprinting method. Two access points were set as reference points which transmit beacons every 100 ms. Fingerprints were collected every 1 m in a grid with the size of $10 \times 4$ m. Radio images were created from 20 samples from the first access point and 20 samples from the second access point. An example of a fingerprint radio image is shown in Fig. 95.

In total 44 fingerprints were created. Then we take fingerprints on random places in the area and estimate the position of the mobile station using the k-NN algorithm. Neighbors were selected by the correlation coefficient $r$

$$r = \frac{\sum_m \sum_n (A_{mn} - \overline{A})(B_{mn} - \overline{A})}{\sqrt{(\sum_m \sum_n (A_{mn} - \overline{A})^2)(\sum_m \sum_n (B_{mn} - \overline{B})^2)}} \qquad (56)$$

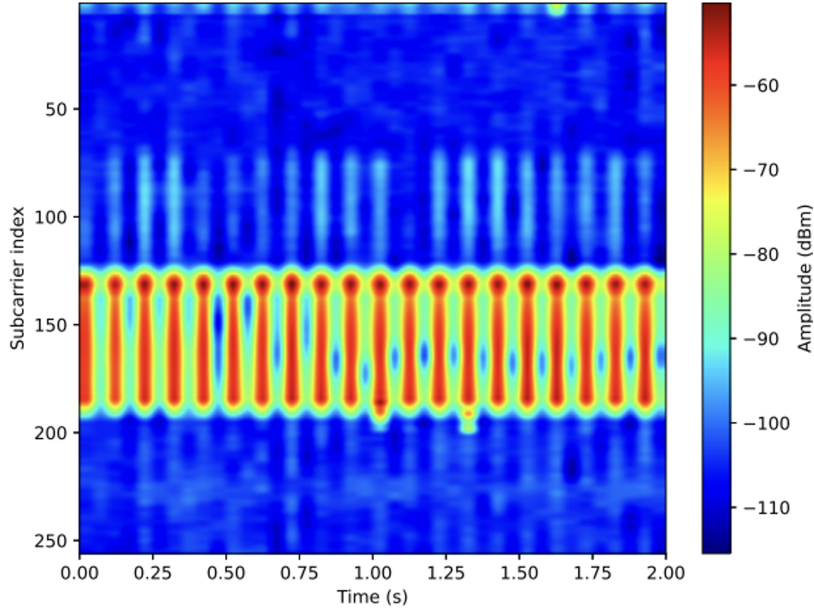

Figure 95: Example of radio image fingerprint.

calculated for each fingerprint $A$ stored in the database and measured fingerprint $B$ in real-time. Results are shown in Tab. 8.

Table 8: Accuracy using k-NN a CSI radio images.

| **Number of neighbours** | 2 | 3 | 4 | 5 |
|---|---|---|---|---|
| **Accuracy [m]** | 1.2 | 1.20 | 0.90 | 1.02 |

The utilization of CSI presents an opportunity across diverse sectors and applications, ranging from healthcare to various industries and localization services. The potential of CSI in these fields emphasizes its importance as a valuable asset for driving innovation, enhancing operational workflows, and delivering improved services across a diverse array of industries and applications.

### 3.5.2 WiFi Channel State Information: Model, Sensing parameters, and Tools

MIMO transmission and beamforming have been introduced in many wireless communication systems, including the WiFi system. Therefore, a CSI frame has been included in the PHY of WiFi systems since the IEEE 802.11n standard, allowing the WiFi device to perform spatial multiplexing and diversity techniques. Generally, CSI contains radio propagation information in terms of a complex-valued channel frequency response between two WiFi node antennas on each OFDM subcarrier. By simply assuming a perfect synchronization between a pair of WiFi nodes, the ideal CSI of the $i$th WiFi OFDM packet measured at time $t_\mathrm{i}$ and at the $k$ subcarrier frequency $f_k$ between the $n$th and $m$th antennas of two WiFi nodes can be modeled [KHST18] by

$$H_{nm}(t_i, f_k) = \underbrace{\sum_{l=1}^{L_\mathrm{S}} \gamma_{lnm}^\mathrm{S}(f_k)\exp\left(-\mathrm{j}\frac{2\pi}{\lambda_k}d_{lnm}^\mathrm{S}\right)}_{H_{nm}^\mathrm{S}(f_k)} + \underbrace{\sum_{p=1}^{P_\mathrm{D}} \alpha_{pnm}^\mathrm{D}(t_i, f_k)\exp\left(-\mathrm{j}\frac{2\pi}{\lambda_k}d_{pnm}^\mathrm{D}(t_i)\right)}_{H_{nm}^\mathrm{D}(t_i, f_k)}, \quad (57)$$

$$d_{pnm}^\mathrm{D}(t_i) = d_{pnm}^\mathrm{D}(t_{i-1}) + \lambda_k \nu_{pnm}(t_i)\Delta t_i \quad (57\mathrm{a})$$

where $H_{nm}^\mathrm{S}(f_k)$ and $H_{nm}^\mathrm{D}(t_i, f_k)$ are static and dynamic components of WiFi CSI, respectively. The former describes a contribution of the WiFi signal propagating through $L_\mathrm{S}$ multipaths and interacts with the static interacting object (IO), whereas the latter describes a contribution of the WiFi signal propagating through $P_D$ multipaths and interacts with dynamic IOs. The total propagation distance of the $l$th static and $p$th dynamic paths are notated by $d^S_{lnm}$ and $d^D_{pnm}(t_i)$, respectively. Here, $\Delta t_i$ is the time interval between adjacent OFDM packets in the time samples $t_i$ and $t_{i-1}$. The wavelength of the $k$th subcarrier and the Doppler shift induced by the moving IO along the $p$th path are defined as $\lambda_k$ and $\nu_{pnm}$, respectively. With the availability of multipath parameters $d^S_{lnm}$, $d^D_{pnm}(t_i)$, and $\nu_{pnm}(t_i)$ embedded in WiFi CSI, many studies have investigated the possibility of WiFi CSI for sensing applications [WGYZ18].

The use of WiFi CSI for sensing can be viewed as an application of bistatic OFDM MIMO radar system. Consequently, the primary sensing parameters include distance, Doppler, and AoA or AoD. From the CSI model in (57), the distance to static and dynamic objects relative to both WiFi nodes can be obtained from $d^S_{lnm}$, $d^D_{pnm}(t_i)$, respectively, while $\nu_{pnm}(t_i)$ provides the radial speed of the sensing objects. Moreover, object positions can also be determined from AoA and AoD which are estimated through spatial processing to the CSI along the $n$th and $m$th element domain of the array. However, since WiFi systems were not originally designed for sensing purposes, the full potential of these sensing parameters embedded in WiFi CSI may not be fully utilized. For example, clock synchronization between two WiFi nodes is required to obtain an absolute range information that is not required for data transmission. WiFi bandwidth is still narrower in comparison with the radar, resulting in poorer range resolution. Only a fewer array antenna elements can physically be installed in the commercial WiFi node such as laptops and access points, which affects the angular resolution. The non-periodic interval transmission of the WiFi packets also impacts the Doppler resolution. These challenges have highlighted the difficulties of WiFi sensing using commercial WiFi chips and therefore led to the form of the IEEE 802.11bf Task Group, the under-developing WiFi standard for sensing [CSL$^+$23].

Typically, PHY layer signaling including CSI cannot be obtained directly from commercial WiFi devices at the time of this report. In the early stage, the work in [PGGP13] used SDR as a WiFi transceiver to extract CSI. Later, CSI-based WiFi sensing has gained much attention following a release of open source software-modified firmware of network interface card (NIC), enabling CSI stream capture directly from commercial WiFi chips such as Linux 802.11n CSI Tool [HHSW11], Atheros CSI Tool [XLL15], and PicoScenes [JLR$^+$22]. These tools can also capture CSI when WiFi nodes operate in injection and monitoring modes, allowing users to manually control the packet transmission interval and the MCS. The measured CSI of the $i$th OFDM packet can be obtained as a complex-valued matrix with dimension of $M \times N \times K$ where $M$ and $N$ are the number of antenna elements of two WiFi nodes and $K$ is the number of OFDM subcarrier.

**Measured CSI Calibration Technique for WiFi Sensing**

Suppose a WiFi packet is sent from the Tx WiFi node and received by the Rx WiFi node. The preamble received from the transmitted signal is generally utilized for the synchronization and extracting of WiFi packet [SFFM99b]. Although synchronization can be optimally obtained, it inevitably introduces additional phase offset and rotation to the CSI since the internal clocks of the Tx and Rx WiFi nodes are not synchronized. Therefore, the measured CSI can be modeled [KHST18] by

$$\hat{H}_{nm}(t_i, f_k) = H_{nm}(t_i, f_k) \exp\left(j\Psi_{nm}(t_i, f_k)\right) + \mathcal{N}_{nm}(t_i, f_k) \tag{58}$$

where $\mathcal{N}_{nm}(t_i, f_k)$ and $\Psi_{nm}(t_i, f_k)$ are measurement noise and the overall phase offset and rotation, respectively. The effect of these undesired phase terms will contaminate the actual CSI phase due to wave propagation, preventing the estimation of the Doppler frequency. The phase offset also results in an unreliable estimation of the propagation distance. The sources of phase offset and rotation due to non-synchronization WiFi chips have been widely discussed throughout the CSI phase-related work [ZZX17]. It can be generally grouped into at least four sources: carrier frequency offset, phase-locked loop offset, sampling frequency offset, and packet boundary delay. Except for the phase-locked loop offset, others are time-varying, which cause temporal phase rotation in the measured CSI phase. In addition to hardware-induced phase offset, spatial mapping and cyclic shift diversity [KHST18] also contributed to the undesired phase offset to measured CSI. Specifically, the latter also induces phase offset among antenna elements, which may affect the accuracy of AoA and AoD estimation.

To handle this unwanted phase offset and rotation due to non-synchronization WiFi chips, various techniques have been introduced in the literature. In general, their methods could be divided into four approaches in the following.

- **Calibration by CSI phase removal** One of the widely used and straightforward calibration methods is to utilize only the CSI amplitude. Although Doppler reside in the phase term, as analytically investigated in [WLS$^+$17], the contribution of Doppler frequency can also be manifested in the CSI amplitude in terms of multipath fading. However, both positive and negative Doppler shifts cannot be distinguished due to a real-valued CSI amplitude, and distance information cannot be obtained in this calibration method.

    **Calibration with linear phase fitting** This is another well-implemented signal processing method [SRCM12b]. Basically, it assumes that the undesired phase offset and rotation are linearly and dominantly distorted by the measured CSI phase. After removing this linear phase offset, the phase residual should contain a temporal rotation due to the Doppler frequency. The drawback of this approach is that a partial CSI phase information including those due to Doppler frequency is also removed. Nevertheless, it can be directly utilized to detect a simple motion through the temporal phase profile. Due to its empirical approach, the sensing application may be limited and cannot be generalized to a wider range of scenarios. Same as in the previous calibration method, distance information cannot be acquired.

- **Calibration with measured CSI from another antenna element** The concept of this method stems from the assumption that both received antenna elements generated by the same WiFi ship should experience the same undesired phase rotation. The first approach of this technique [QWZ$^+$17] applies a cross-correlation between the CSI measured from two nearby antenna elements to produce the calibrated CSI. Analytical investigation showed that the cross-correlation can mitigate the undesired phase rotation and thus recover the Doppler frequency from the measured CSI. However, the issue with this approach is obviously the image of Doppler frequency. The more robust approach was later proposed in [ZWX$^+$19] by normalizing the measured CSI with that of another antenna element. Analytical investigation and experiment have proven that the Doppler frequency can be recovered without the image issue. This technique is applicable when the static components are significantly stronger than the dynamic component and the two antenna elements should be as close to each other as possible. Since the Doppler frequency can be recovered, this calibration technique has recently gained a lot of attention for the motion sensing application. Again, this calibration method cannot provide reliable distance information.

- **Calibration with back-to-back CSI from another antenna port** This calibration method uses the same concept as the previous. However, instead of using measured CSI from two antenna elements, the back-to-back calibration [KHST18] introduces a wired connection between antenna ports of the Tx and Rx WiFi nodes. The CSI measured in the wired channel is called back-to-back (B2B) CSI, which has a known channel properties: a single-path channel with propagation distance proportional to the cable length. The calibrated CSI is computed by normalizing the measured CSI with the B2B CSI. This calibration technique can recover not only the Doppler frequency, but also the propagation distance [KST22] that resides in the measured CSI. However, the drawback of this technique is the dedicated cable connecting between two WiFi nodes as well as an power attenuator to prevent the power saturation in the measured CSI due to the shorter dynamic range of the WiFi chip. Due to hardware modification requirements, this calibration technique may be suitable as a low-cost WiFi-based channel sounder for the validation of the motion sensing model.

### 3.5.3 Human Activity recognition using WiFi Signals

In WiFi sensing, the bistatic deployment, where the transmitter and receiver are spatially well-separated, is usually considered for human activity recognition (HAR).

The field of wireless sensing for HAR in indoor environments has been significantly developed in recent years. It involves measuring wireless channel characteristics using existing wireless networks, such as WiFi networks, to sense environmental changes in the surrounding area of the network. On the one hand, this field is taking advantage of the vast worldwide deployment of wireless communication networks and, on the other hand, increasingly advanced DL techniques.

Moreover, HAR with wireless signals offers advantages over the other most common techniques of HAR: image processing or tracking through wearable devices. Preserving user privacy and the signal propagating through opaque surfaces are the main advantages of CSI-based HAR over image processing. The user-free interaction is a key advantage compared to HAR based on wearable devices.

Due to the ubiquity of WiFi systems in indoor environments, the WiFi networks provide an excellent infrastructure for wireless sensing. The RSSI and the CSI are measurements commonly used for sensing and tracking in HAR. RSSI can offer some valuable information as received power is affected by the environment, but this single value is not as meaningful as the CSI. Therefore, in recent years, CSI data have been widely used [YSL+18]. The CSI is the estimation of the wireless communication link's CFR performed by the receiver. The CSI consists of the complex values (amplitude and phase) of the baseband equivalent of the CFR at each subcarrier in an OFDM symbol. While the CSI amplitude provides a reasonably accurate estimate of the CFR amplitude, the estimated CSI phase contains uncertainties that hinder its use in many HAR applications and theoretical developments. Correcting these phase uncertainties in the frequency and time domains is a complex task [SFFM99a], so many proposals in wireless sensing choose to work exclusively with amplitude [BLM+22, BAR22, IOYG23, SDSEV18].

In this section the applicability of the proposed phase processing in [DSE+23] in HAR is discussed. In this context, the main contributions are described below:

- The method of CSI phase processing is evaluated with a genuine CNN using data from a new HAR dataset recorded using one Raspberry Pi (RPI) as transmitter and three as receivers using 80 MHz bandwidth measurements. Results are obtained for activity recognition and people counting.

- In [DSE+23], the phase processing method was tested for TL using Prototypical Network with the same data and features that appear in [BAR22]. In this work, a different TL strategy is applied, such as fine-tuning, to validate the model in more few-shot learning (FSL) scenarios and prove that this phase processing method can be applied with different TL strategies, in this case, exploring no feature extraction.

- In addition, a CSI data fusion is proposed from three receivers, based on synchronizing their CSI traces employing the timestamps obtained with the Nexmon CSI Tool. Using synchronized data for the same measure can produce a more robust transferability between days.

**Channel state information and phase processing**

In OFDM-based technologies, such as WiFi, the received signal in the frequency domain can be modeled as it is described by equations 53, 54. Most of the work on HAR using CSI data only works with the amplitude because the phase contains several offsets in time and frequency whose estimation is a complex task.

In this work, we must consider three main types of estimation errors [SFFM99a] affecting the phase: SFO, symbol time offset (STO), and CFO. First, SFO is due to the mismatch between the transmitter's and the receiver's local oscillators. This lack of synchronization generates a time shift of the received signal with respect to the transmitted signal. As the local oscillators remain stable over a short time, the SFO is usually treated as a constant. In the second place, STO occurs because the receiver detects the beginning of the symbols by correlation operation and signal power calculation. Due to imperfections in the algorithms, this process introduces a random time shift. Finally, CFO occurs because the transmitter and the receiver center frequencies are not perfectly synchronized. Note that the data demodulation process is not affected by these three offsets but are essential when working with CSI for HAR classification in indoor environments.

Therefore, due to hardware instability, the frequency offset cannot be entirely determined, and this residual offset causes a non-negligible error in the phase. Mathematically, the measured phase $\theta_{s,k}$ at the $k$-th subcarrier of the $s$-th CSI frame can be expressed as:

$$\theta_{s,k} = \varphi_{s,k} + 2\pi \frac{m_k}{N} \cdot \Delta t + \gamma + Z \tag{59}$$

where $\varphi_{s,k}$ is the actual phase, $\Delta t$ is the time lag due to SFO and STO, $m_k$ is the subcarrier index of the $k$th subcarrier, $N$ is the discrete Fourier transform size for the OFDM generation, $\gamma$ is the unknown phase offset due to CFO, and $Z$ is the measurement noise.

Given the phase vector $\boldsymbol{\theta}_s$ of the $s$-th symbol, a linear regression is computed to remove linear offsets. So, the linear regression slope (i.e., $a_s$) of the measured phase can be expressed as:

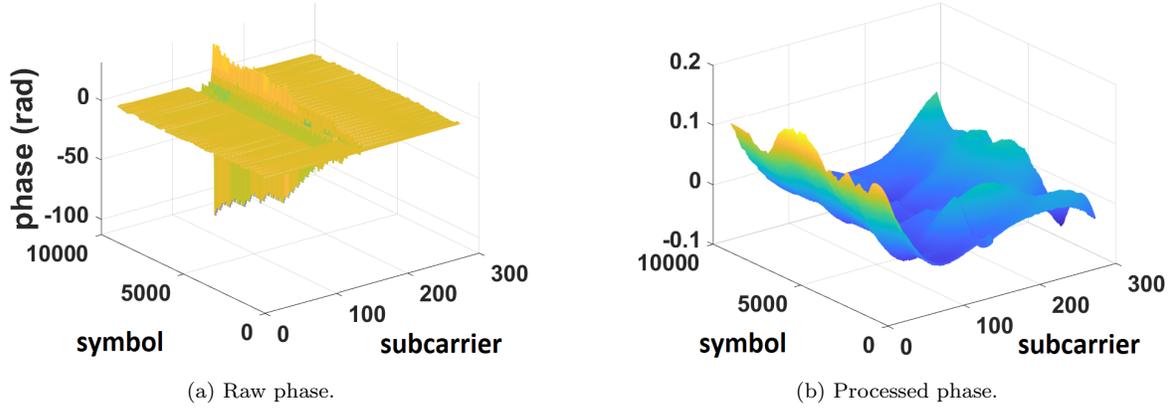

(a) Raw phase.  (b) Processed phase.

Figure 96: Comparison between the raw phase and the processed phase. Image extracted from [DSE+23]
.

$$a_s = \frac{\sum_{k=1}^{K} \left(\theta_{s,k} - \bar{\boldsymbol{\theta}}_s\right)\left(k - \bar{k}\right)}{\sum_{k=1}^{K} \left(k - \bar{k}\right)^2} \tag{60}$$

where $\bar{\boldsymbol{\theta}}_s$ is the averaged phase of the symbol $s$.

Therefore, the angle of the slope $a_s$ with respect to the horizontal axis is

$$\beta_s = \arctan a_s \tag{61}$$

and the pair $(k, \theta_{s,k})$ is rotated by $\beta_s$ degrees, obtaining the rotated phase $\phi_{s,k}$. Let $\boldsymbol{\phi}_s$ be the rotated phase vector of the $s$-th symbol formed by each $\phi_{s,k}$ value, then, the calibrated phase for each $k$ subcarrier and $s$ symbol $\widehat{\phi}_{s,k}$ is given by:

$$\widehat{\phi}_{s,k} = k \cdot \sin \alpha_s + \theta_{s,k} \cdot \cos \alpha_s - \bar{\boldsymbol{\phi}}_s \tag{62}$$

where $\bar{\boldsymbol{\phi}}_s$ is the averaged phase for the symbol $s$.

Once the linear correction has been performed, we apply the Savitzky-Golay filter in the time domain [Sch11, LJW17, YJS+21] of the phase to ensure time continuity. Finally, to rebuild the frequency after time filtering, we establish that the difference between the phase of two adjacent subcarriers $k$ and $k-1$ at a symbol $s$ cannot be greater than a pre-defined threshold $\gamma_s$. Otherwise, we make that $\widehat{\phi}_{s,k} - \widehat{\phi}_{s,k-1} = \gamma_s$. This threshold is defined as the sum of the mean and the standard deviation of the rotated phase at each symbol and can be used assuming $\widehat{\phi}_{s,k} - \widehat{\phi}_{s,k-1}$ as a Gaussian distribution gauss. An example of this phase processing method can be seen in the Fig. 96. The aforementioned phase processing method can be consulted in more detail in [DSE+23].

**Data acquisition**

To evaluate the phase processing method with FSL, a HAR dataset presented in this section has been recorded. This dataset contains CSI data related to activity recognition and people counting. In the activity recognition subset, people perform the following activities: jumping, standing, walking, and empty. In the people counting subset, people walk freely around the room.

The dataset has been recorded using four RPI and the Nexmon CSI Tool software [GSLH19] to extract the CSI data. Nexmon CSI Tool also provides the timestamp for each symbol, and it will be used for a synchronization process, as described in the next section. Three RPIs are in listening mode and are used as receivers (Rx). Ethernet cables connect these three RPI to the same router that sends the information from RPI to a laptop, using an ethernet cable. The transmitter (Tx) is another RPI constantly sending a large file via file transfer protocol (FTP) to another server. The CSIs were calculated at a rate of 1 KHz, in the 5 GHz band, for 80 MHz BW, with 256 subcarriers. However, after eliminating guard, pilot and empty subcarriers, the final number of subcarriers is $K = 240$. In this sense, each Rx picks up a different number of frames due to different distances to the Tx, the direction of each antenna, human activity, and elements in the room.

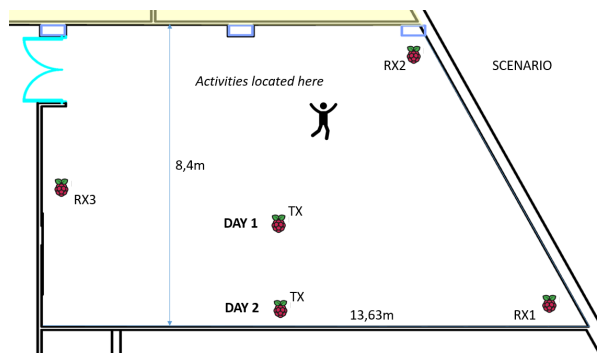

Figure 97: Room for experimental measurements

The scenario is located in the basement of the School of Engineering of Bilbao, as isolated as possible from other WiFi signals. It is a room like the one shown in Fig. 97. The recordings were made over two days, changing the transmitter's position to change the propagation channel. Four men and one woman, aged from 32 to 60, participated in the activities.

**Activity recognition**

Each person performs each activity alone in the room for one minute. The activities are performed consecutively, with a thirty-second rest period between them. Participants receive no instructions or guidelines on performing the activity to simulate a situation as naturally as possible. The activities are always performed at or around the same point, located between Tx and Rx2.

**People counting**

Five recordings are made each day. In each recording, we start with four people walking freely around the room and another resting outside the room, and every minute one person leaves the room. Thirty seconds are given to leave the room each time. In each recording, the order of the people leaving the room changes. The number of people recorded in each realization ranges from four to zero, and their combination is never repeated.

**Deep Learning model - CSI-based inputs**

The CSI data are grouped into sets of 50 symbols to consider a sufficient time interval for environmental changes, and phase and amplitude are obtained from complex values. Therefore, the dimensions of the inputs of this network are $(K, 50, 2)$, where $K$ is the number of subcarriers and the two channels of the CNN is due to phase and amplitude. This way, the input obtained is comparable to a two-color image. Around 30.000 and 60.000 symbols are recorded each minute, depending on RPI and the lost packets, so an image recording frequency between 5 and 10 images per second, respectively, can be provided. The model can be fed with data from several Rx on different days. If the model is trained with data from one Rx, the number of subcarriers is $K = 240$. However, data from the three Rxs can be used jointly by training the model to classify on days, in which case the number of subcarriers is $K_{\text{sync}} = K \cdot 3 = 720$, as is shown in the following.

**Synchronized fusion of CSI data**

In order to add diversity to the model input, data fusion can be performed. Since each symbol has been assigned one timestamp through Nexmon Tool, data from different Rx can be first synchronized. This synchronization causes the Rx with the least symbols to force the number of symbols on the other two Rx. Considering $S_{\text{RX}}$ as the amount of symbols collected by each Rx for the same activity simultaneously, and supposing $S_{\text{RX1}} < S_{\text{RX3}} < S_{\text{RX2}}$, the amount of data synchronized by each RPI $S_{\text{RX}}^{\text{sync}}$ is $S_{\text{RX1}} = S_{\text{RX2}}^{\text{sync}} = S_{\text{RX3}}^{\text{sync}}$. Due to timestamps, it is possible to find for the $i$-th symbol $s_{i\text{RX1}}$ from the smallest dataset (in the example RX1), a symbol quasi-synchronized with it from each of the other Rxs, i.e., $s_{i\text{RX2}}^{\text{sync}}$ and $s_{i\text{RX3}}^{\text{sync}}$, such that their timestamp differences are as small as possible.

Therefore, using $S_{\text{RX1}}$, $S_{\text{RX2}}^{\text{sync}}$, and $S_{\text{RX3}}^{\text{sync}}$ combined, we obtain a CSI matrix with shape ($K_{\text{sync}}$, $S_{\text{RX1}}$) with $K_{\text{sync}} = K \cdot 3 = 720$. A scheme of this example is depicted in Fig. 98.

**Convolutional Neural Network**

In this work, we have employed the CNN represented in Fig. 99. This network consists of a two-channel input layer of size ($K$, 50, 2) or ($K_{\text{sync}}$, 50, 2) and three two-dimensional convolutional layers with 64, 32, and 32 neurons with three max-pooling layers between them. One batch-normalization layer is placed before the second convolutional layer. Behind the convolutional layers is a flattened layer to vectorize the output. Then, there are three fully connected layers, the first with 32 neurons, the second with 16, and the last one with the number of classification classes. A dropout layer, with rate = 0.2, is situated between the first and the second fully connected layers. As an activation function, we have used the mish activation function [Mis20].

**Transfer learning**

Fine-tuning is a TL technique that transfers the knowledge acquired from a DL model trained with one dataset to another, fixing the weights of some model layers. In FSL, the knowledge is acquired in an extensive dataset and transferred to classify another dataset with a few labeled samples. For this purpose, the trained weights of the deepest layers are kept fixed, while the last layers are re-trained with the examples of the new dataset. For our case, the convolutional layer weights, corresponding to the first six layers, have been frozen, while fully-connected layer weights have been re-trained.

To test the effect of fine-tuning, we first save the trained model with all the data from the large dataset. Since our data are all extensive, we have to simulate a dataset with little data. To do this, we load the trained model and retrain the unfrozen layers with a low percentage of the data from the second dataset; in this case, it will be 5 %, which corresponds with a record of three seconds out of a minute. The test and validation sets are each half of the remaining data.

**Results**

In this section, the results obtained for each test are presented. Excepting results in Table 9, the rest were obtained using amplitude and phase together. As mentioned above, 5 % of data is used in the TL strategy to obtain the few-sampled dataset. All datasets are separated into three subsets: train, test, and validation. Stratified shuffle split cross-validation with 5 iterations are used in the training step.

First, Table 9 compares the accuracy obtained using amplitude, phase or both for activity recognition and people counting. The results are presented using the raw phase, linear correction (LC), and our processing method. The model has been trained, tested, and validated in these results with the same dataset. Training has been done with 70 % of data, while testing and validation sets with 15 % of data each of one. We can see that the proposed phase method achieves higher accuracy compared to the raw phase, the phase with LC, and the amplitude, regardless of recognizing activity or counting people. These results confirm those obtained in [DSE$^+$23] and indicate that the proposal is a robust phase processing method for HAR.

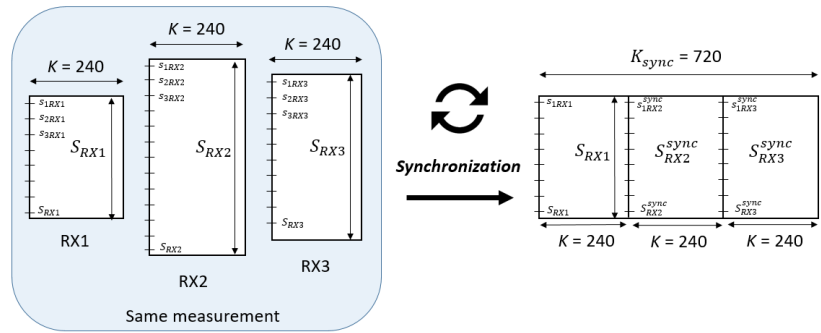

Figure 98: Scheme of CSIs synchronization for the three Rxs.

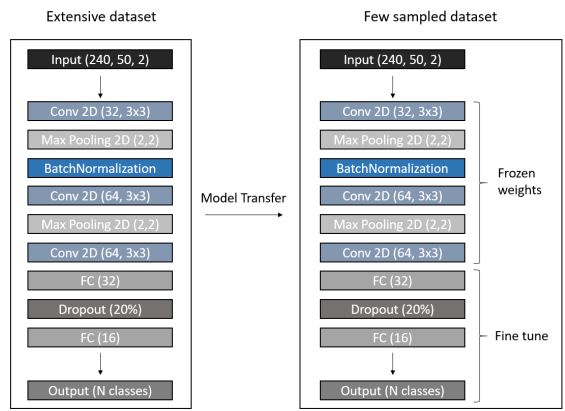

Figure 99: Fine-tuning scheme for one Rx used in this work. In the few sampled dataset, convolutional layers weights are frozen while the fully-connected layers are retrained. For synchronized data from three Rxs, the input layer shape would be (720, 50, 2).

Table 9: Accuracy values (%) for each $RYDZ$, classifying by activity and people counting for each phase stage.

| RX | Variable | Raw | | LC | | Ours | |
|---|---|---|---|---|---|---|---|
| | | Act | Count | Act | Count | Act | Count |
| R1D1 | abs | 70 | 71 | 70 | 71 | 70 | 71 |
| | phase | 23 | 23 | 57 | 52 | **99** | **95** |
| | abs+phase | 56 | 42 | 56 | 58 | 95 | 93 |
| R2D1 | abs | 74 | 77 | 74 | 77 | 74 | 77 |
| | phase | 30 | 21 | 35 | 35 | **98** | 88 |
| | abs+phase | 49 | 30 | 49 | 32 | **98** | **95** |
| R3D1 | abs | 71 | 68 | 71 | 68 | 71 | 68 |
| | phase | 26 | 23 | 43 | 29 | **99** | 98 |
| | abs+phase | 44 | 25 | 44 | 32 | 97 | **99** |
| R1D2 | abs | 67 | 73 | 67 | 73 | 67 | 73 |
| | phase | 21 | 21 | 38 | 30 | **99** | **99** |
| | abs+phase | 34 | 28 | 37 | 33 | 96 | 98 |
| R2D2 | abs | 65 | 60 | 65 | 60 | 65 | 60 |
| | phase | 22 | 22 | 52 | 32 | **98** | 80 |
| | abs+phase | 37 | 27 | 57 | 29 | **98** | **83** |
| R3D2 | abs | 68 | 73 | 68 | 73 | 68 | 73 |
| | phase | 25 | 25 | 45 | 30 | **99** | **99** |
| | abs+phase | 38 | 29 | 58 | 42 | 97 | **99** |

Secondly, the results of fine-tuning between different RPI and different days are shown in Tables 10a and 10b for activity recognition and people counting, respectively. In both cases, the last row, "*REF*", refers to the accuracy training the model from scratch on the same dataset with 5 % of the data, and serves as a reference to see the improvement produced by fine-tuning. The rest values show the accuracy obtained by training and testing in the "RX train" receiver with 75 % and 25 % of data, respectively, and applying fine-tuning to 5 % of data in the "RX validation" receiver. For activity recognition, fine-tuning of a pre-trained model improves the performance by around 10 %, with respect to a model trained from scratch with the 5 % of data. For people counting, there is a more significant variation in the improvement produced by fine-tuning. It is an improvement that varies from 20 % to 5 % or 15 % to 3 %, depending on the Rx trained. One can observe that CSIs from different Rxs on different days can offer more information to the model than that obtained on the same day. For example, for activities, training with R3D2 is the one that offers the best results when classifying data from the three Rxs on Day 1, whereas for validation on Day 2, two of the three best results are obtained with Day 1 Rx training. Regarding counting people, models trained with the Day 1 Rxs offer better results for both days. Based on the results, we can see that fine-tuning of a pre-trained model improves the performance when dealing with few-sampled scenarios. However, the improvement difference can vary considerably.

| RX train | RX Validation | | | | | | RX train | RX Validation | | | | | |
|---|---|---|---|---|---|---|---|---|---|---|---|---|---|
| | R1D1 | R2D1 | R3D1 | R1D2 | R2D2 | R3D2 | | R1D1 | R2D1 | R3D1 | R1D2 | R2D2 | R3D2 |
| R1D1 | - | 90 | 75 | **82** | 70 | 74 | R1D1 | - | **84** | 72 | 76 | **50** | 64 |
| R2D1 | 88 | - | 75 | 79 | 70 | **76** | R2D1 | 68 | - | **73** | 67 | 42 | 55 |
| R3D1 | 87 | 87 | - | 81 | 72 | 75 | R3D1 | **80** | 82 | - | **80** | 46 | **66** |
| R1D2 | 90 | 90 | **77** | - | 70 | 73 | R1D2 | 67 | 75 | 70 | - | 43 | 54 |
| R2D2 | **91** | 90 | 74 | 78 | - | 72 | R2D2 | 73 | 74 | **73** | 63 | - | 58 |
| R3D2 | **91** | **91** | **77** | 80 | **73** | - | R3D2 | 68 | 81 | 70 | 73 | **50** | - |
| REF | 86 | 82 | 67 | 70 | 60 | 66 | REF | 65 | 66 | 68 | 58 | 40 | 48 |
| (a) | | | | | | | (b) | | | | | | |

Table 10: (a) Activity recognition. Accuracy values (%) for fine-tuning between each $RYDZ$, using the processed phase. (b) People counting. Accuracy values (%) for fine-tuning between each $RYDZ$, using the processed phase.

Concerning the use of combined data from the three receivers each day, the results are presented in Table 11, where the "REF" column is obtained as previously explained. Again, we achieve higher accuracy values by fine-tuning activity recognition and people counting. This table also compares results from the receivers individually or synchronizing using the timestamps described above. When TL is not carried out at REF column, we can observe that there is no improvement in using one method over the other. However, when fine-tuning is employed, there is an improvement in the performance results by using data fusion. For activity recognition, the improvement is 16 % over the reference and 8 % over non-synchronization. For people counting, the improvement is 15 % regarding the reference and 5 % regarding non-synchronization.

Table 11: Accuracy values (%) for fine-tuning between days, using the processed phase and 5 % of data in the train.

| Train / Val | Input | Activities | | Counting | |
|---|---|---|---|---|---|
| | | Acc | REF | Acc | REF |
| Day1 / Day2 | No sync | 66 | 55 | 70 | 58 |
| | Sync | **75** | 58 | **75** | 55 |
| Day2 / Day1 | No sync | 77 | 71 | 80 | 73 |
| | Sync | **86** | 69 | **85** | 75 |

**Conclusions**

This work deals with CSI-based HAR in few-sampled scenarios. In this context, a previosly-proposed phase processing technique of CSI has been analyzed with a fine-tuning strategy of a CNN model in few-sampled scenarios. Additionally, a data fusion method from different receivers has been proposed to improve classification for activity recognition and people counting in few-shot environments.

Results have verified that the processed CSI phase is a robust measurement for HAR since it can offer a higher accuracy performance than that achieved with the amplitude. Concerning few-sampled scenarios, fine-tuning strategy has improve accuracy in all cases.

### 3.5.4 WiGig-based Joint Multi-Person Positioning and Respiration Sensing

WiFi sensing is a technology that leverages sensing capabilities inherent in WiFi communication systems. It has attracted substantial interest because it is cost-efficient. WiFi sensing can be employed for conducting various tasks, including localization, people counting, vital signs monitoring, and gesture recognition, *etc.* [DRHN+24, MZW19, KNH17, KGPH15]. These capabilities offer promising solutions for future smart cities and homes [ZY20].

The main use cases in smart homes, such as human localization and breath rate (BR) estimation, have been widely studied by using low-frequency WiFi under legacy standards [KJBK15, AHY15, ZY20]. However, it is worth noting that low-frequency WiFi systems are naturally such as up to 40 MHz in IEEE 802.11n [11n09] and up to 160 MHz in IEEE 802.11ax [11a21a]. Consequently, an insufficient bandwidth degrades range resolution and localization accuracy, making multi-person sensing (MPS) tasks challenging.

For the MPS task, the key lies in extracting patterns from CSI and matching them for each person. Previous approaches, such as blind source separation (BSS) using independent component analysis (ICA) [ZY20], cannot identify and match the respiration patterns of multiple individuals because of the limited angular resolution. Moreover, these methods often require certain prior knowledge, such as the number of persons in the room. A way to match estimated patterns to individual targets is to use beamforming generated by antenna arrays [GZY+23]. However, this method is proposed based on legacy WiFi standards, which require an extra device for beam management.

While WiFi devices work in mmWave bands (WiGig), it can increase the estimation accuracy given the wide bandwidth and high sensitivity on Doppler shift [AGPHG20]. For example, studies in [PLM+23] and [PLA+24] have achieved decimeter-level positioning using the IEEE 802.11ay standard [11a21b]. However, [PLM+23] operates in a monostatic configuration, where the transmitter and receiver are co-located. This configuration does not align with typical communication scenarios. On the other hand, [PLA+24] introduced a bistatic tracking system but relied on the assumption that the relative positions of the transmitter and receiver are known. This assumption, however, limits real-world implementations, particularly in scenarios where the transmitter and receiver may be subject to movement or require reconfiguration. Such dynamics can alter the geometric parameters critical for accurate sensing.

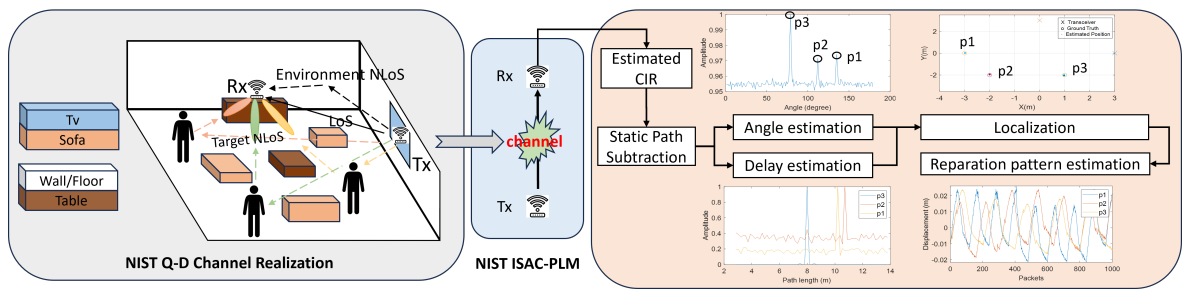

Figure 100: System Overview for Multi-person Positioning and Respiration Sensing[XCL+23, XCMP24]

**Technical Contributions**

To address the aforementioned limitations, this paper introduces a new methodology for multi-person localization and BR estimation utilizing WiGig, without relying on prior knowledge. Fig. 100 displays a systems overview.

Our simulations are based on the National Institute of Standards and Technology (NIST) Q-D Channel Realization Software and NIST Integrated Sensing and Communication Physical Layer Model [BRSG22], an open-sourced end-to-end simulation platform that supports IEEE 802.11ay standard. In our system design, we consider the multi-person sensing scenario in a $7 \times 7 \times 3$ m living room with furniture. The transmitter and receiver are placed in distinct locations. Given the high attenuation properties of mmWave signals, this indoor environment is ray-traced with the LOS path and first-order reflection paths. Our system considers the receiver equipped with a 64-phased antenna array, while the transmitter utilizes an omnidirectional antenna. This configuration enables the receiver to sense the CIR from specific angles defined in the codebook while allowing the transmitted signal to be broadcast in all directions. Our approach utilizes the training field (TRN) and beam scanning capabilities specified in IEEE 802.11ay to determine the relative positions of the transmitter and targets in relation to the receiver.

The principal contributions of this paper are summarized as follows:

- We propose a multi-person sensing framework using IEEE 802.11ay that achieves centimeter-level localization of multiple targets by exploiting CIR estimation and beam sweeping functionalities.

- We present an algorithm for multi-target respiration estimation within a bistatic radar setup by utilizing the acquired location information. This method distinctly identifies and matches breathing patterns to individuals at varying locations, ensuring accurate identification of even identical breathing patterns among different human targets.

- We investigate the impact of package sampling rate and physical blockage on sensing capability and validate our findings through numerical analysis. These insights are valuable for designing practical sensing systems using the IEEE 802.11ay standard.

- We introduce a prior-knowledge-free method for multi-person localization and respiration rate estimation using IEEE 802.11ay in an indoor environment. We utilize the beamforming functionalities specified in IEEE 802.11ay to estimate the transmitter's position, thereby aiding in target localization.

- To enhance the fidelity of the simulation, we model the chest target as a surface instead of points. This augmentation serves to replicate the complexities inherent to practical systems, where multiple reflection delays and angle responses are manifest for each distinct target because of scattering points of the human body.

- We employ the density-based spatial clustering of applications with noise (DBSCAN) algorithm to associate scatterers with their respective targets, enhancing the robustness of BR estimation.

**Discussion and Future Work**

Future research should explore methods to overcome practical challenges such as NLOS conditions, mobility, and multilink operations.

1) *NLOS conditions:* Our current method, which does not require prior knowledge, depends on estimating the LOS path for effective target sensing. However, in practical scenarios, the LOS might be obstructed, posing challenges in localizing the transmitter. Future studies could explore techniques for enabling prior-knowledge-free sensing in environments where NLOS conditions prevail.

2) *Mobility:* Existing research on wireless vital signs sensing generally assumes that human subjects remain stationary during experiments. However, mobility can distort the detected vital signs signals in dynamic settings where the target or surrounding objects are in motion. Identifying effective strategies to address mobility will be crucial for advancing future wireless vital signs sensing technologies.

3) *Integrate sensing and communication performance analysis:* In this work, we introduced an MPS approach that uses the IEEE 802.11y standard for localization and sensing vital signs. Future research should investigate the combined dynamics of sensing and communication by examining the fundamental trade-offs involved, such as the balance between sensing accuracy and communication efficiency. Specifically, research should focus on the impact of sensing performance on accuracy and communication performance in terms of bit error rate or spectral efficiency.

4) *Multilink data fusion:* Given the widespread deployment of WiFi devices, the potential for multiple devices within a single environment is high. It is anticipated that WiGig will become similarly prevalent. A significant area for future research is the fusion of data across multiple links, which could enhance accuracy in estimations. Furthermore, merging data from various viewpoints can better address challenges such as occlusions.

### 3.5.5 Radiolocalization in the cellular systems

The development of modern cellular systems like the fourth generation, Long Term Evolution (4G-LTE) and 5G-NR is clearly focused on increasing the network capacity and enabling a wide range of services. Thus can be clearly pointed out in first development of the 6G generation. It is evident that wide group of customers demand two measurable parameters i.e. high speed data transmission and low latency.

Locating the terminal using cellular network radio interfaces like the LTE, 5G-NR or narrowband internet of things (NB-IoT) seems to be a natural process, especially in the context of environment awareness. Besides the laboratory and prepared environments UE localization still may be challenging considering technical limitations of commercial networks. It seems fully justified to conduct further research regarding the accuracy of UE position estimation taking into account the real and emulated signals and data processing mechanisms implemented in the receiving path of the cellular terminal.

The 3GPP technical specification of the 4G-LTE, 5G-NR and NB-IoT interfaces defines main downlink reference signals used for time and frequency synchronization, determining the cell identifier (Cell ID) and estimating the state of the radio channel. The initial stage of network sensing process is based on detecting the primary synchronization signal (PSS), narrowband primary synchronization signal (NPSS) (for NB-IoT) and secondary synchronization signal (SSS), narrowband secondary synchronization signal (NSSS) (for NB-IoT) signals. These signals are detected by any 4G and 5G terminal during cell attachment process and also during normal operation. It must be noticed that PSS/NPSS and SSS/NSSS signals are different for each radio interface, e.g. different pseudo-random sequence, but their application purpose is common [BY04a].

In all three mentioned radio interfaces it is possible to allocate in the radio subframe signals dedicated to the localization purposes, i.e. the PRS (Positioning Reference Signal), narrowband positioning reference signal (NPRS) for NB-IoT, which take the form of dedicated complex symbols allocated over subcarriers in predefined time pattern [rGPP20], [OL20], [YL19]. Nevertheless, observing the operators' procedures in the 4G and 5G networks, the emission of non-obligatory signals is usually not enabled in order to allocate resource elements to the needs of the physical downlink shared channel. A similar situation occurs in all the mentioned radio interface, where the emission of the PRS/NPRS signal is unfavorable from a business point of view. Moreover, e.g. for the NB-IoT the technical specification imposes the creation of the so-called muted frames in which, apart from the emission of synchronization signals, there is no emission in the narrowband physical downlink shared channel (NPDSCH). Consequently, to localize the UE other signals already present in the radio frames must be used to estimate the position, in this case the PSS/NPSS signals broadcasted periodically in each radio frame were selected [rGPP20], [PR23a].

**Localization in telemetry systems using NB-IoT radio interface**

The NB-IoT radio interface, firstly announced in $13^{th}$ release of 3GPP specification, is dedicated for realization telecommunication services, i.e. in stationary conditions and in difficult propagation conditions, with a bit rate up to 250 kb/s in the uplink. The introduced modifications to the 4G-LTE, both in the physical layer and in higher layers, enabled its usage in environments with very difficult radio signal propagation conditions. It allowed to obtain the maximum coupling loss (maximum coupling loss (MCL)) parameter at the level of 164 dB [rGPP20], [OL20], [IR16], which is more that in second generation (2G) network. This radio interface is still considered as coexisting alongside the 4G or 5G-NR networks, even if a never one, the 5th Generation Reduced Capacity (RedCap), is also proposed. The RedCap is assumed to be used in e.g. IoT and telemetry systems but its physical signals are not different in assumptions comparing to the 5G-NR radio interface [rGPP24].

In presented research, estimation of the UE position is done by using the NPRS signal detection time, where the detection method is based on the cross-correlation. Thus, this process is conceptually identical and combined with time and frequency synchronization operation. The locally generated reference NPSS signal was cross-correlated with the received radio signal while introducing hypotheses about the present frequency deviation, e.g. with coarse step of 100 KHz and a fine step in the range of ±15 kHz. This process made it possible to estimate the frequency deviation of the receiver relative to the evolved node B base station (eNodeB), its compensation, and to estimate the radio signal detection time [XL17]. The estimated reception times for each eNodeB were the input data for the position estimation algorithm based on the TDoA method and the Foy algorithm [Sad18], [HLS19].

During the research it was required to have samples of radio signals with a controlled influence of the radio channel to reliably assess the process of detecting reference signals during the position estimation process. It was decided to generate a set of test signals using the Rohde&Schwarz CMW500 radio communication tester. Using the NB-IoT radio interface emulation module, a radio signal was generated for the Standalone operation mode, modeling three channel profiles, extended pedestrian A model-5 (EPA-5), extended typical urban (ETU-1) and AWGN. For each of the channel profiles, changes were made to the SNR value, where reducing the SNR value was based on adding filtered Gaussian noise to the radio signal with power spectral density of −60 dBm/15 kHz. The generated radio signal in the B2 band was recorded using the created laboratory testbed based on the USRP

Ettus-Research-X310 device which used the native sampling frequency for the NB-IoT interface, i.e. 1.92 MHz. The developed software was used to record and analyze the samples, both in terms of time and frequency synchronization and terminal position estimation [PR23a].

The main goal of the conducted research was to verify the accuracy of estimating the user's terminal position based on detection the NPSS signals, not intended for location purposes. The NB-IoT signals were generated in a way to emulate a rectangular test area with dimensions of 1500 m and 3000 m, with three eNodeB stations and uniformly distributed test points (test point (TP)) with a spacing of 100 m.

The accuracy of the user terminal position estimation was verified based on the measurement data assigned to each TP. These data were the radio signal propagation times from each eNodeB to the TP (test signals from three eNodeBs to each TP were the different fragments of the recorded radio signal) rounded to an integer multiple of the sampling period $t_{sp} = 1/30.72$ MHz. This reflects the number of samples n by which the test signals corresponding to one radio frame of the LTE signal were delayed (19200 + n). It is worth adding that the method of generating the NB-IoT signals in real eNodeBs will not allow in many devices to control time dependencies in the radio interface with a higher resolution than the one adopted in the presented research [OL20].

Numerical analysis of position estimation accuracy was carried out for the signals with different SNR ratio and different emulated radio channel models (AWGN, ETU-1, EPA-5). Fig. 101 shows the error CDFs of the determined position estimates for the AWGN profile and three selected values of the SNR parameter: [16 dB, 8 dB, −2 dB]. For a good quality signal (SNR 16 dB), the use of an additional approximation [PR23a] [Sad18] (designation A, without approximation: WoA) makes it possible to reduce the position estimation error from 258 m to 81 m in 90% of cases, and the standard deviation $\sigma$ from 61 m to 21 m. For low-quality signals (e.g. SNR −2 dB), a significant degradation of position estimation accuracy is visible (687 m for 90 % of cases, $\sigma = 159$) and there are cases when the Foy algorithm does not converge. For the presented case, in four percent of the analyzed measurement data, the estimated location estimates were more than 1000 m away from the actual location of the TP.

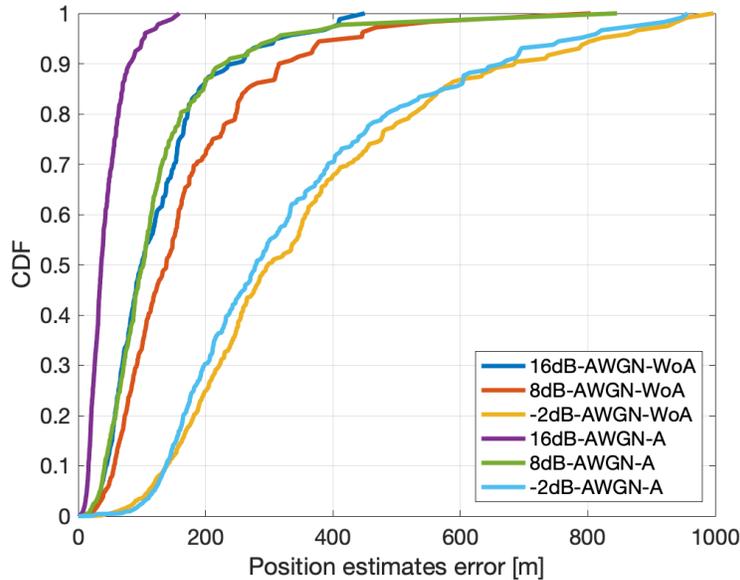

Figure 101: CDFs of position estimates errors for the AWGN channel profile [PR23a]

In the next phase of the research, the accuracy of the position estimation was verified, when the ETU-1 and EPA-5 models were used to emulate the influence of the radio channel. Fig. 102 shows error distributions for a signal with an SNR of 8 dB. In this case, the position estimation accuracy increased from 390 m ($\sigma = 91.03$ m) to 210 m (at $\sigma = 60.46$ m) and 405 m ($\sigma = 112.18$ m) to 370 m ($\sigma = 98.44$ m) for 90 % of cases and the ETU-1 and EPA-5 profiles, respectively.

In the last phase of the research, the degree of degradation of position estimation accuracy was verified by introducing non-ideal time synchronization of the eNodeB stations. The degree of non-synchronization of each eNodeB was modeled with the normal distribution with the mean value $\mu = 0$ and $\sigma^2 = 2.25 \cdot 10^{-12}$, which corresponds to the degree of synchronization of the eNodeB station at the level of 1.5 µs (S1 case) and the normal distribution with parameters $\mu = 0$ and $\sigma^2 = 1 \cdot 10^{-14}$ which corresponds to the degree of synchronization of the eNodeB station at the level of 0.1 µs (case S2). The degree of synchronization of the eNodeB station at the level of 1.5 µs is the expected value for cells with a radius of up to 3 Km, while the assumed value of 0.1 µs is expected for the LTE-Advanced (LTE-A) network [Goe16].

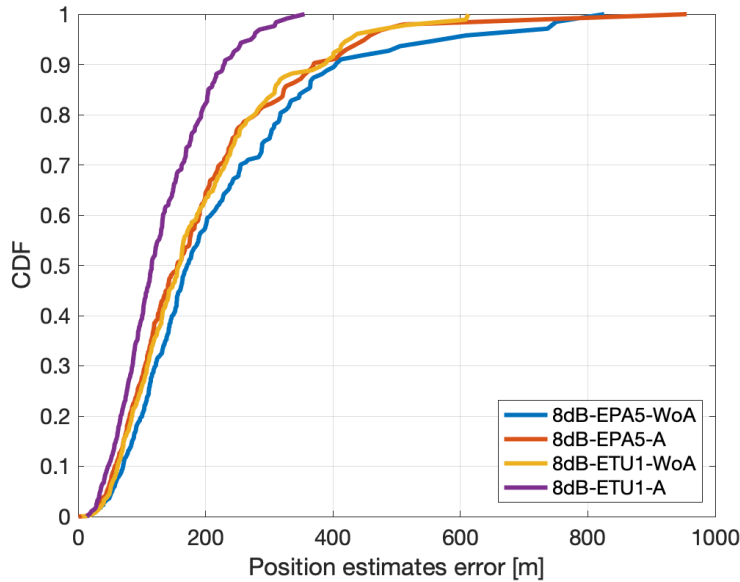

Figure 102: CDFs of the position estimates errors for the ETU-1 and EPA-5 channel profiles [PR23a]

CDFs of position estimates errors for the analyzed two cases (S1, S2) of different degrees of non-ideal synchronization of the eNodeB station are shown in Fig. 103. It is visible that with a higher degree of non-synchronized eNodeB, the additional approximation of the maximum of the correlation function makes it possible to reduce the position estimation error, the difference between the cases S1 and S2 is - respectively 822 m and 286 m for 90 % of cases and the ETU-1 channel profile. However, it is worth pointing that the real eNodeBs often have a high degree of repeatability of time synchronization deviations, which makes it possible to introduce estimated corrections for individual stations.

**Localization in 5G-NR networks**

The observed development of 5G-NR 5G standalone (SA) networks, enabling private ones, gives a flexible possibility to carry out experiments representing realistic scenarios. In works related to the problem of localizing the UE position in various environments and use cases are widely analyzed. Nevertheless, there is still the need to investigate localization performance not only via simulation studies but also by measurements carried out in actual propagation environments, with a noticeable influence of the radio channel. In some works, authors investigated the ability of UE localization even using only one gNodeB, which involves a complex analysis of the channel impulse response, being sensitive to changes in the environment state and multipath propagation. Nevertheless, in most cases it is assumed that UE will detect dedicated positional signals from at least three gNodeBs to perform a two-dimensional localization [MK22], [Tra22], [dPRRLSSG18], [3GP18c].

The localization of the user terminal in 5G networks is based on the similar manner as in 4G-LTE or NB-IoT, in the meaning if method and used signals. In presented studies the TDoA approach, along with the Foy algorithm was used [Sad18]. Similarly to problems desribed in previous subsection, in most of the 5G networks the UE cannot be localized using dedicated PRS signals but other ones, e.g. the PSS, named as signals of opportunity. To extend the research filed in this studies, the assumed major goal was to investigate the influence of the environment type on the accuracy of position estimation (i.e., not the influence of the measurement location or the gNodeB geometry). The registered signals

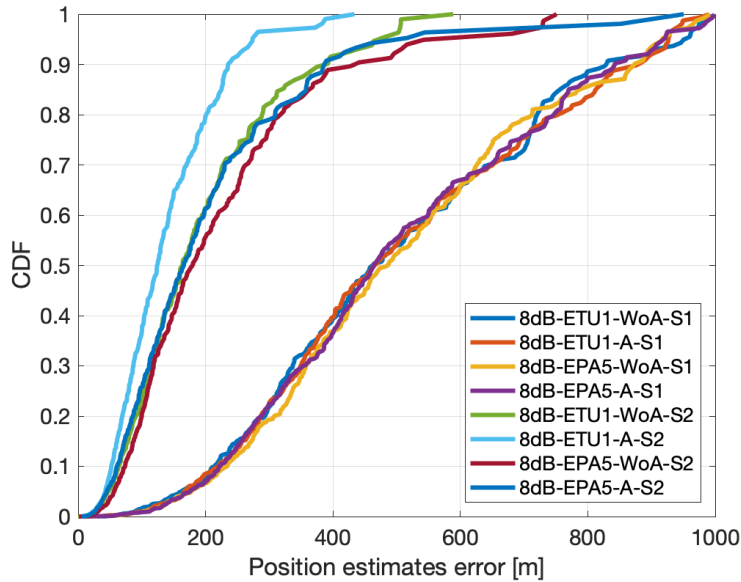

Figure 103: CDFs of the position estimates errors for the ETU-1 and EPA-5 type channel profiles with the influence of imperfect synchronization of the eNodeB station [PR23a]

from different gNodeBs were processed by adding delaying samples [PR23a]. As it was described in previous subsection, this operation emulated the location of the UE on the test area in this case with an optimum, triangular gNodeB geometry. Therefore, it was possible to investigate the accuracy of position estimation influenced by non-ideal signal detection caused by the influence of the environment type.

A square test area with dimensions of $700 \times 700\,\text{m}^2$ was emulated, in which 400 test points were placed, with 35 m spacing along the axes. This area was located inside an equilateral triangle created by the gNodeBs deployment. The accuracy of calculated position estimates was evaluated by calculating the RMSE between the estimated positions and the reference ones, and by the corresponding standard deviation [PR24].

In order to obtain the localization of a UE involving signals from a commercial network, a flexible testbed was designed and used. The SDR platform, based on the USRP X310, along with a PC-class computer was used to obtain samples of the radio signals, and further process them using the developed software. The 5G-NR signal was sampled at 30.72 MHz, i.e., four times higher than the required minimum to analyze the 5G-NR synchronization block, containing the PSS and SSS signals [PR24].

The 5G-NR downlink signal was recorded stationary in three locations corresponding to different types of environments, as typical localization of users, i.e., outdoor, light-indoor and deep-indoor. The network operated in TDD mode on 2.59455 GHz center frequency with 50 MHz of bandwidth. The metrics taken for differentiating these environments were the ratio between the LOS and the NLOS components of the signal, represented by the Rice K-factor ($K$) [ITU21], [JW20], [EIA21], [PR24]. The $K$ values for three analyzed environments have values of 14.0 dB, $-0.7$ dB, $-13.2$ dB respectively for outdoor, light-indoor and deep-indoor.

For the outdoor environment an almost total superposition of the estimated locations with the reference ones was achieved, being the best result: the RMSE with a median of 28 m and the 90 % percentile of 39 m, together with a standard deviation of 11 m. It enables a good use of the approach, since TP are separated 35 m in either vertical or horizontal directions. Moreover, the shape of the test area is maintained, with no significant deformations caused, e.g., by improper detection of the PSS signals, enabling the Foy algorithm to converge in all cases. This is the environment with a strong LOS and a relatively low delay spread [PR24].

For the light-indoor environment, a noticeable increase in the position estimation error was achieved. The RMSE is 640 m for the 90 % percentile and 322 m for median, with a standard deviation of 165 m, which are much higher than the grid resolution of 35 m, hence, not enabling the use of the approach

[PR24].

The analysis of the calculation process shows that this phenomenon for the light-indoor case is caused by the improper detection of the PSS signal. When the LOS component is not dominant, one of the multipath components is selected as the dominant one during the PSS signal detection, which ends up not being a continuous and uniform process across all positions, hence, the high estimation error and the large variation. The estimated signal detection time is influenced by the propagation delay for this dominant component. In many cases, it causes an alternate selection of different multipath components, which is also observed in the CDF presented in Figure 104. It is worthwhile mentioning that a low delay spread does not cause unrealistic input data for the TDoA method and Foy algorithm which converges for a very large percentage of cases. The estimated positions must be further analyzed so that they can be discarded if, e.g., there is large displacement between nearby test points [PR24].

In the results for the deep-indoor environment an intermediate values for RMSE can be observed compared to the two previous ones: 342 m and 229 m for the 90 % and 50 % percentiles, with a standard deviation of 102 m. However, there is a strong increase in the number of test points where the Foy algorithm did not converge, up to almost 33 %. The main cause for these results is the additional propagation loss observed in this environment, since all signal components were in NLOS, which significantly disturbs the PSS signal detection [14]. Nevertheless, most of the multipath components were attenuated, which eliminates the problem of detecting reflected signals as the dominant ones (as in the light-indoor case). Still, a significant influence of the propagation delay can be observed. The CDFs for all the analyzed cases are presented in Fig. 104 [PR24].

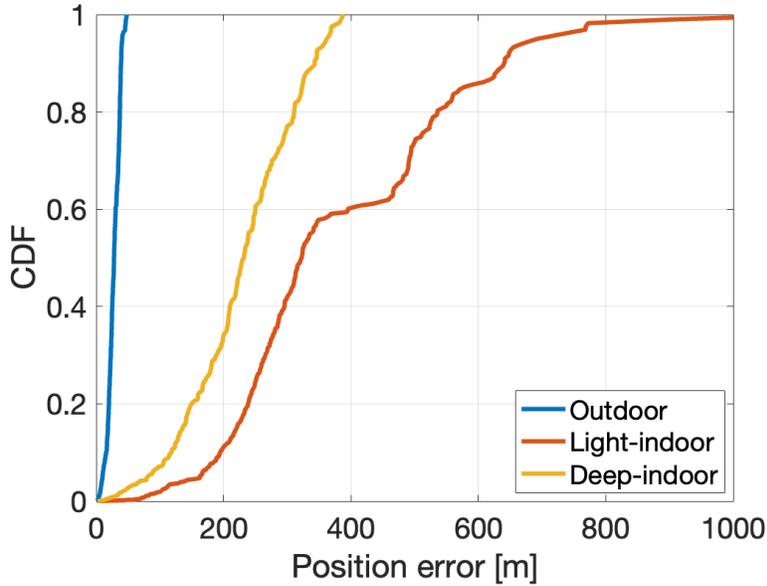

Figure 104: CDFs of the position estimates errors for the three analyzed environments [PR24]

### 3.5.6 Sensing and Localization in IoT Networks

The integration of localization functions into IoT wireless communication is driven by the increasing demand for context-aware applications across various sectors, enhancing operational efficiency, safety, user experience, and resource management. ILAC enables IoT devices not only to communicate data but also to understand and report their position in space, therby opening up new dimensions for automation, monitoring, and interaction within physical environments [KSL+]. Wireless sensing enables devices to gather information about the surrounding environment or objects within it by analyzing variations in the properties of radio frequency (RF) signals. Due to the interaction of the signal with the environment, changes in the signal's power, frequency, and phase can be observed, which reflect the characteristics of the surroundings. Consequently various IoT protocols have begun to focus on the integration, or incorporation, of diverse localization techniques into their protocol specifications.

BLE is widely adopted in IoT devices due to its low power consumption, high-data-rate connectivity, and compatibility with smartphones and wearables. BLE supports basic localization functionalities with the help of RSSI measurements, allowing fingerprint-based and proximity-based localization [MRL+]. A significant advancement in BLE capabilities for localization services was introduced with the release of Bluetooth 5.1, which allows devices to estimate the AoA and determine the direction from which a signal arrives at a receiver [PAGY]. Recent directions in BLE standardization aim to further enhance the accuracy of localization by adding the channel-sounding feature alongside the communication protocol. This feature uses multi-carrier phase difference (MCPD) ranging to analyze the radio signal's phase information for precise distance estimation between devices, even in challenging multipath environments.

Conversely, for IoT scenarios that require low-power, low-data-rate connectivity with potentially large numbers of devices over a relatively large area, alternative IoT protocols are used. These typically include ZigBee, Thread, IPv6 over low-power wireless personal area networks (6LoWPAN), and Wireless highway addressable remote transducer (HART) which are based upon the IEEE 802.15.4 standard. Central to the standard is the adoption of the time slotted channel hopping (TSCH) protocol in the IEEE 802.15.4e-2016 amendment, which enables energy-efficient communication among wireless nodes and offers high reliability, low latency, and scalability [VWC+].

The sensing and localization in IoT networks are currently limited to the estimation of the radio wave direction of arrival and the distance estimation between two devices.

A phase information of the signal can be used to determine the direction from which this signal arrives, a technique known as the AoA estimation. With the use of an antenna array, where antenna elements are separated by less than half of signal's wavelength, the same signal is received at slightly different times and hence different phases. The measured phase difference observed across the array is used for AoA estimation with high-resolution algorithms such as MUSIC, push different data algorithm (PDDA), ESPRIT etc.

The method of acquiring the phases at different antennas depends on the architecture of the receiver and the antenna configuration. IoT devices and corresponding radios are typically designed to be low cost, low power, and low complexity, which results in the implementation of a single RF signal chain. One common approach, to be able to estimate AoA, is the use of RF switches to connect the single RF chain to multiple antennas. This setup allows the device to switch between antennas sequentially, capturing the signal's phase from each antenna at different times. Different antenna configurations are available, such as uniform linear array (ULA), uniform circular array (UCA), and uniform rectangular array (URA), and a variety of switching patterns that provide application-oriented features [SJSH23].

Although switching antenna arrays represent a cost-effective solution, they introduce a time delay between measurements from each antenna, which can produce large errors in case of CFO between the devices or a SFO at the receiver. However, advanced signal processing techniques can calibrate for these offsets and accurately estimate the direction of the signal [MSK+23].

**Estimation of distance between two devices**

The phase difference between the received and transmitted signal, measured at multiple frequencies, can be used to estimate the distance between the devices. Specifically, a technique called MCPD analyzes the RF channel characteristics by observing how the phase of a signal at multiple frequencies changes with distance. Phase-based ranging (PBR) has gained a lot of attention since it can provide more accurate distance estimation than techniques based on received signal strength, even in challenging multipath environments.

In the MCPD scheme, one device acts as the initiator and the other as the reflector. The initiator begins the channel estimation process by transmitting an unmodulated CW signal at a specific frequency. Upon receiving the signal, the reflector measures the phase of the incoming RF signal relative to its LO. To eliminate initial unknown phase offsets in the transmitted signal, the roles are reversed. The reflector transmits a CW signal back to the initiator, where the phase information is extracted by comparing the received signal with initiator's LO. To increase the maximum unambiguous range (UR) and overcome limitations like the ambiguity in phase measurement (since phase wraps around $2\pi$), the devices extend the apparent bandwidth by repeating the phase measurements at several frequencies. The frequency step between the measurements determines the UR, while the total bandwidth used determines the resolution of the system.

As long as the time interval between measurement at the initiator and the reflector is very short, being within the channel coherent time, the initiator can construct the two-way channel frequency response (TW-CFR) from the obtained phase measurements. The TW-CFR directly correlates to the distance between the devices and can be used for sensing algorithms. Two-way notation denotes that the measured signal went in both directions and with channel reconstruction methods it can be transformed into one-way channel frequency response (OW-CFR). The method of obtaining phase samples can be either uniform [SW] or non-uniform [OGSA] and it can include weighting of the phase samples [WRZM]. Once the CFR of a link is acquired, spectral analysis techniques can be employed to estimate the distance between the devices. A variety of algorithms are available for this purpose, including slope-based [PBH], FFT-based [SRW], MUSIC-based [BRGD], and ML-based [VMDADT] algorithms.

**Integration of localization functionality into the IEEE 802.15.4 TSCH**

In this subsection an IEEE 802.15.4-2016 TSCH protocol modification is presented, in which the integration of localization functionality alongside the communication process is proposed. With the enhancement, the node can send a packet to a target device and estimate both the AoA and the distance estimates to the receiver node, which can be used for localization purposes. The method was developed considering the findings from the proposed methods for TSCH protocol enhancement with MCPD distance estimation [MGHJb] and AoA estimation . The method was chosen to assess the performance of hybrid localization and involves modifications only to the PHY and MAC layers of the IEEE 802.15.4 standard.

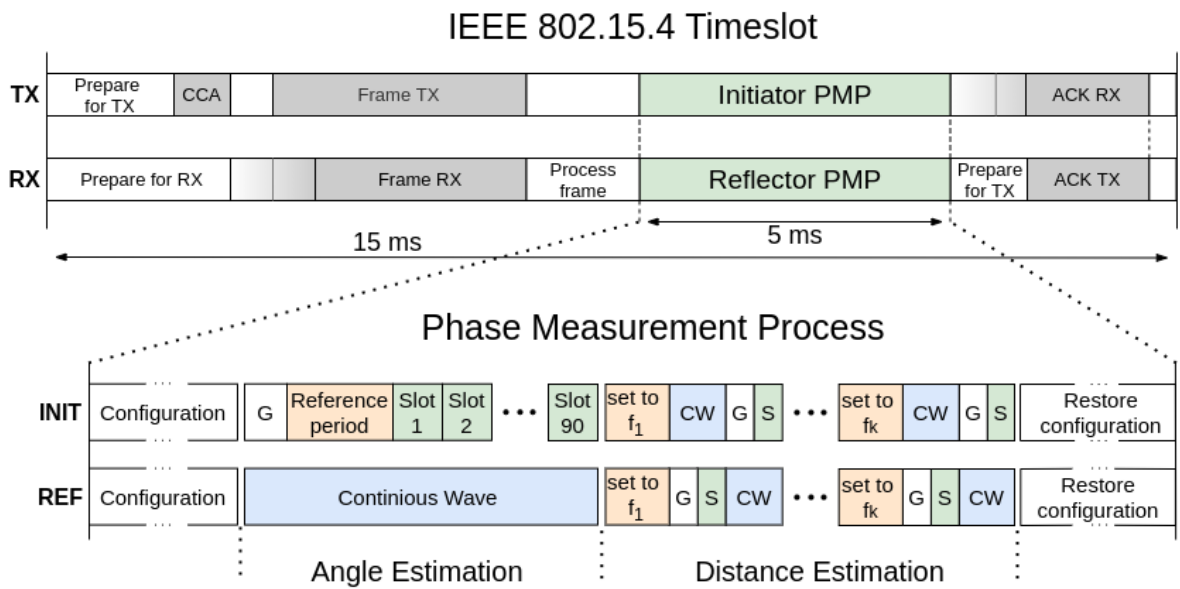

Figure 105: Integration of Phase Measurement Process into IEEE 802.15.4 TSCH timeslot.

To add a localization function to the protocol, the communication process is briefly interrupted so that the devices can measure the characteristics of the RF channel within the procedure called phase measurement process (PMP). The TSCH timeslot is extended with an additional interval, which allocates the required time for PMP, as shown in Fig. 105. Since TSCH by definition builds a synchronized network, both devices start with the channel measurement at the same time. Leveraging the TSCH protocol further, PMP is positioned after the data packet exchange, allowing the measured phase information to be returned to the initiator by the reflector in the form of an information element, as part of the acknowledgment exchange. The approach defines several steps within the PMP:

(1) **Radio configuration**: as the phase measurement is not a part of typical radio operation during communication, a time interval is provided for radio configuration.

(2) **Angle estimation**. The PMP angle estimation differs depending on the type of device. For the reflector, a time interval is defined during which it emits a CW on the same frequency previously used to send the data packet. For the initiator, a first predefined guard interval (marked with G in Fig. 105) is established, followed by the reference period. The reference period enables the reflector to sample up to 8 phase samples on a reference antenna, which can be used for CFO estimation. Afterward, 90 slots are allocated for measuring the phase using various antenna elements, following a predetermined switching pattern. To avoid any phase distortion that might occur due to changes in the signal path during the measurement, enough time is allocated in each slot to sample the phase and to switch to the next antenna element.

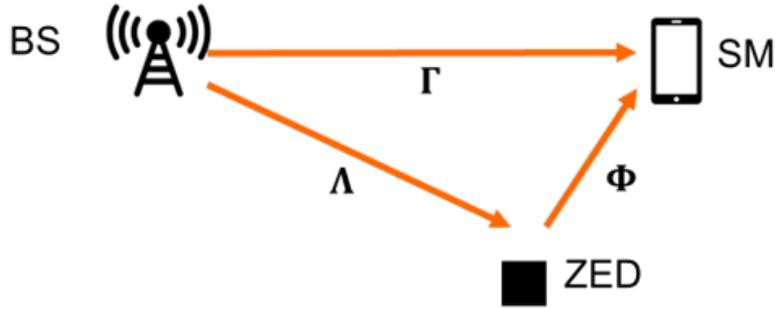

Figure 106: Channel model from the basestation directly to the smartphone and through the zero-energy-devices (ZED)

(3) **Distance estimation**: The PMP for distance estimation consists of short events that are repeated on multiple frequencies to obtain phase response from the entire available RF band. In the event, first, a time interval is provided for both devices to configure their LO to the desired frequency. Afterward, events differ based on the role of the device. For the reflector, a time interval is provided to emit the CW (marked $CW$ in Fig. 105), to wait for a guard time and to sample the phase (marked $S$ in Fig. 105). For the reflector, the time intervals are reversed, so that it can wait for the guard time, sample the phase and afterwards emit the CW. This scheme allows the devices to obtain phase measurement of a channel both ways, sequentially obtaining TW-CFR from multiple frequencies while maintaining synchronization with each other.

(4) **Radio re-configuration**: At the end of PMP, a short period is provided to restore the radio configuration, so that the devices can continue with the communication process.

The proposed generic solution, which applies to any radio, has been modified for the currently available low-cost off-the-shelf radios that offer a phase sampling rate of 8 μs. The intervals shown in Fig. 105 were selected accordingly, and the timeslot was extended from the default 10 μs to 15 μs. Although less time is available for communication, sensing and localization functionalities can be achieved using the same hardware.

### 3.5.7 Integrated Localization and Communication (ILAC) thanks to Ambient Backscatters

Indoor self-localization of smartphones (SMs) can still be improved in terms of cost and energy efficiency. As GPS based localization only works properly outdoor [ZM08], many alternatives have been studied and even commercialized based on terrestrial wireless networks. Indeed, wireless network nodes with known locations can be used to locate the SM, either through the analysis of downlink signals by the SM or through the analysis of uplink signals by the network side. In practice, 5G cellular networks macro BSs [YRC22], 5G small cells [BWC22], WiFi access points [XWP17] or BLE [Wan13][LBM20] have been successfully tested, with several meters of accuracy. However, they require a strong density of nodes and/or highly complex radio finger-printing schemes. Unfortunately, all these nodes need power supply, and thus, their deployment cannot be massive. Recently, the first European Flagship Project Hexa-X I on the future 6G of networks has set as a goal, to provide digital services (including localization) sustainably [ea21] and introduced a new type of device called crowd-detectable ZED (CD-ZED) [ea23]. A CD-ZED, or simply ZED, is a self-powered device that harvests ambient energy to power itself and that backscatters (reflects) ambient waves from the BSs of the cellular network to communicate with a SM [DTPHdR22]. In practice, the ZED switches between a transparent state and a reflective state to transmit a bit '0' or a bit '1'. A SM closed by, detects the ZED message by analyzing variations in the received signal issued from the BS. Very recently, several material proofs-of-concept and experiments have been successfully conducted with 4G base stations as sources of ambient waves (as 4G is a mature ambient network), and pilot-based detection at the SM side [JLPH23, JL23, PN23]. Furthermore, [DTPHdR22] presents a ZED prototype that powers itself with solar and indoor light energy, stores some energy in a small battery, and keeps transmitting data night and day. Finally, such types of devices are currently being studied at the 3GPP for potential introduction in future mobile networks standards, under the umbrella concept of "Ambient Internet-of-Things" (A-IoT) [3GP23b]. A novel localization method is possible for indoor relying on ZEDs beacons, instead of active beacons. Such method is expected to be more sustainable. In the following, we present a first implementation, where 4G is used as an ambient source of signal, as it is the most mature network today. However, similar studies can be done with 5G or beyond standards, enabling the integration of the localization of devices through the communication radiowaves from basestations.

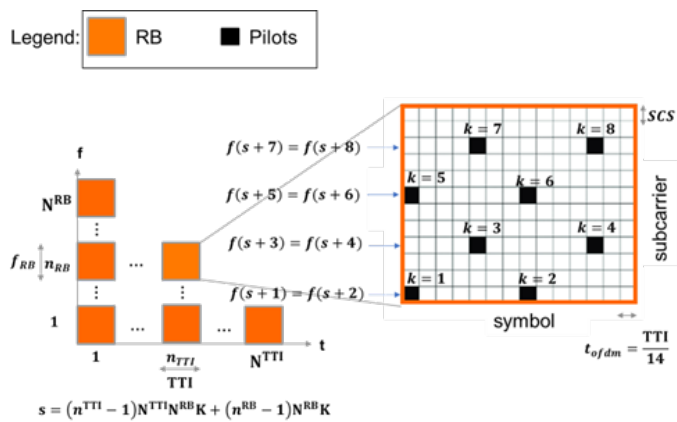

Figure 107: Position of pilots in a PRBs and a transmission time interval (TTI)

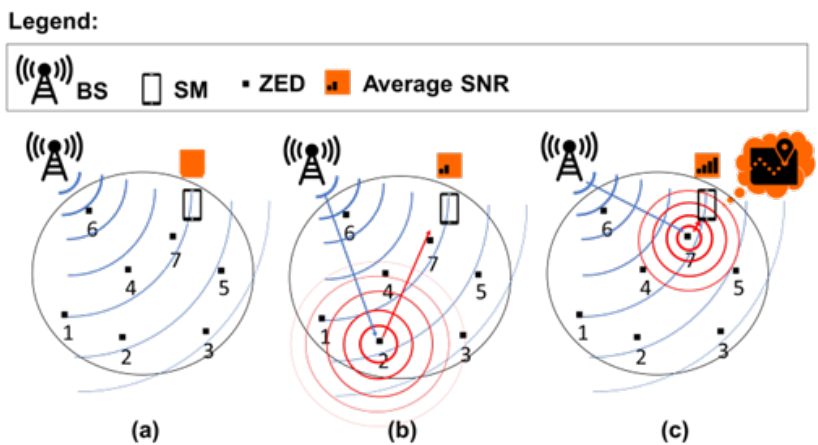

Figure 108: Localization scheme principle : (a) SM monitors the received signal from the BS and detects no ZED ; (b) SM detects $ZED2$ with a low SNR ; (c) SM detects $ZED7$ with the strongest SNR, thus as the closest one, and deduces its location on the map.

**ZED-to-SM communication model**

Fig. 106 illustrates the main principle of the ZED-based system. We consider the communication between a ZED and a SM illuminated by an ambient 4G BS transmitting an OFDM waveform over $N_{\mathrm{RB}}$ PRBs, every sub-frame of TTI duration [3GP19], as illustrated in Fig. 107. The BS sends K pilot symbols in each PRBs, periodically, every subframe, on standardized locations in time and frequency (i.e. on pre-defined sub-carriers and OFDM symbols, inside the PRBs). Without loss of generality, we assume that all pilot symbols have the same constant magnitude. As in [DTPHdR22], the ZED transmits bit $b = 0$ by being in a 'transparent mode' where it is transparent to ambient waves from the BS. Reversely, the ZED transmits bit $b = 1$ by being in a 'reflecting mode' where it backscatters (reflects in all directions) ambient waves from the BS. In our current study, the ZED transmits one bit during one period of $N^{\text{TTI}}$ consecutive subframes. Finally, the ZED sends a data sequence of $N^b$ successive bits. The SM monitors the pilots from the BS, estimates the downlink propagation channel, and searches for the ZED message in the channel estimates. Therefore we suppose that the communication system between the ZED and SM is composed of two phases: a learning and synchronization phase; and a data communication phase (communication meaning here transmission of the identifier of the ZED). We assume that the channel is constant during the learning and synchronization phase, and the whole communication phase.

**ZED-based localization system model**

In Fig. 108, we illustrate the ZED-based localization scheme. $N_{\text{ZED}}$ ZEDs are deployed inside a building. Each ZED is equipped with a unique identification number coded into a bit-sequence, and its precise position on the map is recorded. An SM inside the building is assumed to have the map of ZEDs. Additionally, the building is in the coverage of a BS. ZEDs are assumed to transmit their identification bit-sequence periodically, but at different times, to avoid collisions. The SM monitors the received pilots from the BS, detects, successively, the messages of ZEDs nearby, and computes their respective SNRs. Then, the SM determines the ZED with the strongest SNR and considers it as the closest one. Finally, the SM locates itself on the map at the same location as the selected ZED. Fig. 108-a) illustrates the waves transmitted by the BS alone, when the ZEDs near the SM are inactive. Fig. 108-b) illustrates the case where the SM detects *ZED*2 with a weak SNR. Fig. 108-c) illustrates the case where the SM detects *ZED*7 with the strongest SNR and deduces its location on the map.

**Performance evaluation methodology**

- Deployment scenario and simulation assumptions

  We consider a real existing building under the coverage of a real BS from Orange 4G commercial network [Car], both illustrated in Fig. 109a. The BS has three azimuth orientations: 50°, 180° and 300° and transmits in the LTE band 852 MHz-862 MHz. The frequencies f(n) of pilot symbol n are calculated based on the 4G pilot standardized Time-Frequency pattern [3GP19] illustrated by Fig. 107, where $f_{\text{RB}}$ is the frequency bandwidth of a PRB, SCS is the subcarrier spacing, and $t_{\text{ofdm}}$ is the duration of an OFDM symbol. For each PRB there is a total of $K = 8$ pilot symbols. Various numbers of TTIs $N_{\text{TTI}}$ could be considered. Detailed values of all parameters could be found in [Vil24]

- Raytracing-based channel model

  In order to evaluate the performance of the proposed scheme with a realistic propagation channel model, we use the Orange internal ray-tracing simulation tool STARLIGHT [DTPHR22] to generate the complex channel gain of each propagation path (ray) of the multipath channel, at the central carrier frequency. This channel gain already includes the BS antenna gain and the effect of the antenna diagram, for LTE antenna port 0 [3GP19], and the SM antenna gain (which is assumed to be 0 dBi).

**Example of simulation results**

Fig. 109b illustrates heatmaps indicating the number of detected ZEDs by the SM relative to its true position, focusing on scenarios where the detection probability $P_{\text{d}}$ is greater than 0.9. The analysis considers false alarm probabilities $P_{\text{fa}}$ set at $0.1, 10^{-3}, 10^{-5}$ and $10^{-8}$. Each color on the heatmap represents to the number of detected ZEDs. White areas indicate no ZED with a $P_{\text{d}}$ greater than 0.9 detected at that position within the building. The impact of varying $P_{\text{d}}$ is evident in the heatmap, with a decrease in the number of detected ZEDs as $P_{\text{d}}$ decreases.

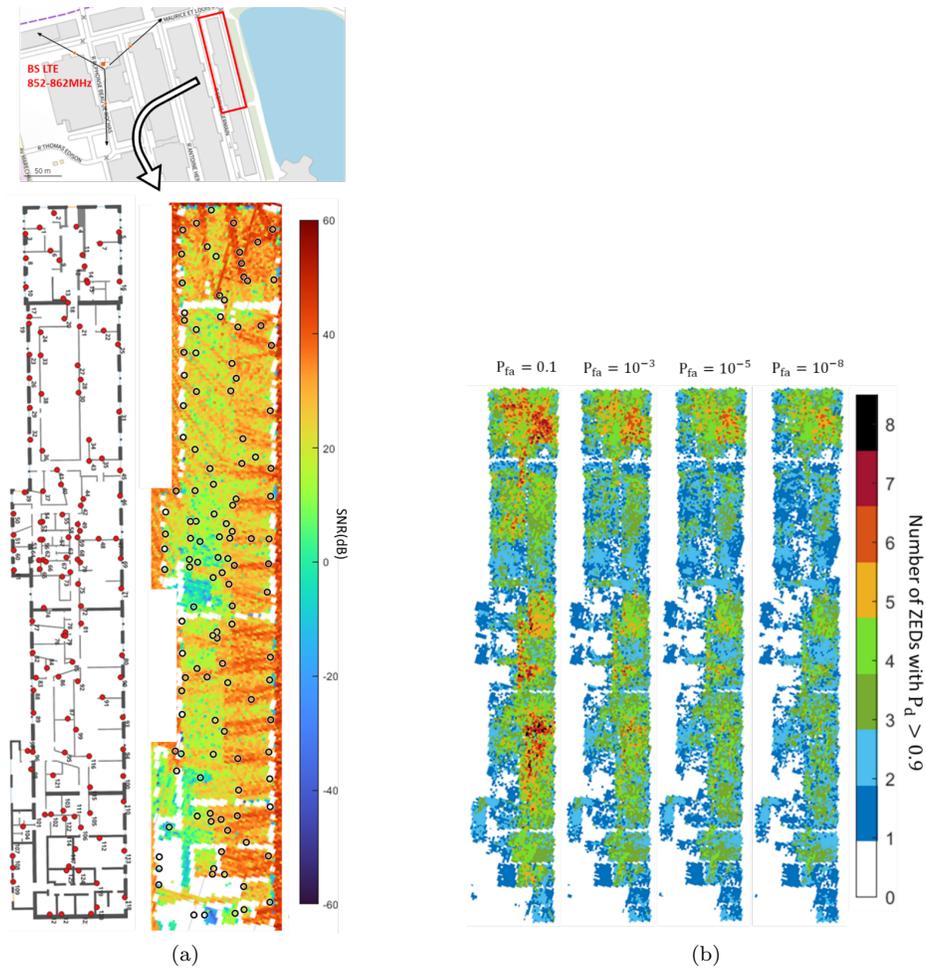

Figure 109: (a) Commercial 4G BS and building location, map of ZEDs deployed in the building, BS coverage inside the building expressed in SNR (dB) and (b) Heatmap simulated with $P_d<0.9$, for $P_{fa}=0.1, 10^{-3}, 10^{-5}$ and $10^{-8}$.

# 4  Acronyms

**2D** two-dimensional

**2G** second generation

**3D** three-dimensional

**3G** third generation

**3GPP** Third Generation Partnership Project

**4D** four-dimensional

**4G-LTE** fourth generation, Long Term Evolution

**4G** fourth generation

**5G-NR** 5G new radio

**5G** fifth generation

**6G** sixth generation

**6LoWPAN** IPv6 over low-power wireless personal area networks

**AB** adaptive boosting

**ACE** antenna coupling errors

**ADAS** advanced driver assistance system

**ADC** analog-to-digital converter

**ADMM** alternative direction method of multipliers

**AED** automorphism ensemble decoding

**AES** advanced encryption system

**AFDM** affine frequency division multiplexing

**AGC** automatic gain control

**AI** artificial intelligence

**AM** amplitude modulation

**AoA** angle of arrival

**AoD** angle of departure

**AP** access point

**AR** augmented reality

**AUD** active user detection

**AWGN** additive white Gaussian noise

**B2B** back-to-back

**BBHT** Boyer, Brassard, Høyer and Tapp

**BER** bit error rate

**BiRa** bistatic radar

**BLE** Bluetooth low energy

**BLER** block error rate

**BM** beam management

**BOK** binary orthogonal keying

**BP** belief propagation

**BPSK** binary phase shift keying

**BR** breath rate

**BS** base stations

**BSS** blind source separation

**BW** bandwidth

**CBRS** citizens broadband radio service

**C-MPC** clutter multipath components

**C-V2X** cellular vehicle-to-Everything

**C-VAE** conditional variational autoencoders

**CC-LDPC** convolutional LDPC code

**CC** convolutionnal code

**CCR** conventional correlation receiver

**CDF** cumulative distribution function

**CDMA** code division multiple access

**CD-ZED** crowd-detectable ZED

**CE-OFDM** constant envelope-orthogonal frequency division multiplexing

**CFAR** constant false alarm rate

**CFO** carrier frequency offset

**CFR** channel frequency response

**CFmMIMO** cell-free massive MIMO

**CIR** channel impulse response

**CNN** convolutional neural network

**CNSU** check Node self-update

**CoMP** coordinated multi point

**COTS** commercial-of-the-shelf

**CP** cyclic prefix

**CPE** common phase error

**CPU** central processing unit

**CR** coding rate

**C-RAN** cloud-RAN

**CRB** Cramér-Rao bound

**CRC** cyclic redundancy check

**CRLB** Cramér-Rao lower bound

**CS** compressed sensing

**CSI-RS** channel state information - reference signal

**CSI** channel state information

**CSS** chirp spread spectrum

**CUDA** compute unified device architecture

**CW** continuous wave

**D-MIMO** distributed MIMO

**DA** data association

**DAFT** discrete affine Fourier transform

**DAS** distributed antenna systems

**DBS** drone base station

**DBSCAN** density-based spatial clustering of applications with noise

**DC** direct current

**DD** delay-Doppler

**DDMC** delay-Doppler multicarrier

**DDOP** delay-Doppler orthogonal pulses

**DFnT** discrete Fresnel transform

**DFT-s-OFDM** DFT-spread-OFDM

**DFT** discrete Fourier transform

**DHA** Dürr-Høyer algorithm

**DL** deep learning

**DmMIMO** distributed massive MIMO

**DMnT** discrete Mellin transform

**DMPSK** differential M-ary phase shift keying

**DNN** deep neural network

**DoA** direction of arrival

**DoS** denial of service

**DQN** deep Q-Networks

**DRL** deep reinforcement learning

**DT** decision tree

**DTA** decision tree adaboost

**DTR** decision tree regression

**DZT** discrete Zak transform

**EBTa** Euro bicyclist target adult

**ECC** error-correcting code

**EcC** elliptic-curve cryptography

**EDC** efficient detection criterion

**EKF** extended Kalman filter

**EM** expectation-maximization

**eMBB** enhanced mobile broadband

**EMM** exponential mixture models

**eNodeB** evolved node B base station

**EPA** extended pedestrian A

**EPA-5** extended pedestrian A model-5

**EPTa** Euro pedestrian target adult

**ESPRIT** estimation of signal parameters via rotational invariant techniques

**ET** extremely randomized trees

**ETU-1** extended typical urban

**Euro-NCAP** European- new car assessment programme

**FBMC** filter bank multicarrier

**FCC** Federal Communication Commission

**FDM** frequency division multiplex

**FEC** forward error correction

**FEKO** Feldberechnung bei Körpern mit beliebiger Oberfläche

**FER** frame error rate

**FFT** fast Fourier transform

**FG** factor graph

**FHSS** frequency-hopping spread spectrum

**FIR** finite impulse response

**FLRC** fast long range communication

**FM-OFDM** frequency modulated-orthogonal frequency division multiplexing

**FM** frequency modulation

**FMCW** frequency modulated continuous wave

**FPGA** field-programmable gate array

**FOCUSS** focal under-determined system solution

**FR** frequency range

**FSK** frequency shift keying

**FSL** few-shot learning

**FTP** file transfer protocol

**FWS** feedforward signal

**GAN** generative adversarial network

**GAO** Gaussian-approximated Optimization

**GBM** gradient boosting machine

**GBSCM** geometry-based stochastic channel model

**GFSK** Gaussian frequency shift keying

**GMM** Gaussian mixture model

**GMSK** Gaussian minimum shift keying

**GNB** Gaussian Naive Bayes

**GNN** generative neural network

**gNodeB** next-Generation node B

**GNSS** global navigation satellite system

**GPS** global positioning system

**GPU** graphics processing units

**GRAND** "guessing random additive noise decoding"

**GSM** global system for mobile communications

**GSSM** geometry-based stochastic signal models

**H-BER** hierarchical bit error rate

**H-MAC** hierarchical MAC

**HAPS** high-altitude platform systems

**HAR** human activity recognition

**HARQ** hybrid automatic repeat request

**HART** highway addressable remote transducer

**HCC** hybrid concatenated code

**HDF** hierarchical decode and forward

**HPBW** half power beam width

**HRNet** high-resolution network

**HSM** hardware security module

**HW** hardware

**LoS** line-of-sight

**IA** initial access

**IAA** iterative adaptive approach

**IBFD** in-band full-duplex

**ICA** independent component analysis

**ICAS** integrated communication and sensing

**ICI** inter-carrier interference

**IDFT** inverse discrete Fourier transform

**IEEE** Institute of Electrical and Electronics Engineers

**ILAC** integrated localization and communication

**IMU** inertial measurement unit

**IO** interacting object

**IoT** internet of things

**IQ** in-phase and quadrature

**IQI** in-phase/quadrature imbalance

**IP** internet protocole

**IR** incremental redundancy

**ISAC** integrated communication and sensing

**ISFFT** inverse symplectic finite Fourier transform

**ISM** Industrial, Scientific, and Medical

**ITS** intelligent transportation system

**JCAS** joint communication and sensing

**JCS** joint communication sensing

**k-NN** k-Nearest Neighbors

**KPI** key performance indicators

**KPM** key performance metrics

**LA** link adaptation

**LBS** loopback signal

**LC** linear correction

**LDA** linear discriminant analysis

**LDPC** low-density parity-check code

**LEO** low Earth orbit

**LFM** linear frequency modulation

**LNA** low noise amplifier

**LO** local oscillator

**LoRa** long range

**LoRaWAN** long range wide area network

**LOS** line-of-sight

**LPI** low probability of intercept

**LPWAN** low power wide area network

**LR-WPAN** low-rate wireless personal area network

**LS** least squares

**LTA** linear tree adaboost

**LTE-A** LTE-Advanced

**LTE** long term evolution

**LTF** long training field

**MAC** medium access control

**MAP** maximum a posteriori

**MCL** maximum coupling loss

**MCO** Monte-Carlo-based optimization

**MCPD** multi-carrier phase difference

**MCS** modulation and coding scheme

**MDN** mixture density network

**MDS** multidimensional scaling

**MFs** map features

**MI** mutual information

**MIMO** multiple-input multiple-output

**MISO** multiple-input single-output

**ML** machine learning

**MLD** maximum likelihood detection

**MLP** multi-layer perceptron

**mMIMO** massive multiple-input multiple-output

**MMSE** minimum mean square error

**mMTC** massive machine type communications

**mmWave** millimeter wave

**MoA** molecular absorption

**MoS** molecular scattering

**MPC** multipath components

**MPS** multi-person sensing

**MRF** pairwise Markov random fields

**MSA** multi-satellite access

**MSE** mean squared error

**MU** multi user

**MUD** multi user detection

**MUSIC** multiple signal classification

**MVDR** minimum variance distortionless response

**NB-IoT** narrowband internet of things

**NB** non binary

**NBP** nonparametric belief propagation

**NC** non-coherent

**nRT RIC** near real-time RIC

**non-RT RIC** non-real-time RIC

**NI** National Instruments

**NIC** network interface card

**NIST** National Institute of Standards and Technology

**NLOS** non-line-of-sight

**NOMA** non-orthogonal multiple access

**NPDSCH** narrowband physical downlink shared channel

**NPRS** narrowband positioning reference signal

**NPSS** narrowband primary synchronization signal

**NR** new radio

**nRMSE** normalized root mean squared positioning error

**NSSS** narrowband secondary synchronization signal

**NTN** non-terrestrial networks

**O-RAN** open radio access network

**OAI** OpenAirInterface

**O-RU** open-radio unit

**O-DU** open-distributed unit

**O-CU** open-central unit

**O-CU-CP** open-central unit control plane

**O-CU-UP** open-central unit user plane

**OCDM** orthogonal chirp division multiplexing

**ODDM** orthogonal delay-Doppler division multiplexing

**ODSS** orthogonal delay scale space

**OEB** orientation error bound

**OFDM** orthogonal frequency division multiplexing

**OFDMA** orthogonal frequency division multiple access

**OMA** orthogonal multiple access

**OMP** orthogonal matching pursuit

**OOB** out-of-band

**OTFS** orthogonal time frequency space

**OTSM** orthogonal time sequency multiplexing

**OW-CFR** one-way channel frequency response

**PA** power amplifier

**PAL** phase alternating line

**PAPR** peak-to-average power ratio

**PAS** probabilistic amplitude shaping

**PBR** phase-based ranging

**PC5** proximity-based services (ProSe) Communication at 5.9 GHz

**PCC** parallel concatenated code

**PDDA** push different data algorithm

**PDF** probability density function

**PDCP** packet data convergence protocol

**PEB** position error bound

**PF** particle filter

**PHY** physical layer

**PLNC** physical layer network coding

**PLS** physical layer security

**PMP** phase measurement process

**PN** phase noise

**PoC** proof-of-concept

**PRB** physical resource block

**PRS** positioning reference signal

**PRU** positioning reference unit

**PS** point scatterers

**PSD** power spectral density

**PSK** phase shift keying

**PSS** primary synchronization signal

**PT-RS** phase tracking-reference signal

**PUF** physical unclonable function

**QAM** quadrature amplitude modulation

**QAOA** quantum approximate optimization algorithm

**QMSA** quantum minimum searching algorithm

**QoS** quality of service

**PKI** public key infrastructure

**RAN** radio access network

**RIC** RAN intelligent controller

**RBF** rank-based fingerprinting

**RCS** radar cross section

**RedCap** 5th Generation Reduced Capacity

**RF** radio frequency

**RGB** red, green, and blue

**RGBD** red, green, and blue- depth

**RIMAX** radio channel parameter estimation from channel sounding measurements

**RIS** reconfigurable intelligent surfaces

**RL** reinforcement learning

**RLC** radio link control

**RLNC** random linear network coding

**RMSE** root mean square error

**RNN** recurrent neural network

**RPI** Raspberry Pi

**RR** repetition redundancy

**RRC** radio resource control

**RRM** radio resource management

**RSMA** rate splitting multiple access

**RSA** Rivest–Shamir–Adleman

**RSS** received signal strength

**RSSI** received signal strength indicator

**RT** ray tracing

**RTLS** real-time localization system

**RTT** round trip time

**Rx** receiver

**S-MPC** sensing multipath components

**SA** 5G standalone

**SAC** shaping after coding

**SAGE** space alternating generalized EM

**SB-SPS** sensing-based semi-persistent scheduling

**SBC** shaping before coding

**SBL** SR-sparse Bayesian learning

**SC-FDMA** single-carrier frequency division multiple access

**SC-LDPC** spatially-coupled LDPC

**SC** successive cancellation

**SCC** serially concatenated code

**SCI** sidelink control information

**SDN** software-defined networking

**SDM** spatial-division multiplexing

**SDNR** signal to distortion (plus) noise ratio

**SDAP** service data adaption protocol

**SDP** semidefinite programming

**SDR** software-defined radio

**SER** symbol error rate

**SF** spreading factor

**SFFT** symplectic finite Fourier transform

**SFO** sampling frequency offset

**SIC** successive interference cancellation

**SIMO** single-input multiple-output

**SINR** signal-to-interference-plus-noise ratio

**SL-RSRP** sidelink reference signal received power

**SLAM** simultaneous localization and mapping

**SLS** system level simulator

**SM** smartphone

**SMO** service management and orchestration

**SNR** signal-to-noise ratio

**SO-TDMA** self-organized time division multiple access

**SOCP** second-order cone programming

**SOP** signals of opportunity

**SpC** splitting code

**SR** super-resolution

**SS** synchronization signal

**SSB** synchronization Signal Block

**SSS** secondary synchronization signal

**STO** symbol time offset

**SVM** support vector machine

**SWiT** self-supervised wireless transformer

**TBS** transport block size

**TDD** time-division duplexing

**TDM** time-division multiplexing

**TDMA** time-division multiple access

**TDoA** time difference of arrival

**TF** time-frequency

**ThIMo** Thuringian Innovation Center for Mobility

**THR** two-hop relaying

**THz** terahertz

**TL** transfer learning

**ToA** time of arrival

**ToF** time of flight

**TP** test point

**TRN** training field

**TRP** transmission and reception point

**TS** tangential-sphere

**TSCH** time slotted channel hopping

**TTI** transmission time interval

**TV** television

**TW-CFR** two-way channel frequency response

**tx-power** transmission power

**Tx** transmitter

**TxPower** Transmission Power

**UAM** urban air mobility

**UAV** unmanned aerial vehicle

**UCA** uniform circular array

**UE** user equipment

**UL** uplink

**ULA** uniform linear array

**UR** unambiguous range

**URA** uniform rectangular array

**URLLC** ultra-reliable and low-latency communications

**USRP** universal serial radio peripheral

**UWB** ultra-wideband

**V2I** vehicle-to-infrastructure

**V2N** vehicle-to-network

**V2V** vehicle-to-vehicle

**V2X** vehicle-to-everything

**V2P** vehicle-to-pedestrians

**VA** virtual anchor

**VAE** variational autoencoders

**VD** vertex diffraction

**VNF** virtual network functions

**VNA** vector network analyzer

**VR** virtual reality

**VRU** vulnerable road user

**VST** vector signal transceiver

**WGDOP** weighted geometric dilution of precision

**WHT** Walsh-Hadamard transform

**WiFi** wireless fidelity

**WiGig** wireless gigabit

**WiT** wireless transformer

**WL** weak learners

**WLS** weighted least squares

**WRF** worse-than-Rayleigh fading

**XAI** explainable artificial intelligence

**ZED** zero-energy-devices

**ZF** zero-forcing

**ZZLB** Ziv-Zakai lower bound